# Design report of
# the KISS-II facility
# for exploring the origin of uranium

Revised in August 3, 2022
Revised in February 8, 2022
Originally released in December 20, 2021

Editors :
Yutaka Watanabe[a)] and Yoshikazu Hirayama[a)]

Authors :
Takamichi Aoki[b)], Yoshikazu Hirayama[a)], Hironobu Ishiyama[c)], SunChan Jeong[a)],
Sota Kimura[c)], Yasuhiro Makida[d)], Hiroari Miyatake[a)], Momo Mukai[c)], Shunji Nishimura[c)],
Katsuhisa Nishio[e)], Toshitaka Niwase[a)], Tatsuhiko Ogawa[f)], Hiroki Okuno[c)],
Marco Rosenbusch[a)], Peter Schury[a)], Yutaka Watanabe[a)], and Michiharu Wada[a)]

*a) Wako Nuclear Science Center, IPNS, KEK*
*b) Department of Physics, The University of Tokyo*
*c) RIKEN Nishina Center for Accelerator-Based Science, RIKEN*
*d) IPNS, KEK*
*e) Advanced Science Research Center, JAEA*
*f) Nuclear Science and Engineering Center, JAEA*

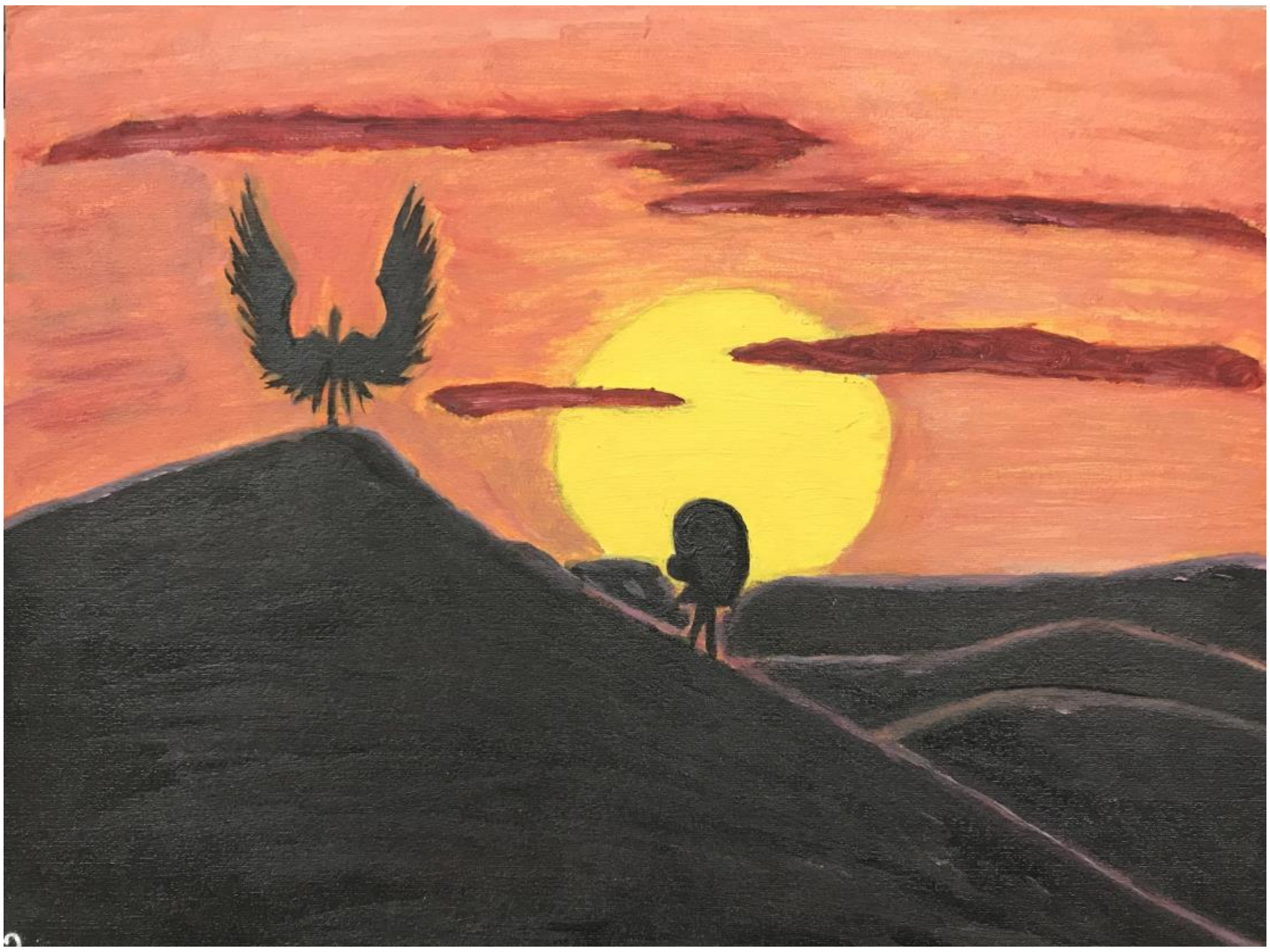

Courtesy of S. NAIMI

# Preface

One of the critical longstanding issues in nuclear physics, noted as a so-called "millennium issue", is the origin of the heavy elements such as platinum and uranium. The r-process hypothesis, generally supported as the process through which heavy elements are formed via explosive rapid neutron capture, was proposed by Burbidge, Burbidge, Fowler, and Hoyle (B2FH: 1957) in the middle of the last century and the originating astrophysical sources and the specific nuclear reaction pathways have not been clarified yet. The study of the origin of heavy elements consists of three pillars: "observation" such as of gravitational waves and other astronomical signals, "experiment" to investigate the characteristics of the nuclei involved, and "theory" to deduce the data and follow the reaction network relevant to the astrophysical source. Many of the nuclei involved in heavy-element synthesis are unidentified, short-lived, neutron-rich nuclei, and experimental data on their masses, half-lives, excited states, decay modes, and reaction rates with neutron etc., are incredibly scarce.

The origin of elements with a high relative abundance around mass number $A = 195$, such as platinum, has been a significant challenge for facilities worldwide, including KISS, which has not been achieved yet. The effort has thus far been hindered by a disconnect between production techniques, separation techniques, and analysis techniques. In recent years, combining the RIBF's in-flight fragmentation reaction separator (BigRIPS) with KISS's gas cell technique and a multi-reflection time-of-flight (MRTOF) mass spectrograph has made it possible to approach the goal.

The ultimate goal is to understand the origin of uranium, the heaviest naturally occurring element on Earth. The nuclei along the pathway to uranium in the r-process are in *"Terra Incognita"*, which no one has been able to reach yet. In principle, as many of these nuclides have more neutrons than $^{238}$U, this region is inaccessible via the in-flight fragmentation reactions and in-flight fission reactions used at the present major facilities worldwide. Therefore, the multi-nucleon transfer (MNT) reaction, which has been studied at KISS, is attracting attention. Like in-flight fission and fragmentation, it simultaneously produces numerous nuclides. Unlike in-flight fission and fragmentation, MNT processes can produce significant yields of nuclides with more neutrons than the target or projectile. However, in contrast to in-flight fission and fragmentation, production via MNT produces RI beams with broad angular distribution and relatively low energies which makes them non-amenable to in-flight separation techniques.

KISS-II would be the first facility to effectively connect production, separation, and analysis of nuclides along the r-process path leading to uranium. This will be accomplished by the use of a large solenoid to collect MNT products while rejecting the intense primary beam, a large helium gas catcher to thermalize the MNT products, and an MRTOF mass spectrograph to perform mass analysis and isobaric purification of subsequent spectroscopic studies. The facility will finally allow us to explore the neutron-rich nuclides in this *Terra Incognita*.

# Contents

# 1. Overview of the KISS-II project

The KISS-II project has been proposed to study the origin of the elements heavier than iron, such as platinum and uranium, presumably produced in the rapid neutron capture process (r-process) in the universe. The ultimate goal of the project is to understand the origin of uranium, heaviest naturally-occurring long-lived element. For the comprehensive understanding of the r-process, the knowledges and collaborative research efforts of theoretical and experimental physicists and astrophysicists, and observational astronomers are crucial. Astrophysical reaction network calculations are the bridge between the nuclear physics and astronomical observation required to understand astrophysical environments relevant to the observed solar r-process abundance pattern. These reaction network calculations require accurate data for various properties, mass, half-life, etc., of nuclides along the r-process path. Unfortunately, as experimental values exist for very few such nuclides, the calculations presently rely on theoretical predictions of these properties for many exotic nuclei.

Along the pathway to uranium and beyond within the r-process, many neutron-rich nuclei, relevant to the cosmic element abundance of $^{238}$U and $^{232}$Th, are located in the *"Terra Incognita"* of the nuclear chart. To fully understand the effects that both their production and their role in so-called fission cycling have on the r-process abundance pattern of lanthanoid elements, experimental studies are highly desirable to determine their various properties of these nuclides, – such as masses, β-decay lifetimes, and decay modes including fissions. Such studies are also important to provide a critical reference for benchmarking various theoretical models of nuclear structure, which can eventually be used to estimate, with improved reliability, the nuclear properties of experimentally unexplored nuclei used in the network calculations. The KISS-II project will focus on measuring, first of all, the nuclear masses as they are one of the most fundamental properties, and performing decay spectroscopic studies such as the lifetime measurements, β–γ spectroscopy, β-delayed fission spectroscopy, and laser spectroscopy. In this way, the KISS-II project will contribute to minimizing the uncertainties originating from a lack of reliable experimental inputs from nuclear physics as much as possible, and eventually to resolve the millennium issues on the cosmic origin of the r-process. The resultant r-process abundance pattern in the network calculation based on nuclear models with highly-reliable predictive powers would help to clarify certain astrophysical models describing explosive events as possible cosmic origin of the r-process.

As an experimental pathway to access such neutron-rich heavy isotopes, we have used the multi-nucleon transfer (MNT) reactions in collisions of heavy projectiles on neutron-rich targets. For detailed discussion on the pattern around the third major peak in the solar r-element abundance, KISS, the ongoing project preceding KISS-II, adopted an MNT reaction system of $^{136}$Xe (projectile) + $^{198}$Pt(target) to produce neutron-rich nuclei relevant to the progenitors forming the peak ($N$=126 waiting-point nuclei). For the β-decay spectroscopy of those isotopes, KISS separates a single species of radioactive nuclei among target-like fragments produced via MNT in the reaction system. This is done by first neutralizing the MNT products using argon gas, extracting the neutral atoms via gas flow, and selectively re-ionizing the desired element by resonant laser excitation. The present system efficiency of ~0.1% has been sufficient for studies closer to stability. Further advancement of the KISS facility will make it possible to reach the progenitors on the third abundance peak. Therefore, the more advanced KISS-II facility (which is proposed as shown in Fig. 1-1) will be able to tolerate primary beam intensity orders of magnitude beyond that currently possible. The efficiency in conversion of the energetic MNT products to a low-energy ion beam will also be improved by orders of magnitude. All-in-all KISS-II will offer a 10 000-fold improvement over KISS.

In order to accomplish this expansion deep into the isotopic horizon, KISS-II will rely on four key devices: a gas-filled superconducting solenoid filter, a large helium gas cell, a variable mass-range separator, and an MRTOF mass spectrograph (MRTOF-MS). The large-bore solenoid (5.5 T, 1.4 m in diameter and 1 m long) will allow an increase in primary beam intensity beyond 1 puA (100-fold beyond the typical intensity at KISS) by collecting the MNT products emitted at large angles to a point beyond the beam dump which will absorb the unreacted primary beam. Near this focus will be located the helium gas cell wherein the MNT products will be stopped by collisions with the helium gas ($P_{\mathrm{He}}$ = 7 kPa at $T$ = 100 K), while remaining charged thanks to the high ionization potential of the helium. A multi-segmented structure is planned for the gas cell to minimize the deleterious effects of induced plasma and allow the ions to be manipulated using RF+DC electric fields. With such a structure, it is estimated that an extraction efficiency exceeding 1% can be achieved, which would represent a 10-fold improvement over the argon gas cell used in KISS. The MRTOF-MS will be an essential device for the KISS-II facility. It can perform precise mass measurements which can be used for particle identification. The high mass resolving power (MRP) approaching $m/\delta m$ = $10^6$ enables achieving the astrophysically necessary mass precision of 100 keV/$c^2$ with the detection of only 10 ions for $A$ ~ 250 nuclei. Moreover, the MRTOF-MS can simultaneously analyze multiple isobar chains, allowing the device to analyze ten or more nuclides simultaneously. In conjunction with the 10-fold improvement in extraction efficiency and the 100-fold increase in primary beam intensity, this constitutes a 10 000-fold improvement in measurement power.

The estimated cost for the construction of the KISS-II facility is 1.5 billion JPY. The construction will be finished by the fourth year following receipt of the construction budget. The offline tests will start from the fifth year, and we can start the experiments from the middle of the fifth year.

At the KISS-II facility, we will perform nuclear spectroscopy around the actinoid region to reveal the mystery of the nucleosynthesis of heavy elements in the universe.

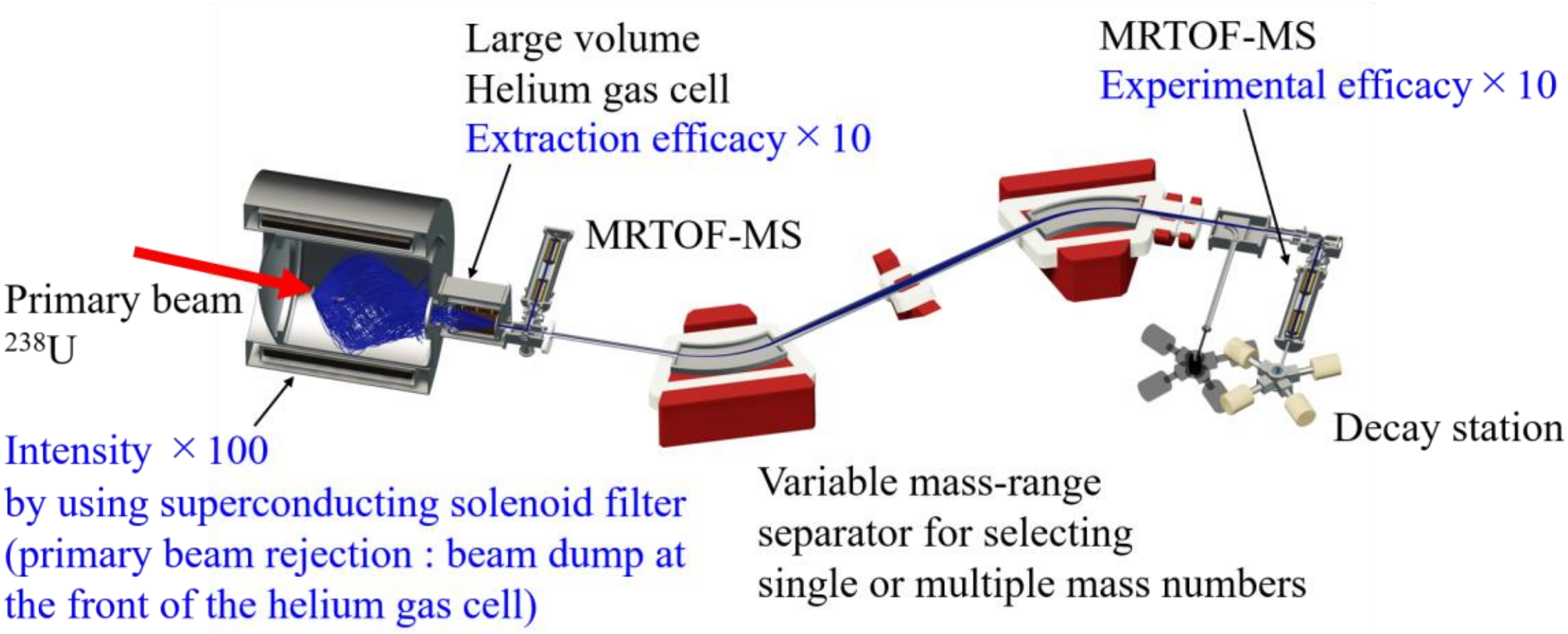

Figure 1-1. Conceptual design of the KISS-II facility.

# 2. Physics motivation

The KISS-II project aims to provide a comprehensive understanding of the rapid neutron capture process (r-process) nucleosynthesis in the universe, which is considered to have produced half of the nuclear abundances among elements heavier than iron. Particularly, all uranium and thorium existing in nature are considered to be remnants of the r-process. Most importantly, because of their long lifetimes, comparable to the age of the universe, the abundances of these two elements observed within the atmospheres of astronomical objects have been used for age estimation. Present limitations in the knowledge of the r-process production ratio for these two elements limits the accuracy of such estimates. Further knowledge of this ratio can provide more strict constraints on the identification of the cosmic origin of astronomical objects.

The present KISS project advances the investigation of nuclear properties relevant to the waiting point nuclei at the $N = 126$ neutron shell closure along the r-process path. Among the various nuclear properties which determine the abundances of remnant nuclei in this region, the neutron separation energies (which can be derived from systematic mass measurements) and β-decay lifetimes of these waiting point nuclei are particularly critical parameters. On the other hand, in the actinoid region, a new decay mode related to fission would play a crucial role for their survival as remnants. Furthermore, fission fragments could enrich lanthanoids, eventually modifying the pattern of the r-element abundances. For the detailed discussion on the r-process abundance pattern, essential for validating the astrophysical conditions of interest, reaction network calculations for the r-process have to be performed.

Because the reaction network calculations involve various properties of nuclei in *terra incognita* of the nuclear chart, the results strongly depend on theoretical predictions of nuclear properties for many unknown exotic nuclei. However, the theoretical calculations of those nuclear parameters, especially for neutron-rich isotopes in and around the actinoid region, have large uncertainties because of complicated single-particle orbitals and nuclear deformations in the midshell nuclear region.

As the reaction rates and decay properties involved in the network calculations depend exponentially on the masses of the relevant nuclei, experimental values for the atomic masses of these exotic nuclides would have a singularly large impact in reducing the uncertainties of the theoretical calculations. The KISS-II facility will advance the frontiers of the experimental mass measurements in and around the neutron-rich actinoid region, and provide significant inputs to theoretical nuclear models. The measurements of β-delayed fission probabilities will make it possible to directly evaluate the probabilities to feed $^{238}U$ and $^{232}Th$. Additionally, lifetime measurements, β–γ spectroscopy, and laser spectroscopy can be performed at the KISS-II facility and will provide various nuclear properties concerning decay schemes, nuclear structures, and collective deformations, which are important for the examination of the theoretical nuclear models.

## 2-1. Nucleosynthesis in the universe

The origin of the existing heavy elements is one of the fundamental questions in natural science. Figure 2-1 shows the solar abundances by mass number, $A$. It indicates characteristic pairs of double-peaked structures observed beyond the iron group ($A > 60$): $^{130}$Te and $^{138}$Ba, and $^{195}$Pt and $^{208}$Pb. The elements heavier than iron are generally considered to have been produced by neutron capture reactions. The existence of a pair of such distinct double-peaked structures indicate that two different neutron capture processes are involved in the production of the majority of elements heavier than iron: a slow process (s-process) and a rapid process (r-process) [2].

The s-process, occurring in a relatively low neutron flux environment, is characterized by neutron capture rates on or below the scale of the β-decay rates of near-stable neutron-rich isotopes. It proceeds as a sequence of neutron capture followed by β-decay, tracing out a pathway along the valley of β-stability and terminating at lead and bismuth. Along the s-process pathway, there exist bottlenecks at the neutron shell closures ($N = 50$, 82, and 126) due to strongly reduced neutron capture cross sections. The stagnation in the s-process at these waiting points results in enhanced production of $^{89}Y_{50}$, $^{138}Ba_{82}$, and $^{208}Pb_{126}$. The major astrophysical site for the s-process has been identified as Asymptotic Giant Branch (AGB) stars based on the presence of the unstable element technetium inferred from observations of their optical spectra.

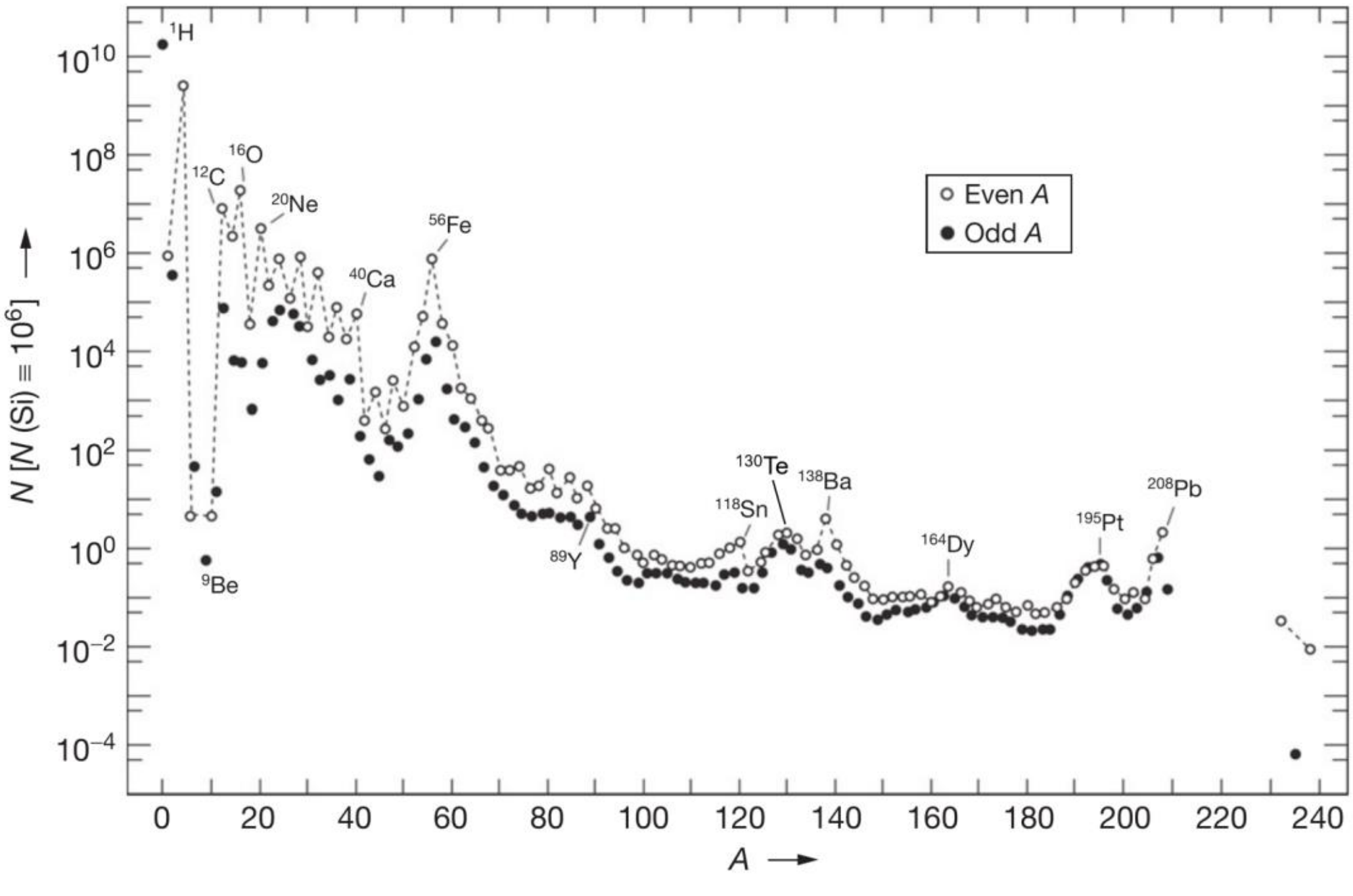

Figure 2-1. Solar abundances by mass number, $A$. Open and solid circles correspond to even and odd masses, respectively. Adapted from Ref. [1].

By contrast, the r-process occurs in an environment with much higher neutron flux and the neutron captures occur much more quickly than the β-decays along most of the r-process pathway; if the neutron flux is great enough the r-process could approach the neutron drip line. Much like in the s-process, the neutron shell closures result in bottlenecks for the r-process. As the r-process extends deep into the neutron-rich nuclides, the freeze-out products formed from β-decay of material in the neutron shell closure bottlenecks when the neutron flux subsides will produce peaks with lower atomic number than the s-process peaks, such as $^{130}Te$ and $^{195}Pt$.

To realize such a high neutron capture rate required by the r-process, an astrophysical environment with very high temperature ($\sim 10^9$ K) and extremely high neutron density ($>10^{20}$ cm$^{-3}$) is required. Under such conditions, the r-process flows through very neutron-rich regions up to actinoid and trans-actinoid elements, leaving uranium and thorium elements among its remnants. Only the r-process can be responsible for the existence of such long-lived actinoid elements, because no other processes can access this neutron-rich region. The astronomical site of the r-process has been a mystery for a long time, because of the lack of comprehensive understanding of the r-process. Ejected layers in a core-collapse supernova explosion had long been considered to be a promising r-process site, which could provide sufficiently high neutron flux. Recently, by the historical discovery of a binary neutron star merger (BNSM) via gravitational wave astronomy, and the subsequent observations of its electromagnetic counterpart, a kilonova [3], the BNSMs have become considered to most likely be the dominant site of the r-process nucleosynthesis. However, it is still not clear whether it can explain total abundances of all r-process elements including the abundance patterns in the very old metal-poor stars.

### 2-1-1. Reaction network calculations

Reaction network calculations of nucleosynthesis are essential to elucidate the comprehensive picture of the r-process. They must treat the density and the temperature of the environment dynamically varying from moment to moment. In the initial phase, with density and temperature high enough to initiate the r-process, the charged-particle reactions produce seed nuclei for the r-process. If a sufficient number of free neutrons remain after charged-particle reactions cease, the r-process can be initiated with the seed nuclei. The r-process involves many reactions and decays: neutron capture reactions, photo-dissociations, β-decays, β-delayed neutron emissions, neutron induced fissions and β-delayed fissions. In dynamical calculations of the reaction network, the relative abundances per unit mass, $Y_{Z,N}$, of the nuclides having the atomic number $Z$ and the neutron number $N$, as a function of time $t$ is obtained by the following differential equation,

$$\begin{aligned}\frac{dY_{Z,N}}{dt} &= \rho N_{\mathrm{A}} Y_{\mathrm{n}} \langle\sigma v\rangle_{Z,N-1}^{(\mathrm{n},\gamma)} Y_{Z,N-1} - \rho N_{\mathrm{A}} Y_{\mathrm{n}} \langle\sigma v\rangle_{Z,N}^{(\mathrm{n},\gamma)} Y_{Z,N} + \lambda_{Z,N+1}^{(\gamma,\mathrm{n})} Y_{Z,N+1} - \lambda_{Z,N}^{(\gamma,\mathrm{n})} Y_{Z,N} \\ &+ \sum_{k_{\mathrm{n}}\geq 0} \lambda_{Z-1,N+1+k_n}^{\beta,k_{\mathrm{n}}} Y_{Z-1,N+1+k_{\mathrm{n}}} - \sum_{k_{\mathrm{n}}\geq 0} \lambda_{Z,N}^{\beta,k_{\mathrm{n}}} Y_{Z,N} \\ &+ \rho N_{\mathrm{A}} Y_{\mathrm{n}} \sum_{Z',N'} \sum_{k_{\mathrm{n}}\geq 0} \langle\sigma v\rangle_{Z',N'}^{\mathrm{nf},Z,N,k_{\mathrm{n}}} Y_{Z',N'} - \rho N_{\mathrm{A}} Y_{\mathrm{n}} \sum_{k_{\mathrm{n}}\geq 0} \langle\sigma v\rangle_{Z,N}^{\mathrm{nf},k_{\mathrm{n}}} Y_{Z,N} \\ &+ \sum_{Z',N'} \sum_{k_{\mathrm{n}}\geq 0} \lambda_{Z',N'}^{\beta\mathrm{df},Z,N,k_{\mathrm{n}}} Y_{Z',N'} - \sum_{k_{\mathrm{n}}\geq 0} \lambda_{Z,N}^{\beta\mathrm{df},k_{\mathrm{n}}} Y_{Z,N}\,, \qquad (2-1)\end{aligned}$$

where $\rho$ and $N_{\mathrm{A}}$ are density at the reaction site and Avogadro's number, $\langle\sigma v\rangle_{Z,N}^{(\mathrm{n},\gamma)}$ and $\langle\sigma v\rangle_{Z,N}^{\mathrm{nf},Z',N',k_{\mathrm{n}}}$ are the product of the cross section and the relative velocity for the neutron capture reaction and the neutron-induced fission producing a fragment $(Z',N')$ with emission of $k_{\mathrm{n}}$ neutrons, respectively, for the nuclide $(Z, N)$ averaged over the velocity distribution, $\lambda_{Z,N}^{(\gamma,\mathrm{n})}$, $\lambda_{Z,N}^{\beta,k_{\mathrm{n}}}$, and $\lambda_{Z,N}^{\beta\mathrm{df},Z',N',k_{\mathrm{n}}}$ are the rate of the photo-dissociation, the β-decay with emission of $k_{\mathrm{n}}$ neutrons, and the β-delayed fission producing a fragment $(Z',N')$ with emission of $k_{\mathrm{n}}$ neutrons, respectively, for the nuclide $(Z, N)$, and $\langle\sigma v\rangle_{Z,N}^{\mathrm{nf},k_{\mathrm{n}}}$ and $\lambda_{Z,N}^{\beta\mathrm{df},k_{\mathrm{n}}}$ are the sum of $\langle\sigma v\rangle_{Z,N}^{\mathrm{nf},Z',N',k_{\mathrm{n}}}$ and $\lambda_{Z,N}^{\beta\mathrm{df},Z',N',k_{\mathrm{n}}}$, respectively, over $Z'$ and $N'$. The first and the second terms in the right-hand side indicate the rate of increase in the abundance of the nuclide $(Z, N)$ by the neutron capture reactions. The third and the fourth terms correspond to the rate of increase concerning photo-dissociations. The fifth and the sixth terms indicate the rate of increase in the abundance by the β-decays. The β-delayed neutron emissions are also considered by summing over the neutron multiplicity, $k_{\mathrm{n}}$. The seventh and the eighth terms are related to the neutron-induced fissions having neutron multiplicity $k_{\mathrm{n}}$. The ninth and the tenth terms are for β-delayed fissions having neutron multiplicity $k_{\mathrm{n}}$. Similarly, the abundance of the neutrons as a function of time $t$, $Y_{\mathrm{n}}$, is obtained from the following differential equation,

$$\frac{dY_{\mathrm{n}}}{dt} = -\rho N_{\mathrm{A}} Y_{\mathrm{n}} \sum_{Z,N} \langle \sigma v \rangle_{Z,N}^{(\mathrm{n},\gamma)} Y_{Z,N} + \sum_{Z,N} \lambda_{Z,N}^{(\gamma,\mathrm{n})} Y_{Z,N} + \sum_{Z,N} \sum_{k_{\mathrm{n}} \geq 1} k_{\mathrm{n}} \lambda_{Z,N}^{\beta k_{\mathrm{n}}} Y_{Z,N}$$
$$+\rho N_{\mathrm{A}} Y_{\mathrm{n}} \sum_{Z,N} \sum_{k_{\mathrm{n}} \geq 1} k_{\mathrm{n}} \langle \sigma v \rangle_{Z,N}^{\mathrm{nf},k_{\mathrm{n}}} Y_{Z,N} - \rho N_{\mathrm{A}} Y_{\mathrm{n}} \sum_{Z,N} \langle \sigma v \rangle_{Z,N}^{\mathrm{nf}} Y_{Z,N}$$
$$+\sum_{Z,N} \sum_{k_{\mathrm{n}} \geq 1} k_{\mathrm{n}} \lambda_{Z,N}^{\beta \mathrm{df},k_{\mathrm{n}}} Y_{Z,N} . \qquad (2-2)$$

Site-independent astrophysical models use the electron abundance, $Y_e$, and the entropy, $S$, as free parameters by assuming adiabatic expansion of a hot matter with the initial radius, $R_0$, density $\rho_0$, temperature, $T_0$, and the velocity, $v_{\mathrm{ex}}$. These differential equations are solved throughout the r-process duration with the density and the temperature varying in each layer of the expanding materials from moment to moment, to simulate the r-process

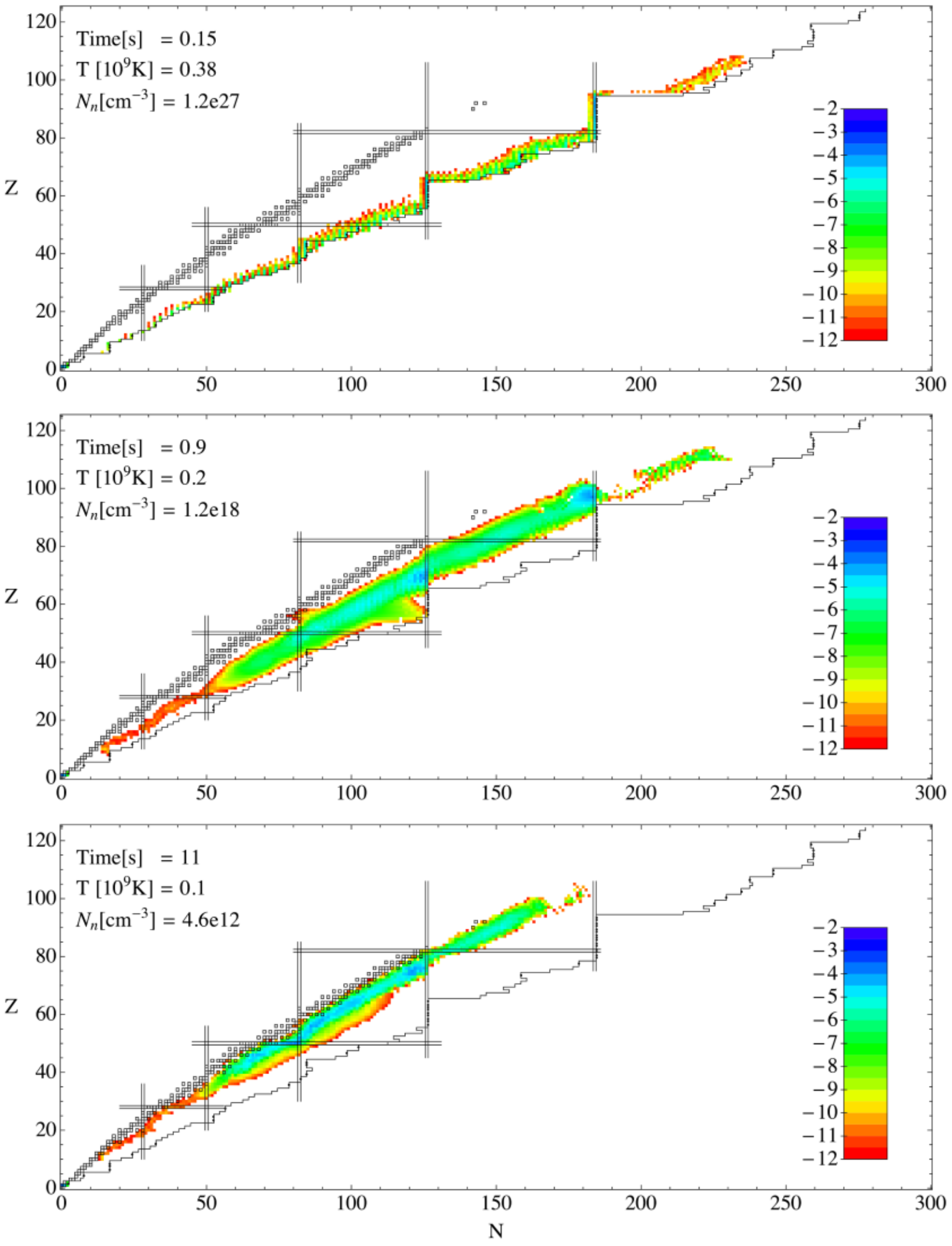

Figure 2-2. The r-process nuclear flow in the 1.30 $M_{\odot}$ − 1.35 $M_{\odot}$ BNSM, color-coded in terms of the mass fraction of nuclei in log scale, at times 0.15 s (top panel), 0.9 s (middle panel) and 11 s (bottom panel) after ejection. Adapted from Ref. [4].

nucleosynthesis.

Figure 2-2 shows the results of the dynamical calculations for the r-process nuclear flow in a 1.30 $M_\odot$ – 1.35 $M_\odot$ BNSM at three different times (0.15 s, 0.9 s and 11 s) after the mass ejection from the BNSM, where $M_\odot$ denotes the solar mass. At the initial stage of the r-process, the flow proceeds along near the neutron drip lines as shown in the top panel. If the neutron density is higher, a larger mass fraction can proceed along the neutron dripline, where the relevant lifetimes are shorter. It makes the timescale of the r-process become shorter and the nucleosynthesis continues to the fissile nuclei, leading to fission cycling. When the duration of the r-process is longer than the fission cycling timescale, several cycles can occur, and the fission fragments capture neutrons to synthesize fissile nuclei again and again. How such fission cycling contributes to the r-process abundance pattern strongly depends on the astrophysical conditions. After some times the temperature falls with the steady decrease in neutron density and the r-process flow shifts toward β-stability as shown in the middle panel. After the neutron density becomes low enough for the neutron capture rate to be dwarfed by the β-decay rate, the r-process is said to undergo freeze out and the nuclei all move toward the valley of β-stability by successive β-decays and possibly β-delayed fissions as shown in the bottom panel.

## 2-1-2. Circumstance to feed uranium and thorium as r-process remnants

As seen in Fig 2-2, after 150 ms (top panel) the r-process can reach as far as $Z \simeq$ 110 and $N \simeq$ 230; by 0.9 s (middle panel) the formation of a major peak around $Z \simeq$ 95 and $N$= 184 can already be seen. When the r-process flows into the region of fissile nuclei, nucleosynthesis proceeds via neutron-induced fissions and β-delayed fissions in addition to β-decays, neutron capture, and photo-dissociations. Figure 2-3 shows theoretical predictions of the fission-barrier heights in the heavy and superheavy mass region [5]. It indicates that superheavy nuclei with $Z$> 110 would be unlikely to be produced even near the neutron drip line due to the significant fission taking place along those isotopic chains under neutron irradiation because of their low fission-barrier heights.

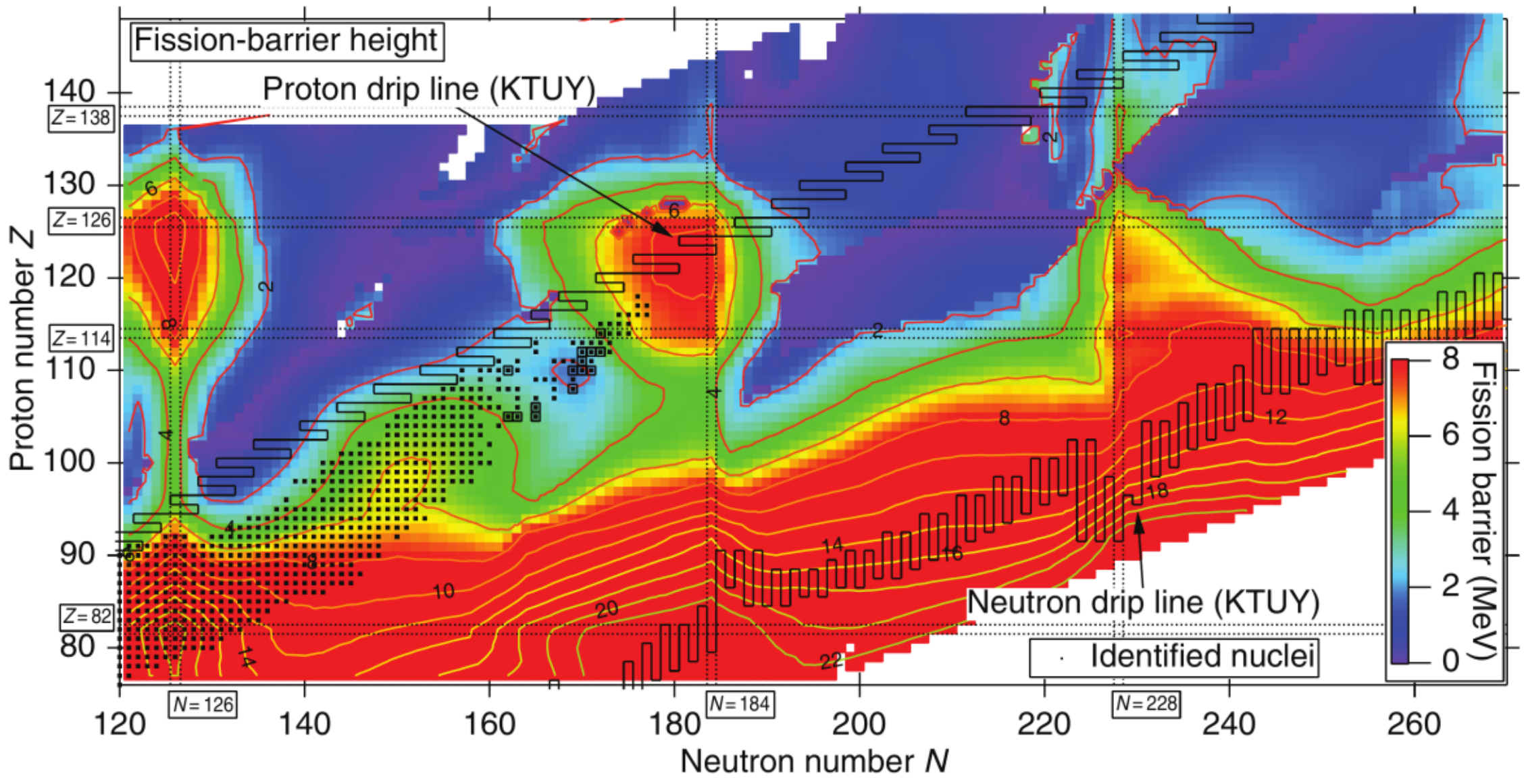

Figure 2-3. Fission-barrier height in the heavy and superheavy mass region. Proton and neutron drip lines from the KTUY mass formula are shown as solid lines. Adapted from Ref. [5].

The existence of significant quantities of long-lived radioactive isotopes $^{232}$Th ($T_{1/2}$ = 14.0 Gyr), $^{235}$U ($T_{1/2}$ = 704 Myr) and $^{238}$U ($T_{1/2}$ = 4.47 Gyr) in nature indicates that the r-process should produce their progenitors, because there are no other possible processes which can access those nuclei. These progenitors undergo a series of β- and α-decays to feed the long-lived thorium and uranium nuclei. Figure 2-4 shows possible β- and α-decay paths feeding $^{232}$Th and $^{238}$U on the nuclear chart, indicated by bold black outlined boxes. The white and black dashed outlined boxes in the figure indicate specific nuclei which are important for their ability to feed $^{232}$Th or $^{238}$U by their α-decays. The nuclides $^{236}$U and $^{240}$Pu produced by β-decays from the r-process progenitors feed $^{232}$Th by α-decays, while $^{244}$Pu, $^{248}$Cm, and $^{252}$Cf can feed $^{240}$U by α-decays, which in turn feeds $^{240}$Pu by two β-decays, which finally α-decays to feed $^{232}$Th. Similarly, $^{238}$U is fed from $^{242}$Pu, $^{246}$Cm, and $^{250}$Cf by α-decays. The β-decay of isobars with $A > 253$ cannot contribute to the production of U or Th because spontaneous fission of $^{254,256}$Cf blocks the production of $^{254,256}$Fm; similar processes block production of other possible α-decay progenitors. Figure 2-5 indicates the integrated β- and α-decay flows [6] after the freeze-out of the r-process in the top and the bottom panels, respectively. The integrated flow $f_x(Z,N)$ $(x = \alpha \text{ or } \beta)$ is expressed as

$$f_x(Z,N) = \sum_i \lambda_{x,i}(Z,N) Y_i(Z,N)(t_{i+1} - t_i) \qquad x = \alpha \text{ or } \beta, \qquad (2-3)$$

where $t_i$ is the time at the time step $i$-th time step in the calculation, and $\lambda_{x,i}(Z,N)$ and $Y_i(Z,N)$ are the rate for the $x$-decay and the abundance of the nuclides $(Z, N)$ at the time $t_i$. The green outlined boxes indicate $^{232}$Th and $^{238}$U, while the black outlined boxes denote nuclei that feed them by α decays, and the dashed diagonal lines represent the most probable β-decay paths feeding them. Approximately 10% of nuclides with $A > 230$ finds its way into either $^{232}$Th or $^{238}$U after 1 Gyr, while the rest populates other short-lived actinoids, continues α decaying to populate lead and bismuth, or simply fissions [6].

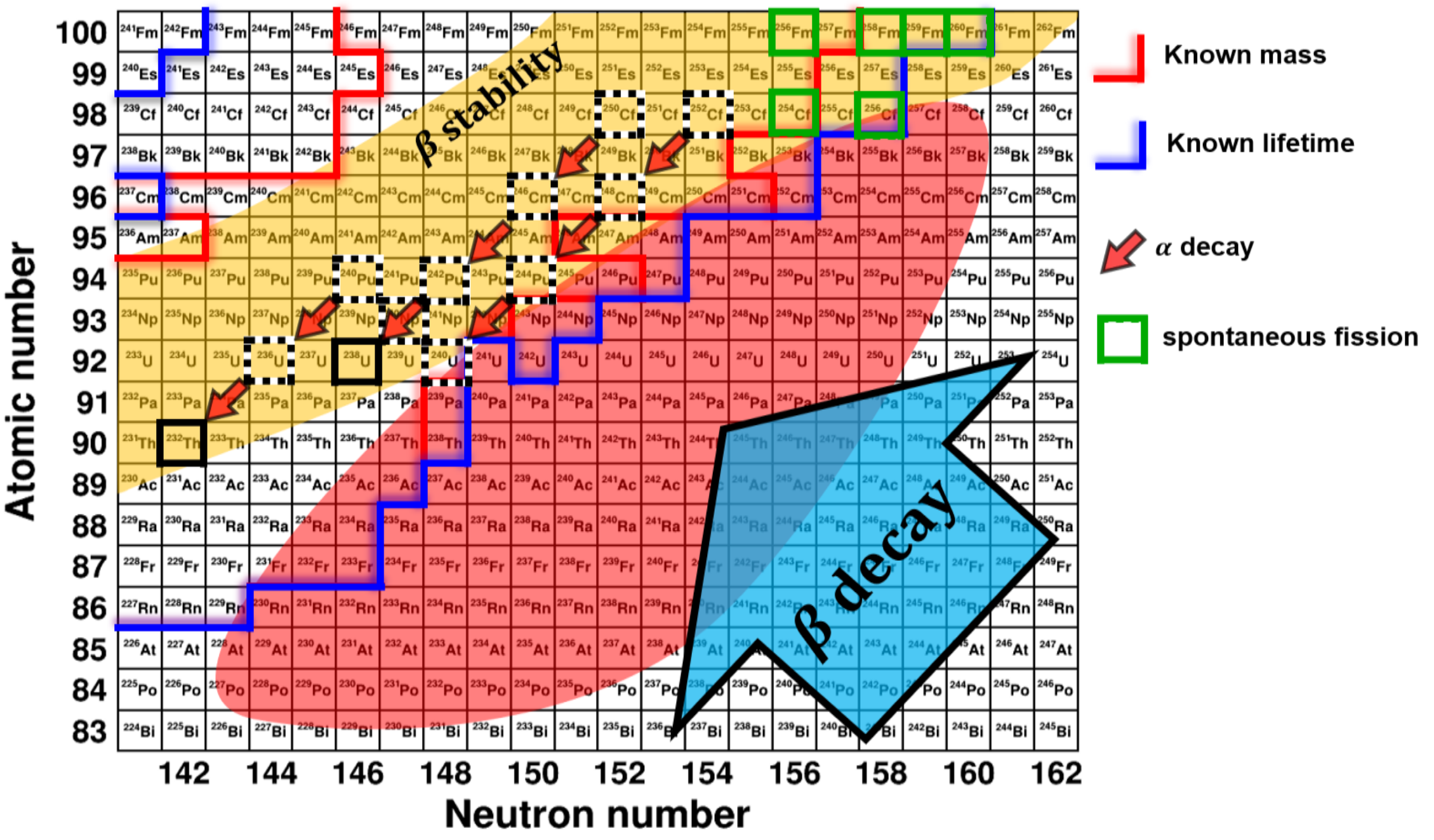

Figure 2-4. Schematic view of the nuclear chart around actinoid elements. Bold black outlined boxes indicate the location of $^{232}$Th and $^{238}$U. The progenitors, produced in the r-process, undergo a series of α- and β-decays to produce $^{232}$Th and $^{238}$U. Red arrows indicate α-decays. Green outlined boxes indicate nuclei which decay by spontaneous fission. Yellow shaded area roughly indicates the valley of the β stability. Blue and red lines indicate boundaries of known lifetimes and masses, respectively. The isotopes inside the red shaded area are the goal of the KISS-II project.

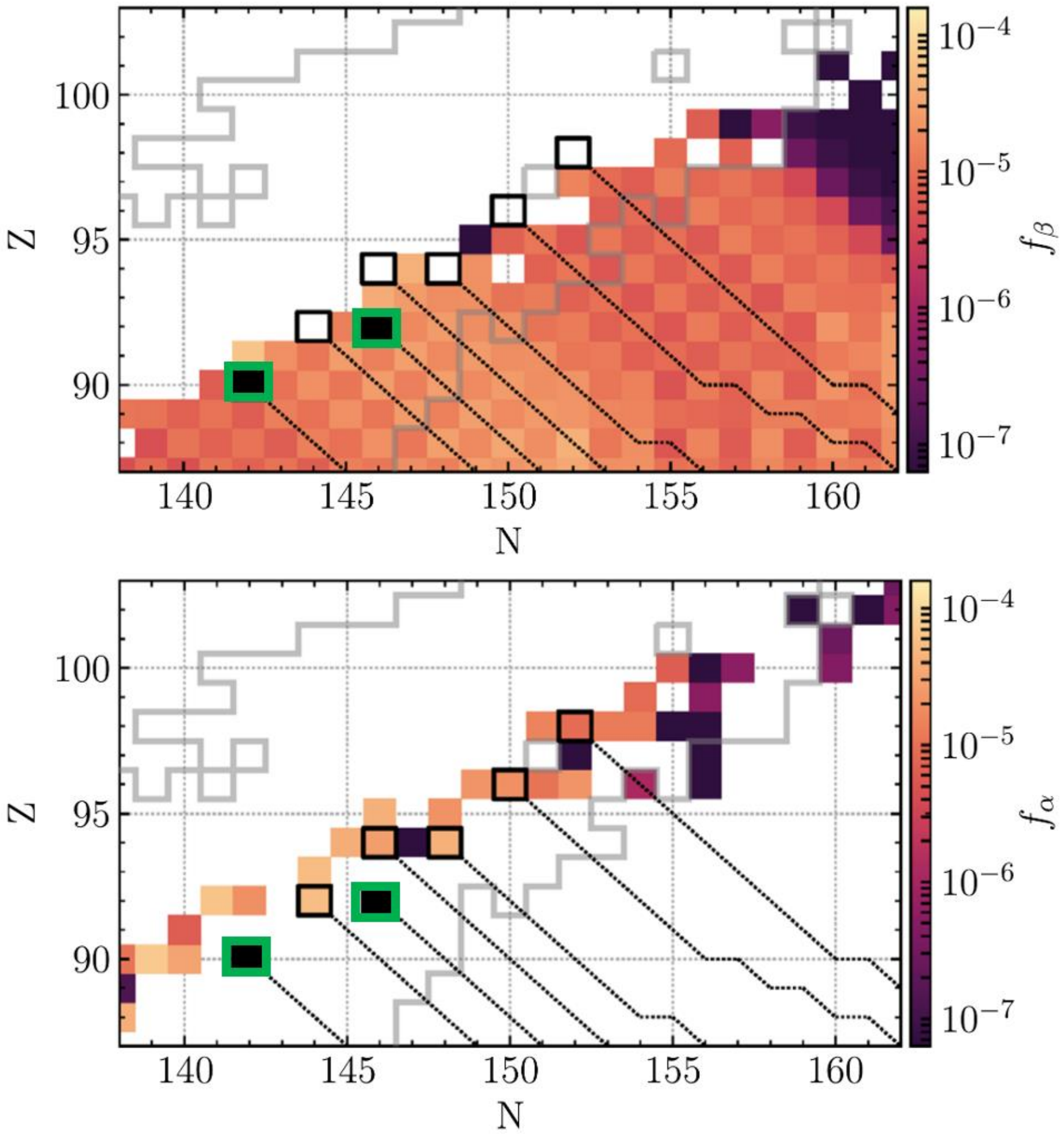

Figure 2-5. Integrated flows at $Y_e = 0.035$ for β-decay (top) and α-decay (bottom). Adapted from Ref. [6].

Thus, a limited range of the nuclides contributes to the abundances of $^{232}$Th and $^{238}$U. Considering the most probable β-decay paths in Fig. 2-5, neutron-rich nuclei with mass number from $A$ = 232 to 252 are important to evaluate those abundances. With KISS-II project we aim to study properties of those nuclei inside the red shaded area in Fig. 2-4. The measurements of β-delayed fission probabilities of those nuclei directly provide the probabilities to feed the thorium and uranium isotopes.

### 2-1-3. Nuclear mass around the actinoid region

The mass is one of the most fundamental parameters of nuclei. It defines the stability of the nuclei, which is the base of all reaction and decay rates including fission appearing in Eqs. (2-1) and (2-2). The red lines in Fig. 2-4 shows the frontiers of experimentally known atomic masses around the actinoid region. In this region, the range of experimentally determined mass values rarely extends more than a few isotopes from the valley of β-stability, due to the difficulty to experimentally produce the neutron-rich nuclei and the indirect mass determination through decay and reaction studies in this region so far. Reliable mass predictions beyond the frontiers are critical for the reaction network calculations. However, any theoretical mass predictions have large uncertainties and there are sizable variances between different models. Figure 2-6 shows the deviation of mass predictions for At isotopes from FRDM2012 mass model [7] among several other common mass models, FRDM95 [8], HFB14 [9], Duflo-Zuker [10], and

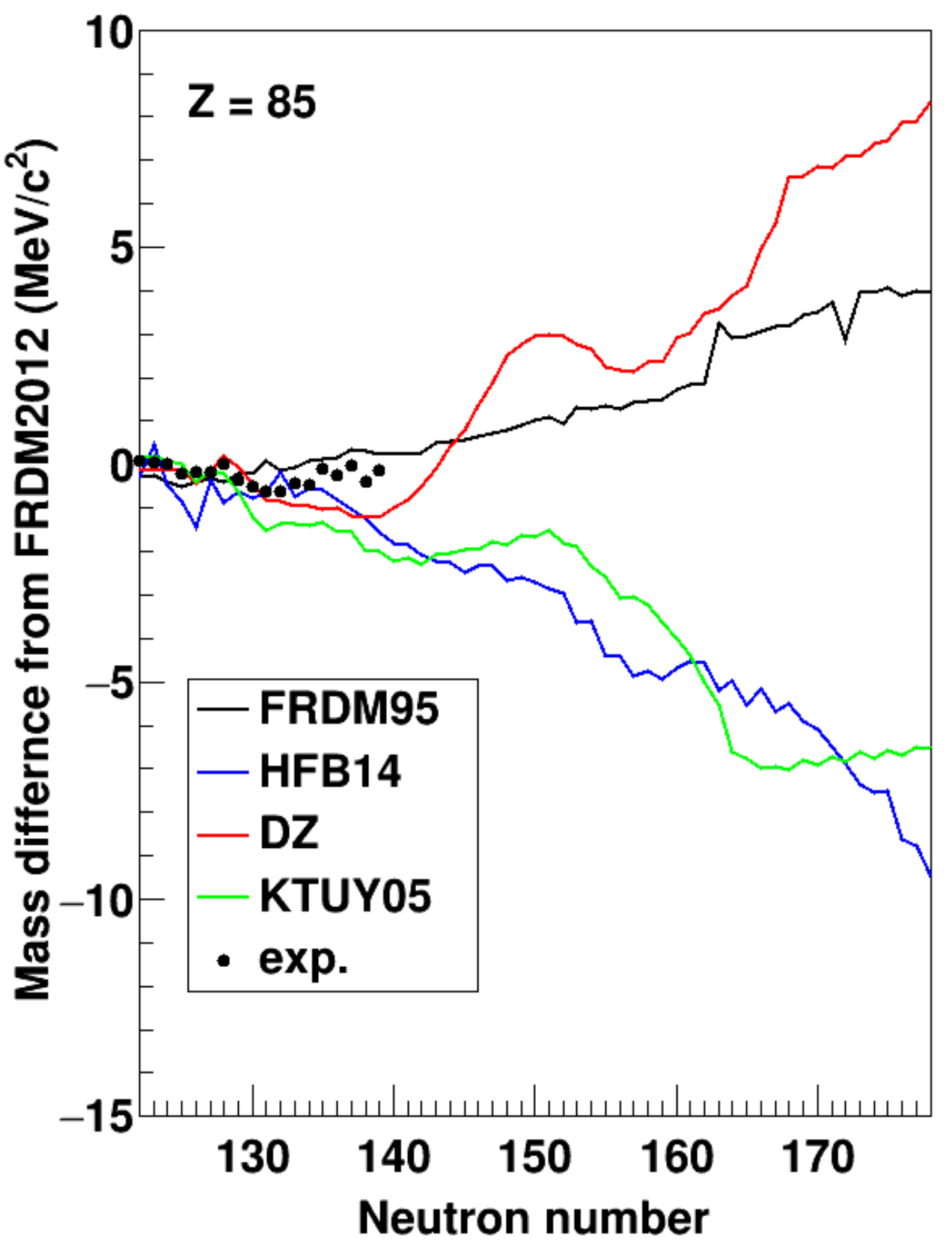

Figure 2-6. Comparison of mass prediction of various theoretical models with experimental data for At isotopes. The ordinate indicates the mass difference from the mass prediction of FRDM2012.

KTUY05 [11], along with experimental data as functions of neutron number for astatine isotopes; the behavior is not atypical. Where experimental data exists, the models all agree within 2 MeV/$c^2$. However, the models quickly diverge beyond the anchoring experimental data, exceeding 10 MeV/$c^2$ at $^{250}$At. Such large variances in mass predictions among different theoretical models result in significant uncertainties in the abundances of $^{238}$U and $^{232}$Th determined by the reaction network calculations. Figure 2-7 shows a comparison of final mass abundance patterns obtained by the reaction network calculations based on various theoretical mass models [12]. It shows a difference in uranium and thorium abundances spanning about half an order of magnitude, and also shows variations in the abundance patterns for r-process elements with $A > 140$. In order to minimize the variations in relative abundances determined by network calculations induced by uncertainties in atomic masses requires that the atomic mass used in the calculation have uncertainties comparable to the thermal energy $k_B T \sim 100$ keV/$c^2$, where $k_B$ is the Boltzmann constant and $T$ = 1 GK is the presumed temperature scale of the r-process site. The KISS-II project will be used to measure nuclear masses to a precision better than 200 keV/$c^2$ for all the nuclides within the red area in Fig. 2-4 – better than 100 keV/$c^2$ in most cases.

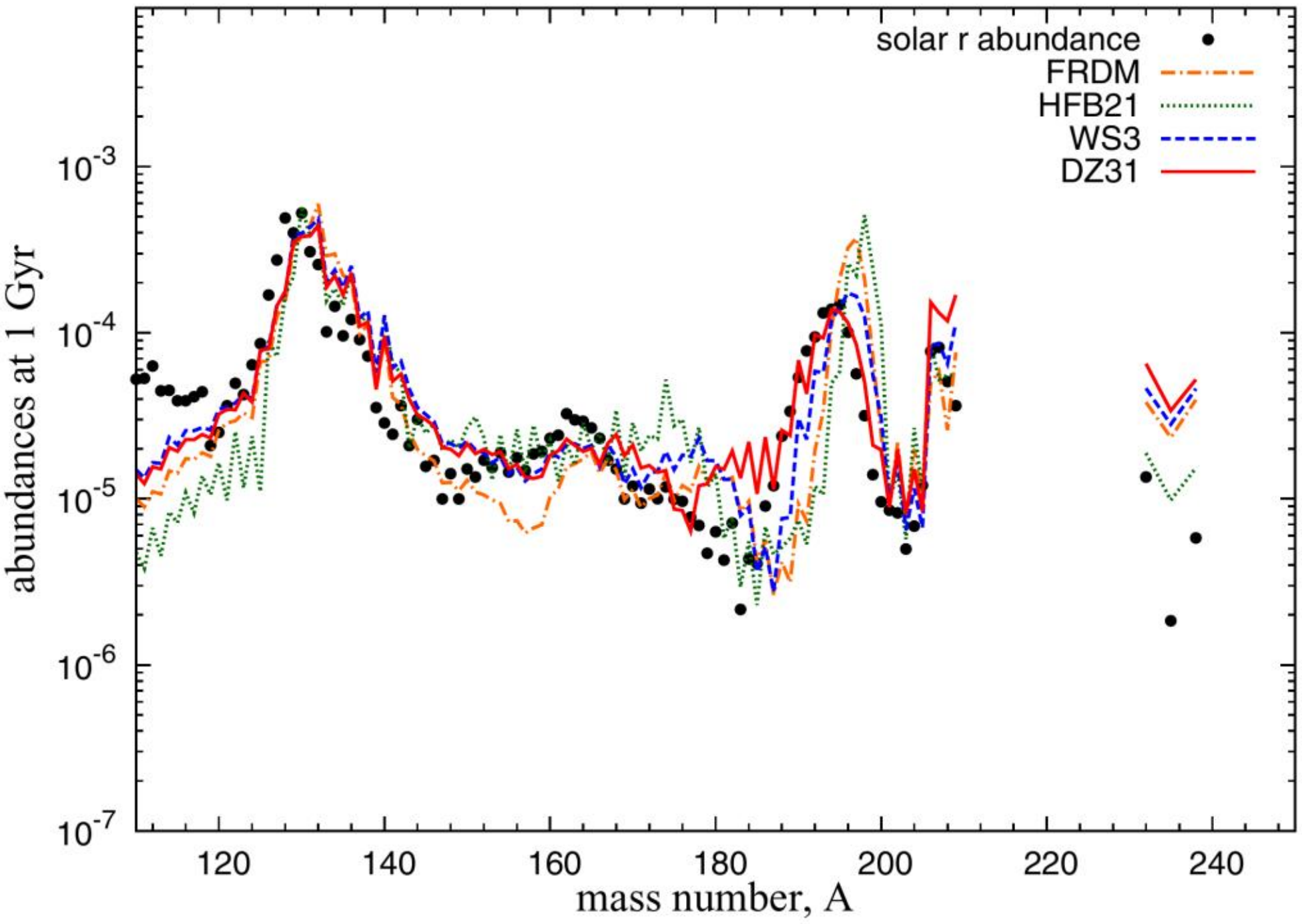

Figure 2-7. Final mass abundance patterns calculated with various theoretical mass models. From Ref. [12].

### 2-1-4. Identification of astronomical site for actinoid elements

$^{254}$Cf decays by spontaneous fission with a 61-day half-life, blocking the flow to $^{238}$U in the freeze-out phase of the r-process as shown in Fig. 2-4. If $^{254}$Cf is produced in kilonovae, which are one of the promising sites for the r-process, its fission is considered to play a significant role in the heating of the kilonovae [13]. Figure 2-8 shows a comparison of the heating rate of the kilonova with the ejected mass $M_{ej}$ = 0.06 $M_\odot$ having velocity $v_{ex}$ = 0.1$c$ with the measured luminosity [14]. The blue and black solid lines indicate the β decay and total heating rates, respectively. The total heating rates include α decays and fissions as well as β decays. The gray shaded area indicates the range of uncertainty owing to the variation of Th/Eu (Eu is typical r-process element) abundance ratios, which indicate the actinoid productivity in the r-process nucleosynthesis. The calculations show a discrepancy between the β-decay and total heating rates after 10 days. The observed luminosities after 40 days, indicated by filled squares, are located between these two curves, indicating contributions of α decays and fission in the luminosity of the cooling system. Such astronomical data could provide definitive evidence for the production of actinoid elements in kilonovae. However, the calculations are still not sufficiently reliable, due to uncertainties in the properties of nuclides which feed $^{254}$Cf, such as lifetimes and decay modes as well as astrophysical models and Th/Eu abundance ratios. The KISS-II facility will support the measurements of lifetimes

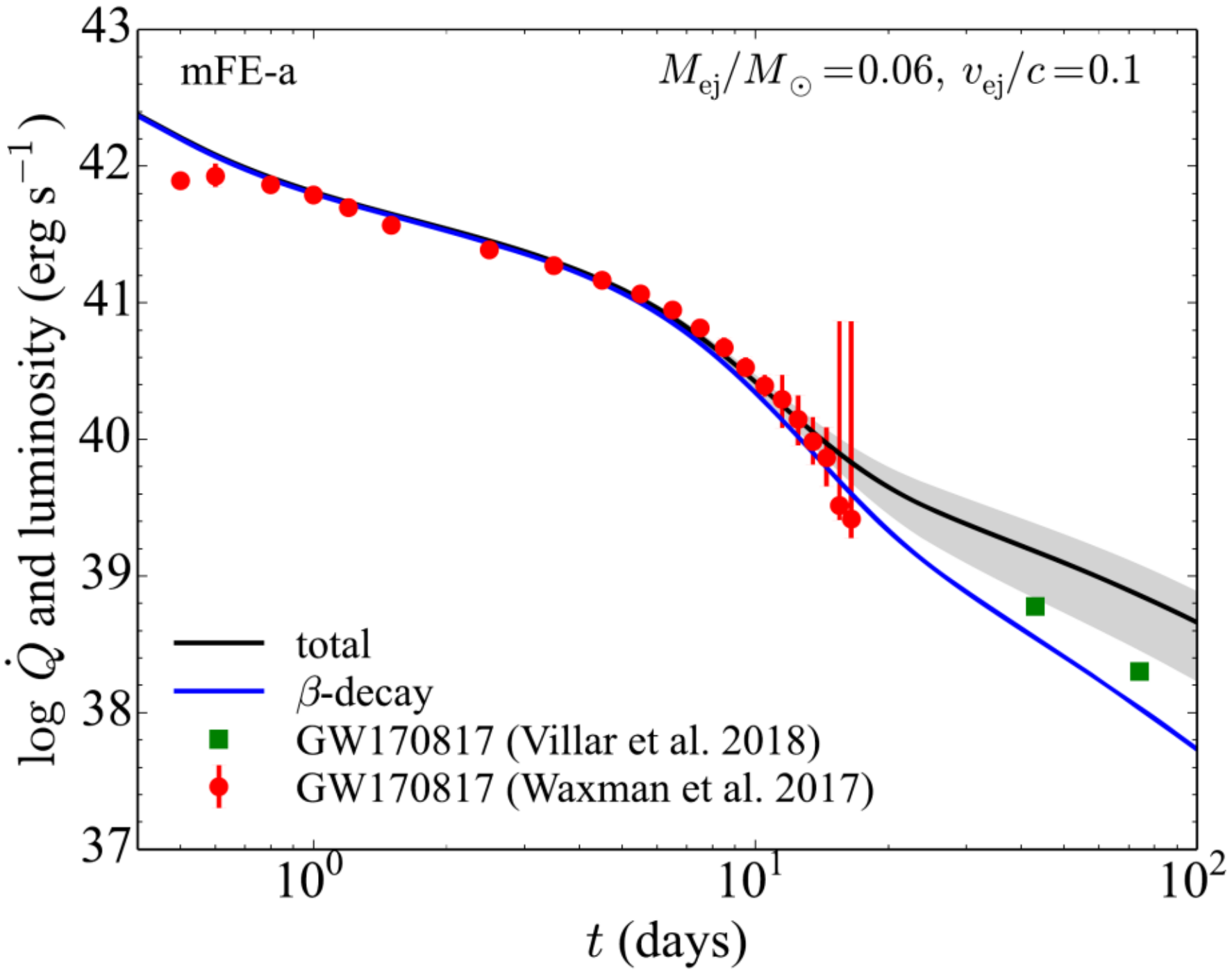

Figure 2-8. Comparison of the heating rate of the kilonova with the eject mass $M_{ej}$ = 0.06 $M_\odot$ and velocity $v_{ex}$ = 0.1$c$ with the measured luminosity (SSS17a/AT2017gfo). From Ref. [14].

and β-delayed fission probabilities for neutron-rich nuclei on the flow path to feed $^{254}$Cf. It will reduce the uncertainties in the calculations of heating by the fissions and α-decays in kilonovae, and may clarify the astronomical site of the actinoid synthesis.

## 2-1-5. Progenitors of r-element abundance peak around $A$ = 195

The assumption of the waiting points and the β-decay steady flows for the r-process leads to the neutron-rich nuclei at the neutron closed shell $N$ = 126 on the r-process path being considered to be progenitors of the nuclides in the r-abundance peak around $A$ = 195. Recent dynamical reaction network calculations involve various nuclei in the vast region around $N$ = 126 depending on the astrophysical environments and the processing time including the freeze-out phase. The masses and β-decay lifetimes of the relevant nuclei are important parameters which have significant impact on the r-process abundances as shown in Fig. 2-9 [15]. However, they are located at the midshell region between $Z$ = 50 and $N$ = 82, and $Z$ = 82 and $N$ = 126, which reveals complicated single-particle levels in combination with a transitional shape evolution making their theoretical predictions difficult. The measurements of those nuclear properties are major subjects at the KISS facility to provide experimental inputs to the theoretical nuclear models for the improvement of their predictability. The $N = 126$ closed shell also acts as the gatekeeper to the actinoid and trans-actinoid region, because the r-process flow has no other means to slow down before terminating when it once passes it. The fission cycling is influential in the formation of the β-decay steady flows and the abundance pattern of lighter species. For the comprehensive understanding of the r-process nucleosynthesis, the nuclear properties of neutron-rich nuclei around $N = 126$ as well as actinoid elements are significant. The KISS-II facility will enable us to directly access those nuclei to study their masses and lifetimes, as well as to perform β-γ spectroscopy and laser spectroscopy, including several neutron-rich nuclei on the $N = 126$ neutron shell closure and beyond.

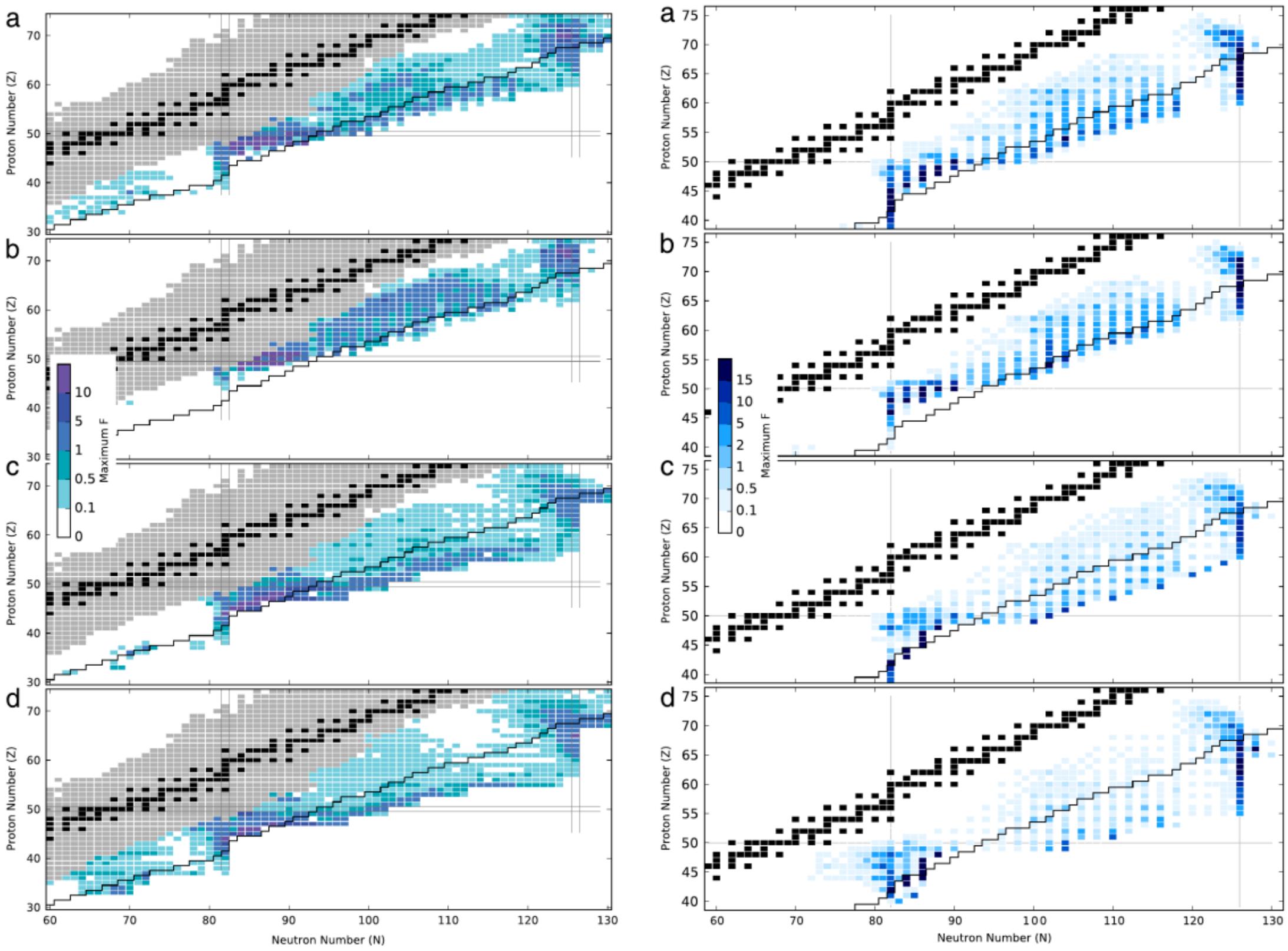

Figure 2-9. Nuclei that significantly impact final r-process abundances in four astrophysical conditions (a) low entropy hot wind, (b) high entropy hot wind, (c) cold wind and (d) neutron star merger. More influent nuclei are shaded darker from the ±500 keV mass uncertainty (left panel) and the factors of 10 variation in individual β-decay rates (right panel). From Ref. [15].

## 2-2. Nuclear properties characteristic to neutron-rich actinoid isotopes

The actinoid elements, which have the atomic numbers larger than 88, reveal characteristic nuclear physical features, owing to many proton and neutron single-particle levels which complicatedly intertwine each other in the deformed nuclear potential, combined with the high Coulomb potential created by so many protons. Figure 2-10 and 2-11 show Nilsson diagrams relevant to the actinoid region. Such plots present proton and neutron single-particle energy levels obtained from momentum-dependent Woods-Saxon potential as a function of quadrupole deformation, $\nu_2$ [16]. The complicated evolution of the single-particle levels both of protons and neutrons in conjunction with the nuclear collective deformation forms sub-shell structures with the deformed magic numbers of 96 and 100 for protons and 134, 142, and 152 for neutrons. The delicate stability of those elements causes their characteristic features of isotopic and isotonic dependence of single-particle energies, nuclear shapes, and nuclear fissions. Investigation of those characteristic features will help to improve the theoretical nuclear models to understand the evolutions of nuclear shells and collective deformations. The theoretical nuclear models play critical roles in the reaction network calculations of the r-process to reveal the origin of uranium and thorium, and, therefore, we need to provide as much experimental input as possible to improve their performance.

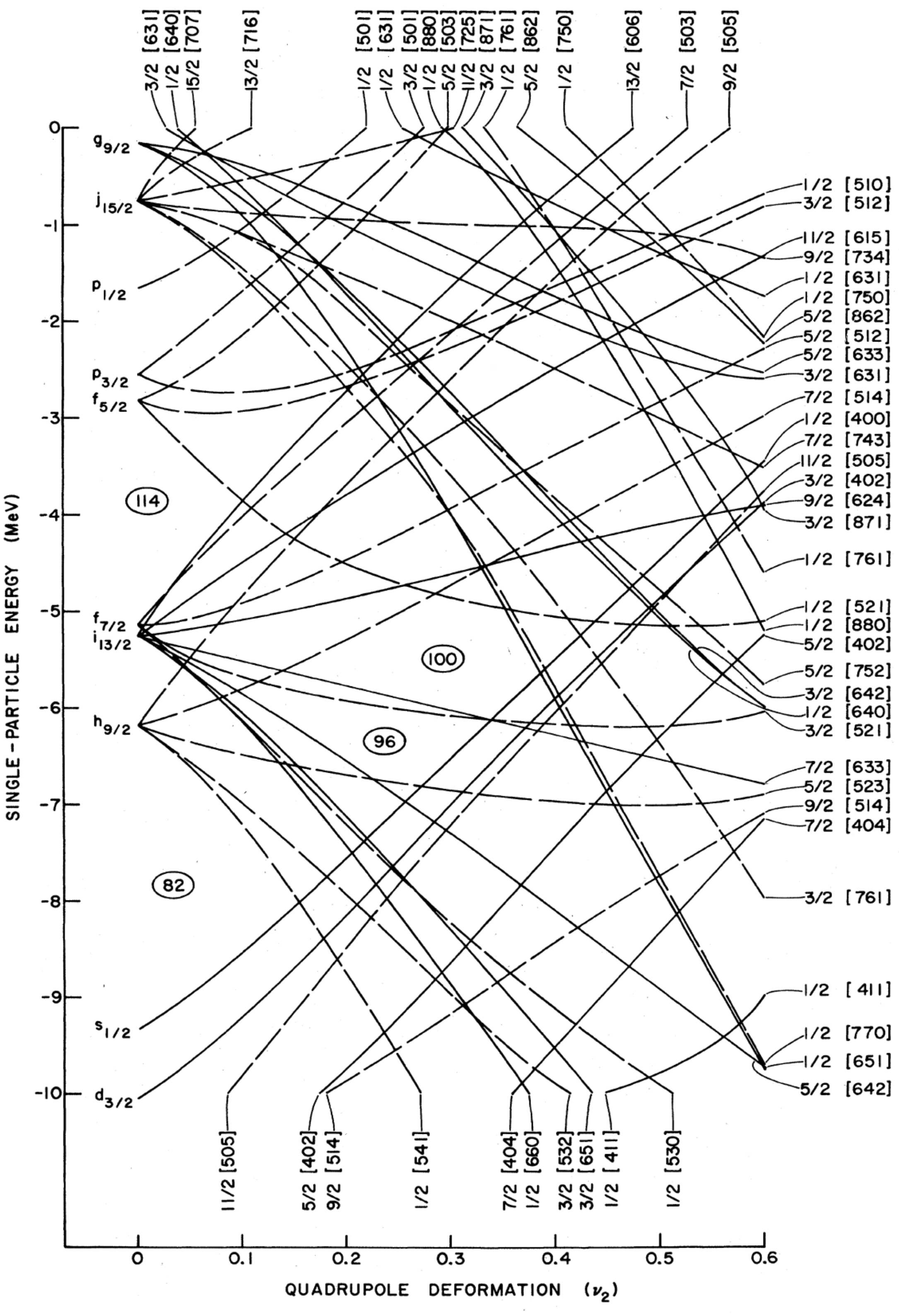

Figure 2-10. Proton single-particle levels around actinoid elements obtained from momentum-dependent Woods-Saxon potential as a function of quadrupole deformation, $\nu_2$. From Ref. [16].

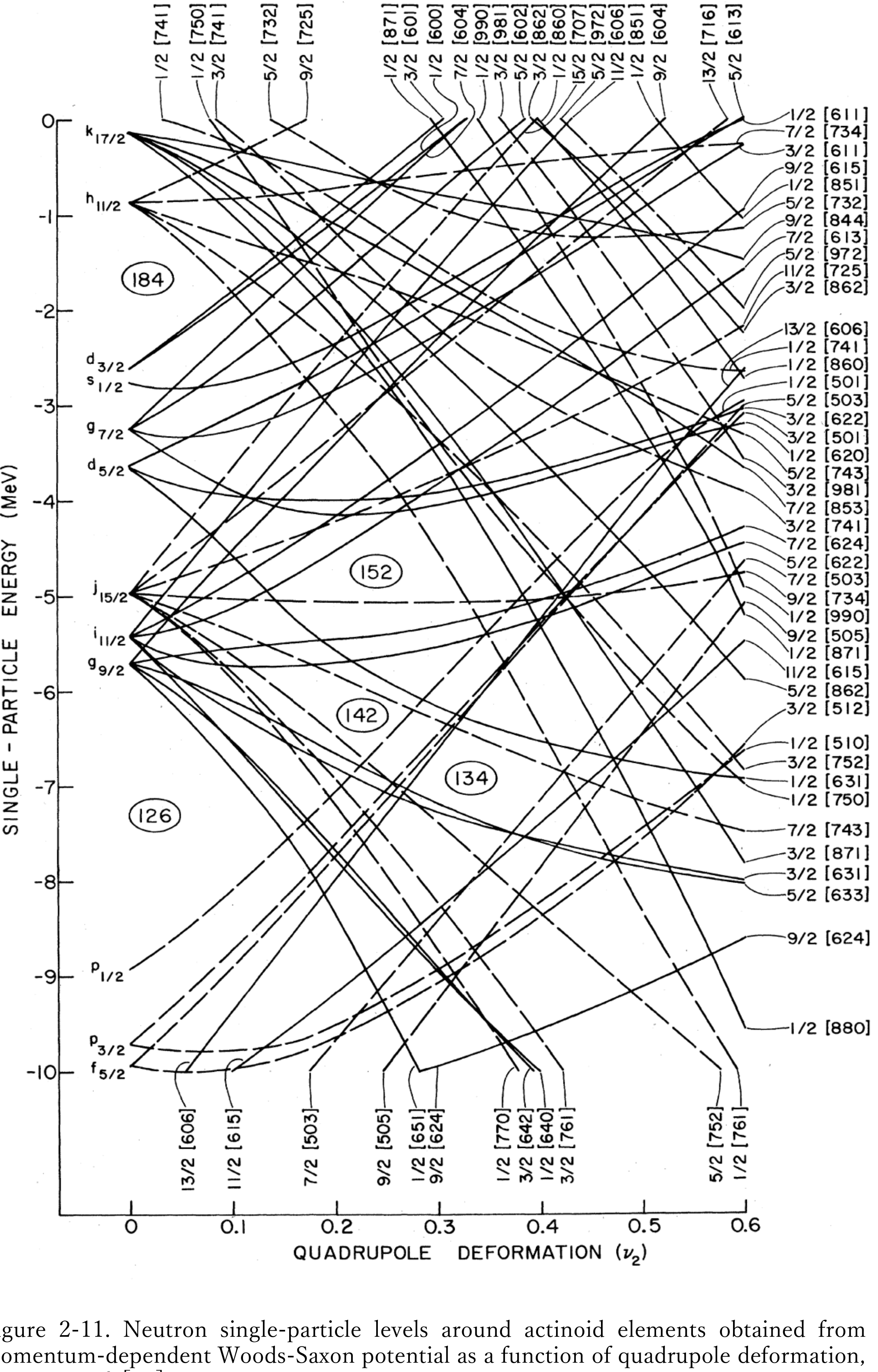

Figure 2-11. Neutron single-particle levels around actinoid elements obtained from momentum-dependent Woods-Saxon potential as a function of quadrupole deformation, $\nu_2$. From Ref. [16].

## 2-2-1. Evolution of single-particle energies

Due to the competition between pairing correlations – that is p – n, n – n, and p – p residual interactions – and collective deformation, the relative energies of the single-particle orbitals, even their ordering, would be varied with increasing number of valence nucleons. Figure 2-12 shows such an example of how the neutron single-particle energies could evolve for the even-$Z$ actinoid isotones having $N = 155$ in the top panel, where different deformation parameters were assumed in the calculation of a macroscopic-microscopic model as shown in the bottom panel [17]. It indicates that the neutron single-particle energies intertwine with each other as the atomic number increases in the delicate balance between the pairing correlation and the collective deformation. Different theoretical models can predict different ordering of the single-particle orbitals because of such delicate energy balance. A similar evolution of the proton single-particle energies could be expected to happen. The nuclear magnetic moment of the ground state of an odd nucleus with a valence nucleon is a good identifier of the spin and parity of the relevant single-particle orbital, because the magnetic property arises mainly from the unpaired nucleon. The measurements of nuclear magnetic moments allow probing of nuclear shell model predictions of the underlying single-particle wave functions. Although the absolute value of the magnetic moment is very sensitive to the variation of the spin-parity and *g*-factor, the corresponding single-particle orbital of the ground state could be uniquely assigned in combination with a microscopic mean-field calculation. Laser spectroscopy makes it possible to identify tmagnetic moments of the ground and isomeric states for neutron-rich nuclei around the actinoid region. The use of β- and/or γ- spectroscopy can directly identify or make constraints on the values of the spin and the parity for the excited states as well as the ground state. They reveal the development of single-particle orbitals along the isotopic and isotonic chains.

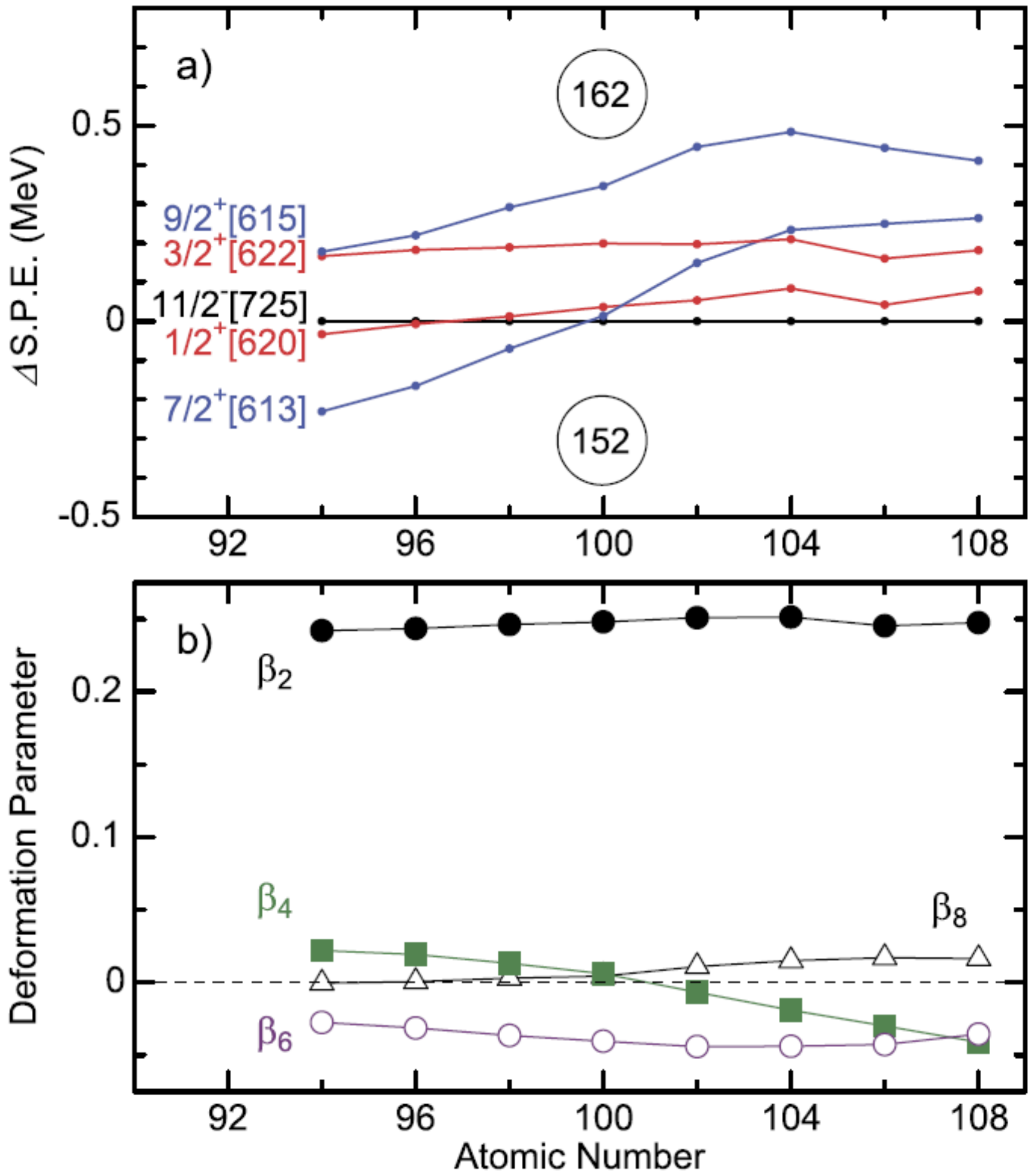

Figure 2-12. (a) Nilsson single-particle energies of $N = 155$ isotones relative to the $11/2^{-}[725]$ energies, calculated with a macroscopic-microscopic model, and (b) deformation parameters used for the calculations of (a). From Ref. [17].

## 2-2-2. Evolution of nuclear shapes

Heavy nuclei located between neighboring closed shells reveal a variety of properties concerning their complex nuclear structures. Around the rare-earth element region, shape transitions from axially symmetric prolate shapes to oblate shapes through triaxial γ softness are known to take place along isotopic chains from the neutron midshell as one approaches the closed shell at $N = 126$ due to interweaving single-particle orbitals in a deformed nuclear potential [18-21]. The behavior of the degree of deformation of the nuclei is of utmost importance to understand the nature of the closed shells. There is a rather well-established deformed region around $N = 152$, which corresponds to the midshell region in the nuclear chart beyond the double-closed-shell nucleus $^{208}$Pb. By the interplay of two intruder orbitals with high total angular momenta in relevant major shells, $6\pi i_{13/2}$ in the proton major shell of $Z = 82 - 114$ and $7\nu j_{15/2}$ in the neutron major shell of $N = 126 - 184$ as shown in Figs. 2-10 and 2-11, the midshell region with $90 \leq Z \leq 110$ and $140 \leq N \leq 160$ would be considered as a transitional region of the deformation degree of nuclei. Its evolution along the isotopic and isotonic chains would provide significant insights towards the island of stability of superheavy nuclei considered to be located at $Z \sim 114$ and $N \sim 184$. The top panel of Fig. 2-13 shows the quadrupole deformation parameters, $\beta_2$, for different even-even isotopes of Th, U, Pu, Cm, and No resulting from the Density Functional Theory (DFT) calculations [22], where the DFT functional was optimized for reproducing the experimental charge radii data. It predicts the maximum in the quadrupole deformation to be around $N = 148$. It suggests the characteristic feature that the central depression of proton density grows with increasing deformation parameter as shown in the inset and the bottom panel of Fig. 2-13. The role of neutron addition beyond $N = 148$ in such evolution should be investigated in a systematic way regarding to the following points: how the deformation parameter evolves beyond the $N \simeq 148$ and how the central depletion of proton density distribution evolves with decreasing deformation parameter. This would reveal the correlation between the collective deformation and the single-particle wave functions. Laser spectroscopy at KISS-II will make it possible to measure the differential nuclear mean-square charge radii and electric quadrupole moments of neutron-rich nuclei in and around the actinoid region. They can be used to deduce the deformation parameters under the assumption of axial symmetry. Application of $\gamma$-spectroscopy can deduce the moment of inertia along the rotational bands, which may reveal the change of nuclear shapes with the excitation energies.

### 2-2-3. Nuclear fission

Fission is a complex nuclear decay characteristic to actinoid and transactinoid elements, which involves an interplay between collective and single-particle effects along the evolution of collective degrees of freedom from a single compound nuclear system to a pair of smaller fragments by overcoming a fission barrier on the nuclear potential energy landscape. The fission barrier height, $B_f$, is an important parameter to formulate the fission probability.

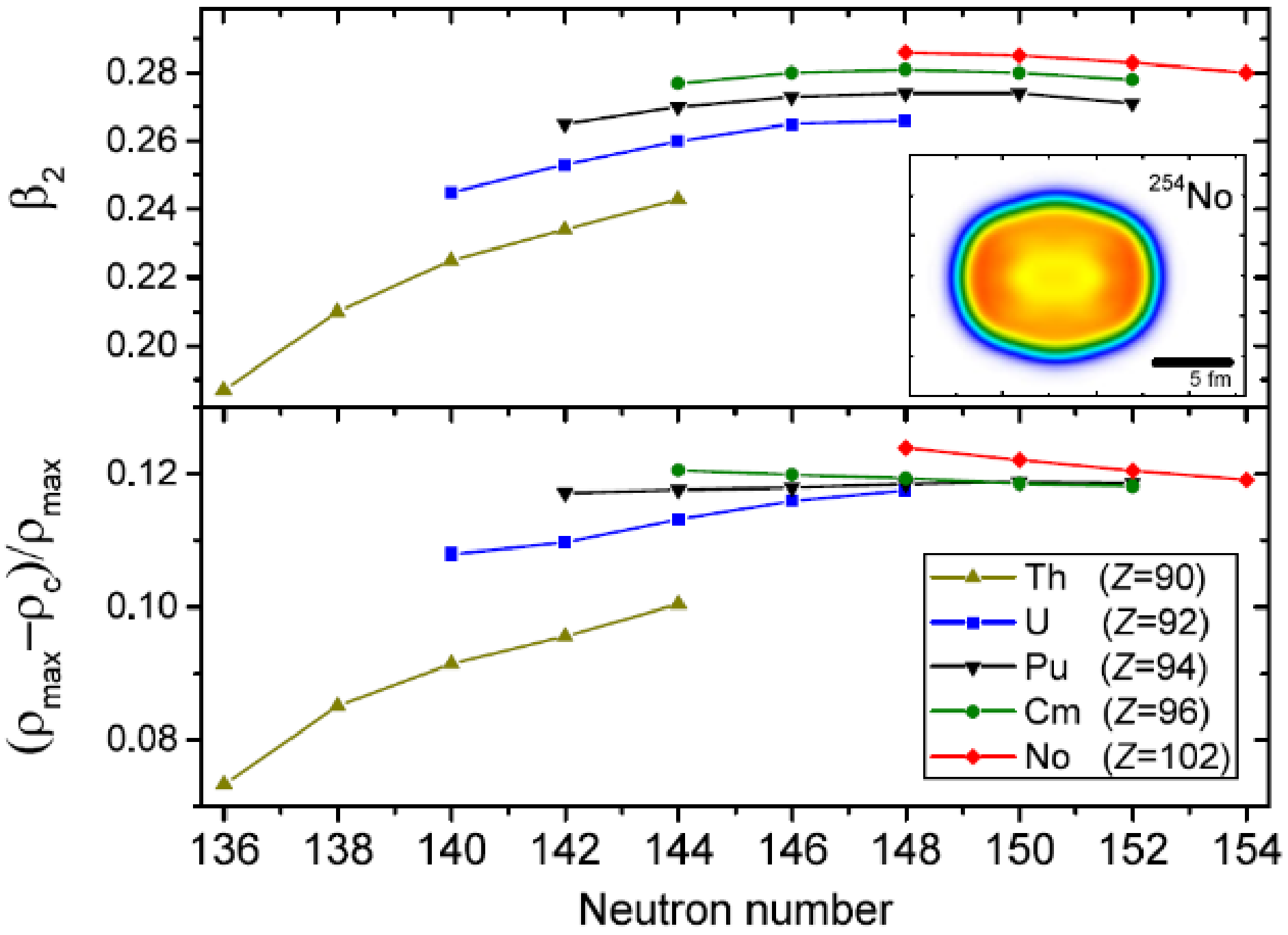

Figure 2-13. Top panel: deformation parameter $\beta_2$ for different even-even isotpes of Th, U, Pu, Cm, and No obtained from the DFT calculations with the UNDEF1 functional. The inset figure shows the calculated proton distributions of $^{254}$No from highest density (red) to low density (blue). Bottom panel: relative depth of the central depletion. From Ref. [22].

Traditional evaluations of fission probabilities rely on multiple-humped fission penetration models with fission barriers of inverted decoupled parabolas. Since their predictive powers are very limited due to a large number of parameters for reasonable fits to experimental cross sections, simple-minded extrapolation of the fitting parameters to experimentally unknown nuclei would be hardly acceptable. As illustrated in Fig. 2-14, where several theoretical calculations of $B_f$ are compared for uranium isotopes, there exist large discrepancies among different calculations at $N > 152$, beyond the existing experimental data indicated by filled circles, although all theoretical predictions reproduce the experimental data for $N < 149$ well.

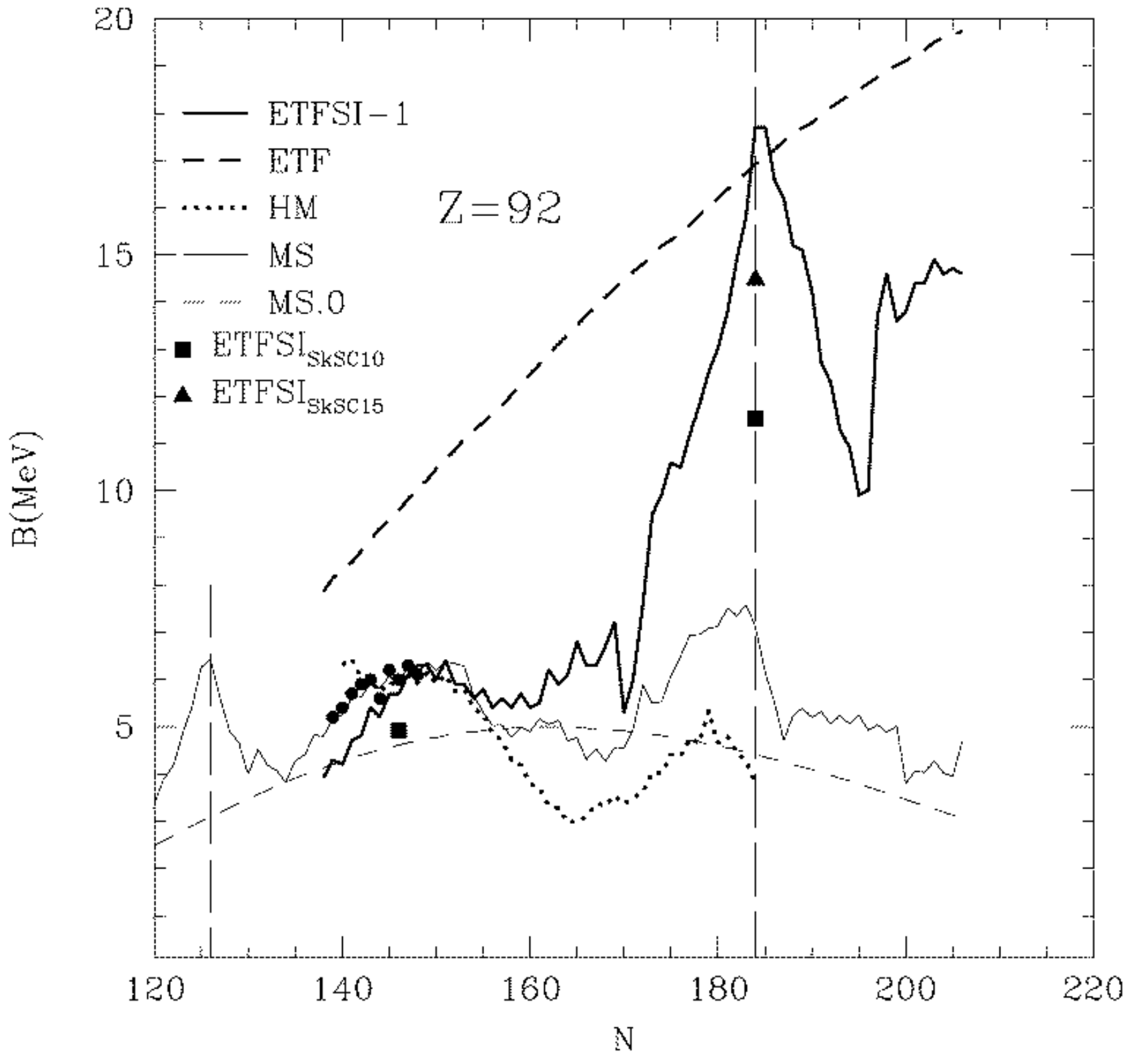

Figure 2-14. Fission-barrier heights for the uranium isotopes, calculated by various models with experimental values indicated by filled circles. From Ref. [23].

Additionally, β-delayed fission (βdf) is another characteristic aspect of the decay modes for actinoid and transactinoid isotopes far from β-stability, in which a parent nucleus first undergoes a β decay to an excited state in the daughter nucleus, which then undergoes fission. Figure 2-15 shows a schematic representation of the βdf process. The excited state in the daughter nucleus fed by the β-decay allows it to overcome the fission barrier, which would be difficult to do from the ground state. As the energy of the populated excited state depends on the β-decay Q-value, $Q_\beta$, and the β-decay strength function, the dynamics of the βdf is strongly affected by the balance between $Q_\beta$ of the parent nucleus and $B_f$ of the daughter nucleus. Under these circumstances, it would be very interesting to experimentally investigate the fission properties of actinoid isotopes across $N = 152$ along their isotopic chains, providing a benchmark for various theoretical models on the market that describe nuclear fission by using different methods of calculation and nuclear forces. The understanding of the shell structures and the collective deformations of actinoid elements from the mass measurements, lifetime measurement, β-γ spectroscopy and laser spectroscopy at KISS-II will reduce the uncertainties of fission barrier heights beyond $N = 152$. The KISS-II facility will also provide opportunities to directly measure the fission barrier heights of neutron-rich actinoid isotopes. The measurements of mass asymmetry for two fission fragments provide hints to understand the potential energy landscape which is affected by the shell structure in the fission dynamics.

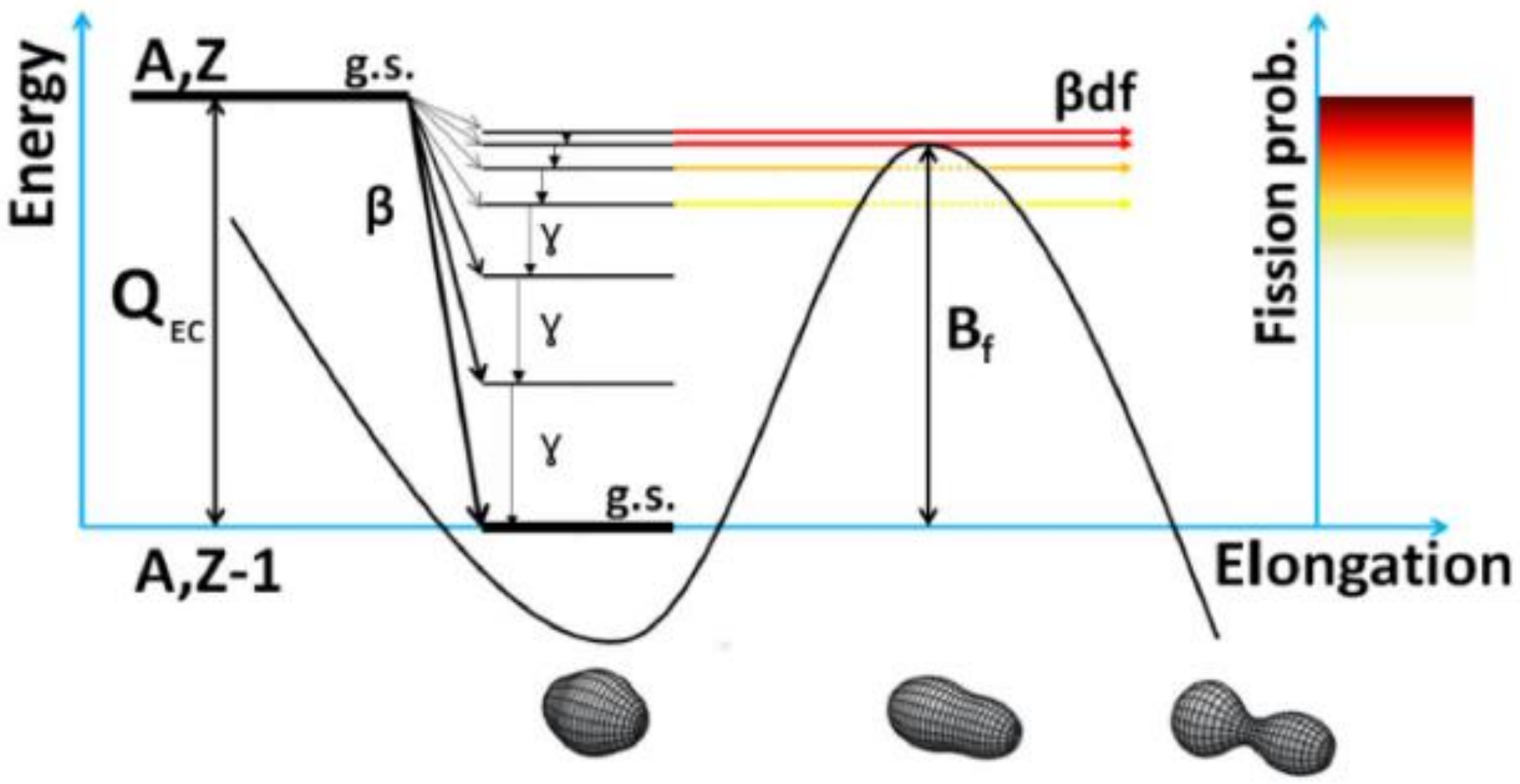

Figure 2-15. Schematic representation of the β-delayed fission process on the neutron-deficient side of the nuclear chart. From Ref. [24].

# 3. Production of neutron-rich nuclei using MNT reactions

Neutron-rich nuclei around the actinoid region can be produced by multinucleon transfer (MNT) reactions [1]. To efficiently produce a given nuclide, the best target would be the heaviest nuclide closest to it on the nuclear chart. However, all actinoids are radioactive. Apart from $^{232}$Th, $^{235}$U, and $^{238}$U – which have long lifetime comparable to the age of the universe – the other actinoids do not exist in nature, apart from sparse decay products of these three long-lived nuclides. Even if we include man-made isotopes based on half-life and available material the possible target materials for reasonable production of neutron-rich actinoids by MNT would still be limited to $^{232}_{90}$Th ($T_{1/2}$ = 14.0 Gyr), $^{238}_{92}$U ($T_{1/2}$ = 4.47 Gyr), $^{243}_{95}$Am ($T_{1/2}$ = 7.36 kyr), and $^{248}_{96}$Cm ($T_{1/2}$ = 348 kyr). Various reaction systems have been studied so far to try to understand the driving force of the MNT reactions and to investigate the most useful reaction systems for producing neutron-rich actinoid isotopes [2-17]. For example, $^{16,18}$O, $^{20,22}$Ne, $^{40,44,48}$Ca, $^{86}$Kr, $^{136}$Xe, and $^{238}$U beams were used with $^{248}$Cm targets to evaluate production cross sections of actinoid isotopes at energies around the Coulomb barriers; the results were evaluated via radiochemical methods [4-8,13]. Figure 3-1 shows production cross sections of $_{98}$Cf isotopes in the MNT reactions from various projectiles with the $^{248}_{96}$Cm target. It indicates that the production yields of the isotopes close to the target nucleus, having only a few transferred nucleons, are less sensitive to the choice of projectile. However, it also reveals that the heavier projectiles show broader isotopic distributions extending to more neutron-rich actinoid isotopes. The use of the heaviest beam, $^{238}U$, would therefore be advantageous to produce the most interesting, neutron-rich isotopes in the actinoid region with the largest cross sections.

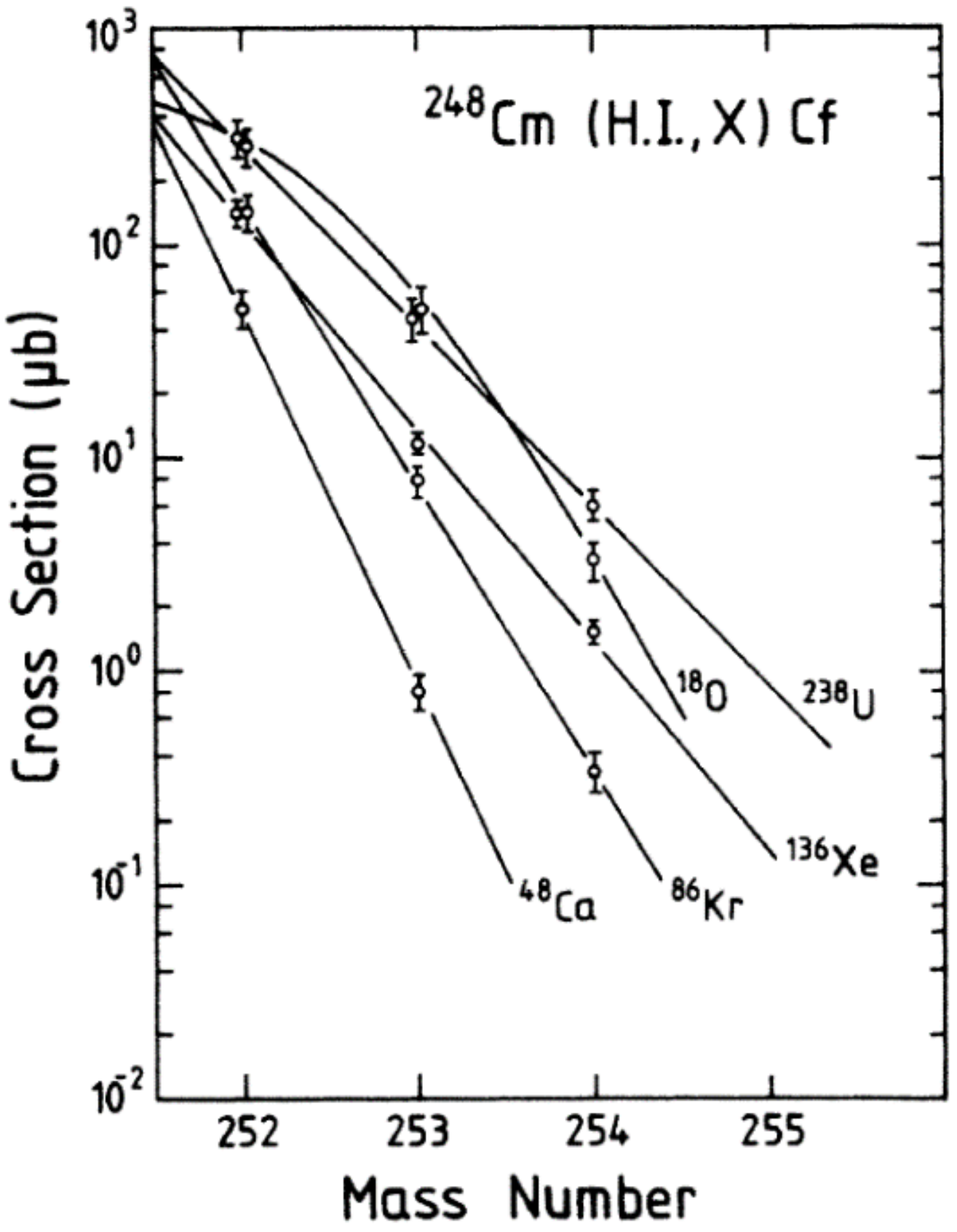

Figure 3-1. Neutron-rich californium yields arising from the reaction of various heavy ions with $^{248}$Cm. From Ref. [8].

The MNT reactions would form the reaction products in excited states from the dissipative collisions, leading to deexcitation by particle emission – γ, neutrons, protons, alphas or even fissions. Figure 3-2 shows cross sections for production of primary fragments – before deexcitation – in the collision of $^{238}U$ + $^{248}Cm$ modeled on the Langevin-type dynamical equations of motion [19-20]. It indicates that this reaction produces neutron-rich nuclei in *terra incognita* with relatively high cross sections – exceeding 1 μb as primary fragments. Figure 3-3 shows the landscape of the excitation energies for the primary fragments in the $^{238}U$ + $^{248}Cm$ collisions at $E_{cm}$ = 750 MeV produced with cross sections larger than 1 μb. It indicates that the neutron-rich reaction products around $^{248}Cm$ have excitation energies up to 40 MeV, which would be high enough for the primary fragments to overcome the fission barrier. Therefore, the surviving nuclei are significantly reduced from the primary fragments by particle evaporations and fission, as shown in Fig. 3-4, where mass distributions of primary fragments and surviving nuclei ($Z \geq 98$) are compared.

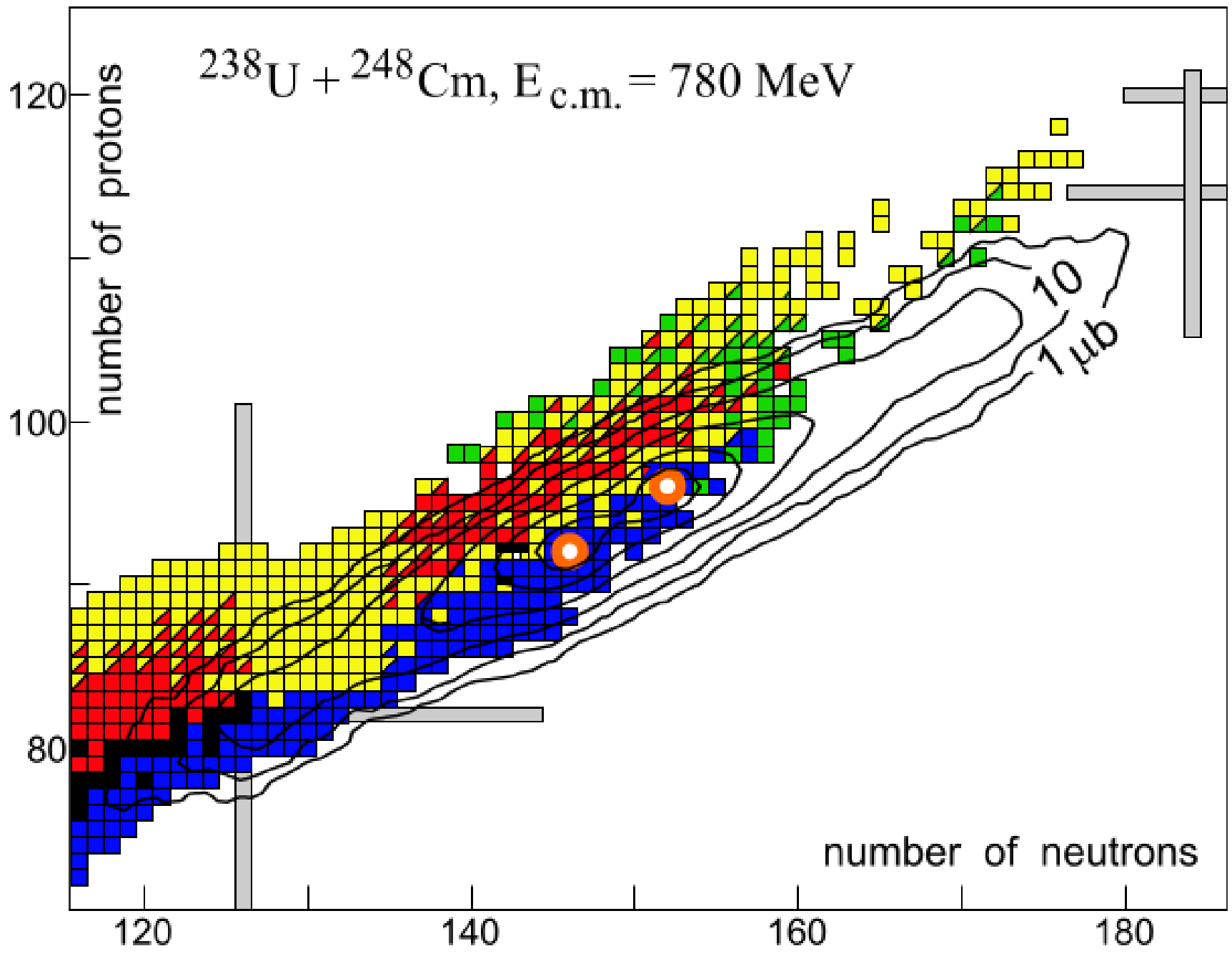

Figure 3-2. Cross sections for production of primary fragments in the collision of $^{238}U$ with $^{248}Cm$ at $E_{cm}$ = 780 MeV. From Ref. [18].

Figure 3-5 compares the experimental isotopic distributions for the reaction products in the reactions $^{238}U$ + $^{238}U$ ($E$ = 7.5$A$ MeV) and $^{238}U$ + $^{248}Cm$ ($E$ = 7.4$A$ MeV) with theoretical

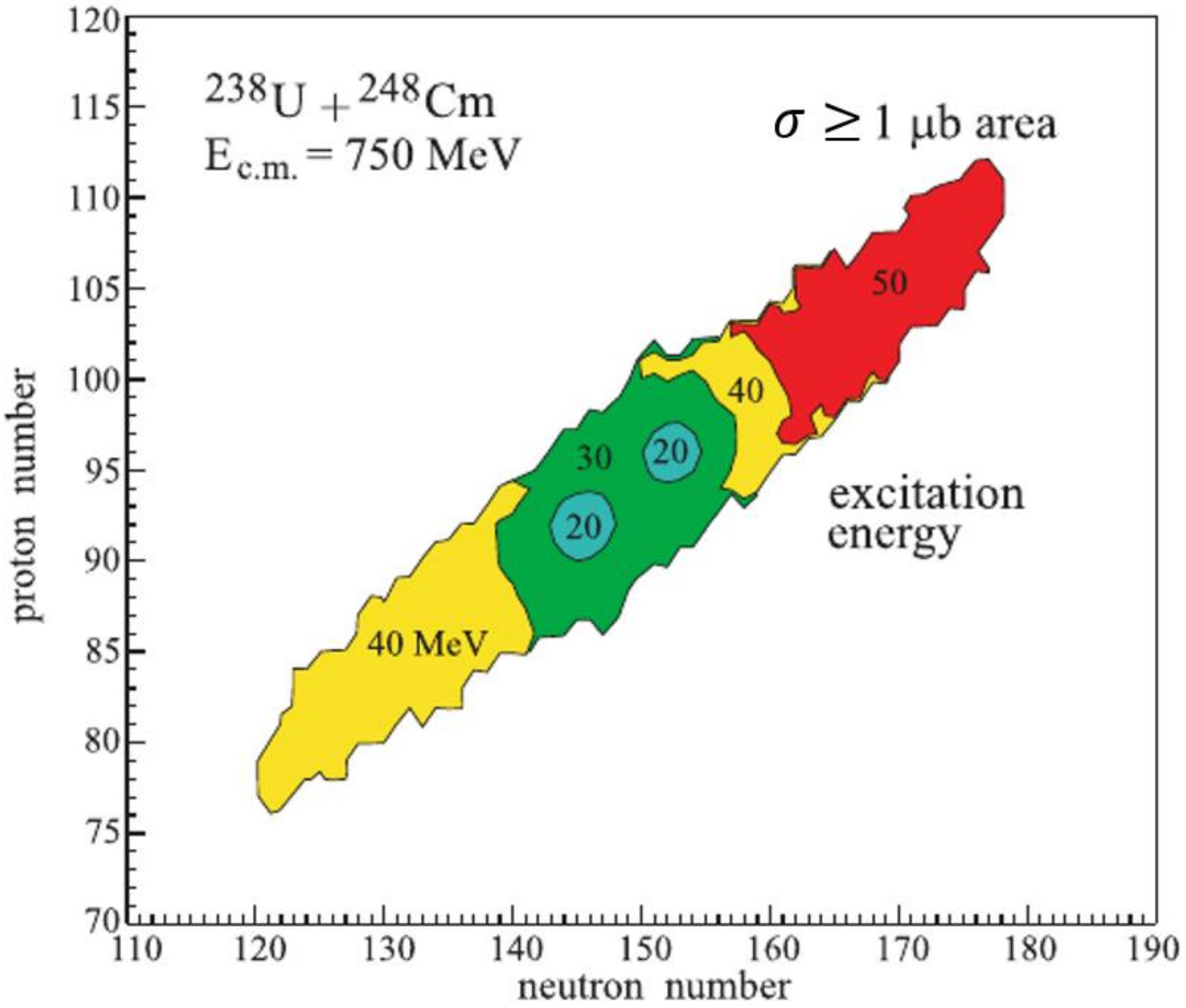

Figure 3-3. Landscape of excitation energy of primary fragments formed in the $^{238}U + ^{248}Cm$ collisions at $E_{cm}$ = 750 MeV with cross sections larger than 1 μb. Adapted from Ref. [21].

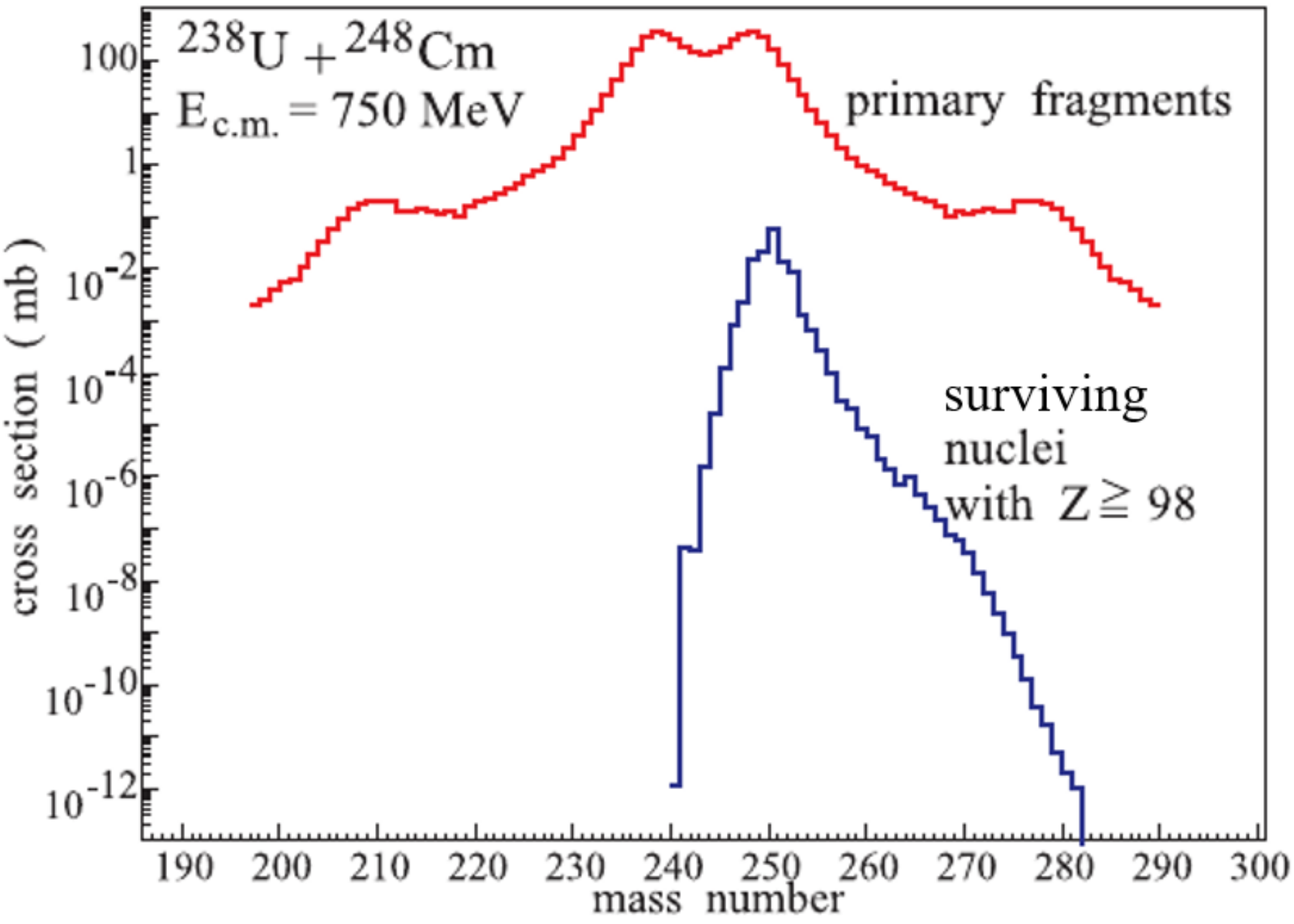

Figure 3-4. Mass distribution of primary fragments (red) and surviving nuclei (blue) formed in the $^{238}U + ^{248}Cm$ collisions at $E_{cm}$ = 750 MeV. Adapted from Ref. [21].

calculations using GRAZING and GRAZING-F codes and a Langevin model. GRAZING is a semiclassical coupled channel program, which was developed from the theory of Ref. [23] to treat the grazing collisions. It includes particle evaporation from the reaction products but does not consider fission. GRAZING-F is a modified version of GRAZING which includes fissions as well as particle evaporations in the secondary process of the reaction products [24]. The figure shows overall agreements between the experimental data and the Langevin model calculations both for $^{238}$U + $^{238}$U and $^{238}$U + $^{248}$Cm. The calculations of GRAZING have a tendency to overestimate the cross sections in $^{238}$U + $^{248}$Cm and to shift the isotopic distributions to the lighter side in $^{238}$U + $^{238}$U. GRAZING-F improves the agreement of the absolute cross sections with the experimental data, especially for $^{238}$U + $^{248}$Cm, yielding cross sections comparable to the Langevin model calculations. However, it must be noted that GRAZING-F systematically *underestimates* the cross sections of the more neutron-rich isotopes compared to experimental data for $^{238}$U + $^{238}$U. The reduction of the cross sections in GRAZING-F compared to GRAZING is about one order of magnitude in the neutron-rich side. Despite the GRAZING code yielding slightly less accurate results than the Langevin model, because the code is available to the public for the calculations of the cross sections and the kinematics in the MNT reactions [25], we will adopt GRAZING calculations – with a uniform 90% loss to presumed fission in the actinoid region – to evaluate the yields in the KISS-II project.

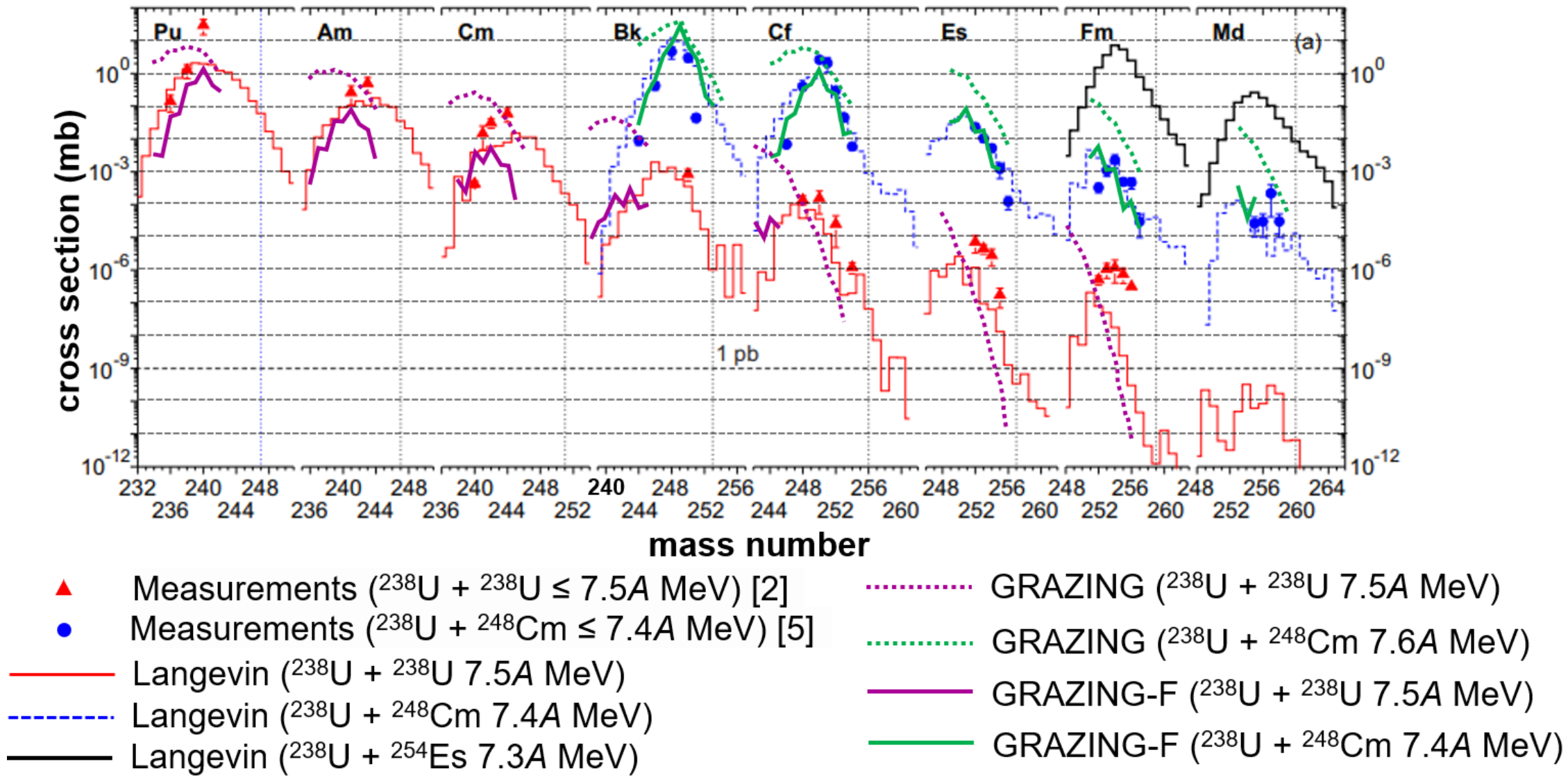

Figure 3-5. Isotopic distributions of final products with $Z$ greater than the target obtained in collisions of actinoids. Triangles and circles indicate the experimental data for $^{238}$U + $^{238}$U [2] and $^{238}$U + $^{248}$Cm [5], respectively. Thin red, dashed blue, and thick black histograms correspond to the results of the Langevin model calculations for the reactions $^{238}$U + $^{238}$U ($E$ = 7.5 MeV/u), $^{238}$U + $^{248}$Cm ($E$ = 7.4$A$ MeV), and $^{238}$U + $^{254}$Es ($E$ = 7.3$A$ MeV), respectively. Dotted and solid purple lines indicate the calculations by GRAZING and GRAZING-F, respectively, for Pu, Am, and Cm in the $^{238}$U + $^{238}$U reaction. Dotted and solid green lines indicate the calculations by GRAZING and GRAZING-F, respectively, for Bk, Cf, Es, Fm, and Md in the $^{238}$U + $^{248}$Cm reaction adapted from Ref. [22].

## 3-1. $^{238}$U + $^{238}$U reaction system: a realistic candidate to access to the neutron-rich actinoid region

The $^{238}$U + $^{238}$U reaction system is a candidate to produce neutron-rich nuclei in and around the actinoid region at KISS-II. Figure 3-6 shows the GRAZING calculated cross sections as functions of projectile energy for production of neutron-rich uranium isotopes as target-like fragments (TFLs) in the $^{238}$U + $^{238}$U reaction system. Above a reaction energy sufficient to overcome the Coulomb barrier, the cross section for an MNT reaction is fairly independent of the reaction energy. However, in the case of the TLFs, when the reaction energy also increases, the minimum angular momentum relevant to the MNT reaction also increases and the energy transfer from the projectile to the target thereby decreases. Thus, the scattering angle of the recoil product becomes larger at higher reaction energies, while the recoil energy becomes smaller. Such kinematic conditions result in the TLFs being liable to stop in the target if the projectile energy is too high. As such, after considering primary beam energy loss in the target along with the TLF energy loss in the target, an optimal primary beam energy and target thickness can be determined. Figure 3-7 demonstrates the GRAZING calculated production yield of $^{250}$U TLFs as a function of target thickness and primary beam energy. For any target thickness, there exists a primary beam energy above which the yield decreases. The target thickness of 13 mg/cm$^2$ with a beam energy of 9.4*A* MeV will be used for KISS-II on consideration of similar functions for other nuclides.

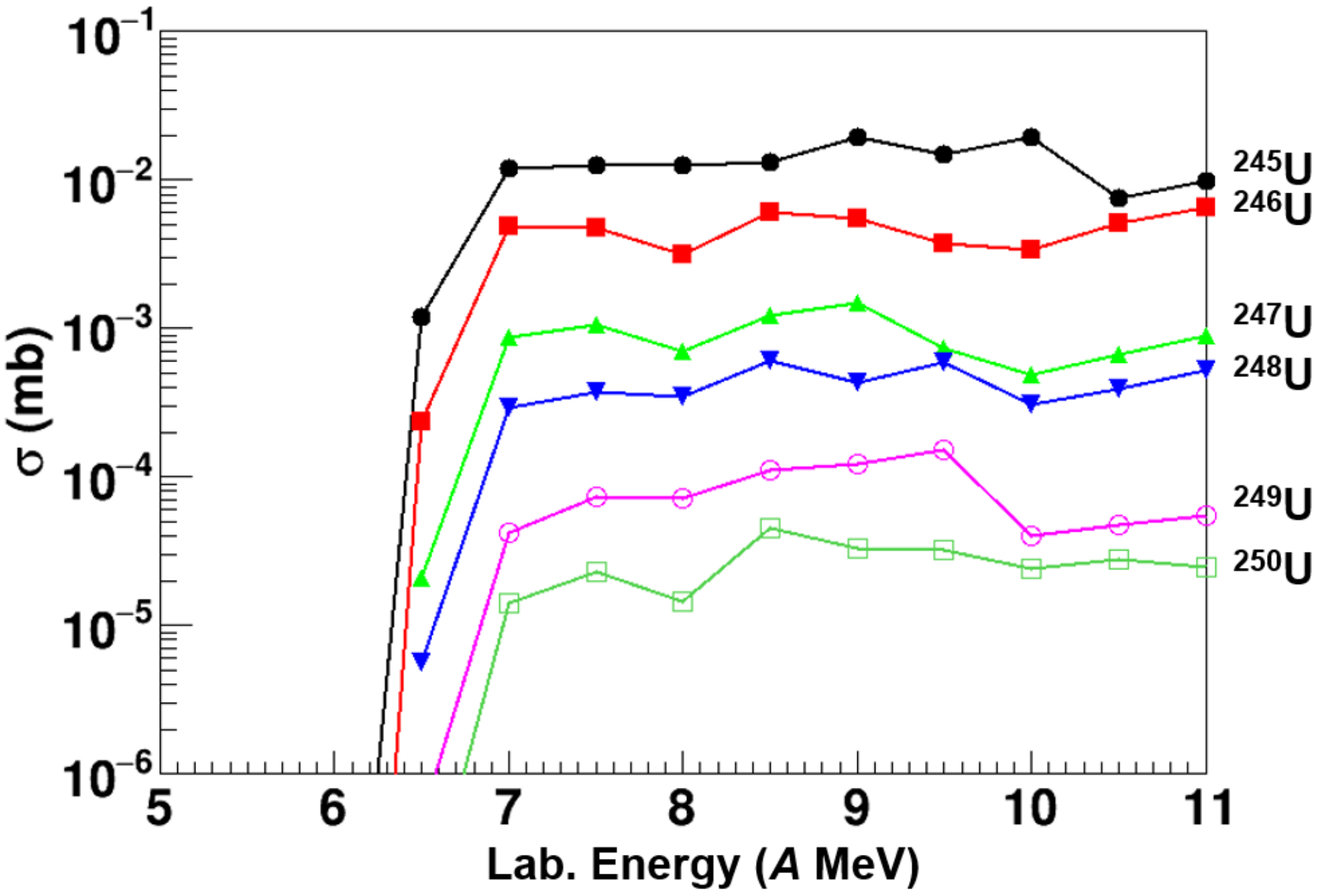

Figure 3-6. GRAZING calculated production cross sections as function of projectile energy for production of uranium isotopes as the TLFs in the $^{238}$U + $^{238}$U reaction system.

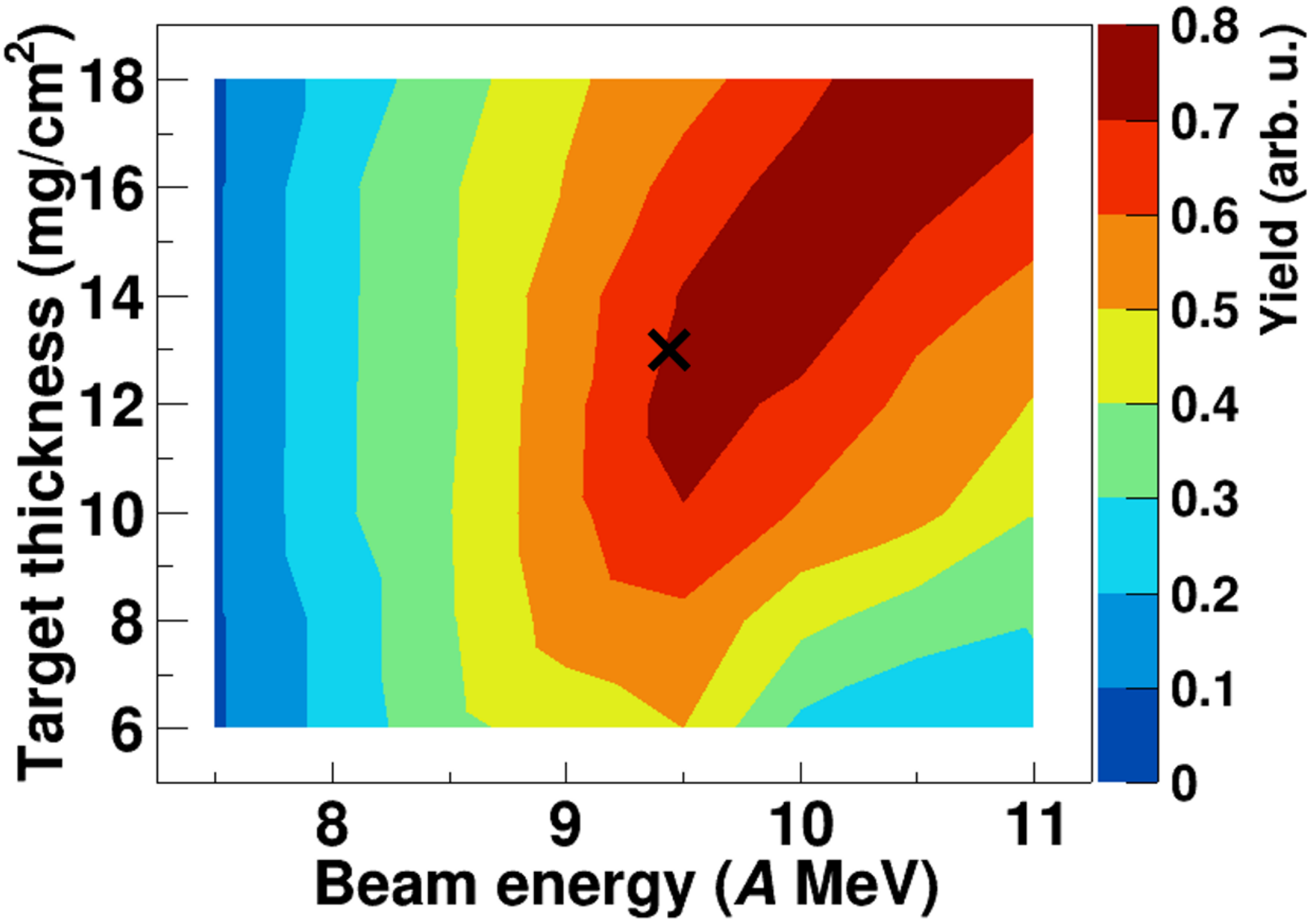

Figure 3-7. Production yield for $^{250}$U TLFs in the $^{238}$U + $^{238}$U reaction system as a function of target thickness and primary beam energy. The calculation takes into account the fraction of TLFs which stop in the target. Cross indicates the energy-thickness values chosen for KISS-II based on consideration of similar function for other nuclides.

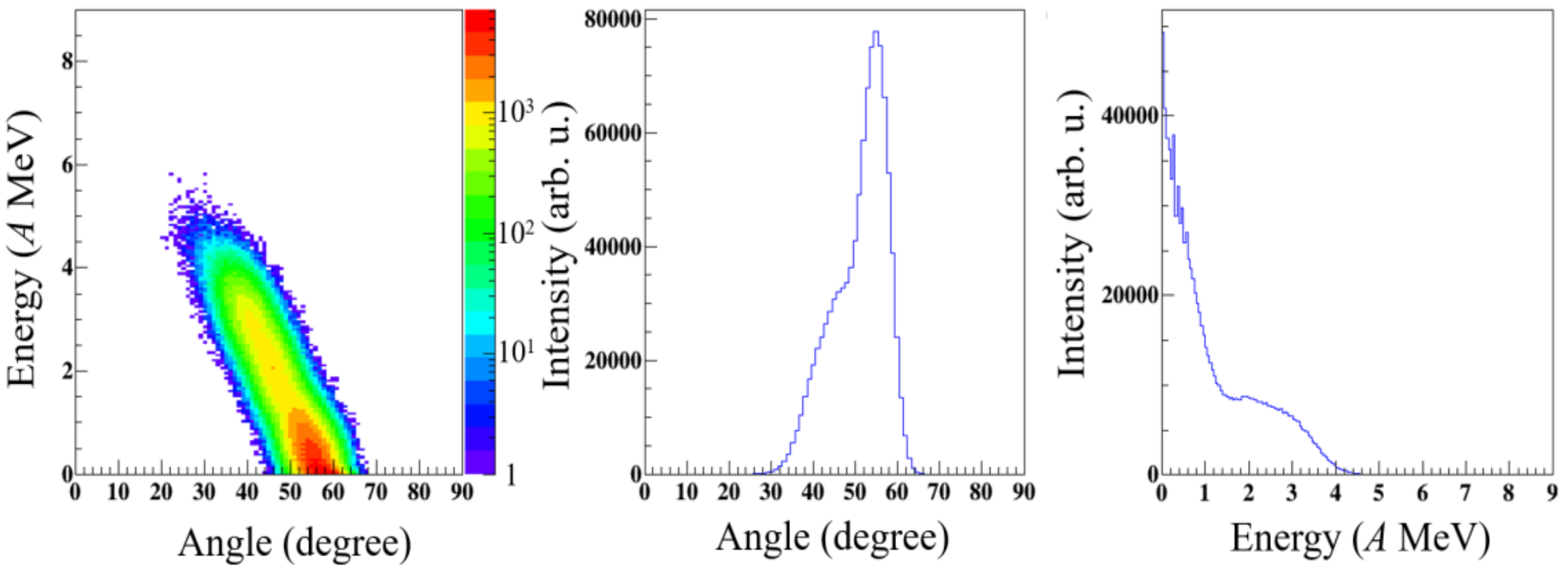

Figure 3-8. Correlation between energies and angles (left), the angle distribution (middle), and the energy distribution (right) for $^{250}$U target-like fragments ejected from the target in the $^{238}$U (9.4*A* MeV) + $^{238}$U (13 mg/cm$^2$) reactions, calculated by GRAZING [25].

Figure 3-8 shows the calculated correlation between energies and angles (left panel), the angle distribution (middle panel), and the energy distribution (right panel) for $^{250}$U TLFs ejected from the target. It shows that the TLFs have large emission angles around the grazing angle - i.e. 30° - 70° - with the energies less than 5$A$ MeV.

Figure 3-9 shows the isotopic distribution of TLFs in this reaction system based on a beam intensity of 1 pμA, and assuming a 90% reduction of yields due to fission. The blue lines indicate the frontier of nuclides whose masses are experimentally known based on AME2020 [26]. The $^{238}$U + $^{238}$U system pushes the research field into a much more neutron-rich region, well- beyond the limits of presently-known masses. The region enclosed by the dashed line corresponds to the red shaded area in Fig. 2-4. The red line indicates the boundary for the yield more than $10^3$ per day, which encompasses the majority of the region of interest having $A \leq 252$. The orange lines indicate the boundary of experimentally discovered nuclides. The figure indicates that more than 110 new nuclides can be accessed at the yield more than $10^3$ per day. Even if the production rates are smaller by two orders of magnitude, more than 60 new nuclides are in the scope.

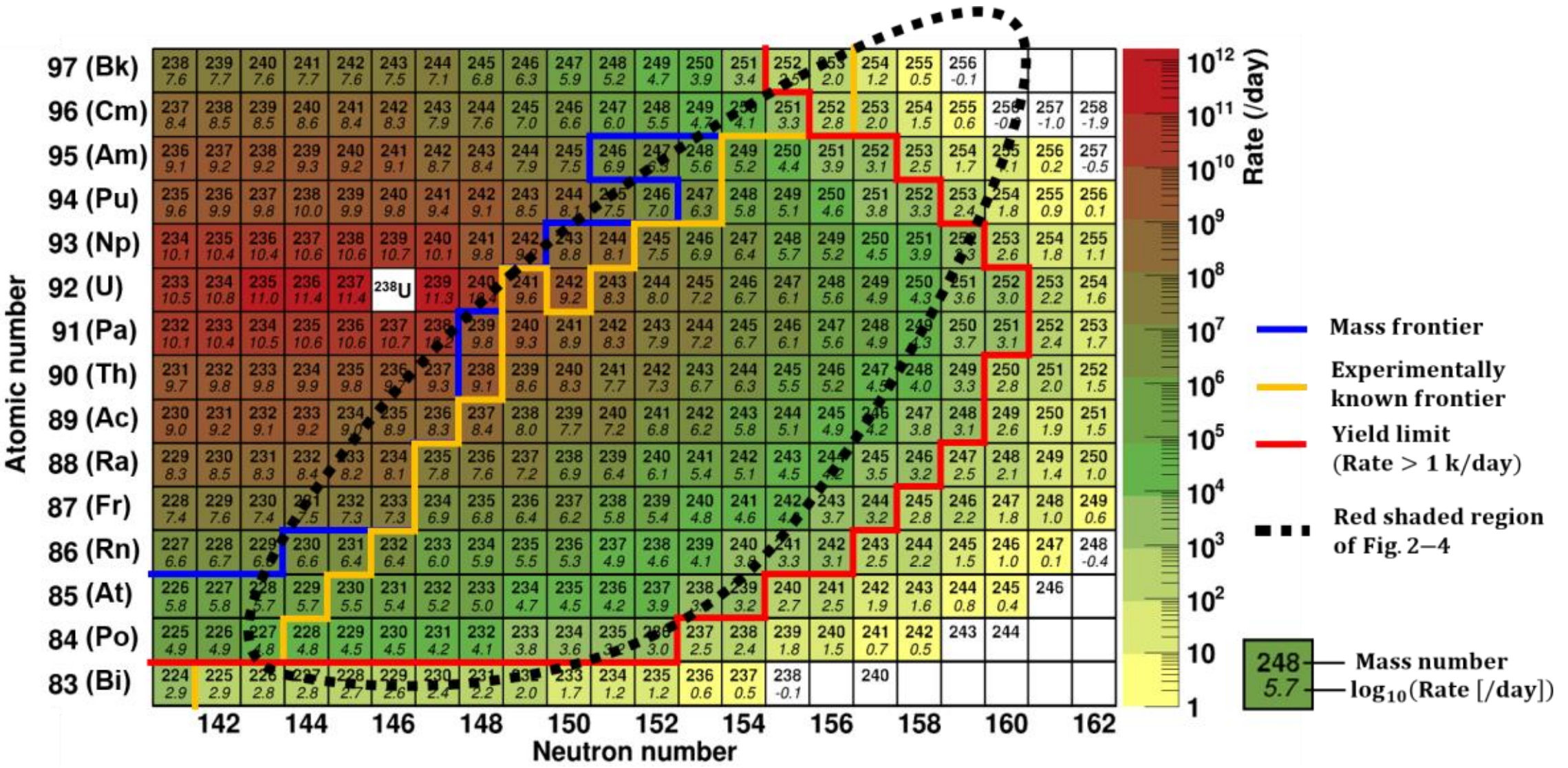

Figure 3-9. Production yield of target-like fragments produced by MNT reactions in the $^{238}$U (9.4$A$ MeV, 1 pμA) + $^{238}$U (13 mg/cm$^2$) reaction system. Expected yields were calculated by GRAZING [23] assuming a 90% reduction from fission. Blue lines indicate the frontier of nuclides with experimentally known masses. Orange lines indicate the boundary of experimentally discovered nuclides. The region enclosed by the dashed line corresponds to the red shaded area in Fig. 2-4. Red lines indicate the boundary for yields larger than $10^3$ per day.

# 3-2. $^{238}$U + $^{243}$Am and $^{248}$Cm reaction systems: candidates for extended access to the neutron-rich actinoid region

Neutron-rich nuclei with larger atomic numbers could be accessed by using the $^{238}$U + $^{243}$Am reaction system. Due to radiation limits stemming from the relatively short half-life ($T_{1/2}$ = 7.36 kyr), the target thickness will be limited to 1/10 that of the previously mentioned $^{238}$U targets, while the available material further limits the number of targets to a wheel 1/10 the circumference of that which would be used for the $^{238}$U targets. The limited number of targets reduces the acceptable primary beam intensity to 0.1 pμA. In spite of this, the enhanced production yield from the heavier target would allow experimental access to heavier and more neutron-rich isotopes than in the $^{238}$U + $^{238}$U reaction system, as shown in Fig. 3-10. The blue lines indicate the frontier of nuclides, whose masses are experimentally known based on AME2020 [26]. The region enclosed by the dashed line corresponds to the red shaded area in Fig. 2-4. The red line indicates the boundary for the yield more than $10^3$ per day, which covers the region of interest up to $A = 256$. The orange lines indicate the boundary of experimentally discovered nuclides. Using the $^{243}$Am target, as many as 20 new nuclides beyond those available from the $^{238}$U + $^{238}$U system can be accessed at the yield more than $10^3$ per day.

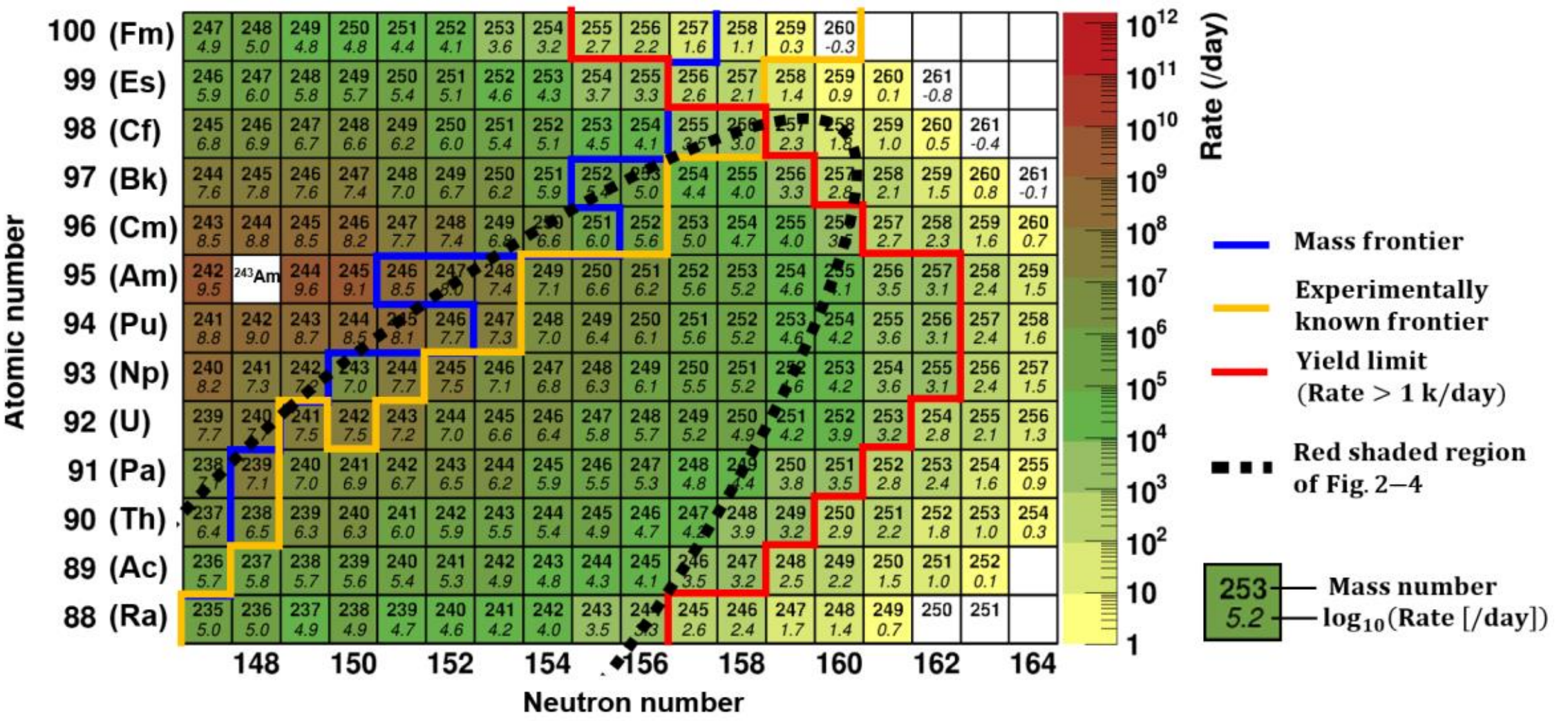

Figure 3-10. Production yields of target-like fragments produced by MNT reactions in the $^{238}$U (7.2$A$ MeV, 0.1 pμA) + $^{243}$Am (1.3 mg/cm$^2$) reaction system. Expected yields were calculated by GRAZING [25] assuming a 90% reduction from fission. Blue lines indicate the frontier of nuclides with experimentally known masses. Orange lines indicate the boundary of experimentally discovered nuclides. Dotted line indicates the red shaded area in Fig. 2-4. Red line indicates the boundary for yields larger than $10^3$ per day.

If we were to use the $^{248}$Cm target with the $^{238}$U beam, further heavy neutron-rich actinoid nuclides could be accessed. Figure 3-10 shows isotopic distribution of TLFs in this reaction system based on a beam intensity of 1 pμA, and assuming a 90% reduction of yields due to fission. The blue lines indicate the frontier of nuclides whose masses are experimentally known based on AME2020 [26]. The region enclosed by the dashed line corresponds to the red shaded area in Fig. 2-4. The red line indicates the boundary for the yield more than $10^3$ per day, which reaches nuclides with $A$ = 261. The orange lines indicate the boundary of experimentally discovered nuclides. The $^{248}$Cm target makes it possible to access more than 40 new nuclides at the yield more than $10^3$ per day beyond those available from the $^{238}$U and $^{243}$Am targets.

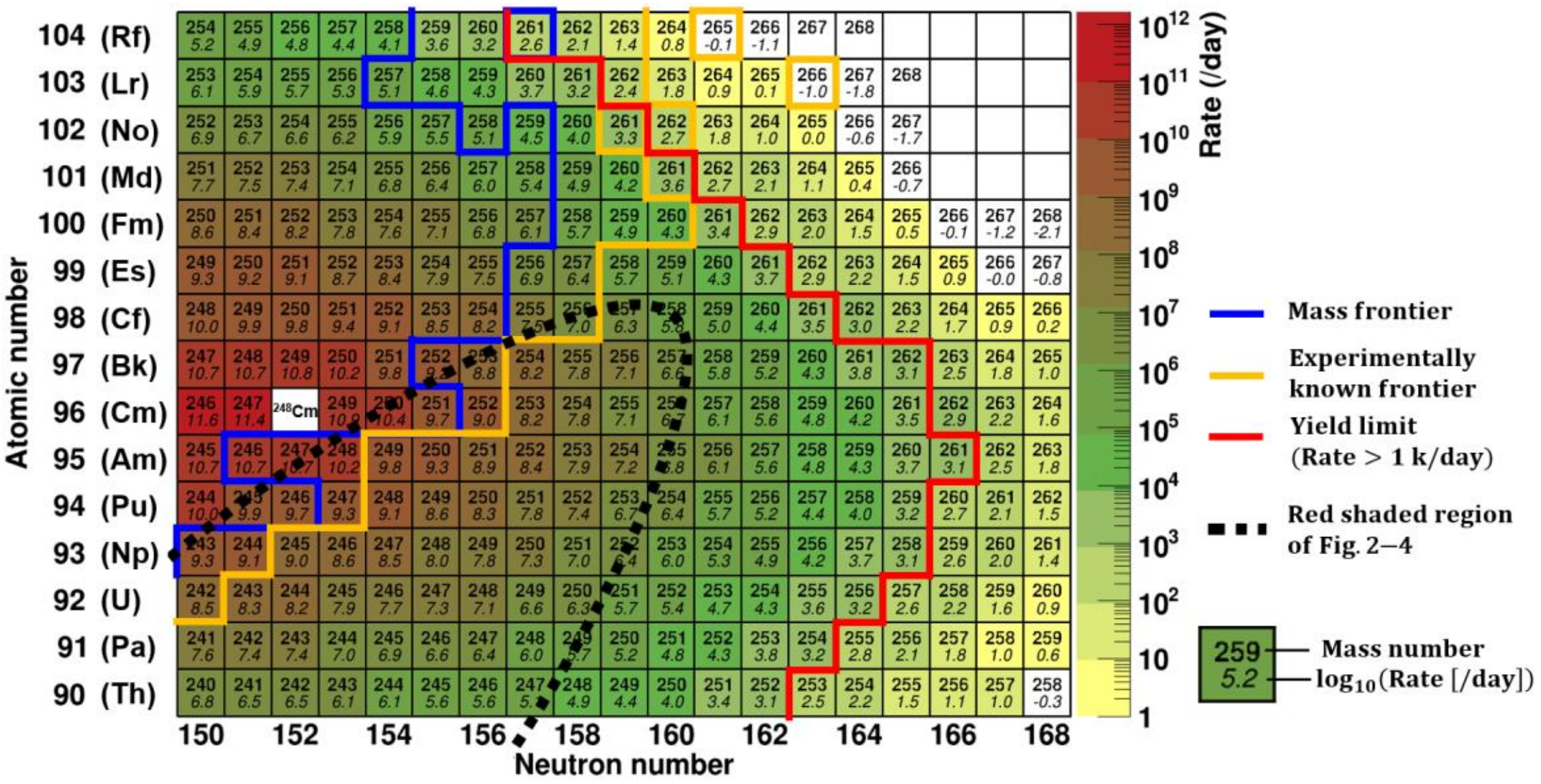

Figure 3-11. Production yields of target-like fragments produced by MNT reactions in the $^{238}$U (9.2$A$ MeV, 1 pμA) + $^{248}$Cm (13 mg/cm$^2$) rection system. Expected yields were calculated by GRAZING [23] assuming a 90% reduction from fission. Blue lines indicate the frontier of nuclides with experimentally known masses. Orange lines indicate the boundary of experimentally discovered nuclides. Dotted line indicates the red shaded area in Fig. 2-4. Red line indicates the boundary for yields larger than $10^3$ per day.

## 3-3. $^{238}$U + $^{198}$Pt reaction system; production of neutron-rich nuclei around $N$ = 126

The present KISS pursues the β-γ spectroscopy, the laser spectroscopy and the mass and lifetime measurements of the neutron-rich nuclei around $N$ = 126 for investigation of the astrophysical origin of gold and platinum elements, using the nuclear production by the MNT reactions in the $^{136}$Xe + $^{198}$Pt reaction system. Because of the limits of the beam intensities and the extraction efficiencies of the present KISS, it is difficult to directly reach the r-process path. Figure 3-12 shows isotopic distribution of TLFs produced by the MNT reactions between the $^{238}$U (9.0*A* MeV, 1 pμA) + $^{198}$Pt (13 mg/cm$^2$) calculated by the GRAZING model [25]. The blue lines indicate the frontier of nuclides, whose masses are experimentally known from AME 2020 [26]. The red line indicates the boundary for the yield more than $10^3$ per day. The orange lines indicate the boundary of experimentally discovered nuclides. The use of the intense $^{238}$U beam with the $^{198}$Pt target and the effective collection of reaction products at KISS-II make the playground of the research expand to more neutron-rich region around the neutron closed shell $N$ = 126 on the nuclear chart.

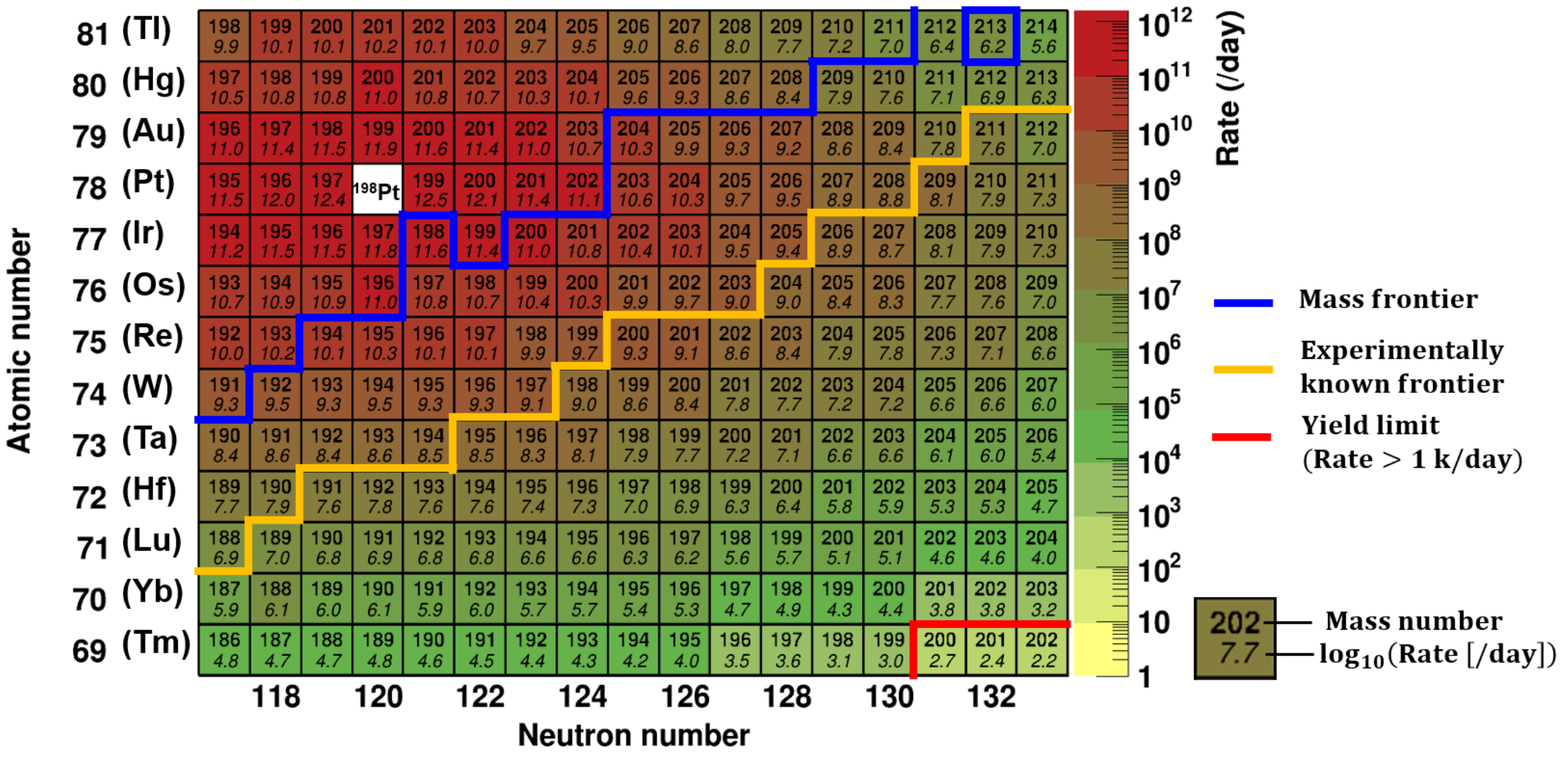

Figure 3-12. Production yields of target-like fragments produced by MNT reactions in the $^{238}$U (9.0*A* MeV, 1pμA) + $^{198}$Pt (13 mg/cm$^2$). Expected yields were calculated by using GRAZING [25]. Blue lines indicate the frontier of nuclides with experimentally known masses. Orange lines indicate the boundary of experimentally discovered nuclides. Red lines indicate the boundary for yields larger than $10^3$ per day.

# 4. Configuration of the KISS-II facility

The KISS facility [1-3] was developed by the KEK Wako Nuclear Science Center (WNSC) and has been used to perform nuclear spectroscopy experiments such as decay spectroscopy, laser ionization spectroscopy, and mass measurements [4-11] of nuclides near $N = 126$ to understand the astrophysical environment for the formation of the third peak in the observed solar abundance pattern. However, limited yields of more neutron-rich nuclei have restricted these studies from reaching $N = 126$ nuclides. In order to extend the experimental range and promote new activities not only in the vicinity of $N = 126$ but also into the actinoid region where the r-process would terminate, much higher yields are essential.

To overcome the present limitation, we propose to construct the KISS-II facility which will comprise four key devices: a superconducting solenoid filter, a helium gas cell, a variable mass-range separator, and a multi-reflection time of flight mass spectrograph (MRTOF-MS). It is envisioned that the KISS-II facility will achieve experimental efficacy 10 000 times higher than the original KISS facility.

In this chapter we will introduce each device, providing details of the design, operation, and purpose of the various components. Due to its unique complexity, however, a detailed description of the solenoid filter will be reserved for Chapter 6

## 4-1. Efficacy gain over the original KISS facility

The original KISS facility relies on neutralization in an argon-filled gas cell, followed by selective re-ionization of neutral atoms that were transported by gas flow to a laser interaction region. The combined efficiency of these processes has been found to be $\varepsilon \sim 0.1\%$. A medium-resolution ($A/\Delta A \sim 900$) magnetic separator has been used to limit the mass range of ions transported to the experimental devices to a single isobar set. This had the benefit of providing strong selectivity to allow decay studies of a specific nuclide, while disposing of any other isotopes of that nuclide which might have been re-ionized.

In the time since the original KISS facility was designed, there have been numerous technical advances which would improve experimental efficacy. These new technologies have been leveraged so that every part of the KISS-II facility has been designed to enhance the experimental efficacy compared to the original KISS facility.

### 4-1-1. Superconducting solenoid filter – efficacy gain: 100-fold

The maximum permissible primary beam intensity has been limited to 10 pnA by the present KISS argon gas-cell installed. To increase the available primary beam intensity by more than two orders of magnitudes, to more than 1 pμA (up to 2 pμA), the KISS-II facility will rely on a gas-filled magnetic solenoid filter. The gas-filled solenoid filter will efficiently transport the ions produced in MNT reactions – by suppressing the transport of unwanted elastic particles – along trajectories largely determined by $A/Z^{1/3}$ due to the charge-equilibrium condition resulting from collisions with helium atoms. The details of the helium gas selection for the gas-filled solenoid filter are discussed in Sec. 6-1-4-1.

A water-cooled Faraday cup will be placed inside the bore of the solenoid to stop the intense primary beam. A set of stopper plates will similarly be installed to suppress elastically scattered particles. The magnetic field will guide the MNT reaction products around these barriers and efficiently transport them to a next-generation, multi-segmented helium gas cell. The details of the target and filter – Faraday cup and stopper plates – will be presented in Sec. 4-3, while the details of the 5.5 T solenoid magnet will be given in Chapter 6.

### 4-1-2. Next-generation, multi-segmented helium gas cell – efficacy gain: 10-fold

The reaction products transported through the solenoid filter will be implanted in a helium gas cell. The energetic ions will enter the gas cell in a highly-charged state. Through collisions with helium atoms the ions will lose energy and eventually stop in the gas cell. A combination of static and dynamic electric fields can then be used to transport the ions out of the gas cell and into vacuum. During this process, the ions will quickly discharge through interactions with the helium, eventually reaching a singly, doubly, or triply charged state; due to the high ionization potential of helium, the ions should never neutralize.

In the stopping process, each incoming ion produces a large number of helium ions. In the proposed system, reaction products (mainly elastic particles) will enter the gas cell with a rate on the order of $10^8$ particle-per-second (pps), creating an intense non-neutral plasma. Such a plasma will strongly shield ions from the applied electric fields necessary to transport the ions into vacuum.

Since dynamic RF electric fields better penetrate the plasma produced in the gas cell, we have long used them to overcome these space charge effects. In this extreme case, however, even the RF field penetration may have a rather limited penetration depth. To overcome this, we propose to construct a gas cell using 20 equally spaced RF carpets constructed in a substrate-free configuration. In this way, despite the plasma-induced electric field attenuation, the ions will generally be sufficiently near the electrodes to experience the effects of the applied electric fields and can thereby be efficiently transported through and extracted from the gas cell.

In Sec. 4-4 we provide a detailed discussion of the gas cell's design and operation. Furthermore, simulations are presented which support an estimated gas cell efficiency of $\varepsilon \sim 1\%$, a 10-fold gain over the present KISS gas cell.

### 4-1-3. Variable mass-range separator

As previously mentioned, the original KISS facility utilizes a magnetic separator with a fixed resolving power of $A/\Delta A \sim 900$ to ensure all nuclei delivered to the experimental devices belong to a single set of isobars. However, by making use of advanced detector technologies it would be possible to analyze several nuclides – across multiple isobar chains – which would greatly enhance the experimental efficacy of the KISS-II facility. To that end, the facility is envisioned to make use of a variable mass-range separator for selecting single or multiple mass numbers. The expected mass-resolution and mass-dispersion are $A/\Delta A \sim 1000$ and 20 mm/%, respectively. The proposed mass-separator will be able to transport nuclides with 5 different mass-numbers. By changing a horizontal slit position, the nuclide with the desired single mass-number will be able to be selected and transported to detectors for precision spectroscopy such as β/γ-decay spectroscopy, β-delayed fission or β-delayed neutron multiplicity measurements, which could be difficult to perform with wide-spectrum cocktail beams even using the most advanced detector schemes. In other experiments – such as laser ionization spectroscopy and mass spectrometry measurements – the separator will be operated in a wide mass-number mode (by opening the slit fully) to allow multiple isobaric chains to be simultaneously delivered to the experimental stations. The variable mass-range separator will be discussed in Sec. 4-6.

### 4-1-4. Multi-reflection time-of-flight mass spectrograph – efficacy gain: 10-fold

The MRTOF-MS will be an essential device for both mass measurements and particle identification at the KISS-II facility. It consists of an advanced suite of low-pressure buffer gas filled RF ion traps, a pair of electrostatic ion mirrors separated by a field-free drift region, and an ion-impact detector. The ion traps allow concomitant measurement [12, 13] of analyte and reference ions with virtually no duty cycle loss. After cooling in the ion traps, ions are ejected and captured between the two ion mirrors. They reflect between the ion mirrors for a predetermined duration before being released to travel to the ion-impact detector. The time between ejection from the RF ion trap and detection of ions striking the ion-impact detector defines the time-of-flight (TOF) from which an ion's mass-to-charge ratio can be inferred.

While the achievable mass resolving power continues to increase, the present typical mass resolving power (MRP) of m/δm~1 000 000 would, even for the heaviest nuclides ($A > 250$), yield the astrophysically mandated mass precision $\delta$ m = 100 keV/$c^2$ with a mere six detected ions. Furthermore, the MRTOF-MS can be operated with a wide mass bandwidth to simultaneously analyze several nuclides across five or more isobaric chains [13]. Using the variable mass-range separator to facilitate this capability will yield a 10-fold gain in experimental efficacy.

In recent years we have found success in embedding nuclear decay detectors inside the ion-impact detector to facilitate correlated TOF-decay spectroscopy. At present we have demonstrated success with α-decay [14,15], and are presently beginning tests with β-decays. It is envisioned to further extend the technique to γ rays for the KISS-II facility. This technique, which is described in more detail in Sec. 4-7, will allow simultaneous decay spectroscopy of cocktail beams comprising up to 10 or more nuclides by correlating the decays with prior ion-implantations.

# 4-2. Overview of the proposed KISS-II facility layout

Figure 4-1 shows a schematic overview of the present KISS and proposed KISS-II facilities spanning the E2 and E3 experimental rooms in RIKEN RIBF facility. An intense primary beam of $^{238}$U with a fixed energy of 10.75$A$ MeV can be delivered to the E2 room from the RIKEN Ring Cyclotron (RRC). In the E2 room, the KISS facility's production target station and argon gas cell based laser ion source is located on the E2B beam line; it has been under development and in operation since 2011. The low-energy beams of MNT products – produced in reaction systems such as $^{136}$Xe primary beam impinging upon a $^{198}$Pt target – are transported to the neighboring E3 experimental room where decay and laser spectroscopy and mass measurements are performed. We plan to install the solenoid filter and gas cell for the KISS-II facility on the E2A beam line in the E2 room. They will not interfere with the KISS facility on the E2B beam line, as demonstrated in Fig. 4-1.

Figure 4-2 shows a more detailed overview of the proposed KISS-II facility. The superconducting solenoid is envisioned to have a field strength of 5.5 T, with a 1.4 m coil diameter bore and 1 m length. The solenoid will be filled with helium gas at a typical pressure of 1 kPa for efficiently transporting the reaction products to a helium gas cell with suppressing the elastic particles of primary beam and the target. For the use of intense (up to 2 pμA) primary beam, the primary beam line will have a differential pumping system.

A target station will be located inside the solenoid. The target station will consist of two synchronized rotatable wheels – one for energy degraders and another for the targets. The design of the target station has been carefully considered to accommodate the maximum possible primary beam intensity while also minimizing the transfer of elastically scattered particles.

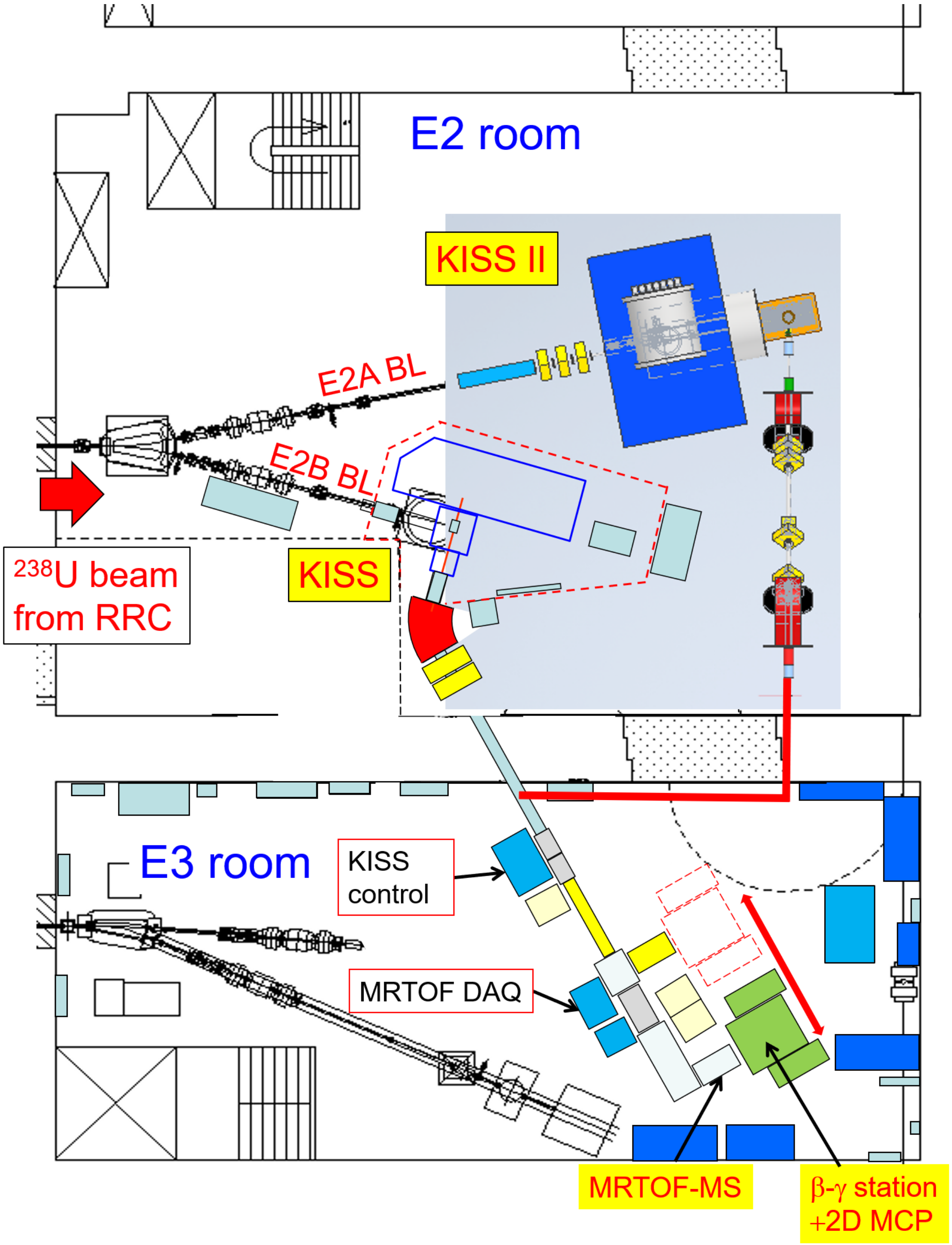

Figure 4-1. Schematic top overview of the KISS and KISS-II facilities installed in the E2 and E3 experimental rooms. The red dotted line around E2B beam line (BL) indicates a high-voltage platform of the KISS laser-ion-source based on the argon gas cell. The solenoid filter's large yoke is indicated by the blue area on the E2A BL. Most of the detectors and experimental devices are installed in the neighboring E3 room.

Target-like MNT reaction products emitted from the target will pass through the bore of the solenoid and be collected in a next-generation, helium-filled gas cell. A series of carefully designed adjustable barriers inside the solenoid's bore will strongly suppress both elastically scattered and unreacted primary beams to reduce the deleterious effects they would have on the operation of the gas cell.

The low-energy ions extracted from the gas cell will be transported by an RF multipole ion guide and accumulated in a buffer gas filled linear Paul trap to produce a pulsed ion beam. The pulsed ion-beam with the electrovalency of $q$ can either be accelerated to $2q$ keV for transfer to an MRTOF-MS located in the E2 room, or it can pass through an electrostatic deflector and be accelerated to $50q$ keV for transfer to the E3 experimental hall. The acceleration will be performed by use of pulsed drift tubes. These techniques are well-established within the KEK WNSC group.

The $50q$ keV beam will pass through the variable mass-range separator before continuing downstream. The beam will pass through a planned collinear laser ionization spectroscopy line before merging with the original KISS facility beam line and travelling to an MRTOF-MS or decay station for mass spectroscopy, decay spectroscopy (measurements of half-lives of nuclear states, β-delayed γ-rays, β-delayed neutrons, fission probabilities, fission branching ratios, and fission barrier heights ($B_f$) etc.), or decay-correlated mass spectroscopy. The collinear laser ionization spectroscopy beam line will include an optional neutralizer for laser ionization spectroscopy.

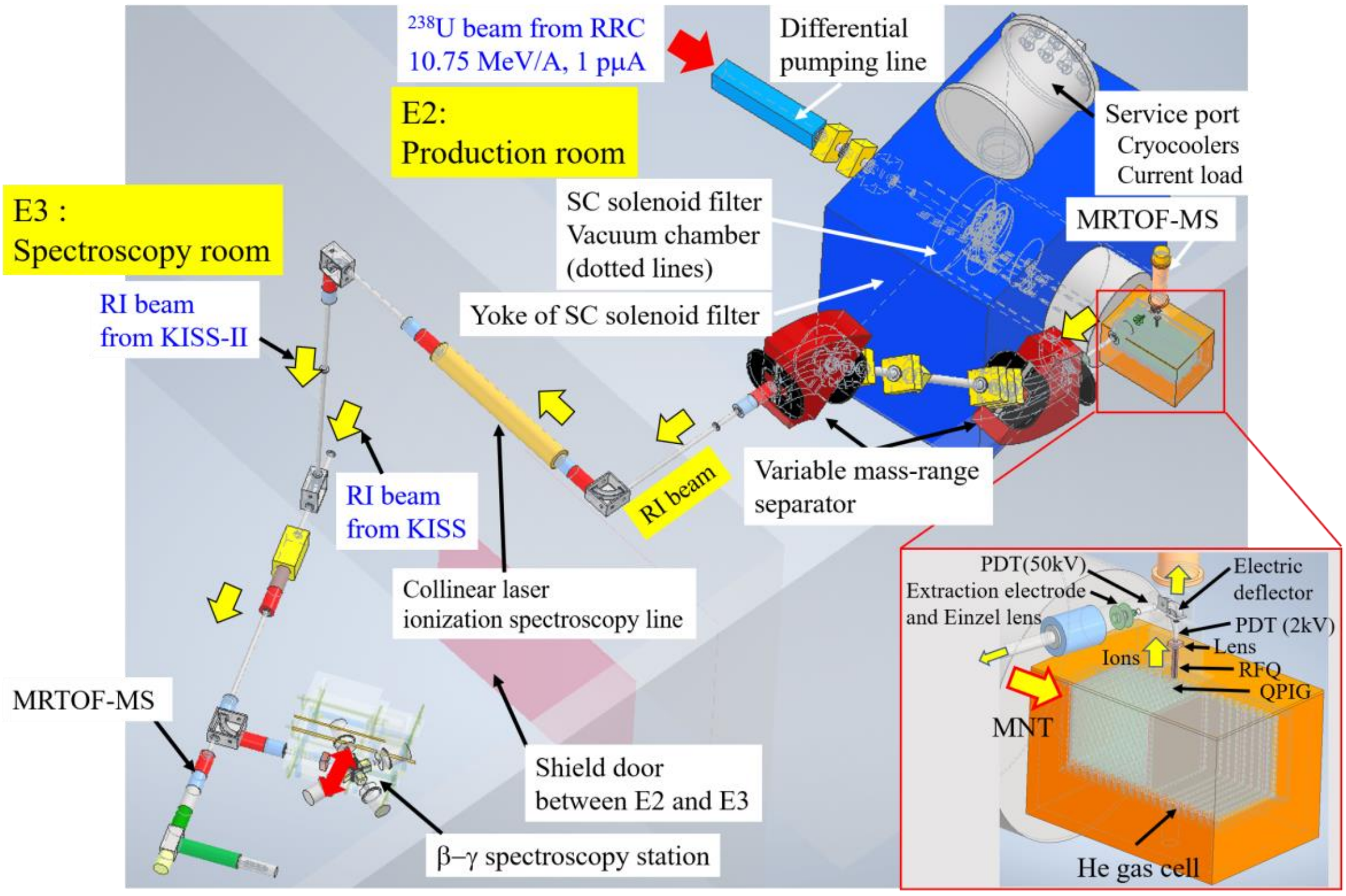

Figure 4-2. Schematic drawing of the KISS-II facility. The MNT production target station, solenoid filter, and helium gas cell are installed in the E2 room. Most detectors and experimental are installed at the neighboring experimental room E3. The laser devices (not shown) are installed in the J3 room underneath E2.

## 4-3. Production target and solenoid filter components

Heavy unstable nuclei will be produced by the MNT reactions induced when $^{238}$U beams impinge upon targets of e.g., $^{238}$U or $^{198}$Pt. Production rates have been reported in Sec. 3-1.

Figure 4-3 shows the calculated energy and angular distributions for $^{250}$U and $^{200}$W, which are characteristic of MNT products in the heavy actinoid region and near $N$ = 126, respectively. In these simulations, the primary beam energy and the target thickness were optimized to maximize the production rate. Under such conditions the MNT products will be emitted from the production target with large emission angles, near the grazing angle – i.e., 30° – 70° . Thanks to this characteristic angular distribution the solenoid can be used to transport the MNT products around a central beam stopper and refocus them into the gas cell

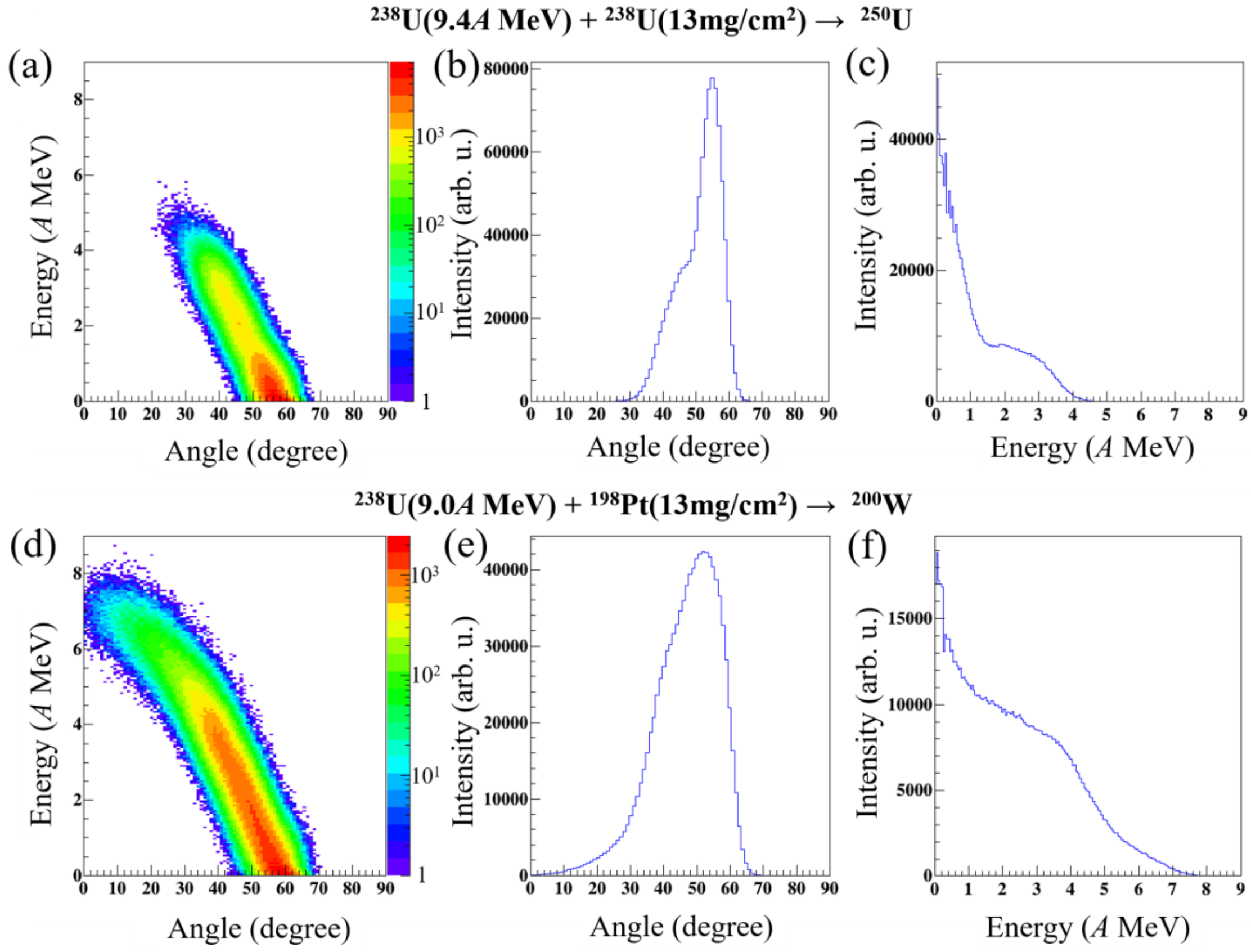

Figure 4-3. (a) and (d) Angle-energy correlation for $^{250}$U and $^{200}$W, respectively. (b) and (e) Laboratory frame angular distributions for $^{250}$U and $^{200}$W, respectively. (c) and (f) Energy distributions for $^{250}$U and $^{200}$W, respectively. All data calculated using the GRAZING code.

while the primary beam is fully rejected by the beam stopper. In this way the space charge within the gas stopper can be maintained at a tolerable level while efficiently collecting the MNT products.

Figure 4-4 shows simulated trajectories for $^{250}$U (top, blue) and $^{200}$W (bottom, orange) passing through the solenoid filter as determined using the GEANT4 toolkit [16], accounting for the kinematics of $^{250}$U and $^{200}$W shown in Fig. 4-3, the geometry of the solenoid filter, and charge-exchange interactions [17] with the 1 kPa helium gas in the solenoid filter. The design of the rotating production target station, the beam stoppers for both projectile and elastically scattered beams – including the Faraday cup (FC) – and the helium gas cell were guided by these simulations.

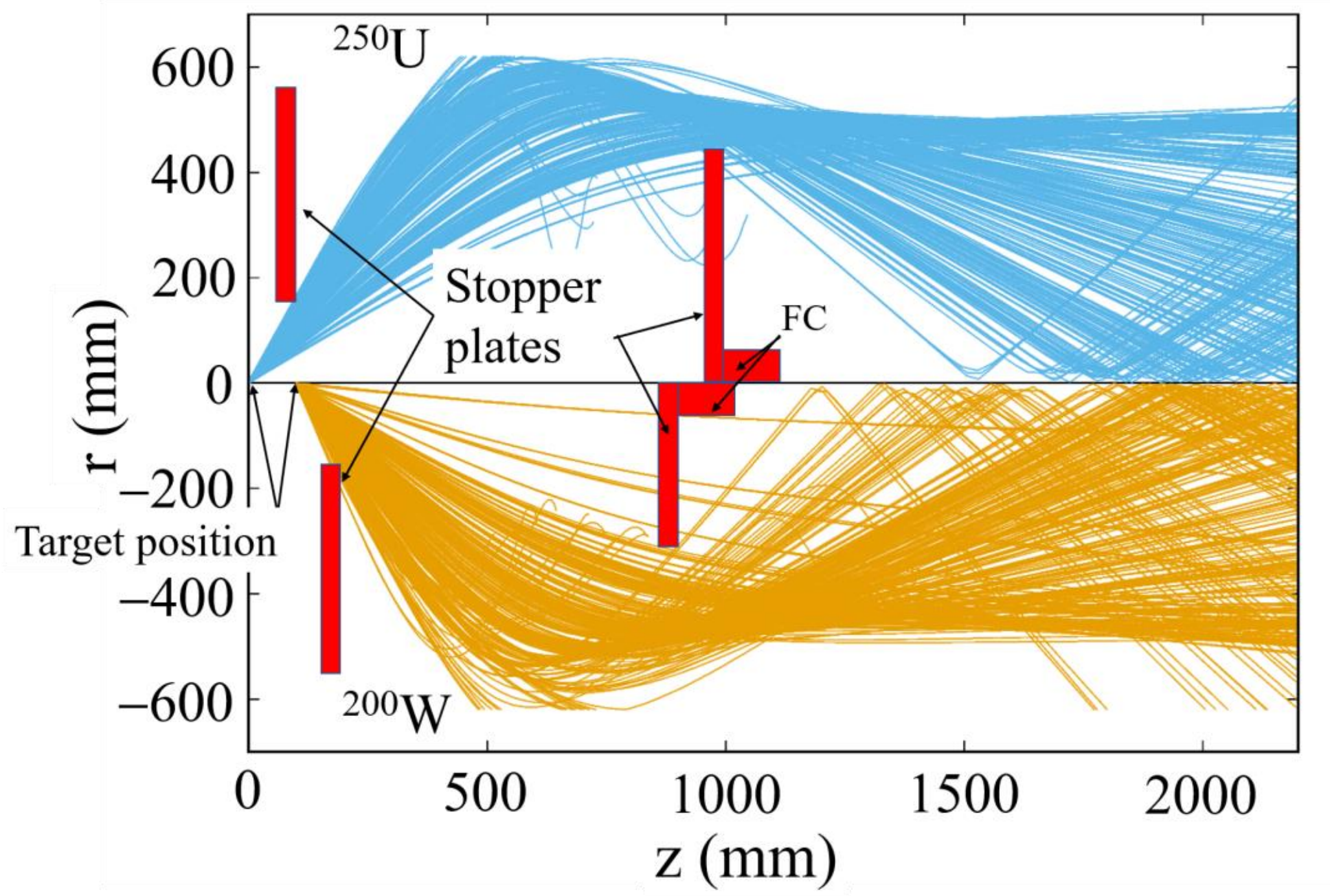

Figure 4-4. Simulated trajectories in the solenoid filter for $^{250}$U (top, blue) and $^{200}$W (bottom, orange). According to the reaction system, the target and stopper plate positions have been optimized to maximize the available intensities of each unstable nuclide. The stopper plates and Faraday cup (FC) were not included in the simulations.

### 4-3-1. Design of the production target station

The KISS-II facility will initially produce neutron-rich nuclei using $^{238}$U and $^{198}$Pt targets with thicknesses of 13 mg/cm$^2$. The optimal energies for production of neutron-rich nuclides by MNT reactions with a $^{238}$U primary beam are 9.4*A* MeV and 9.0*A* MeV for the $^{238}$U and $^{198}$Pt targets, respectively. Since the energy of the beam delivered by the RRC is fixed at 10.75*A* MeV, energy degraders will be needed to achieve the desired energy-on-target. To this end multi-layered titanium foils will be used, with each ~3 μm thick layer degrading the projectile energy by ~0.5*A* MeV.

At the 1 pμA intensity for the primary beam intensity, due to the primary beam's energy loss the titanium foil degraders will need to sustain a power dissipation of 320 W ~ 420 W, while the targets will experience ~700 W of heating power. To avoid thermal damage under such high heating power requires rotating degrader and target wheels. Based on the experience of the RIKEN SHE group, to avoid runaway heating will require 500 mm diameter wheels rotating at 1000 rpm. It is assumed that the 1 kPa helium pressure in the target region will help with cooling.

To reduce the power density the beam will also be defocused to a σ = 3 mm Gaussian profile covering most of the target. In order to fine tune the focus, it is foreseen to include insertable aluminum oxide beam viewers at the front of the degrader and target wheels.

Metallic targets will be used in the initial operational phase of KISS-II. The maximum primary beam intensity will be constrained by the melting points of uranium (1405 K), platinum (2042 K), and titanium (1939 K). The helium gas in the solenoid filter should assist in the cooling of the targets and degrader as the wheel rotates, allowing initial use of up to 1 puA intensity for the $^{238}$U primary beam. Later, after development of appropriate chemical targets – e.g., uranium carbide (melting point: 2620 K), uranium oxide (melting point: 3140 K), etc. – the primary beam intensity could be increased beyond 1 puA.

The wheels will accommodate 32 segments, each 40 mm long and 20 mm high. Elastically scattered beam is a large concern for efficient operation of the helium gas cell. To that end the degrader and target wheels will be constructed as shown in Fig. 4-5 (a), with a tag at the end of each spoke to trigger a photo-sensor that will allow the primary beam to be rapidly turned off and on to avoid both heating the wheel support structure and producing extraneous scattered particles. A detailed discussion of the structure and its operations can be found in Refs. [2, 18].

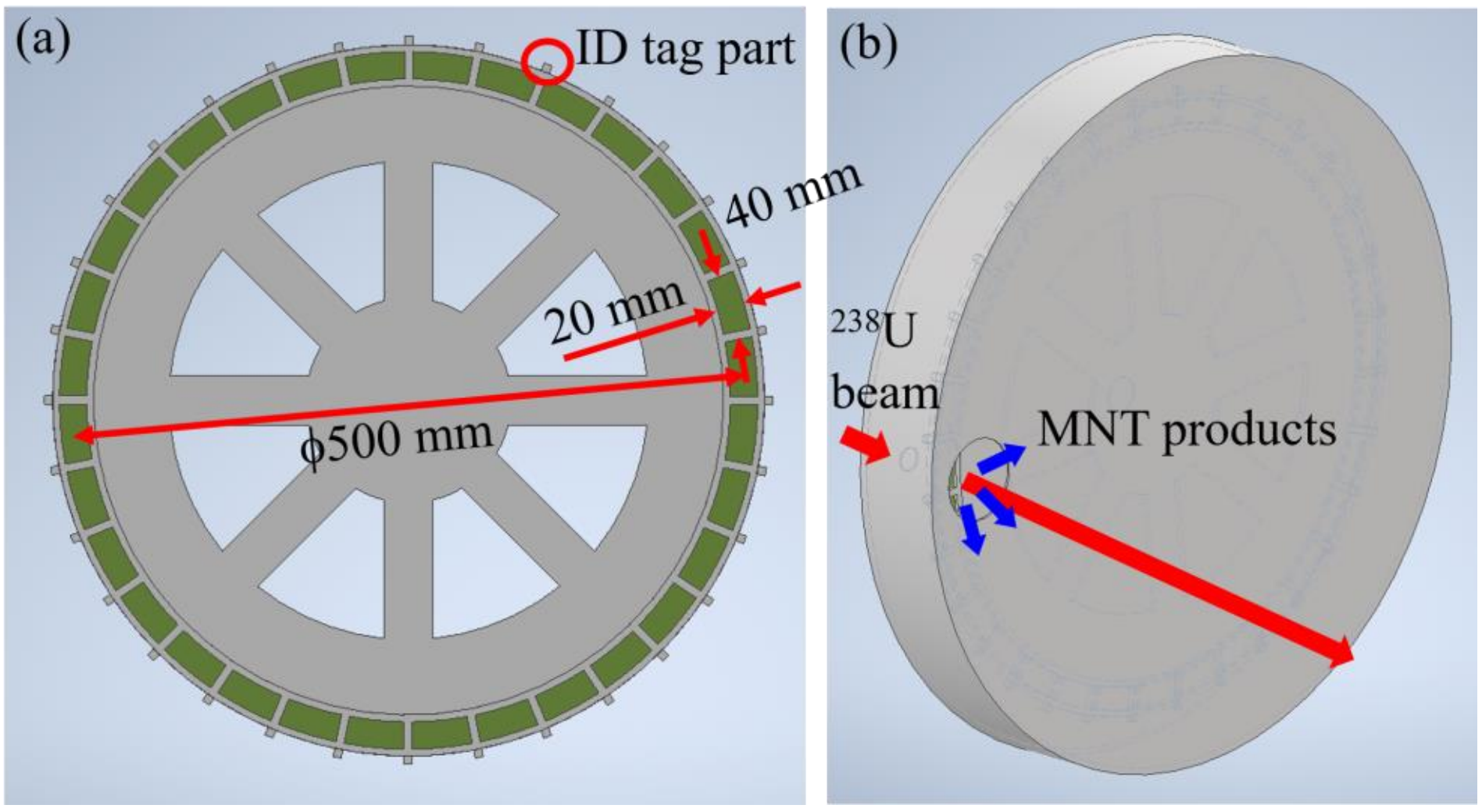

Figure 4-5. (a) Schematic design of the rotating wheel to be used for target and energy degraders. (b) Target enclosure showing the large aperture for the transmission of MNT products from the target.

A collimator will be installed between the energy degrader wheel and the target wheel, as shown in Fig. 4-6. This will also further reduce the extraneous elastically scattered particles from reaching the gas cell.

To avoid excessive contamination in the case of one or more target segments experiencing a catastrophic thermal failure, the target wheel is enclosed as shown in Figs. 4-5(b) and 4-6. The upstream side of the enclosure offers a small entrance aperture for the collimated primary beam to pass through, while the downstream side (see Fig. 4-5(b)) offers a larger aperture for the MNT products and unreacted primary beam to pass through. If necessary, it is foreseen to be able to cool the enclosure via either a cooling loop or Peltier element.

To simplify maintenance, the motor used to rotate the degrader and target wheels should be installed on the air-side of the system rather than in the vacuum chamber. This will be accomplished by coupling the rotation shaft to the motor via a magnetic fluid seal. The motor and magnetic fluid seal cannot be operated in an external magnetic field exceeding 10 mT. Thus, the shaft must be rather long, as indicated in Fig. 4-6, to provide sufficient shielding from the solenoid magnet's field.

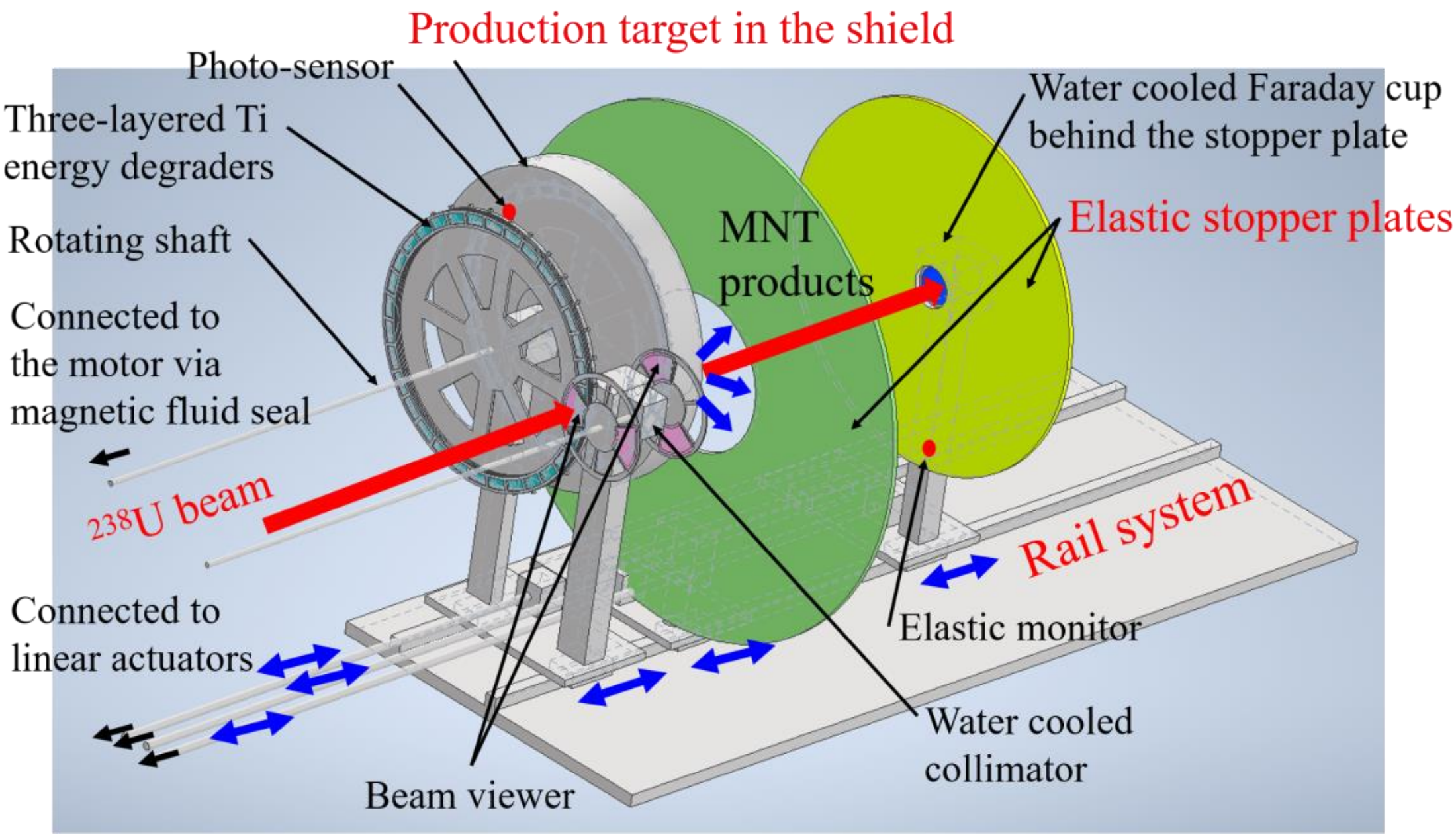

Figure 4-6. Schematic view of the proposed target station and solenoid beam stopper inserts. The three components are mounted on a rail system; a set of three linear actuators allows remote control of the relative positions of these three components

As implied by Fig. 4-4, the ideal location of target within the solenoid is dependent on the choice of target and the neutron-rich nuclide to be studied. To facilitate placing the target at the ideal position, the target station will be mounted on a rail system as shown in Fig. 4-6, allowing the target position to easily be moved along the beam axis.

## 4-3-2. Design of beam stopper plates

The solenoid filter will make use of a pair of plates to suppress the elastically scattered beam, one of which will be mounted on a Faraday cup beam stopper that collects the unreacted primary beam. For optimal performance, as indicated in Fig. 4-4, the location of each of these filter components needs to be adjusted based on the choice of target and the neutron-rich nuclide to be studied. As shown in Fig. 4-6, this is accomplished by mounting them on the same rail system as the target station; all three devices on the rail have an independent linear actuator to remotely adjust their positions.

A 10.75*A* MeV beam of $^{238}$U has a power of 2.6 kW/pμA, most of which must be dissipated by the Faraday cup beam stopper. As such, the device will require water cooling. The precise mechanical design for the Faraday cup and its cooling loop will be determined based on the technical support of the RIKEN and KEK accelerator groups. It will be desirable to minimize the transverse cross section of the Faraday cup support structure to limit the fraction of beam it might block.

As shown in Fig. 4-6, a compact detector ("elastic monitor") will be mounted on the second beam stopper plate. This will provide the capacity to monitor the health of the targets. The relative energy and intensity of the elastically scattered particles will vary with the target thickness. Should one target fail it would be quickly identifiable by a sudden shift upward in the energy of the elastically scattered particles. Ideally, however, imminent target failure could be predicted well-ahead of a catastrophic failure by a slow and steady upward shift in the elastically scattered particle energy, indicating erosion of the target material.

The optimal dimensions of the two plates for suppressing the elastically scattered beam have been determined via the previously mentioned GEANT4 simulations. The simulations indicate that the first plate should be 1000 mm in diameter with a 300 mm aperture (which is Iris type for further fine adjustment), while the second plate should be 800 mm in diameter. With such dimensions, GEANT simulations such as those shown in Fig. 4-4 indicate that roughly 66% of target-like fragments can pass through the solenoid filter. Prior to each experiment, the ideal position of each component can be determined by GEANT4 simulation and precise optimum can quickly be determined online through adjustment of the positions via the linear actuators.

### 4-3-3. Differential pumping scheme of the primary beam line

Differential pumping scheme is essential for connecting high-vacuum primary beam line to the gas-filled region in the solenoid filter. To use the intense primary beam of $^{238}$U (typical beam spot ~ 2mm in FWHM) for the production of unstable nuclei without any window materials on the beam line, a differential pumping line is installed. We need to reduce the helium gas pressure from 1 kPa to less than $1.5 \times 10^{-4}$ Pa for transporting the primary beam to the solenoid filter. Figure 4-7 shows a schematic drawing of the primary beam line with a differential pumping. The differential pumping line (about 2 m length) will consist of five sections, and then a triplet magnetic quadrupole (MQ) will be installed to make a defocused primary beam spot at the target position. Each section will be connected by an orifice with 15 mm in diameter and 50 mm in length to limit the conductance. Each section will be pumped by a turbo-molecular pump (TMP), screw pump, or dry booster pump as shown in Fig. 4-7 to fulfill the required pressure. The estimated pressures in each section are also shown in Fig. 4-7. A primary beam profile monitor and another FC are installed at the upstream of the TMQ to check the primary beam transport through the orifices. By applying the configuration, we will be able to achieve less than $1.5 \times 10^{-4}$ Pa for the beam transport.

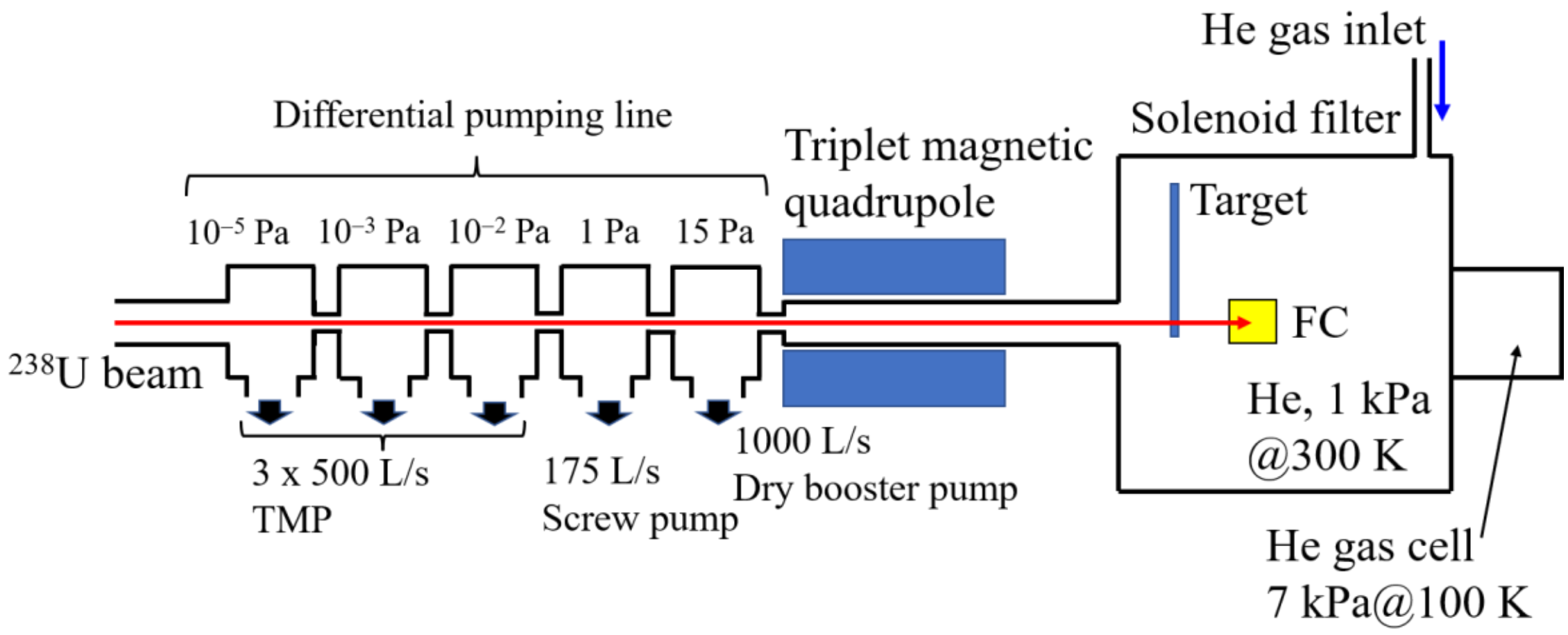

Figure 4-7. Schematic drawing of the primary beam line with a differential pumping system.

## 4-4. Multi-segmented gas cell

After passing through the magnetic solenoid filter, the ions will still have kinetic energies up to 7*A* MeV as noted in Fig. 4-3 earlier. The performance of high-precision spectroscopy, however, requires ions which can be easily manipulated, stopped, or stored – preferably with low axial and transverse emittance. Gas stopping cells have proven to be the most effective way to accomplish this transformation in beam properties.

As described in Sec. 4-1-2, the basic principle of operation for a helium gas cell is that high-energy ions will undergo collisions with helium atoms in a high-pressure environment, loosing energy through ionizing collisions which create $He^{+}$-$e^{-}$ pairs. Stopping an ion of $A$ = 250 which enters the gas cell with an energy of 1$A$ MeV will produce $10^7$ such pairs. The large asymmetry in mobilities results in the electrons quickly being removed by the application of the electric field that is required to guide the stopped ion out of the gas cell, while the $He^{+}$ produce a non-neutral plasma.

Within the KEK WNSC group there is a long history of using radiofrequency (RF) fields to penetrate the plasma and guide the stopped ions. Until this point, however, we have generally operated with heavy ion implantation rates limited to be around $10^4$ pps [19]. Under the space-charge condition induced at such rates, the ions could be guided along the length of the gas cell by a combination of RF and low-frequency (or even static) electric fields. In these helium gas cells built within the WNSC and SLOWRI groups to date, a so-called RF carpet has been used to confine ions above a plane. The carpets have been composed of many wires, parallel or concentric, on a printed circuit board. Adjacent wires have RF signals (typically 50 $V_{pp}$ ~100 $V_{pp}$ at 6 MHz ~ 10 MHz) applied which are 180° out of phase, which produces a pseudo-potential barrier to the ions. Superimposed on the RF signals have either been an electrostatic gradient or a set of low-frequency (10 kHz ~ 100 kHz) signals. When the low-frequency signal applied to adjacent wires is 90° out of phase it produces a travelling wave that pulls the ions with it. Most importantly, when considering the plasma-induced shielding, is the need for a super-imposed electrostatic "push field" to press ions onto the RF carpet. If the plasma shields ions from the push field or the low-frequency travelling wave field, the RF carpet will not be able to effectively transport the ions.

As such, the intense beam anticipated at the KISS-II facility will necessitate a next-generation gas cell design. To mitigate the effects of the plasma in shielding ions from external electric fields, two major departures will be made from our traditional designs. First, we plan to construct a multi-segmented, RF wire gas cell with a large stopping volume (height/width/depth = 500/500/1000 mm) as indicated in Fig. 4-8. Each RF wire carpet unit will be separated from its neighbor by 50 mm and operate like an independent gas cell. The tight spacing between these wire carpets will greatly reduce the plasma-induced electric field attenuation. The active stopping length of the device will be 1 m.

Secondly, rather than superimposing a low-frequency travelling wave field on top of the RF pseudo-potential barrier, we will use an RF travelling wave (RFTW) technique, whereby the high-voltage RF signals applied to a wire will be 90 degrees out of phase those applied to neighboring wires. This technique has been demonstrated to provide an extraction efficiency of $\varepsilon_{HeGC}$ = 10% using an intense ($10^7$ pps) beam of high-energy (~150$A$ MeV) of $^{47}K$ ($T_{1/2}$ = 17.5 s) [20]. As the ion beam energy will be lower at the KISS-II facility – thereby resulting in fewer $He^+$ for each stopped ion – and the RF carpet pseudo-potential barrier weakens for lighter ions, we confidently assume we will be able to achieve $\varepsilon_{HeGC}$ = 10% for extraction of ions stopping within any given RF wire carpet unit.

As indicated in Fig. 4-8, the gas cell will have two rooms. The RF wire carpet units will occupy the first room. The ions from each of the units will be transferred to the second room. The second room via PCB-based RFTW carpets (1st RF carpet) will contain a long PCB-based RFTW carpet (2nd RF carpet) to collate the ions from the 20 RF wire carpet units and finally extract them to vacuum as a single, continuous, low-energy ion beam.

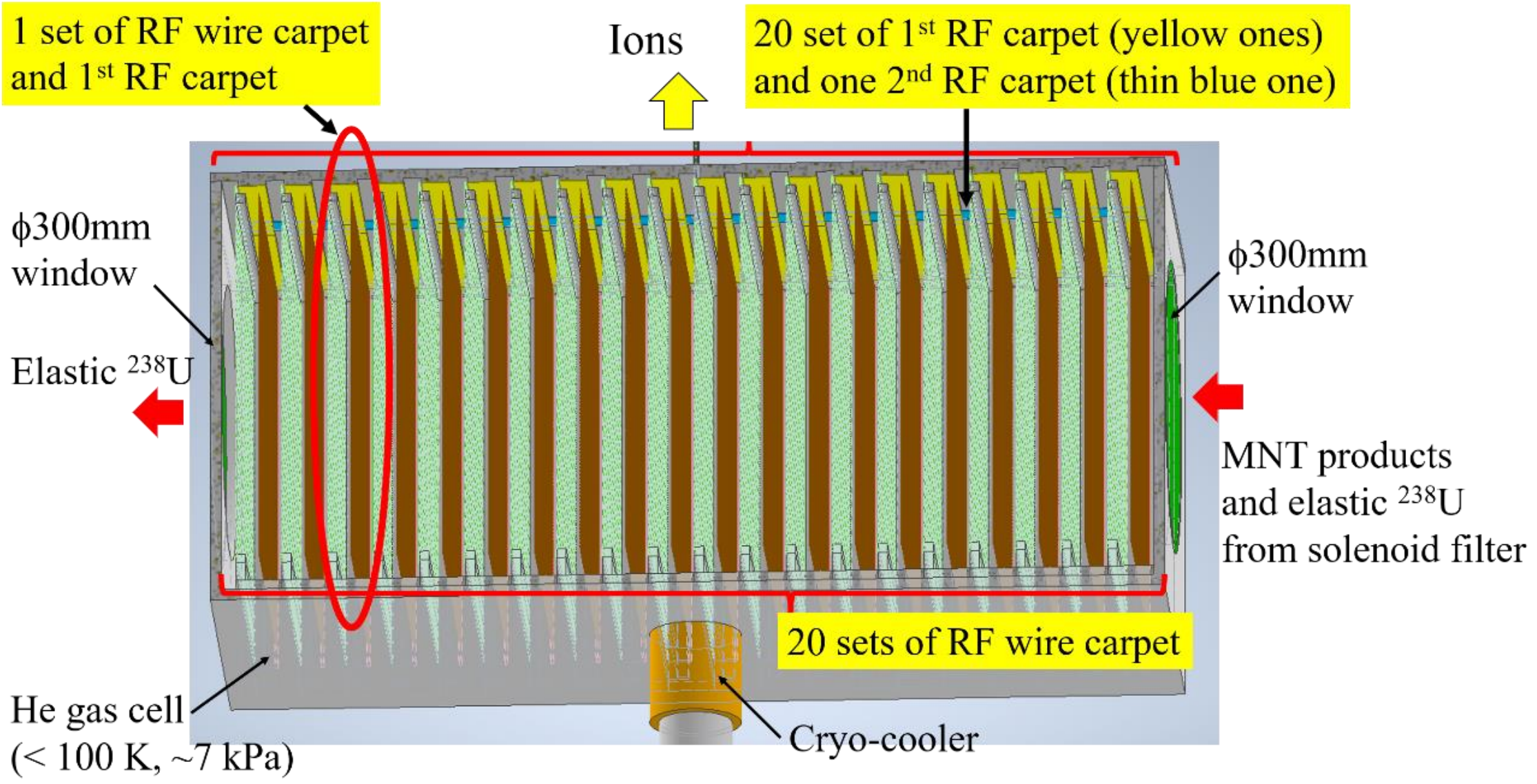

Figure 4-8. Schematic overview of the proposed multi-segmented, RF wire carpet gas cell. MNT products will mostly come to rest within one of 20 RF wire carpet units, while the more energetic elastic particles will tend to pass through without stopping. The ions extracted from each of the RF wire carpet units will be collated by an RF carpet in a second room and extracted as a single, continuous, low-energy ion beam.

The gas cell chamber will be mounted on a cryocooler to allow operation at $T < 100$ K, thereby enhancing the cleanliness and greatly reducing the risk of the MNT product ions being lost to neutralization or chemical reactions with contaminant such as $O_2$ or $C_nH_mX$ [21]. This will further allow the gas cell to be operated at 7 kPa while achieving the same gas density as 20 kPa at room temperature, greatly reducing the time between helium bottle changes.

The gas cell will feature thin windows – 10 μm polyimide or 6 μm Havar – supported by honeycomb grids [22]. By using thin windows at both the entrance and exit, portions of the beam which are not stopped in the gas can be handled by an external beam stopper, rather than degrading the system performance by ablating surface contaminants from an in-gas-cell beam stopper. Additionally, by placing e.g., a plastic scintillation detector after the exit window, we can monitor the stopping condition and make fine adjustments by varying the gas pressure. The windows are presently foreseen to be 300 mm in diameter, limited by the available sizes of polyimide or Havar foils. This will limit the acceptance to roughly 50% of the beam passing through the solenoid filter.

### 4-4-1. RF wire carpet unit

A schematic overview of an RF wire carpet unit is provided in Fig. 4-9. Each unit will consist of three active components: a RF wire carpet (brown), a pair of wire grids (green) to generate the electrostatic push field, and the 1$^{st}$ RF carpet. The RF wire carpet will be made using 50 μm-diameter beryllium-copper (BeCu) wire with 600 μm center-to-center pitch. The wire grids will use thicker wire with wider pitch; the wires will run perpendicular to the RF carpet wires. Both will have an area of 500 mm × 500 mm.

A 2$^{nd}$ RF carpet will be mounted at one end of each assembly. It will consist of a set of concentric circular segments as shown in Fig. 4-9 (c), printed on a polyimide or Teflon substrate and will have an active area of 30 mm x 600 mm.

Each of the RF carpets will use RF signals of 2 MHz ~ 4 MHz with amplitudes up to 100 $V_{pp}$. The signals will be applied in a travelling wave configuration – the signal applied to adjacent wires will differ in phase by 90$^{o}$. The RF wire carpet will have a negative DC bias relative to the push field grids; the 1$^{st}$ RF carpet will similarly have a negative DC bias relative to the RF wire carpet. Ions which stop between the two push field grids will be held between the RF wire carpet and push field grid by a balance of the DC gradient and the RF pseudo-potential force. The RF travelling wave will then drag the ion toward the 1st RF carpet. The DC gradient between the two carpets will cause the 1$^{st}$ RF carpet's RF travelling wave to draw the ions toward a 0.65 mm central aperture.

A combination of gas flow and electric fields will draw the ions through the aperture and into a second chamber common to all the RF wire carpet units. The ions from each unit will collect on a long (30 mm × 1000 mm) 2$^{nd}$ RF carpet – of similar design to the one in each RF wire carpet unit – located at the rear of the second chamber. The traveling wave RF pseudo-potential will drag ions to the central aperture of this 2$^{nd}$ RF carpet, from which the ions will be transferred to a lower-pressure region. The mean extraction time is expected to be a few 100 ms based on the measured extraction time of a similar helium gas cell [21].

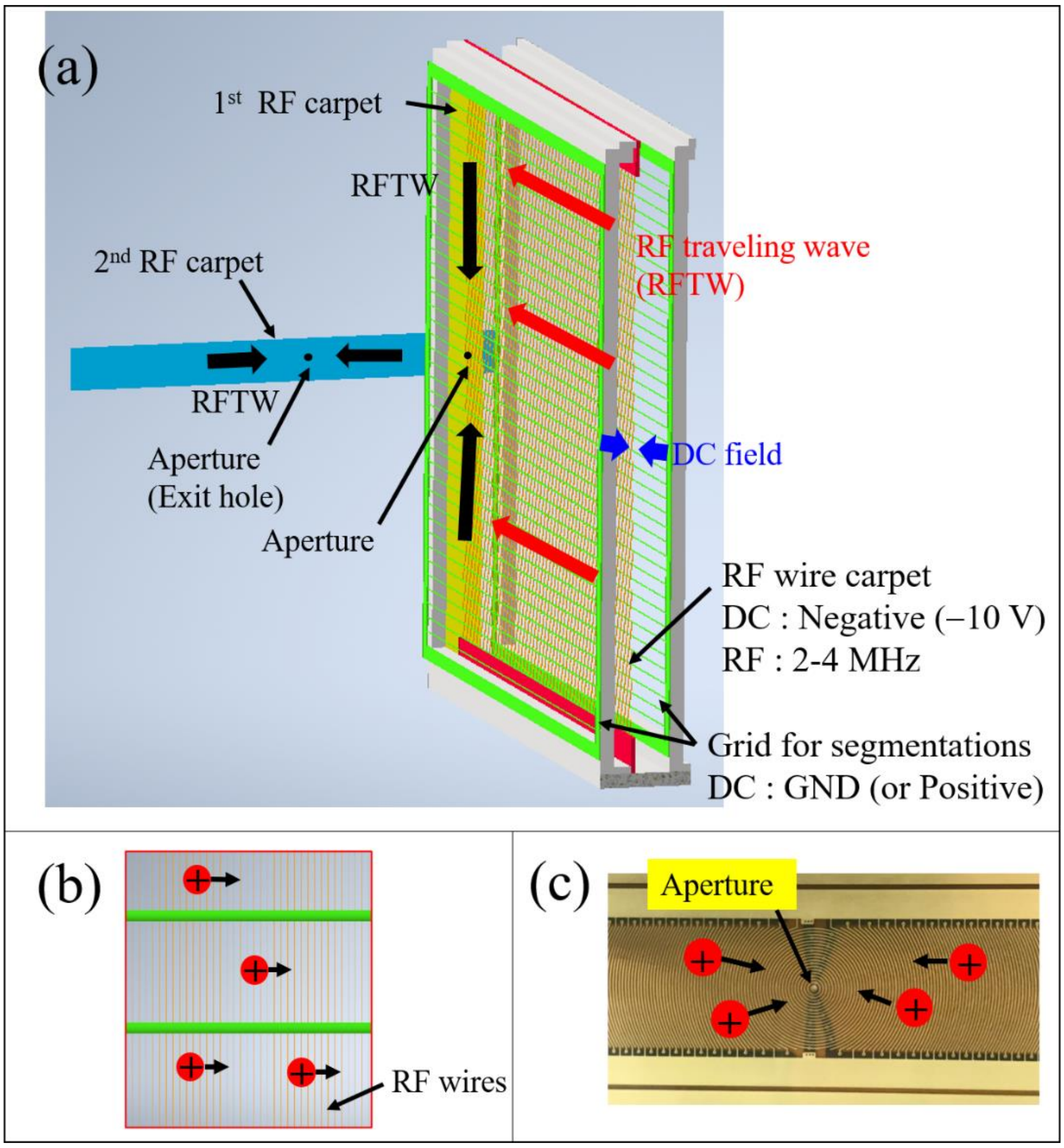

Figure 4-9. (a) Detailed schematic overview of the RF wire carpet unit assembly. (b) Detail of wire carpet (brown) and push field grid (green) showing the direction of travel for positive ions. (c) Detail of the 1st and 2nd RF carpets showing how ions will be drawn to a small aperture for transfer to the second chamber.

## 4-4-2. Results of simulated performance

The design of the gas cell was motivated by extensive simulations. The most salient task, in terms of the design, was the simulation of the ion stopping profile in the gas cell. These calculations were vital to determine the appropriate length of the gas cell system. A comparison of simulation results for target-like fragments (TLF) of $^{250}U$ and elastically-scattered $^{238}U$ can be seen in Fig. 4-10.

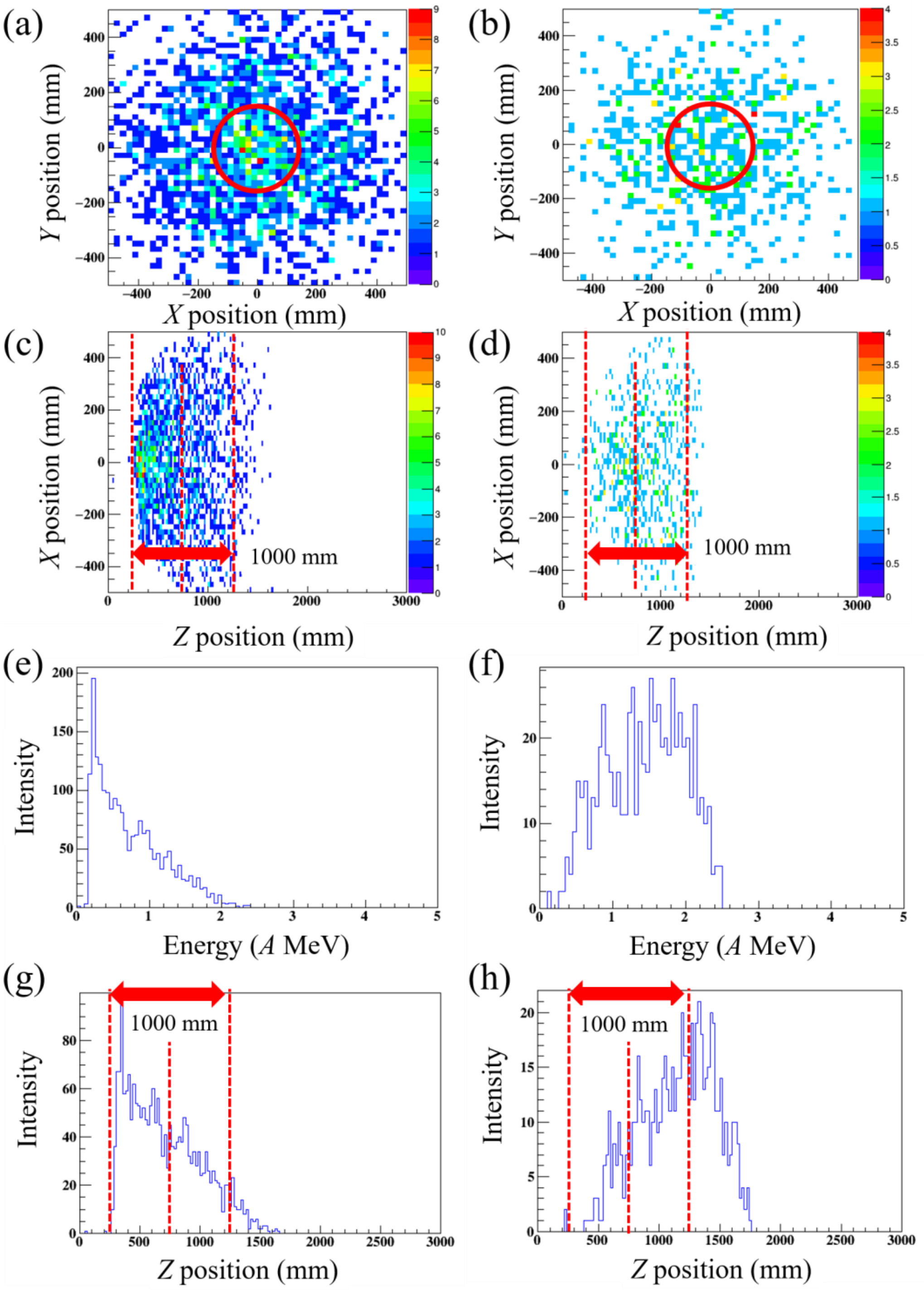

Figure 4-10. Results of simulated stopping distributions in helium gas (7kPa, 100 K) for characteristic target-like fragments (TLF) and elastically scattered particles from the target. The transverse stopping distributions for $^{250}$U (TLF) and $^{238}$U (elastic) are similarly shown in (a) and (b), while the axial stopping distributions are shown in (c) and (d). The red circles in (a) and (b) indicate the outline of the entrance window. The energy profiles for $^{250}$U (TLF) and $^{238}$U (elastic) before passing through a 10 μm polyimide window are shown in (e) and (f), respectively. The axial stopping profiles for $^{250}$U (TLF) and $^{238}$U (elastic) are shown in (g) and (h), respectively, indicating that the elastic particles stop deeper in the gas cell, on average, than the TLFs

The simulations make clear that the elastic particles come to rest deeper inside the gas cell than the TLFs. Of the ions reaching the end of the solenoid filter, about 50% of $^{250}$U TLFs and $^{238}$U elastic particles were able to penetrate through the 10 μm polyimide window with the diameter 300 mm. Of those ions which were able to penetrate the window, 42% of $^{250}$U TLFs came to rest within the 1 m length of the proposed gas cell, while about 10% of $^{238}$U elastic particles did.

Even accounting for the solenoid filter, it is expected that as many as $1.5 \times 10^8$ pps elastic particles will pass through the gas cell entrance window when using a $^{238}$U primary beam with 1 pμA intensity. As such, on the order of $10^6$ pps elastic particles could be extracted from the gas cell. As will be explained in Sec. 4-5, this could hamper the system efficiency by overfilling the ion traps used to convert the continuous beam extracted from the gas cell into a pulsed beam.

An additional benefit of the multi-segmented approach to the gas cell is that it can be used to reduce the elastic-to-TLF ratio. Based on the stopping profiles shown in Figs. 4-10 (g) and (h), further simulations were made with consideration of the 20 wire RF carpet units which will compose the gas cell. These simulations accounted for ion losses arising from the 90% transmission of ions passing through each unit due to collisions with the RF carpet or push field wires. The results can be seen in Fig. 4-11.

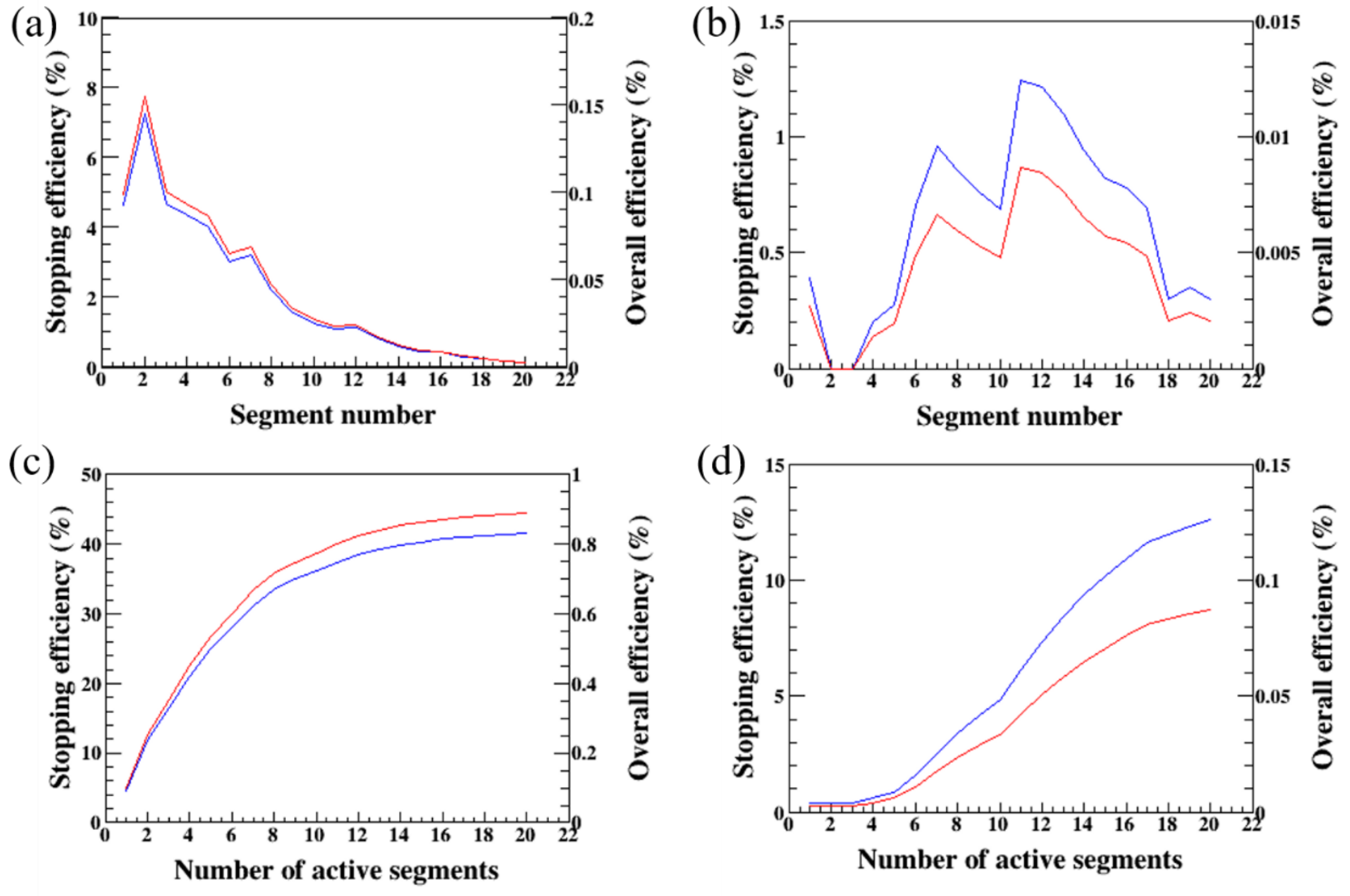

Figure 4-11. Results of simulated stopping and extraction accounting for the 20 wire RF carpet units. The fraction of the incoming $^{250}$U TLFs and elastic $^{238}$U particles which can be stopped and extracted from each of the units is shown in (a) and (b), respectively. Here, red and blue lines indicate stopping efficiency and overall efficiency (see Table 4-1 for the definition). The integrated fraction of incoming $^{250}$U TLFs and elastic $^{238}$U particles, respectively, which can be stopped and extracted as a function of the gas cell depth in terms of wire RF carpet units is shown in (c) and (d).

Based on this calculation, were the extracted elastic particles to overfill the ion traps and suppress the measurement efficiency of the desired neutron-rich nuclides, we could selectively turn off some wire RF carpet units. By turning off the last 10 units, the elastic particles would be suppressed by more than a factor of two, while the TLFs would only be reduced by ~10%. The efficiencies determined by this simulation are tabulated in Table 4-1.

Table 4-1. Calculated efficiency estimates for various types of particles emitted in the angle range of 30° -70° from the MNT target. The solenoid efficiency ($\varepsilon_{solenoid}$) is the fraction of particles emitted from the target which pass through to the end of the solenoid filter. The implantation efficiency ($\varepsilon_{implant}$) is the fraction of the ions that reached the end of the solenoid filter which then pass through the 300 mm-diameter gas cell window. The stopping efficiency ($\varepsilon_{stopping}$) is the fraction of ions successfully passing through the gas window which then come to rest within the 1 m length of the gas cell, excluding ions which impinge upon some part of the RF carpet structure. The transport efficiency ($\varepsilon_{HeGC}$) is the fraction of stopped ions which are successfully extracted from the gas cell. Numbers in parentheses indicate values when only the first 10 wire RF carpet units are used. Overall efficiency ($\varepsilon_{ovarall}$) is defined as a multiplication of all above efficiencies.

| Components | $\varepsilon_{solenoid}$ | $\varepsilon_{implant}$ | $\varepsilon_{stop}$ | $\varepsilon_{HeGC}$ | $\varepsilon_{overall} = \varepsilon_{solenoid} \cdot \varepsilon_{implant} \cdot \varepsilon_{stop} \cdot \varepsilon_{HeGC}$ |
|---|---|---|---|---|---|
| MNT($^{250}$U,TLF) | 0.66 | 0.32 | 0.42(0.37) | 0.10 | 0.009(0.008) |
| MNT($^{250}$U,PLF) | 0.97 | 0.02 | - | 0.10 | - |
| Elastic of projectile ($^{238}$U) | 0.90 | 0.08 | 0.13(0.06) | 0.10 | 0.0009(0.0004) |
| Elastic of target ($^{238}$U) | 0.33 | 0.21 | 0.11(0.04) | 0.10 | 0.0008(0.0005) |

# 4-5. Pulsed ion beam production

The ions are extracted from the gas cell as a very low-energy (~10$q$ eV) continuous ion beam. However, to transport such low-energy ions from the gas cell in the E2 room to the E3 experimental hall would be slow and inefficient. To utilize efficient ion optical transport, the ions should be accelerated to 50$q$ keV. This could be accomplished by floating the gas cell and solenoid filter on a high-voltage platform, but that proves to be dangerous, complex, and expensive. Rather, the KISS-II facility will leverage the deep well of experience with RF ion trapping techniques within the KEK WNSC and SLOWRI groups.

As shown in the inset image of Fig. 4-2, the ions extracted from the gas cell will initially be transported by a compact radiofrequency quadrupole ion guide (QPIG) and will then accumulate in a linear Paul trap (RFQ) for 1 ms ~ 10 ms. The RFQ will be filled with a dilute helium gas ($P_{\mathrm{He}}$ = 0.1 Pa ~ 1 Pa) to cool the ions to form a compact ion bunch. The ions will then be ejected from the RFQ as a brilliant ion pulse and transferred downstream.

As such an ion trap can typically hold ~$10^4$ ions before Coulomb interactions degrade their performance, if the ion trap is operated on a 100 Hz cycle it will be able to accommodate on the order of $10^6$ ions/s delivered from the gas cell. Reducing the cycle time to operate at 1 kHz will limit the ion cooling and slightly degrade performance, but allow the trap to accommodate $10^7$ ions/s. In the case of exceedingly high extraction rates producing low efficiency due to the space charge limit, the RFQ could be replaced to a two-gap buncher system [24] operated by sawtooth waves.

The brilliant ion pulse ejected from the RFQ will initially be accelerated to $2q$ keV, then either sent to an MRTOF-MS installed nearby the gas cell or pass through an electrostatic deflector and be further accelerated up to $50q$ keV. The accelerations will be performed by use of pulsed drift tubes. The first pulsed drift tube will be 100 mm long with a 20 mm inner diameter. The second pulsed drift tube will be 100 mm long with a 30 mm inner diameter.

When the ion pulse enters the drift tube it is typically biased to the same potential as the beam line elements before and after. When the ion pulse is in the drift tube – preferably well-centered, to limit fringe-field effects – the bias applied to the drift tube is quickly (typically $\Delta t << 1\mu s$) raised by a $\Delta V$. During the change in the bias, the trajectories of ions inside the tube are unaffected. However, when the ions reach the end of the tube, they accelerate by $\Delta V$. The technique provides a cost-effective and compact means to accelerate pulsed ion beams with low emittance and small beam waist.

After passing through the second pulsed drift tube, the ions will be transported to a variable mass-range separator via a series of standard ion optical elements.

## 4-6. Variable mass-range separator

As mentioned in Sec. 4-1-4, the KISS-II facility should be capable of providing beams of individual isobaric chains for cases of high-precision spectroscopy, including β-delayed neutron emission probabilities ($P_{\mathrm{n}}$) and fission branching ratios and barrier height ($B_f$). However, for most nuclear spectroscopy measurements, the experimental efficacy would be greatly enhanced by an ability to provide a wider mass bandwidth. To that end, a variable mass-range separator will be implemented.

Figure 4-12 shows a schematic overview of the proposed variable mass-range separator, which will have mass resolving powers of $A/\Delta A \sim 1000$. The desired number of mass numbers will be able to be selected by using a slit system located at mass-dispersive plane. The mass separator will consist of two 45° dipole magnets (DM) with $\rho$ = 1.3 m, a doublet electric quadrupoles (DEQ), single magnetic quadrupole (SMQ), two pairs of doublet magnetic quadrupoles (DMQ), and a slit system to select the number of mass numbers for the experiments, and position-sensitive 2D-MCP to tune the beam profile and position. The two dipole magnets, two DMQ and SMQ are almost symmetrically installed to realize a wide-band mass number transport.

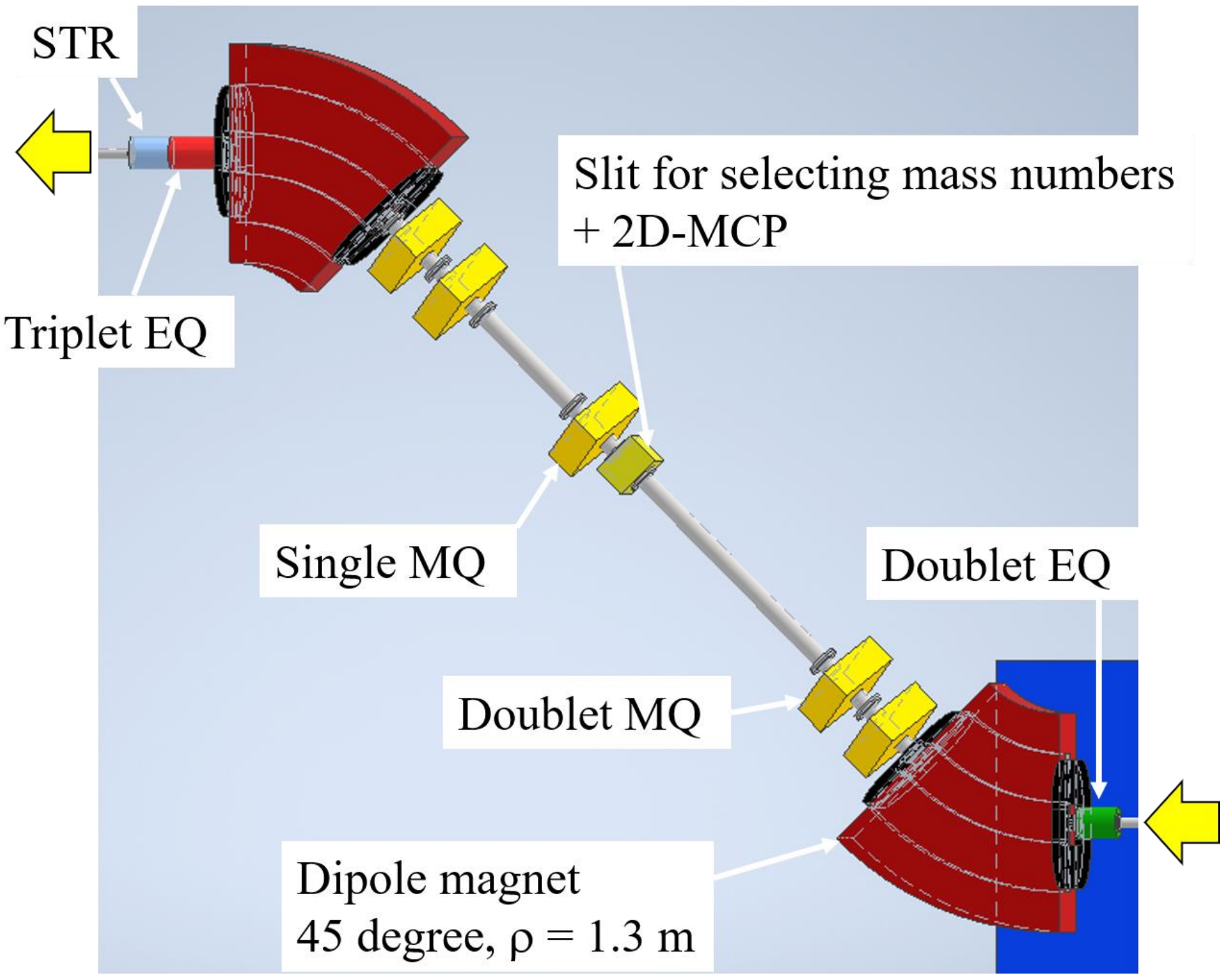

Figure 4-12. Schematic overview of the proposed variable mass-range separator.

Figure 4-13 shows an expected beam trajectories of (a) vertical (Y) and (b) horizontal (X) directions, (c) horizontal and (d) vertical profiles and (e) the corresponding contour plot of a beam with mass-number $A-2$ at the mass-dispersive plane, as an example, calculated by using a GIOS code [26] with the assumption of reasonable emittance of 10$\pi$ mm·mrad. The expected mass-dispersion will be 20 mm/%, and the mass-separator will be able to transport a cocktail beam with 5 different mass numbers by applying the horizontal slit width of ~ 50 mm. While by applying the slit width of ~ 6 mm, single mass number can be selected. If a comb-shaped slit were installed, the mass separator can transport nuclides with required mass-numbers to the downstream for nuclear spectroscopy.

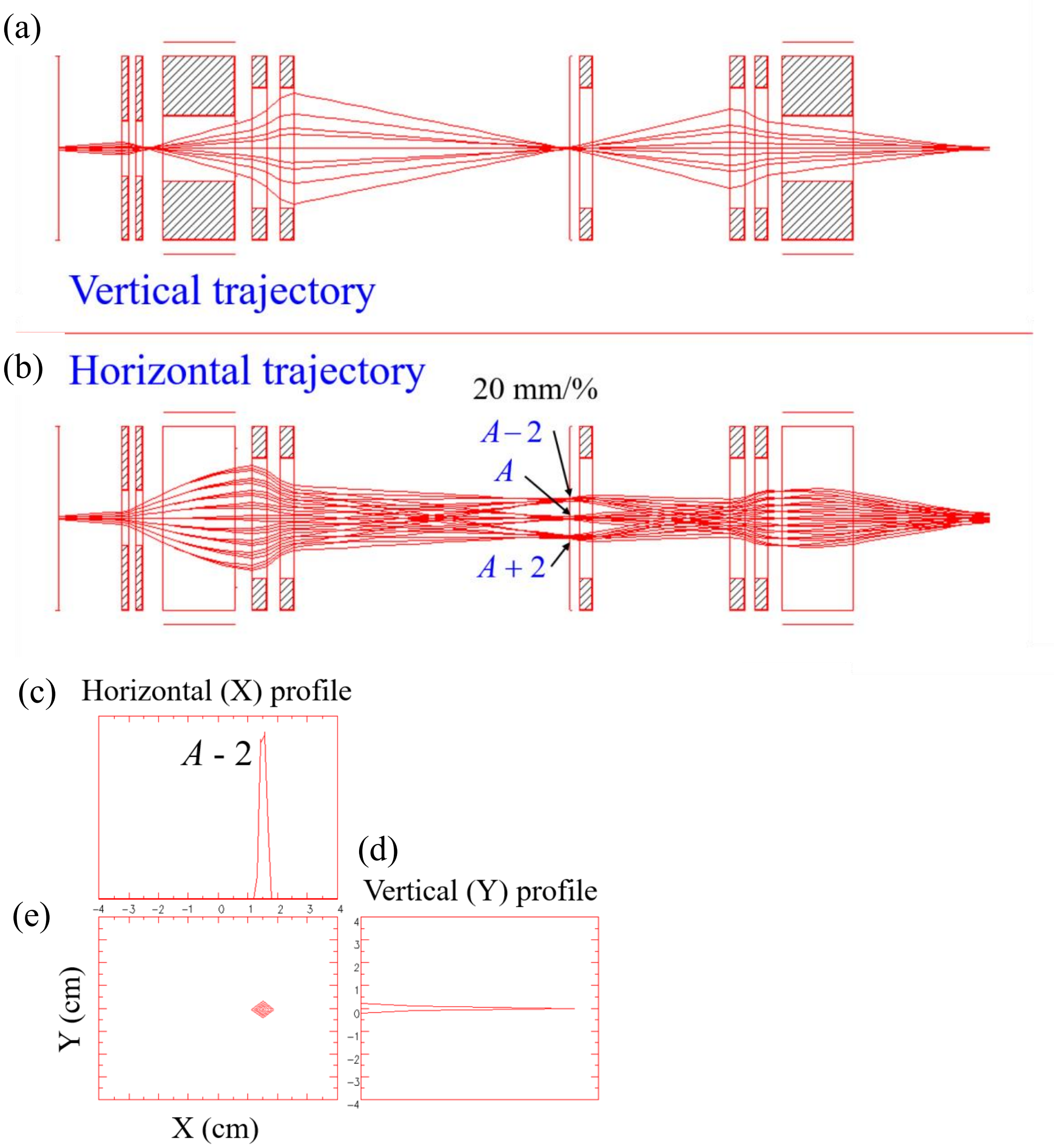

Figure 4-13. Simulation results of (a) the vertical and (b) horizontal beam trajectories, (c) horizontal and (d) vertical profiles and (e) the corresponding contour plot of a beam with mass-number $A-2$ at the mass-dispersive plane by using the GIOS code [26]. Here, we assumed the beam with $A = 250$ and the energy of 50 keV.

## 4-7. KISS-II beam line

Figure 4-14 shows the proposed layout for the beam line to transport ions from the variable mass-range separator to the various experiment devices for nuclear spectroscopic work in the E3 experimental hall. After the variable mass-range separator, the beam profile can be inspected at the exit of the collinear laser spectroscopy line (this device will be similar to CRIS [27,28] at ISOLDE) where a position-sensitive 2D-MCP to monitor the beam profile and position will be installed.

For experiments other than laser spectroscopy, the collinear laser ionization spectroscopy line can be operated as a high-efficiency ion optical element to transport the beam to downstream experimental stations located after the beam is merged with that of the original KISS facility. After the merger, a second beam monitor station will be available to confirm the beam position and profile.

Following the second beam monitor station the beam line branches and the beam can be deflected to the β-γ decay spectroscopy station or travel straight to the mass spectrometry station. The mass spectrometry station is preceded by a low-pressure (~0.2 kPa) helium-filled gas cell ion beam cooler that efficiently converts the $20q$ keV ~$50q$ keV ion beams from the KISS or KISS-II facilities into low-energy beam amenable to ion trap experiments. The line ends with the MRTOF-MS. For advanced decay-correlated mass spectroscopy experiments, the decay station can be moved, using a rail system, to allow the MRTOF-MS to operate in conjunction with it.

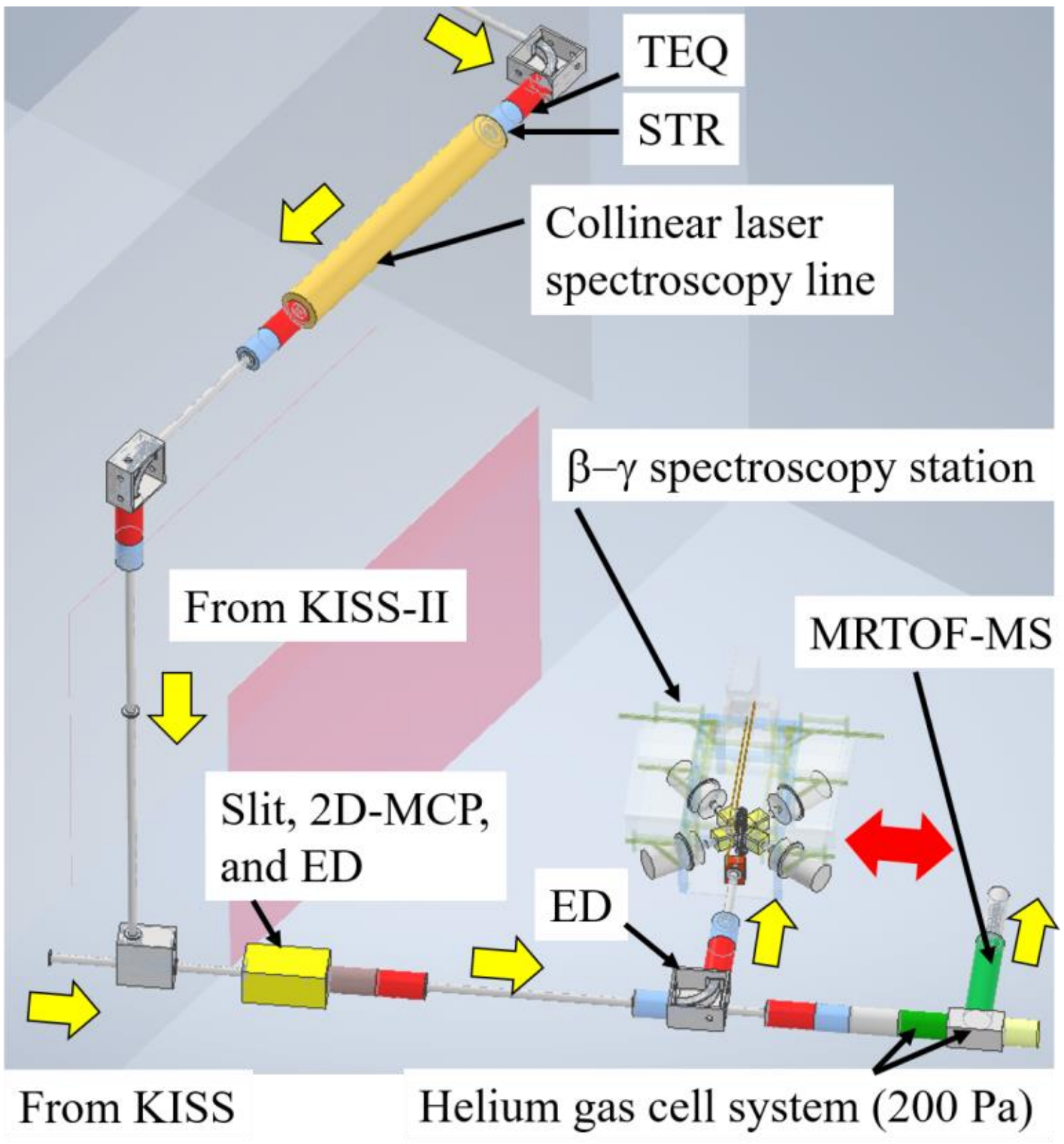

Figure 4-14. Schematic overview of the beam line after the variable mass-range separator and detailing the initially-planned experimental equipment. The decay station and mass spectroscopy beam line will be available to both KISS and KISS-II facilities.

## 4-8. MRTOF for advanced decay spectroscopy techniques

Figure 4-15 shows a schematic overview of the MRTOF-MS station [25]. Pulses of ions from the ion trap suite are injected into the MRTOF. This is accomplished by momentarily lowering the potential applied to the first electrode ("injection endcap"). The ions then reflect between a pair of ion mirrors. After a sufficient number of reflections, the potential applied to the final electrode ("ejection endcap") is lowered to allow ions to leave the reflection chamber. The ions will then impact a TOF detector to measure the total flight time between the ions' ejection from the ion trap until being implanted on the detector. Alternatively, the detector can be replaced with a pulsed drift tube to accelerate a purified ion ensemble for transport to other detector stations.

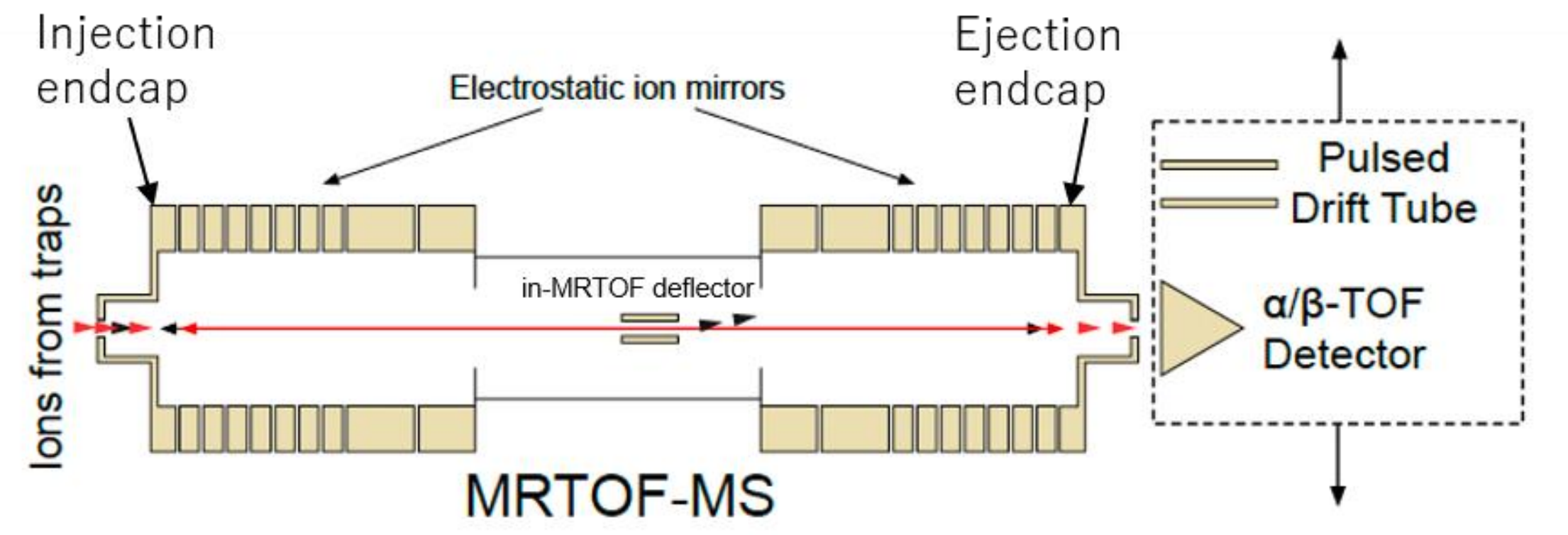

Figure 4-15. Schematic overview of the MRTOF-MS. Ions reflect between two ion mirrors. An in-MRTOF deflector can selectively remove ion species to produce a purified ion ensemble. After leaving the MRTOF-MS ions will impinge upon a TOF detector or be post-accelerated using a pulsed drift tube.

An in-MRTOF deflector in the field-free drift region between the two ion mirrors can be pulsed to remove ions with undesirable mass-to-charge ratios. Figure 4-16 demonstrates the benefit of this ion rejection capability for MRTOF-MS measurements. This technique will enable suppression of unwanted ions for performing precise nuclear spectroscopy via mass, decay, and laser spectroscopy at the KISS-II facility.

The MRTOF-MS can presently achieve a mass resolving power approaching $m/\delta m = 10^6$, as shown in Fig. 4-17. The achievable mass resolving power is progressively improving with advancements in electronic switches and ever more stable power supplies. With the present maximum achieved mass resolving power of $m/\delta m = 10^6$ the MRTOF could fully resolve isomeric states with excitation energies above 250 keV for $A$ = 250 nuclides. Figure 4-18 shows the TOF spectrum of an admixture of $^{134g}Sb$ and $^{134m}Sb$ ($E_x$ = 279 keV) in which the ground and isomeric states are clearly identified. Consequently, it would be possible to use the in-MRTOF deflector to provide isomerically pure beams in such cases. Moreover, with a mass resolving power of $m/\delta m = 10^6$ it would be possible to achieve a relative mass precision of $\delta m/m = 3\times10^{-7}$ from a mere 10 detected ions; with 100 detected ions we could achieve $\delta m/m = 10^{-7}$.

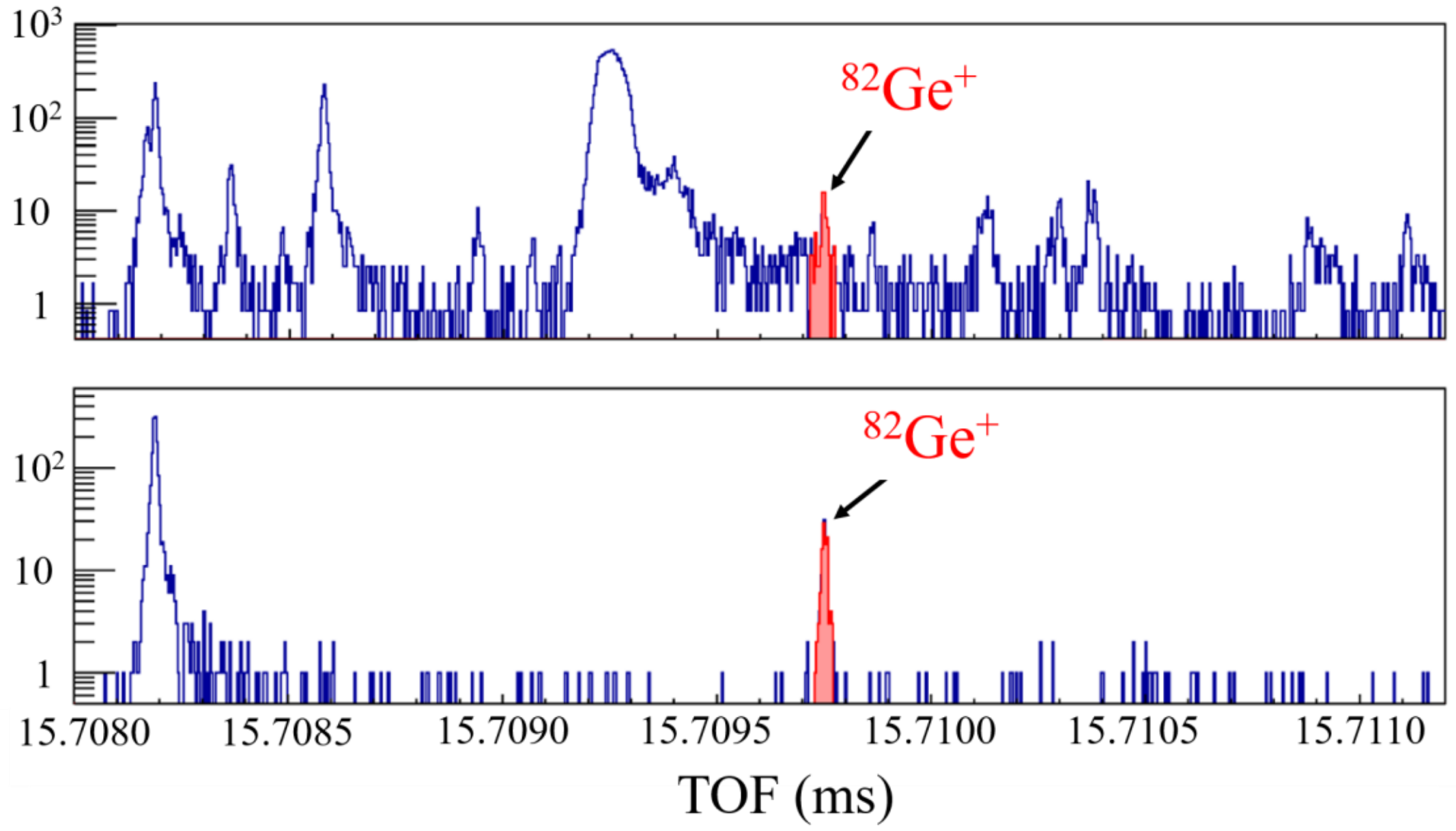

Figure 4-16. Demonstration of the utility of in-MRTOF deflector inside the MRTOF-MS. Some of the injection mirror electrodes were used as a deflector to remove unwanted ions in the test experiment. Upper and lower figures show the TOF spectra with and without the use of ion rejection mode. The red marked ions are $^{82}Ge^+$. After removing the unwanted ions, the $^{82}Ge$ is easily identified.

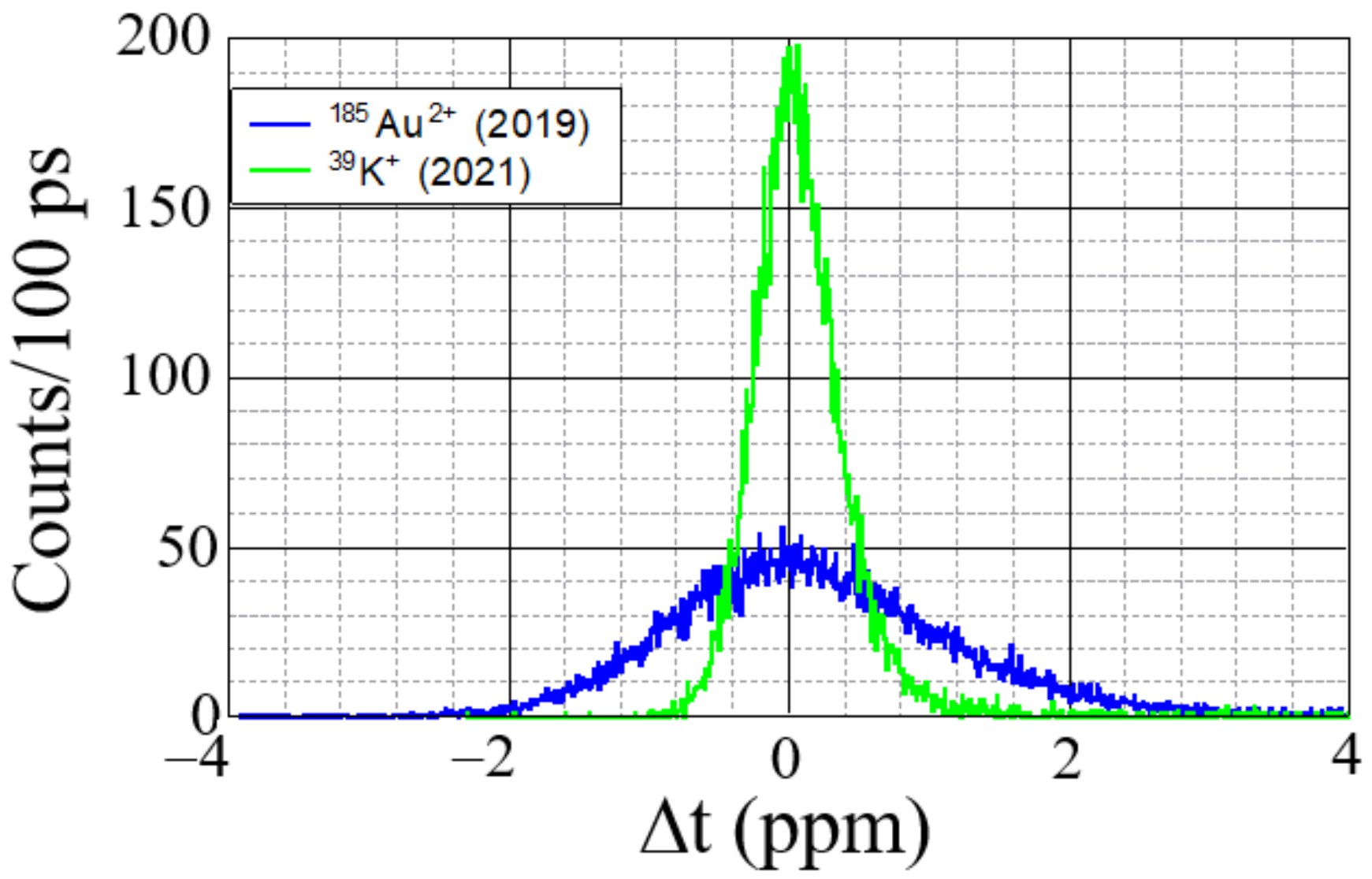

Figure 4-17. Comparison of highest achieved mass resolving power ($m/\delta m$ = 250 000) in 2019 and highest achieved mass resolving power by first half of 2021 ($m/\delta m$ = 800 000). Each peak has a similar number of ions. With improved tuning, longer times-of-flight, and better power supply stability it will be possible to achieve $m/\delta m$ > 1 000 000 in the near future.

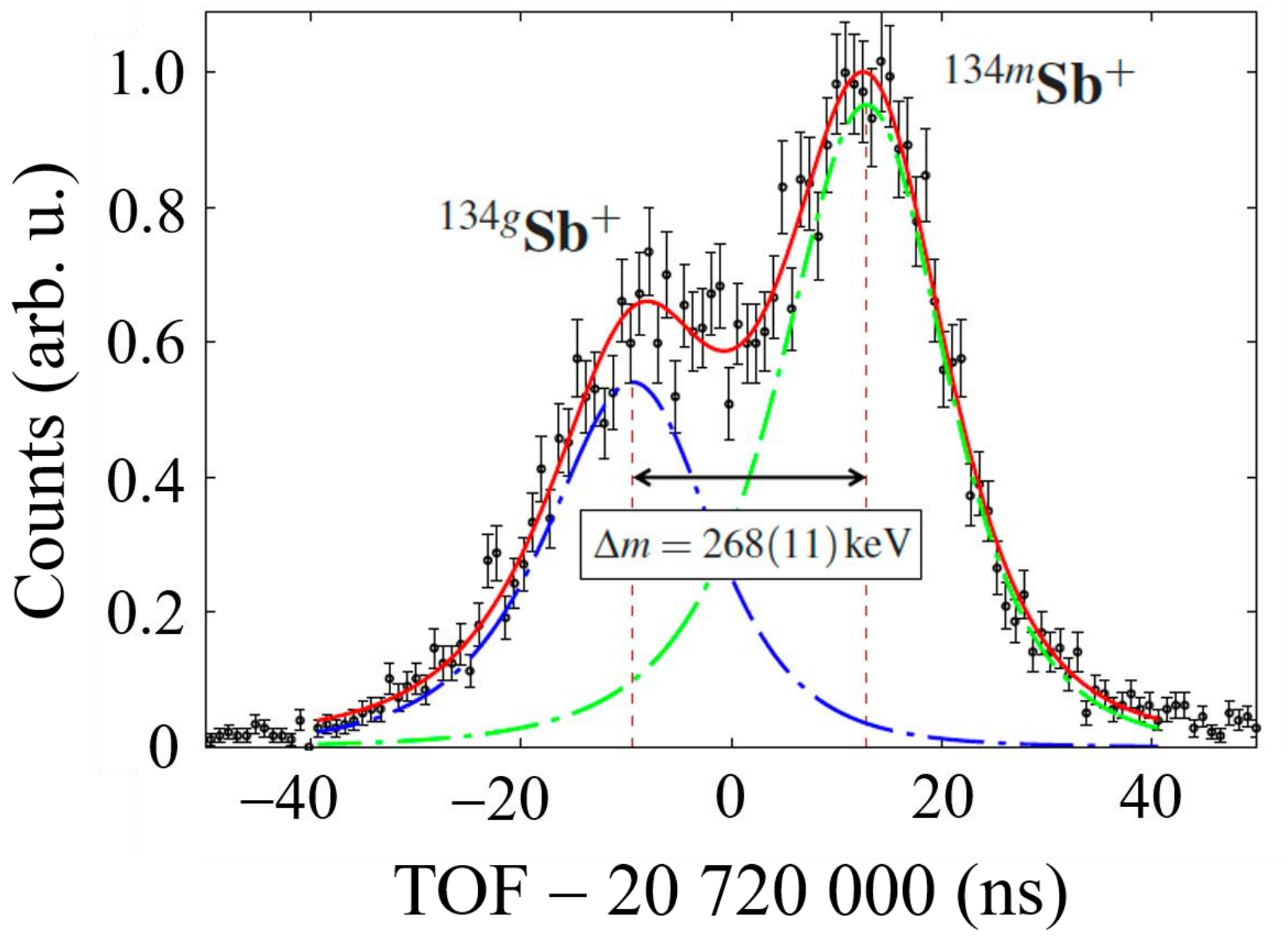

Figure 4-18. TOF spectrum demonstrating the ability to resolve $^{134g}$Sb and $^{134m}$Sb ($E_x$ = 279 keV).

## 4-8-1. Correlated mass and decay spectroscopy

We have successfully developed a silicon detector for measuring α-decays, embedded in the TOF detector [15]. Development of a similar two-layer detector for β-decays is currently under development. Such detectors allow TOF correlated decay studies, allowing a very clean means to measure half-lives while also allowing high-confidence in precision mass measurements of very low-yield species. This could be particularly useful in in the actinoid region.

As mention in Sec. 4-7, the decay station can be moved to allow the MRTOF-MS to deliver ions to it. This will allow the MRTOF-MS to be used for future β- and γ-decay correlated mass measurements as well, as will be discussed in Sec. 5-1-3.

## 4-8-2. Wide-band mass measurement

By operating the MRTOF-MS as a wide-band mass analyzer we can gain an experimental efficacy factor of more than 10. An example of such a result is shown in Figure 4-19. During an online experiment performed using the MRTOF connected to GARIS-II, five isobar chains were delivered to the MRTOF and simultaneously analyzed. A total of 15 nuclides could be measured at one time. When applied to neutron-rich MNT products, this will be a powerful tool for rapid identification in addition to precision mass spectrometer for nuclear physics. Moreover, when operating with low-yield isobar chains this technique can be combined with the correlated decay spectroscopy discussed in Sec. 4-8-1 to simultaneously perform half-life measurements among many other possible spectroscopic studies.

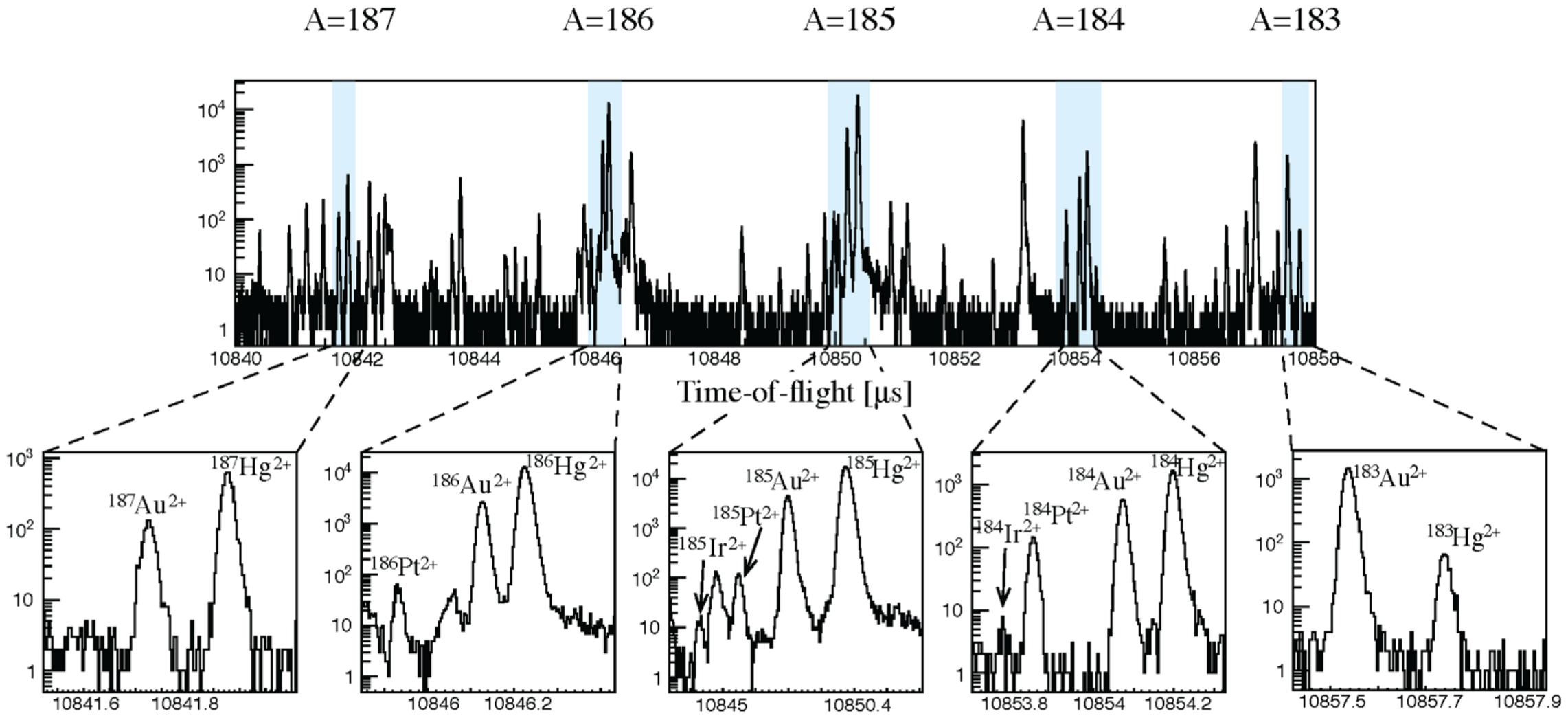

Figure 4-19. Wide-band mass measurement of the nuclei around $A$ = 184 produced by a fusion reaction at the GARIS-II facility in RIKEN.

# 5. Experimental setup for the nuclear spectroscopy

The proposed KISS-II facility should improve on the original KISS facility's capability for performing β-γ decay spectroscopy, half-life ($T_{1/2}$) determinations, and measuring atomic masses while also expanding the capabilities to include measuring β-delayed neutron emission probabilities ($P_{\mathrm{n}}$) and fission branching ratios and barrier height ($B_f$) which are of the astrophysical interest. Moreover, we will perform these spectroscopic works to study exotic nuclear structures which are desirable to improve theoretical models in the heavy regions in the vicinity of $N = 126$ and of actinoid isotopes. From the physics motivation, we prepare three experimental setups. One is for β–γ decay spectroscopy station to measure $T_{1/2}$, $P_{\mathrm{n}}$, and $B_f$. One is for MRTOF-MS to measure masses (already introduced in Sec. 4-8), and another one is for collinear laser resonance ionization spectroscopy (CLRIS station). The details of the new decay-spectroscopy station and the CLRIS station are described in the following sections.

## 5-1. β–γ decay spectroscopy

Figure 5-1 shows the detector station [1] which has been used for β-γ spectroscopy at the present KISS facility consisting of a tape transport system, MSPGC [2,3] for β-ray, X-ray, and

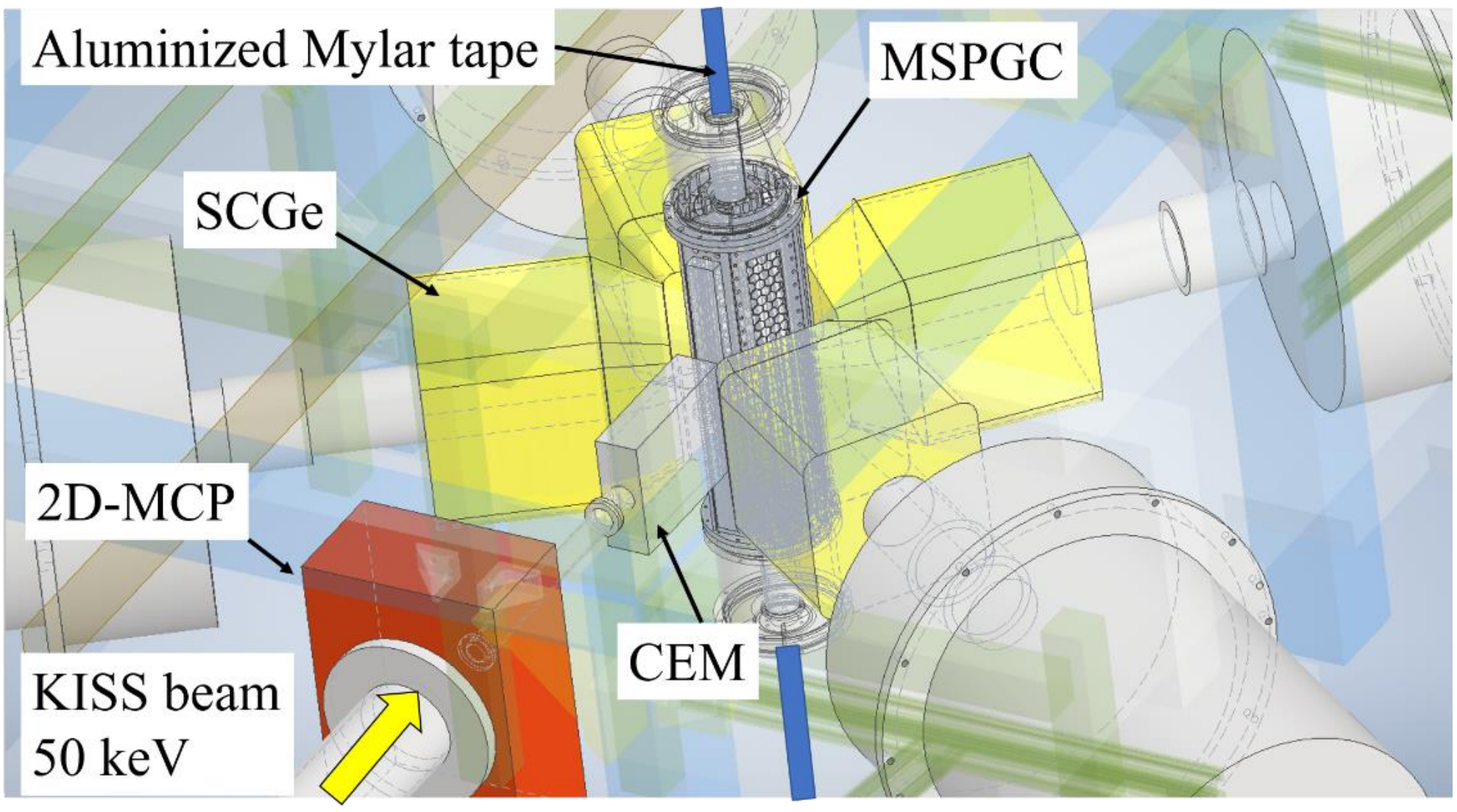

Figure 5-1. Detector setup for the β–γ decay spectroscopy. According to the physics motivation, we can arrange the detector setup. A channel electron multiplier (CEM) is used for KISS beam tuning.

conversion-electron detections, and super clover Ge detectors for γ- and X-ray energy measurements. The setup has been used for laser ionization spectroscopy [4-6] as well as many decay spectroscopy [7-11] at KISS. The detectors are covered by cosmic-ray veto counters made from plastic scintillators to reject background events. The detector station is installed on a rail system to enable it to be able to be moved to the end of the MRTOF-MS for advanced spectroscopic studies. We can perform more precise decay spectroscopy of the ground and isomeric states by the addition of particle and state identification performed by the MRTOF-MS.

By modifying the flexible detector configuration, we can measure nanosecond-order $T_{1/2}$ isomers by adding $LaBr_3$ detectors, $P_n$ values by adding $^3$He gas counters for the neutron detection, and $B_f$ values by adding multi-wire proportional chamber (MWPC) for the detection of fission fragments.

### 5-1-1. Gas counter for β-ray detection

We have successfully developed a compact multi-segmented proportional gas counter (MSPGC) system [2,3], having high efficiency and low background rate, for β-γ decay spectroscopy. The background rate was suppressed down to 0.11(1) cps, and the detection efficiency was 32% for β-ray with $Q_β$ = 1 MeV. As shown in Fig. 5-2, the MSPGC consists of 2 layers, each having 16 proportional counters (32 in total), cylindrically arranged around an aluminized Mylar tape in which the unstable nuclei are implanted. Each counter is pressurized to 100 kPa with P10 gas to facilitate signal detection. In the first implementation of the MSPGC, we used BeCu wire as an anode wire (indicated by red dots inside the blue trapezoids in Fig. 5-2 (b)) and applied two-dimensional hit-pattern analysis to distinguish β-ray, X-ray, and internal conversion electron events from environmental background events [3]. This identification provided a great advantage to the study of long-lived isomer transitions [7-11]. Figure 5-3 shows the measured γ-ray energy spectra of $^{199g}$Pt and $^{199m}$Pt decays as a typical example of the measurements that could be performed with the initial MSPGC implementation. In this example, by gating on events with the multiplicity $M \geq 1$ (which includes $M = 1$ events such as X-ray or internal conversion electron associated with the isomer transition, see Refs [2,3] for an extended explanation of hit multiplicity), we clearly observed 392 keV γ-rays (Fig. 5-3(b)) from the internal transition of isomer to the ground state. In the case of $M = 2$ (one telescope), higher energetic β-rays are dominant, and, therefore, isomer transition peak disappears as shown in Fig 5-3 (c). In order to perform decay spectroscopy of more neutron-rich nuclides, with their lower yields, required a further suppression of the background rate. To that end, we developed three-dimensionally tracking (3DT) MSGPC. In the 3DT-MSGPC, the BeCu wires were replaced by resistive carbon wires in order to determine the axial position of each event and thereby enable three-dimensional tracking of the β-ray trajectory emitted from the implantation area on the tape by measuring the vertical hit positions of telescope gas counter as shown in Fig. 5-4 (a). In the 3DT-MSPGC, the vertical positions can be evaluated by using charge information measured at both the ends of each anode wire. A vertical position resolution of 3.5 mm in FWHM, leading to a background rate of 0.020(3) cps has been achieved [3] by implementation of 3D tracking.

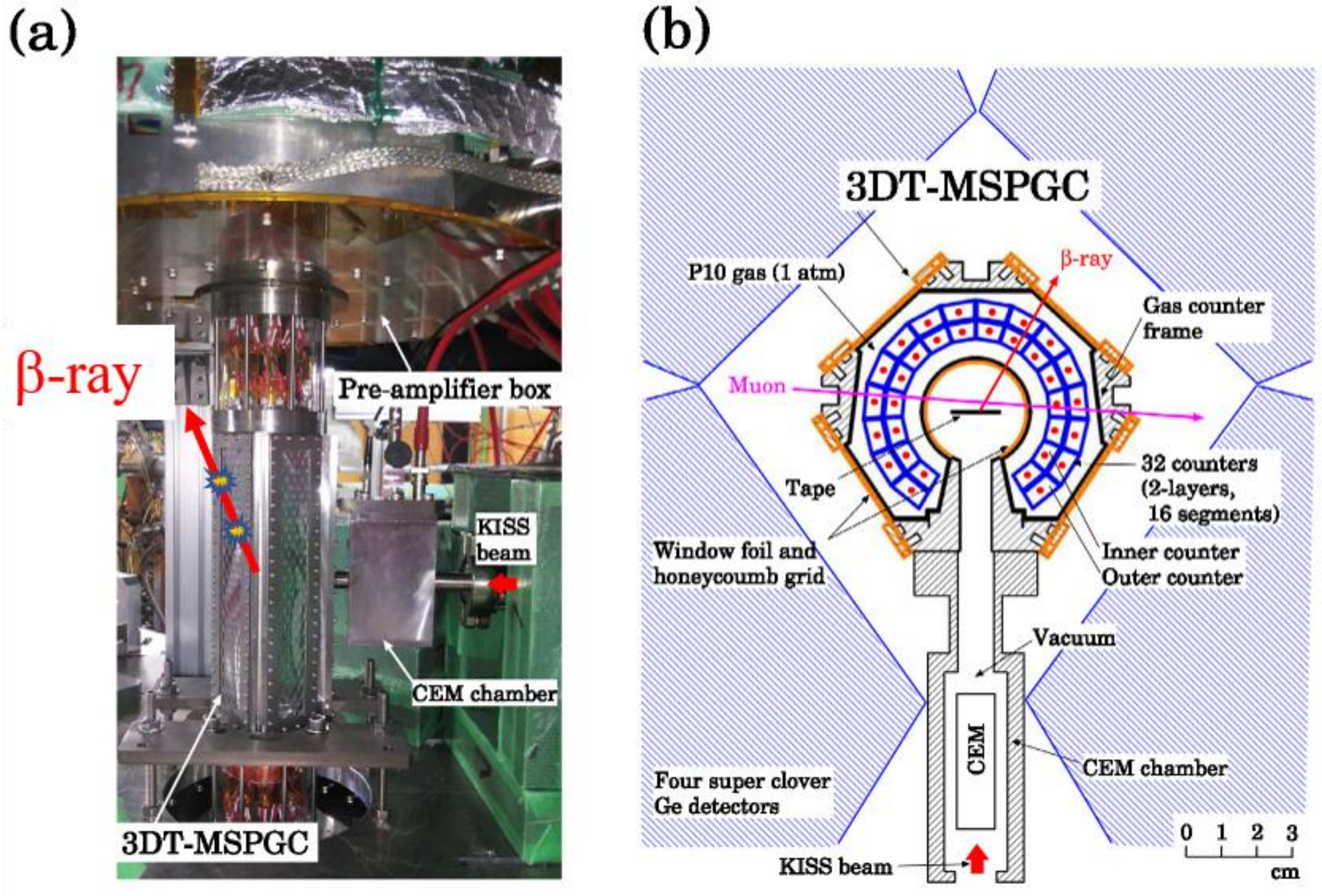

Figure 5-2. (a) Photo of the 3DT-MSPGC, and (b) cross-sectional view of the 3DT-MSPGC and related KISS detector setup. Red broad arrows indicate KISS beam. Blue trapezoids represent the MSPGC proportional counters. Orange areas indicate the window foil and honeycomb grid structure used to contain the P10 gas.

Further reducing the background rate to below 0.01 cps will require improving the position resolution to better than 2.0 mm FWHM. To accomplish this a new concept for the 3DT-MSPGC is planned for the KISS-II facility. The new 3DT-MSPGC will employ a multi-wire gas proportional counter (MWPC) with delay lines [12,13] to determine the hit positions of β-rays in the gas counter. In Ref. [12,13], a gas counter with planar geometry was used to achieve less than 0.5 mm position resolution in FWHM. Figure 5-4 provides a conceptual explanation of how we will perform the position measurements by using the delay lines. Timing information obtained at both the ends of the anode wires and cathode strips are measured, and the avalanche position can be determined from the measured timing differences between the two signals. The trajectory of charged particles can be reconstructed by using the two two-dimensional positions detected in a two-layer MWPC.

Based on this technique, we plan to construct new 3DT-MSPGC as shown in Fig. 5-5. The new 3DT-MSPGC will be built as a two-layer MWPC array. The cathode strips will comprise three concentric cylinders of segmented electrodes. Two arrays of anode wires will be strung perpendicularly between three cathode rings. The volume between the innermost and outermost cathode cylinders will be pressured to 100 kPa with P10 gas. The vertical position and azimuthal angle, respectively, of each hit can be determined from the timing information obtained for each cathode strip and anode wire. The cathode strip electrodes will be made by aluminum vapor deposition onto a thin Mylar foil. The MWPC will have an inner diameter of 40 mm and an outer diameter of 58 mm, and will extend 200 mm longitudinally. Both ends of the anode wires and cathode strips will be connected to delay lines. In the new 3DT-MSPGC, large injection angles of β-rays to the MWPCs are expected because of the short distance between the implantation area and the MWPCs necessary for large solid angles around 80% of 4π. Therefore, the position resolution may not achieve the 0.5 mm reported in other studies, and will need to be investigated in future performance tests.

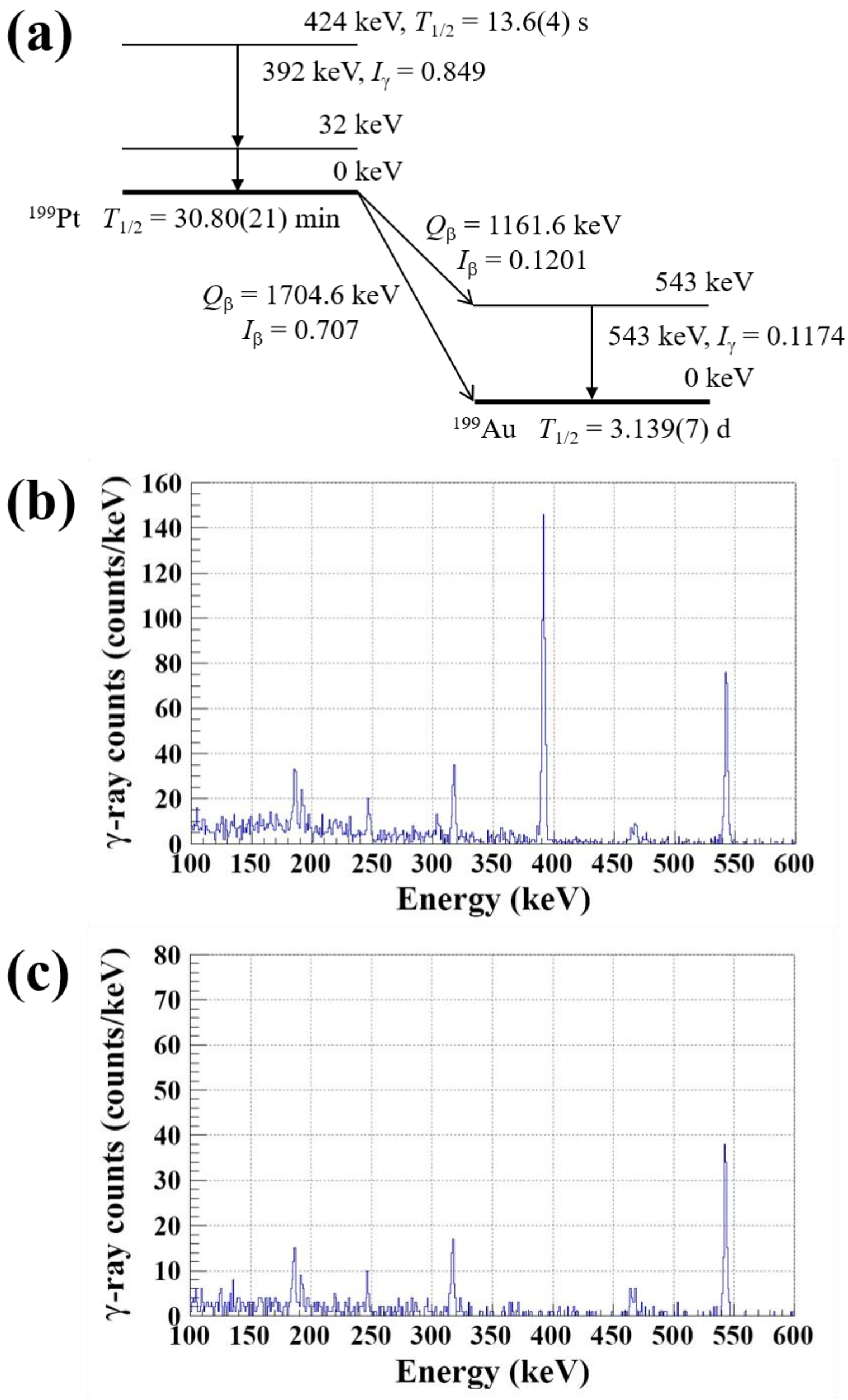

Figure 5-3. (a) Decay scheme of the isomeric state $^{199m}$Pt ($E_x$ = 424 keV, $T_{1/2}$ = 13.6(4) s) and ground state of $^{199g}$Pt ($T_{1/2}$ = 30.80(21) min). Measured $\gamma$-ray energy spectra which coincide with the MSPGC hit patterns of (b) $M \geq 1$ and (c) $M = 2$.

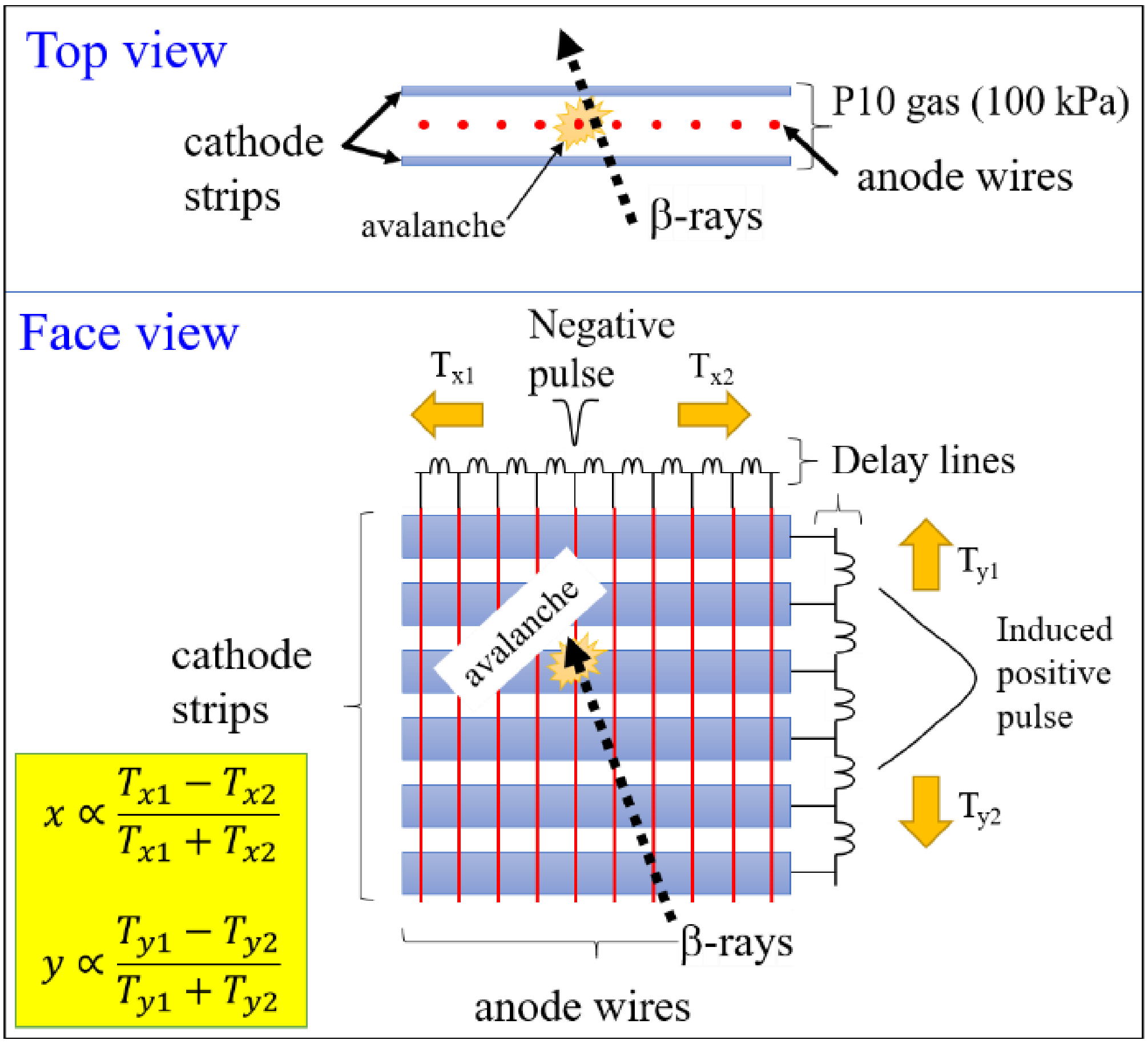

Figure 5-4. Position measurements of MWPC with delay lines.

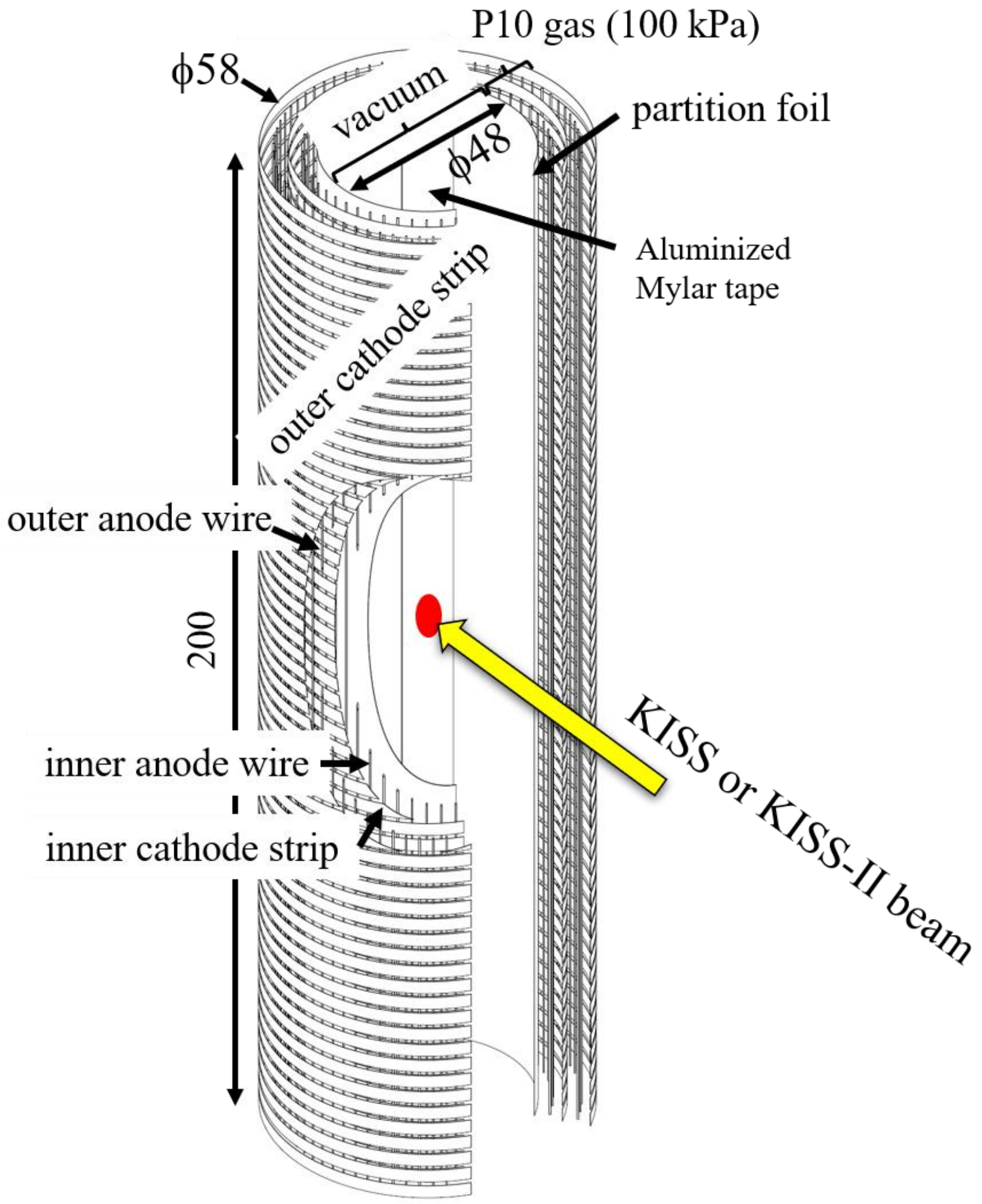

Figure 5-5. Conceptional design of new 3DT-MSPGC by using MWPCs with delay lines.

## 5-1-2. Ge detector for γ-ray detection

The compact geometry of the MSPGC allows for the installation of HPGe detectors within 50 mm of the implantation tape position. Since 2016, the KISS facility was operated with four super clover (SC) HPGe detectors surrounding the MSPGC. The facility initially used four SC HPGe detectors on loan from the Institute for Basic Science in South Korea under a collaborative agreement with KEK. From September 2020 until May 2021 another four SC HPGe, on loan from the Institute of Modern Physics (IMP) of the Chinese Academy of Sciences, were used under a collaborative agreement between KEK and IMP.

The detection efficiency of SC HPGe exceeds that of standard HPGe detectors by approximately 160%. The detection efficiency (e.g., about 10% for 1 MeV γ-ray) of the four

SC HPGe detectors were measured by using $^{152}$Eu and $^{133}$Ba sources at the KISS facility. Unfortunately, SC HPGe detectors are prohibitively expensive. As a compromise, we plan to install eight coaxial HPGe detectors at the decay station for the KISS-II facility. Each of these will have a detection efficiency 60% that of standard HPGe detectors, resulting 75% of the previous detection efficiency.

### 5-1-3. Decay spectroscopy in conjunction with MRTOF-MS

We have successfully performed β–γ spectroscopy [7-11] of the nuclei approaching the $N = 126$ region at KISS. In these spectroscopic works, there were always contaminants from non-resonantly ionized isobars because the resolving power of the KISS dipole magnet ($A/\Delta A$ = 900) was insufficient to suppress isobaric contaminants. We could occasionally perform β–γ spectroscopy of these contaminants simultaneously. However, these contaminants usually made it difficult to analyze the decay data.

In order to improve the situation and perform much more precise decay spectroscopy, we will use the β–γ detectors coupled with the MRTOF-MS in a manner similar to that demonstrated by the use of an α-detector embedded within the MRTOF-MS ion implantation detector [16]. The β–γ detectors will be placed just downstream of the MRTOF-MS as shown in Fig. 5-6. A new MCP detector will be installed just upstream of the aluminized-Mylar tape where the RI beam is implanted, and will detect secondary electrons emitted when ions strike the tape to produce the TOF stop signal of the MRTOF-MS. To enhance the number of the secondary electrons, the unstable nuclei ejected from the MRTOF-MS will be post accelerated to $20q$ keV by a pulsed drift tube. Figure 5-7 shows the proposed setup with the use of an annular-type MCP detector, a solenoid coil, and ring electrode. To efficiently transport the secondary electrons, a solenoid coil will be used to produce a magnetic field to compress and guide the electrons. Ion optical simulations have verified that good performance can be achieved with a small number of modestly biased ion-optical elements. Figure 5-8 shows the secondary electron transport as simulated with SIMION 8.1. We expect to be able to transport the electrons with 85% efficiency. The secondary electron detection should add less than 1 ns FWHM to the measured time-of-flight peak width, which will negligibly affect the achievable mass resolving power. From the TOF spectrum, we will be able to identify the nuclei responsible for each decay radiations by using the coincidence measurement between the secondary electrons and the decay radiations.

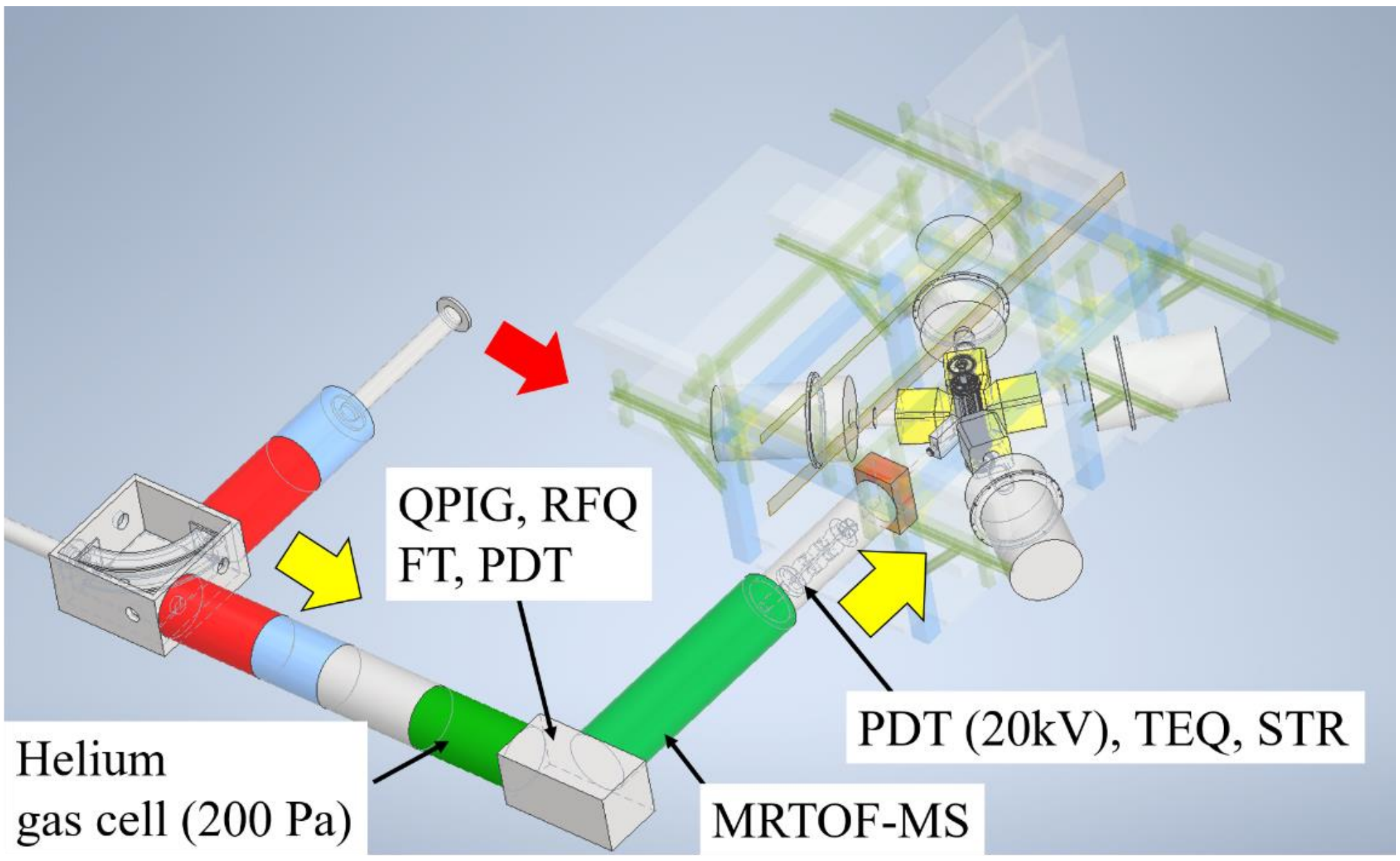

Figure 5-6. β–γ detectors combined with MRTOF-MS. Here, QPIG is a quadrupole ion-guide; RFQ is a radio-frequency quadrupole; FT is a flat trap; PDT is a pulse drift tube; TEQ is a triplet electric quadrupole; STR is a steerer.

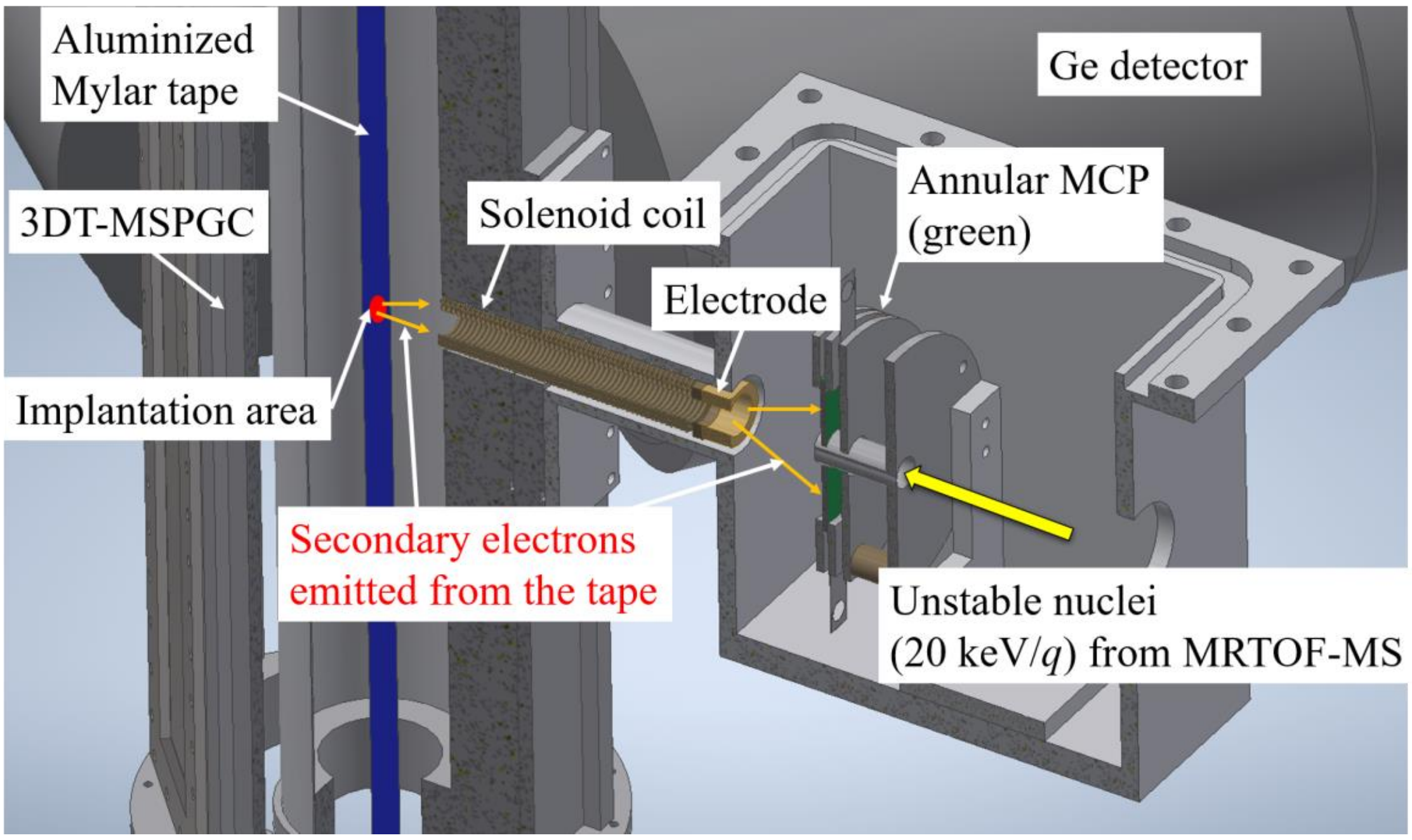

Figure 5-7. Setup for the detection of secondary electrons for the stop signal of MRTOF-MS. The annular-type MCP installed between the tape and MRTOF-MS will detect the secondary electrons emitted when ions from the MRTOF-MS strike the tape.

Figure 5-8. Simulation of the secondary electron transport to the annular-type MCP by using SIMION 8.1. We can transport the secondary electrons with high efficiency of 85% and time resolution of 0.6 ns in FWHM.

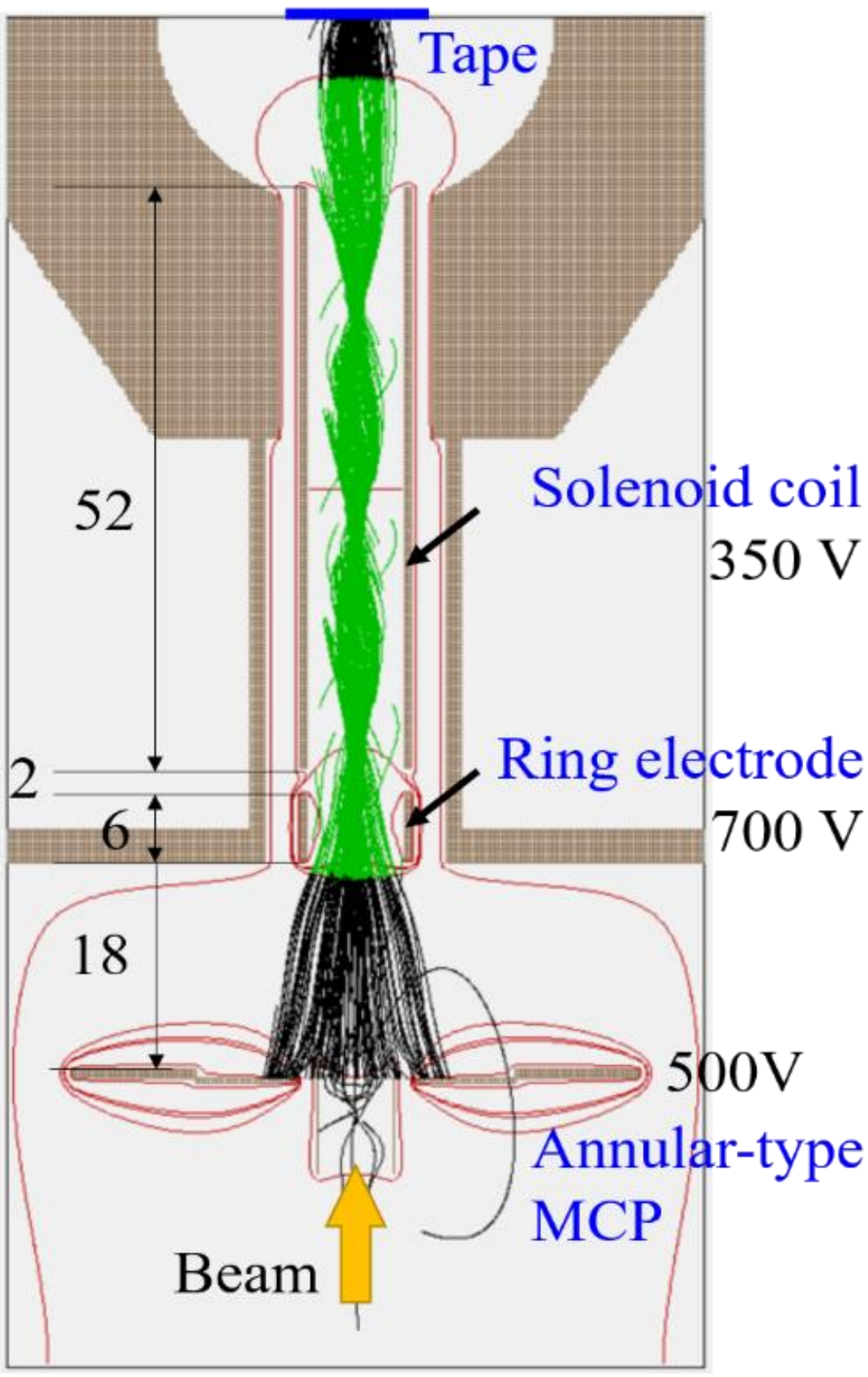

Figure 5-9 shows simulated results of β–γ decay spectroscopy performed in conjunction with the MRTOF-MS for the ground state $^{199g}$Pt and isomer $^{199m}$Pt($E_x$ = 424 keV) based on the known γ-ray energy spectra of $^{199g}$Pt and $^{199m}$Pt [17]. Figure 5-9 (a) shows a correlation between the TOF and measured γ-ray energy. Here, we assumed a mass resolving power of $m/\delta m = 10^6$. We can clearly identify the TOF peaks of $^{199g}$Pt and $^{199m}$Pt as shown in Fig. 5-9 (b) which is the projection of the Fig. 5-9 (a) onto the TOF axis. By gating on respective TOF peaks of $^{199g}$Pt and $^{199m}$Pt, γ-ray energy spectra shown in Figs. 5-9 (c) and (d), respectively, are obtained. Thus, we will perform the β-delayed γ-ray spectroscopy of $^{199g}$Pt and γ-decay spectroscopy of $^{199m}$Pt simultaneously and precisely by MRTOF-MS assisted decay spectroscopy. When we operate the variable mass-range separation system in the wide mass-range mode and utilize the wide-band mass measurement mode of the MRTOF-MS more than 10 nuclides would be transported to the detector system. Therefore, we would be able to simultaneously perform the decay spectroscopy on these nuclides; the analysis can be performed by gating on each TOF peak. This mode of operation will result in high-efficacy utilization of limited online machine time.

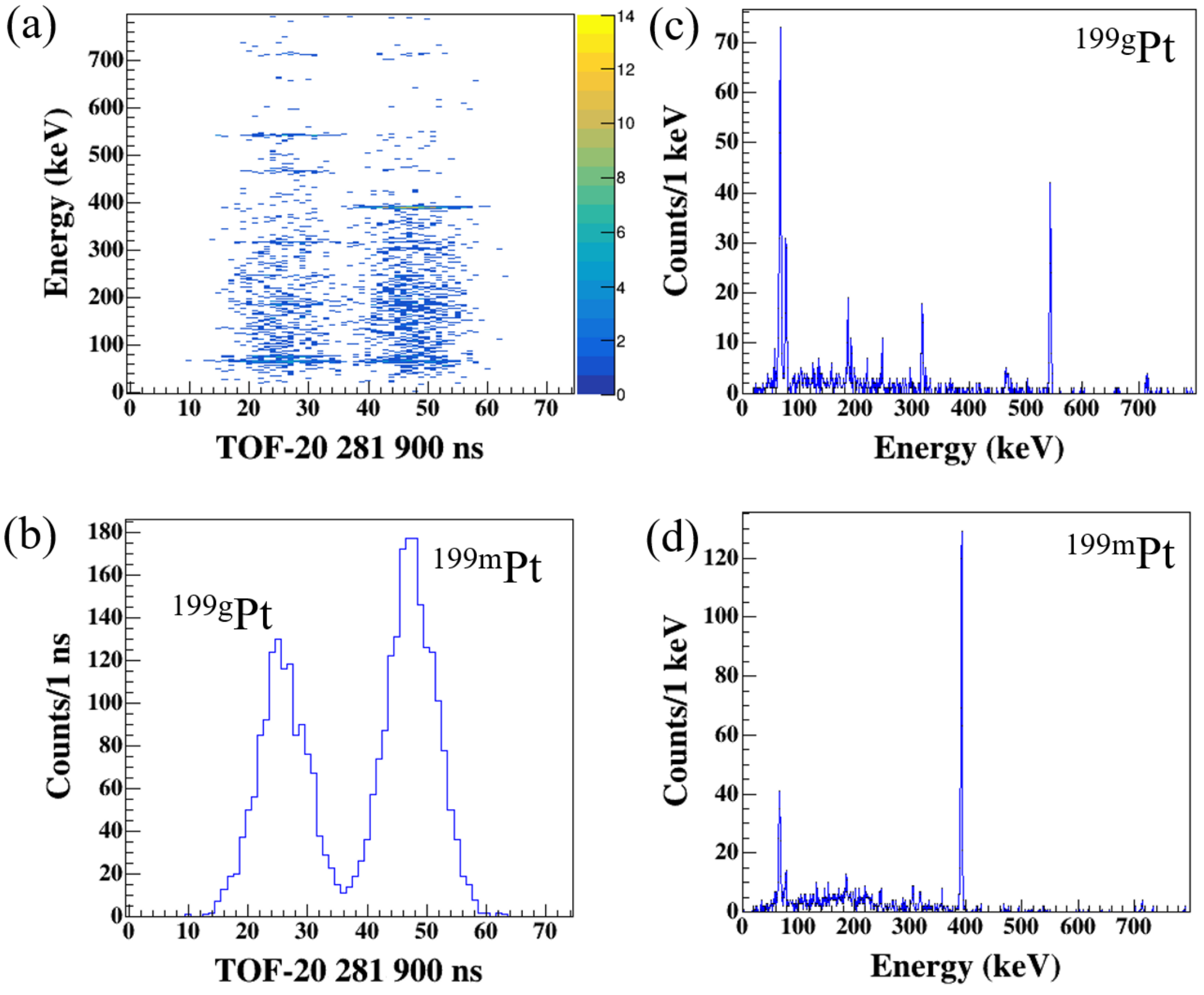

Figure 5-9. Results of simulated β–γ decay spectroscopy performed in conjunction with the MRTOF-MS for an admixture of $^{199g}$Pt and $^{199m}$Pt ($E_x$ = 424 keV) in three-hour run, by accounting for the yields of $^{199g,m}$Pt obtained at the present KISS facility. (a) Simulated TOF and γ-ray energy correlation. (b) Projection of (a) onto the TOF axis. (c) Projection of (a) onto the energy axis, gating on the $^{199g}$Pt TOF peak in (b) to ascertain the β-delayed γ-ray energy spectrum of $^{199g}$Pt. (d) Projection of (a) onto the energy axis, gating on the $^{199m}$Pt TOF peak in (b) to ascertain the γ-ray spectrum from decay of $^{199m}$Pt.

## 5-2. Collinear laser resonance ionization

As previously, noted, the KISS-II facility will employ a helium gas cell for stopping and thermalizing the energetic MNT products. There will essentially be no element-selective ionization process at the helium gas cell system. As one option for suppressing isobaric contamination to enable certain precision spectroscopy studies, a collinear laser resonance ionization beam line will be installed in the E3 experimental hall as shown in Fig. 5-10. This device will work in a manner similar to the Collinear Resonance Ionization Spectroscopy (CRIS) system [18-21] at CERN/ISOLDE.

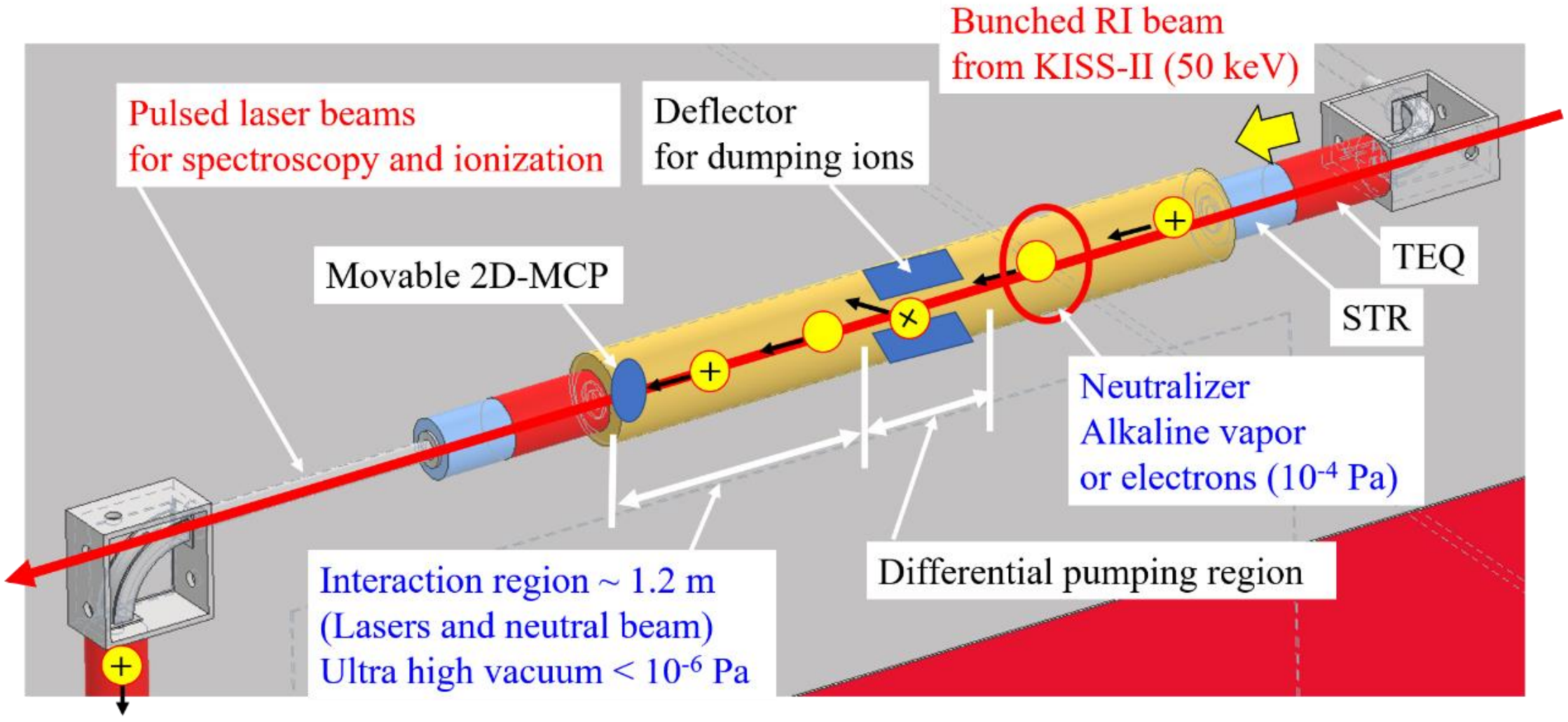

Figure 5-10. Schematic view of collinear laser resonance ionization spectroscopy line proposed to be installed in the E3 room.

The bunched RI beam from the KISS-II facility will be transported to the E3 experimental hall with an energy up to $50q$ keV. By using the TEQ, the STR at the entrance of collinear laser resonance ionization spectroscopy line, and the position sensitive 2D-MCP at the exit of the beam line, the beam profile and position can be adjusted for transporting the beam through the beam line, which has small apertures (typically 6 mm in diameter) for differential pumping.

For the laser ionization spectroscopy, the ion beam can be neutralized by either an alkaline vapor (Na or K) or electrons emitted from a filament. Non-neutralized ion beam will be deflected by an electric deflector installed after the neutralization equipment and be dumped. Then, the bunched neutral beam will be selectively laser-ionized by pulses of excitation (probe) and ionization lasers overlapping in the interaction region (length ~ 1.2 m) which will be under ultra-high vacuum of $P < 10^{-6}$ Pa. Only the laser ionized beam will be transported downstream, and the precise nuclear spectroscopic works such as precision decay spectroscopy, mass measurement, and hyperfine structure (HFS) measurement by detecting the decay radiations and/or counting the number of ions identified at the MRTOF-MS (See Fig. 5-6). The CRIS group has reported achieving the laser spectroscopy efficiency of 1-10% [21], which would be acceptable.

We already have a pulsed dye laser and dye amplifier pumped by intense YAG (355 nm) laser with a repetition rate of 10 kHz in the J3 experimental room located underneath the E2 room. The narrow-band dye lasers (typical bandwidth of 1.5 GHz and 0.1 GHz) can be applied as the excitation (HFS probe) laser, and the intense YAG laser can be used as an ionization laser. This system has been successfully applied to in-gas-jet laser ionization of stable platinum atoms in the KISS argon gas cell system [1], and it can be applied to the collinear laser resonance ionization spectroscopy at the KISS-II facility by some modifications such as the synchronization with the bunched neutral beam and adding elements for laser transport to the E3 room. In the collinear laser ionization spectroscopy, the resolution of HFS spectra can be drastically improved to be less than 200 MHz in FWHM by reducing the Doppler broadening to ~100 MHz by the use of 50$q$ keV beam.

Figure 5-11 shows simulated results of a laser ionization spectroscopy measurement for an admixture of $^{199m,g}$Pt where the MRTOF-MS is used to detect the laser ionized ions. The simulation assumed $^{199g}$Pt to have spin-parity $I^{\pi} = 5/2^{-}$ and a magnetic moment of $\mu = +0.75(8)\ \mu_N$, and for $^{199m}$Pt to have spin-parity $I^{\pi} = 13/2^{+}$, a magnetic moment of $\mu = -0.57(5)\ \mu_N$, and an excitation energy of 424 keV. Figure 5-11 (a) shows a correlation between the simulated TOF and HFS spectrum. Here, we assumed a mass resolving power of

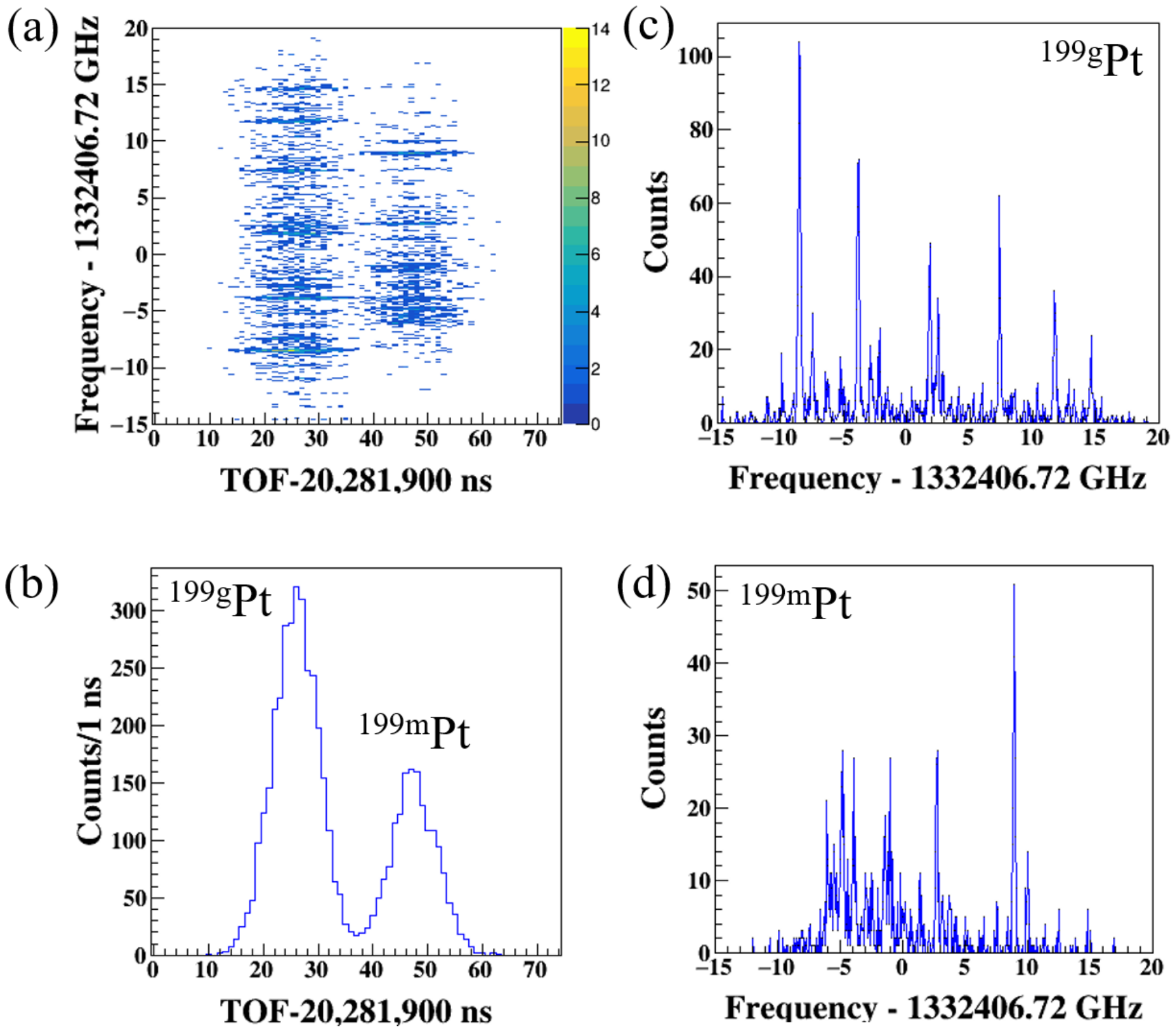

Figure 5-11. Simulation results of laser spectroscopy coupled with the MRTOF-MS for the ground state $^{199g}$Pt and isomer $^{199m}$Pt($E_x$ = 424 keV). (a) correlation spectrum between the simulated TOF and HFS spectrum. (b) TOF spectrum. (c) HFS spectrum of $^{199g}$Pt. (d) HFS spectrum of $^{199m}$Pt.

$m/\delta m = 10^6$. We can clearly resolve the TOF peaks of $^{199g}$Pt and $^{199m}$Pt as shown in Fig. 5-11 (b) which is the projection of 5-11(a) onto the TOF-axis. By gating on the respective TOF peaks of $^{199g}$Pt and $^{199m}$Pt, HFS spectra shown in Figs. 5-11 (c) and (d), respectively, are obtained. Thus, we can perform the laser ionization spectroscopy of $^{199g}$Pt and $^{199m}$Pt simultaneously and precisely by detecting the laser ionized ions with the MRTOF-MS. When we use the mass separation system with wide mass-range mode and wide-band mass measurement mode of the MRTOF-MS, more than 5 nuclides with the same atomic number and different mass-numbers are transported to the MRTOF-MS. Therefore, we can directly measure isotopic shifts with extreme accuracy by scanning through these nuclides with a fixed setting of the laser ionization spectroscopy.

As Fig. 5-11 indicated, we will be able to perform not only purification of the RI beam with the element-selective laser ionization technique when the nuclear spectroscopy work requires such, but also to perform precise laser ionization spectroscopy. Laser spectroscopy is a powerful method to study nuclear structure through magnetic dipole and electric quadrupole moments, the changes of charge radii, and the deformation parameter deduced from the hyperfine structure (HFS) measurements. This method allows performance of spectroscopic works even when the rate of unstable nuclei is as low as a few per second.

## 5-3. Feasibility of various nuclear spectroscopy at KISS-II

Mass measurements, half-life measurements, $(\alpha, \beta, \gamma)$-decay spectroscopy, and laser spectroscopy are the typical nuclear spectroscopic work to be performed at the KISS-II facility. Each of these spectroscopic studies will require their own characteristic minimum beam intensity and purity. Based on the astrophysically required precisions, the general requirements are for a few-day-long run to be able to determine the atomic mass with a precision of about 100 keV/$c^2$, decay half-lives with a relative precision better than 30%, and magnetic dipole moments, quadrupole moments, and isotope shifts with a relative precision of better than 5%, 20%, and 5% to study the nuclear structure. Table 5-1 shows the required minimum rates at the detector position and emitted from the target for four types of spectroscopy work. We assume an ion-extraction efficiency $\varepsilon_{\mathrm{overall}} \sim 0.01$ from the target station through to the variable mass-range separator and an efficiency $\varepsilon_{\mathrm{MRTOF}} = 0.1$ for the MRTOF-MS system, including deceleration and trapping of the $50q$ keV ion beam. For the laser spectroscopy, an efficiency $\varepsilon_{\mathrm{laser}} = 0.1$ is assumed in the estimate of the minimum rate at the target position. Precise γ-decay and laser spectroscopy requires more than 3 orders of magnitude higher rates than that for mass measurements. β–γ decay spectroscopy requires another factor of roughly 300 after accounting for the detection efficiency and typical decay branching-ratio. Complex HFS requires much higher statistics for the laser spectroscopy.

Table 5-1. Required minimum rate at the detector position and target position for mass measurements, decay spectroscopy, and laser spectroscopy.

| Measurement | Required minimum rate at detector position | Required minimum rate at target position |
|---|---|---|
| Mass | a few particles/day | a few $10^3$ particles/day |
| Half-life | 10 particles/day | $10^4$ particles/day |
| $\gamma$-decay spectroscopy | $10^3$ particles/day | $10^7$ particles/day |
| Laser spectroscopy | $10^4$ particles/day | $10^8$ particles/day |

Figure 5-12 and 5-13 show the accessible regions for these spectroscopic works, which is estimated based on the required minimum rate at the target position, based on MNT reaction rates in the system of $^{238}$U+$^{238}$U and $^{238}$U+$^{198}$Pt, respectively. Mass measurements can be performed in a large region systematically. Decay and laser spectroscopy can be performed in the region where many unknown neutron-rich nuclei exist. By improving the efficiencies and measurement techniques, we can eventually extend the accessible regions.

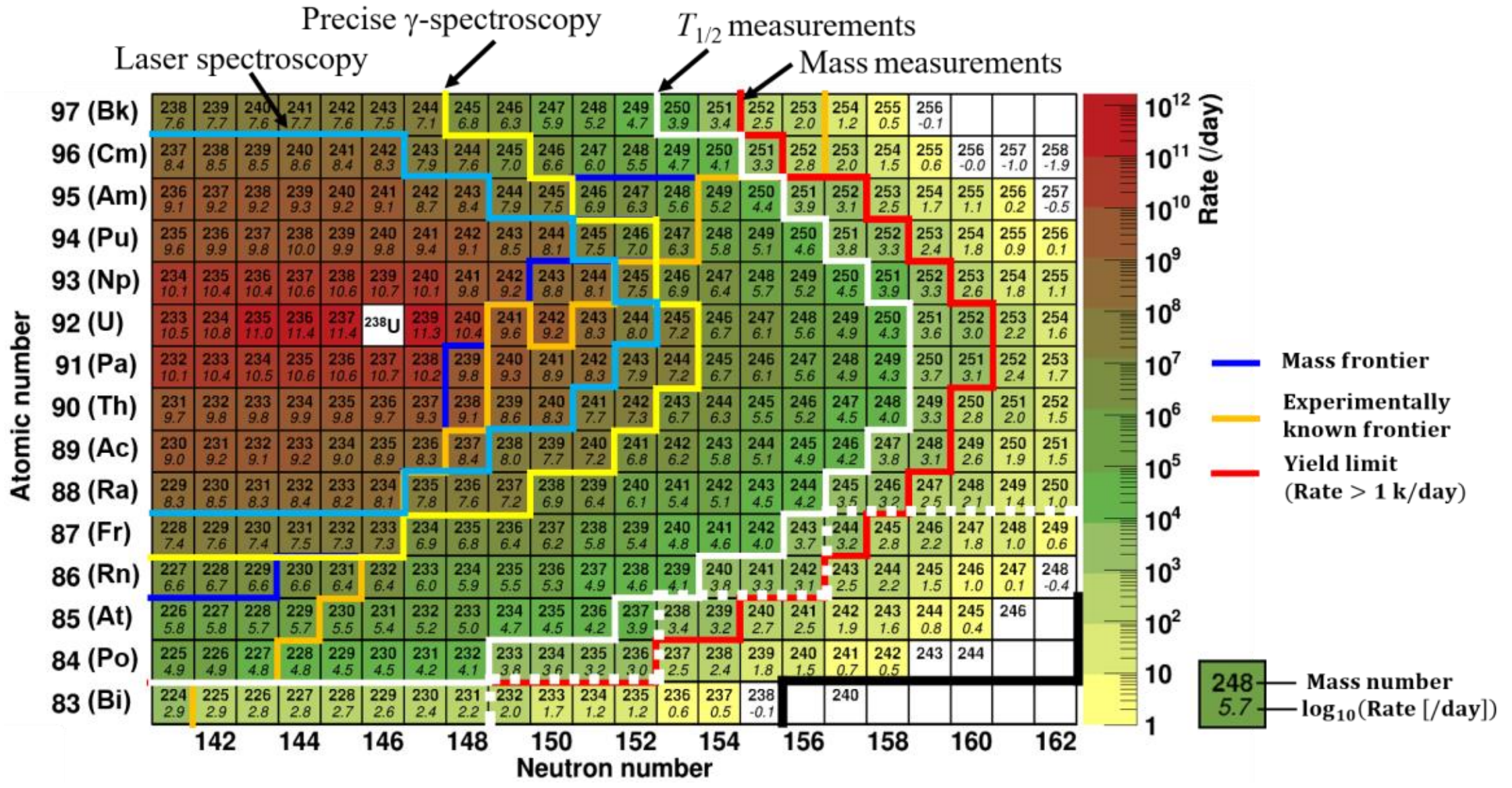

Figure 5-12. Accessible regions for mass measurements (boundary indicated by red line), half-life measurements (orange line), precise γ-decay spectroscopy (yellow line), and laser spectroscopy (blue line) based on MNT reaction products produced by $^{238}$U beam (1 pμA) and $^{238}$U target (13 mg/cm$^2$). White dotted and black lines indicate the bordars of half-lives less than 20 ms and 100 ms, respectively, predicted by KUTY model [22].

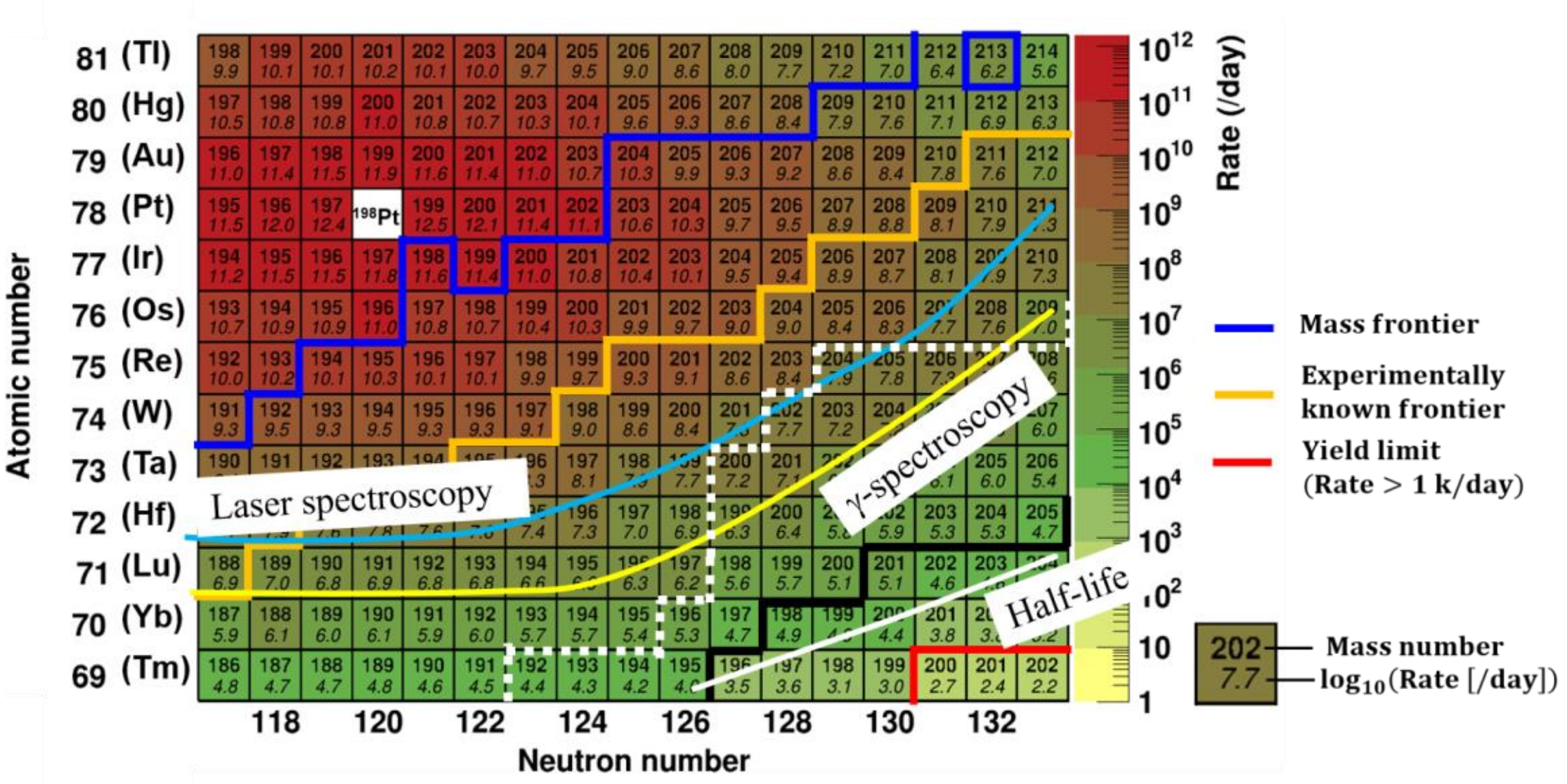

Figure 5-13. Accessible regions for mass measurements, half-life measurements, precise γ-decay spectroscopy, and laser spectroscopy based on MNT reaction products produced by $^{238}$U beam (1 pμA) and $^{198}$Pt target (13 mg/cm$^2$). White dotted and black lines indicate the bordars of half-lives less than 20 ms and 100 ms, respectively, predicted by KUTY model [22].

# 6. Superconducting solenoid filter for KISS-II

The superconducting (SC) solenoid filter will be installed in KISS-II facility. The conceptual design and an expected performance are described in this chapter which consists of four sections. The first section discusses characteristics of the SC solenoid based on a ray-tracing simulation. The second section describes a design of the SC solenoid system including cryostat and yoke. The third section discusses on the fringing field of the solenoid magnet for minimizing the effect to other devices of the KISS-II facility. Finally, radiation damage of the SC coils and radiation safety estimated by using PHITS radiation transportation code are presented in the fourth section.

## 6-1. Expected performance of the SC solenoid filter

In this section, we will discuss an expected performance of the gas-filled SC solenoid filter which will be installed for collecting the reaction products and separating the unwanted particles, e.g., primary beam, elastically scattered particles of the primary beam and production target, and so on. In Sec. 6-1-1, the basic parameters for the gas-filled solenoid filter are discussed for collecting MNT reaction products. Based on the parameters, a simulation for evaluating the performance of the solenoid filter is introduced in Sec. 6-1-2. From the simulation, optimal filling gas and its pressure are evaluated to suppress the unwanted particles with keeping the collection efficiency of the reaction products. The gas pressure dependence of the solenoid filter's resolving power is estimated in Sec. 6-1-3. The collection efficiencies of the MNT reaction products by the solenoid filter and its separation abilities under various reaction conditions are investigated in Sec. 6-1-4. An application of the solenoid filter to fusion reactions is also discussed. The results of these evaluation are summarized in Sec. 6-1-5.

### 6-1-1. Magnetic rigidity and basic parameters

A gas-filled separator with dipole magnetic field has been widely used mainly for studies of (super) heavy nuclei produced through nuclear fusion reactions. It can cancel out the large energy dispersion of fusion evaporation residues via the atomic charge exchange collisions with a dilute buffer gas filled in the magnetic field. Under an appropriate dipole field, evaporation residues recoiling out from the reaction target with large solid angles can be collected on a dispersive focal plane having some resolving power for the mass and atomic number. Thus, this kind of separator with relatively large collection efficiency has played an important role in researches of evaporation residues rarely produced in the fusion reactions.

On the other hand, MNT reaction products have wide energy and angular distributions. So far, efficient collection and purification of target-like fragments (TLFs) of MNT reactions have been difficult with existing experimental techniques. To overcome such difficulty, we propose a gas-filled separator with a solenoid magnetic field. The solenoid field could collect MNT reaction products emitted to any azimuth and polar angles from the target. Thanks to the atomic charge exchanging collision under the gas-filling condition, this type separator would be an ideal device for collecting the MNT products with some separation ability from back ground particles such as elastically scattered ones of the primary beam and the production target.

By assuming the ideal solenoid magnetic field, $\vec{B} = (0,0,B)$, and the vacuum condition, the maximum radius from the solenoid axis, $\rho_{max}$, of any ions emitted from the target on the solenoid axis with velocity $v$, charge $q$, and polar angle in the laboratory frame $\theta_{lab}$, is given by

$$\rho_{\max} = \frac{0.0457}{B}\frac{A(v/v_0)}{q}\sin\theta_{\mathrm{lab}}\ [\mathrm{m}]\,, \qquad (6-1)$$

where $B$ is the magnetic field strength in units of Tesla, $A$ is the atomic mass in atomic mass units, $q$ is the ion's charge in elementary charge units, and $v_0$ is Bohr velocity ($c/137$). The ion repeatedly approaches and leaves the solenoid axis under the ideal condition with a longitudinal distance $z_0$ between target and closest point on the solenoid axis. This can be described by

$$z_0 = \frac{0.143}{B}\frac{A(v/v_0)}{q}\cos\theta_{\mathrm{lab}}\ [\mathrm{m}]\,. \qquad (6-2)$$

Ions moving in a dilute gas will undergo charge exchange reactions with gas materials. If the mean free path of such charge exchange reactions is sufficiently short, the ionic charge exchange collisions will come into equilibrium and the ion charge states can be represented by an equilibrium charge state. The equilibrium charge state $\bar{q}$ is given by the empirical formula:

$$\bar{q} \approx C(v/v_0)Z^{1/3}, \qquad (6-3)$$

where $Z$ is the ion's atomic number and $C$ is a constant. Thus, in the ideal gas-filled solenoid filter, $\rho_{max}$ and $z_0$ depend solely on $Z$, $A$, and $\theta_{lab}$ after substituting the equilibrium charge state $\bar{q}$ for ion charge $q$:

$$\rho_{\max} \approx \frac{0.0457}{C}\frac{1}{B}\frac{A}{Z^{1/3}}\sin\theta_{\mathrm{lab}}\ \ [\mathrm{m}], \qquad (6-4)$$

$$z_0 \approx \frac{0.143}{C}\frac{1}{B}\frac{A}{Z^{1/3}}\cos\theta_{\mathrm{lab}}\ [\mathrm{m}]\,. \qquad (6-5)$$

Considering the case for collecting a TLF, $^{250}$U produced via MNT reactions under the He-gas filled condition, both relations of Eq. (6-4) and Eq. (6-5) are deduced to be

$$\rho_{\max} \approx \frac{3.5}{B} \text{ [m]}, \tag{6-6}$$

$$z_0 \approx \frac{7.6}{B} \text{ [m]}, \tag{6-7}$$

where the constant $C$ is set to 0.6 and $\theta_{\mathrm{lab}} = 55°$ is assumed. Thus, if one considers a solenoid filter having radius 0.6 m, $B \approx 5.5$ T would be required. To satisfy this requirement a superconducting solenoid magnet of length around 1.2 m would be necessary. The conclusion is equally valid for neutron-rich nuclei in the $N = 126$ region up through the superheavy element region, because the $A/Z^{1/3}$ term has only a weak isotopic dependence and the $\theta_{\mathrm{lab}}$ distributions of various MNT reactions show similar features as discussed in the previous sections.

### 6-1-2. Simulation of ionic motions

In order to evaluate the performance of the solenoid filter, a simulation of moving ions in the gas-filled magnetic field was applied [1]. Important processes when considering the gas-filled separator are charge exchange reactions and multiple scattering of ions with filling gas.

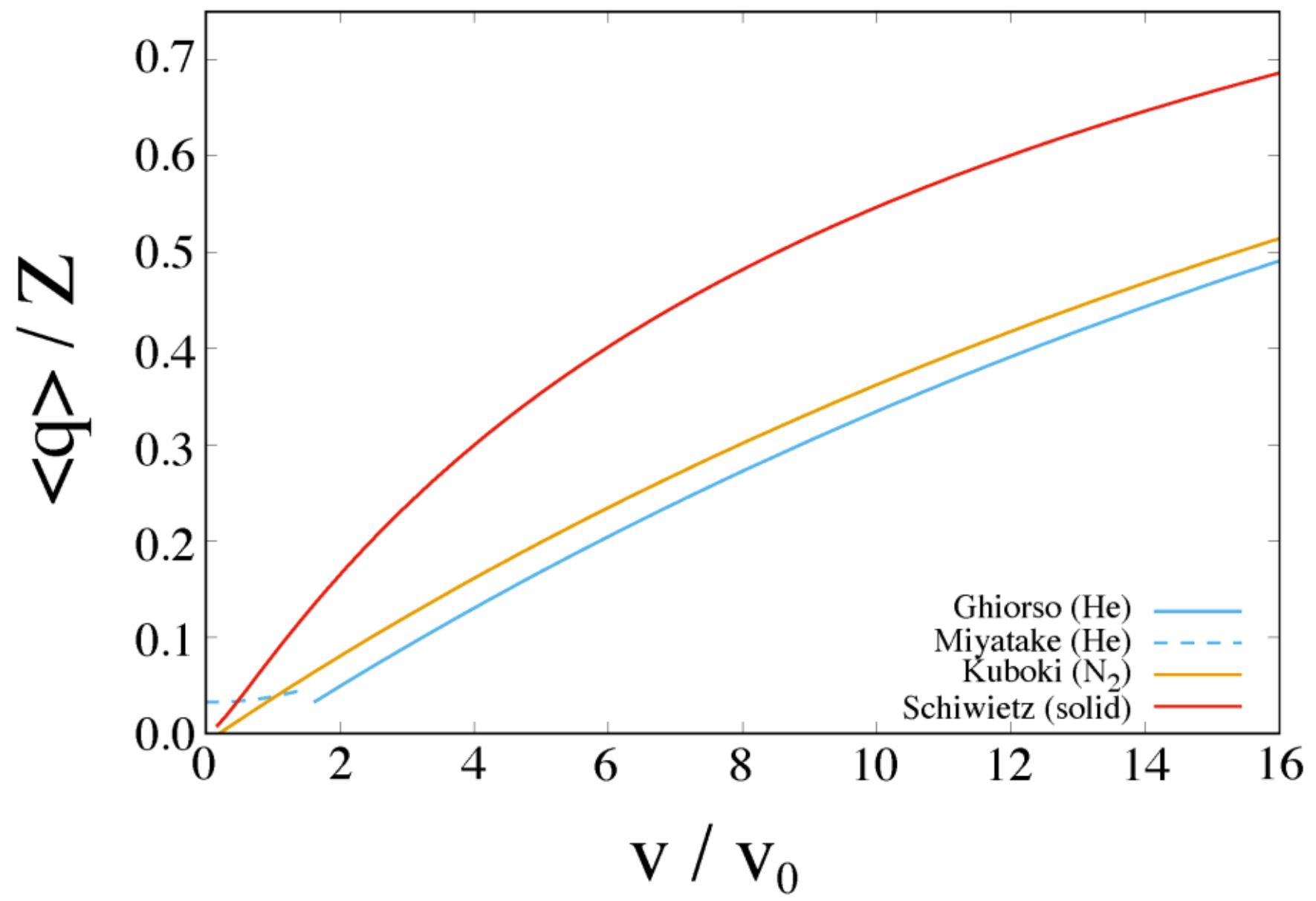

Figure 6-1. Empirical formulae of averaged charge states, $<q>$. Two cyan-colored lines show the formulae in the He gas case; Ghiorso [3] for fast velocity region and Miyatake [6] for the slow region. The $N_2$ gas case is represented by an orange-colored line [7]. A red-colored line corresponds to the solid material case [8].

The simulation code was made based on the simulation toolkit Geant4 [2] and contained both processes. Treatment of charge exchange processes in the simulation followed Ref. [3]. For the multiple scattering process the screened nuclear recoil model [4] was employed. To simulate ion trajectories correctly, appropriate empirical formulae of equilibrium charge states are needed. Two kind of gas species, He and $N_2$, were considered in the simulation. For the He gas case a formula taken from Ref. [5] was used. An empirical formula in Ref. [6] has also adopted for the low velocity region. An empirical formula developed for gas charge strippers [7] was employed in the $N_2$ gas case. Figure 6-1 shows averaged charge state as a function of $v/v_0$ based on several empirical formulae for both gas species. Averaged charge state can be treated as $\bar{q}$, when a charge state equilibrium condition is achieved.

He gas gives slightly lower charge state compared to the $N_2$ gas. The red curve in Fig. 6-1 shows the average charge state as a function of ion velocity after passing through the solid material based on the empirical formula in Ref. [8]. Initial charge distribution of ions just recoiling out the target is described by this formula. And it is noted that relatively high charge state just after the target would affect largely the particle motions both for the low gas pressure circumstance and the vacuum one.

The magnetic field at each ionic orbital point in the simulation was interpolated from the field map data calculated by using POISSON/SUPERFISH code [9]. Results for the calculated magnetic field are shown in Fig. 6-2. Details of design of the superconducting solenoid magnet are discussed in the following sections. Calculation results of GRAZING code were used for describing initial energy and momentum of each TLF ion when simulating

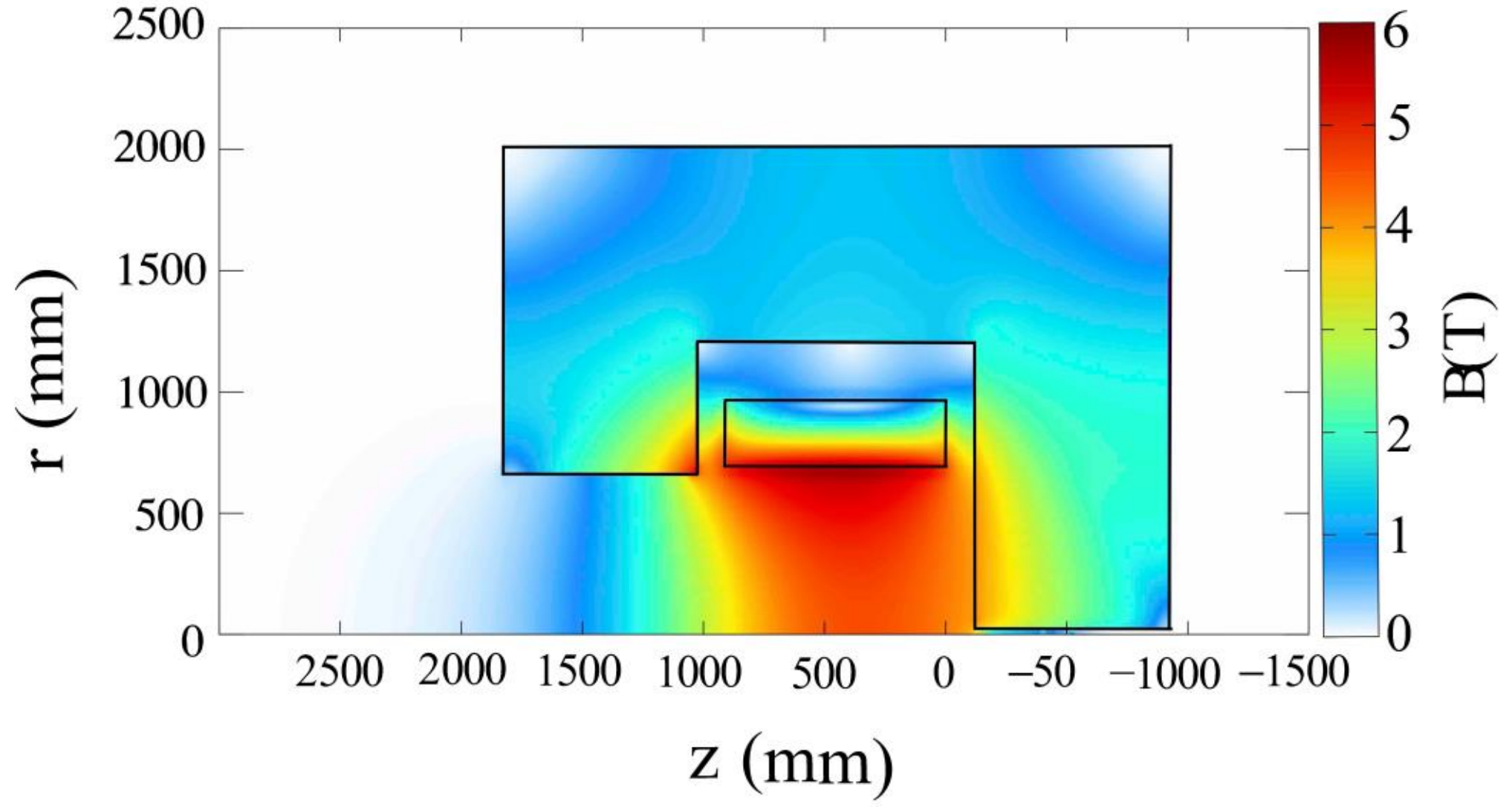

Figure 6-2. Magnetic field of the solenoid filter. Black lines indicate the coil and the yoke. A point of $z = 0$ corresponds to the target position. The maximum field strength is 5.6 T at the coil surface with operational current of 2400 A.

MNT reaction products. For the case of fusion reactions, initial kinematic parameters of reaction products were also determined under the Monte Carlo treatment of the evaporation process of the fusion reactions.

## 6-1-3. Optimization of the gas pressure

A performance of gas-filled separators could be represented by $B\rho$-resolution defined by a magnetic rigidity, $\Delta B\rho/B\rho$, even for the solenoid filter, where there is no momentum dispersive plane. The resolution $\Delta B\rho/B\rho$ is determined by the combination of several components:

$$\left(\frac{\Delta B\rho}{B\rho}\right)_{\mathrm{FWHM}} = 2.35\sqrt{\sigma_{\mathrm{q}}^2(P) + \sigma_{\mathrm{s}}^2(P) + \sigma_{\mathrm{o}}^2}\ [\%]. \qquad (6-8)$$

The first term inside the radical symbol represents dispersion of ion charge states. The ionic charge can be changed frequently via charge exchange reactions during the flight through the separator. The charge states will eventually gather around an equilibrium state if the mean free path between any charge-changing reactions is sufficiently short compared to the total flight length in the magnet. As the filling gas pressure determines the mean free path, $\sigma_q(P)$ is inversely proportional to square root of the gas pressure. The second term, $\sigma_s(P)$, represents the contribution from the multiple scattering process and is proportional to the square root of the gas pressure, in contrast to $\sigma_q(P)$. Other contributions from effects of separator optics, beam size, detector size, etc., are included in the last term $\sigma_o$, which is a minor component compared with both $\sigma_q(P)$ and $\sigma_s(P)$, and is almost independent of the gas pressure. Thus, there must be an optimum gas pressure to minimize $\Delta B\rho/B\rho$ as a function of the gas pressure.

A method to experimentally determine the optimum gas pressure has been established for the common gas-filled separators. However, in the case of the solenoid filter, the same method cannot be applied because it is impossible to measure the $B\rho$-distribution. According to Eq. (6-4), $\rho_{max}$-distributions with $\theta_{lab}$ in the limited region could be applied as an indicator how well the ions of various charge states gather around the equilibrium state. Figure 6-3 shows the simulation results of the gas pressure dependences of the $\rho_{max}$-distribution. TLF ions of $^{250}U$ produced via $^{238}U(^{248}Cm, X)$ reactions were assumed and $\theta_{lab}$ was limited to 48°-52° in the simulation. As a result, the optimum pressures of He and $N_2$ gases were found to be 1 kPa and 0.1 kPa, respectively. The $\rho_{max}$-resolution for the He case is two times better than the $N_2$ case. In addition, weak contribution of the multiple scattering to the $\rho_{max}$-distribution can be also found at the higher pressure than 1 kPa for the He case.

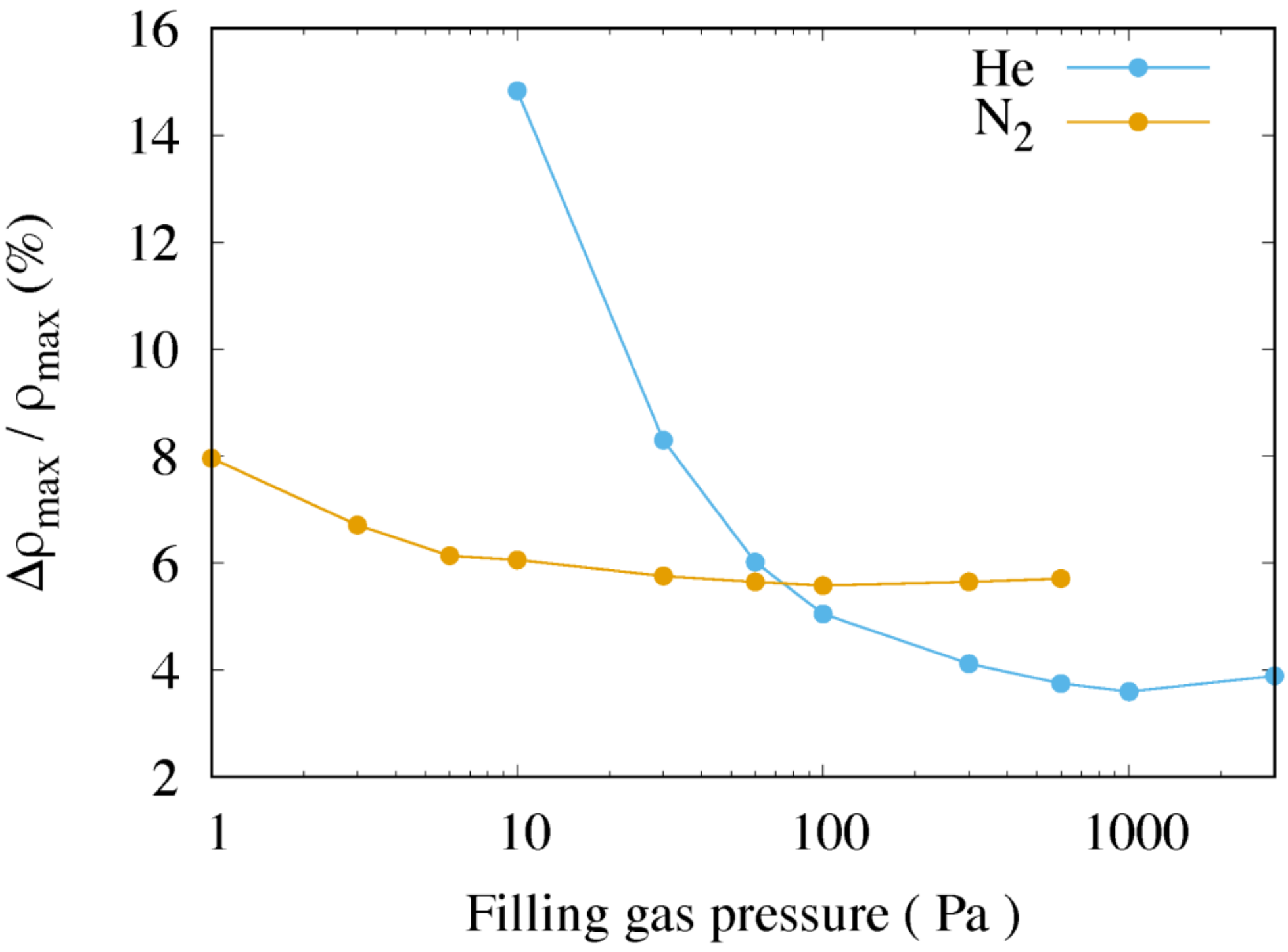

Figure 6-3. $\rho_{\mathrm{max}}$ -distribution as a function of the filling gas pressure. The details are in the text.

### 6-1-4. Collection and separation abilities for MNT and fusion reactions

Collection efficiency of the MNT and fusion reaction products and the separation ability of unwanted ions in the solenoid filter were evaluated. In the simulation, nuclei of interest are collected by inserting a catcher plate as a probe disk. It has sufficient thickness to stop the incoming ions. Unwanted nuclei will be suppressed by two types of the stopper plates which limit the emission angles from the production target as shown inf Fig. 4-4. Optimal positions of the target and catcher plate are determined. A helium gas cell in the actual set-up discussed in Sec. 4-4. will be installed at the position of the catcher plate.

#### 6-1-4-1. Multi-nucleon transfer reaction products

Three TLFs – $^{250}$U via $^{238}$U($^{248}$Cm, *X*), $^{250}$U via $^{238}$U($^{238}$U, *X*), and $^{200}$W via $^{238}$U($^{198}$Pt, *X*) – were considered in the simulation. The expected collection efficiencies for these TLFs are compared to confirm the capabilities of the solenoid filter. The collection efficiencies for $^{250}$U using the 500 mm diameter catcher plate placed at various distance from the target are shown in Fig. 6-4. Differences in the optimum distances for the various filling gas conditions are clearly seen. Vacuum condition where relatively highly-charged states of ions are maintained throughout their trajectories, gives the highest collection efficiencies with the shortest distances. The collection efficiencies with $N_2$ and He gas conditions are not so different compared with the vacuum case, but the optimum distances are significantly different from each other depending on their different equilibrium charge states.

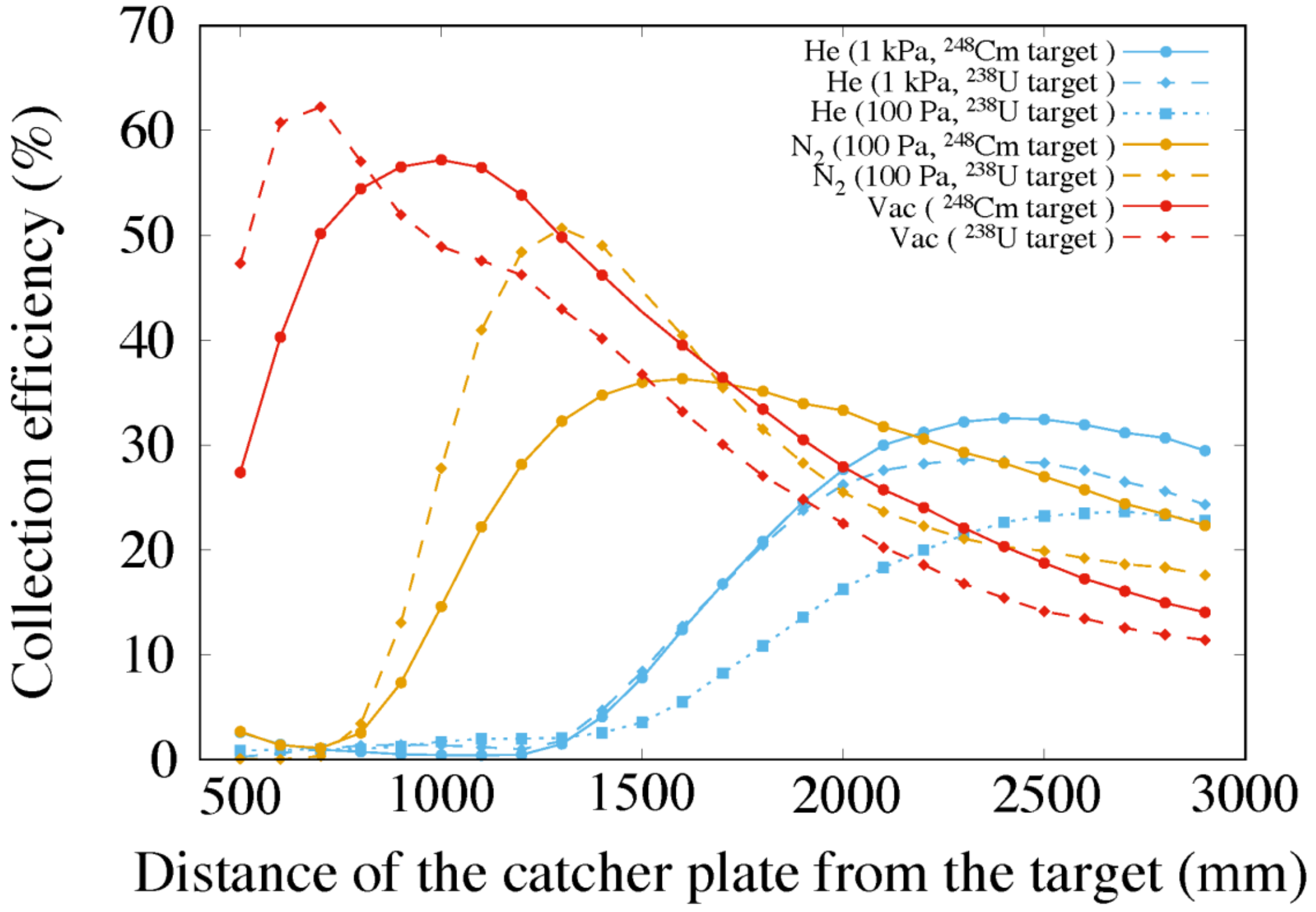

Figure 6-4. Collection efficiencies as a function of the distance from the target. Simulation results by considering the TLFs of $^{250}$U produced via $^{238}$U($^{248}$Cm, $X$) (circle and solid lines) and via $^{238}$U($^{238}$U, $X$) (diamond and square, non-solid lines) are shown. The diameter of the catcher plate was set to be 500 mm.

In Fig. 6-4, the different efficiency profiles between the various reaction systems can also be discussed. Results for the $^{238}$U($^{238}$U, $X$) case show narrow peak profiles compared with those of $^{238}$U($^{248}$Cm, $X$) case. These narrow efficiency profiles are originated from the more localized initial $E$–$\theta_{lab}$ distribution of TLF in the $^{238}$U($^{238}$U, $X$) reaction as indicated in Fig. 6-5.

Figure 6-6 shows the comparisons between the collection efficiencies for two different TLFs: $^{250}$U and $^{200}$W. Weak dependence on $A$ and $Z$ of $A/Z^{1/3}$ is visible in the case of both $N_2$ and He gas conditions, in contrast to vacuum condition. Collection efficiencies around 30% can be achieved in all the reaction system in the present simulation under $N_2$ and He gas conditions.

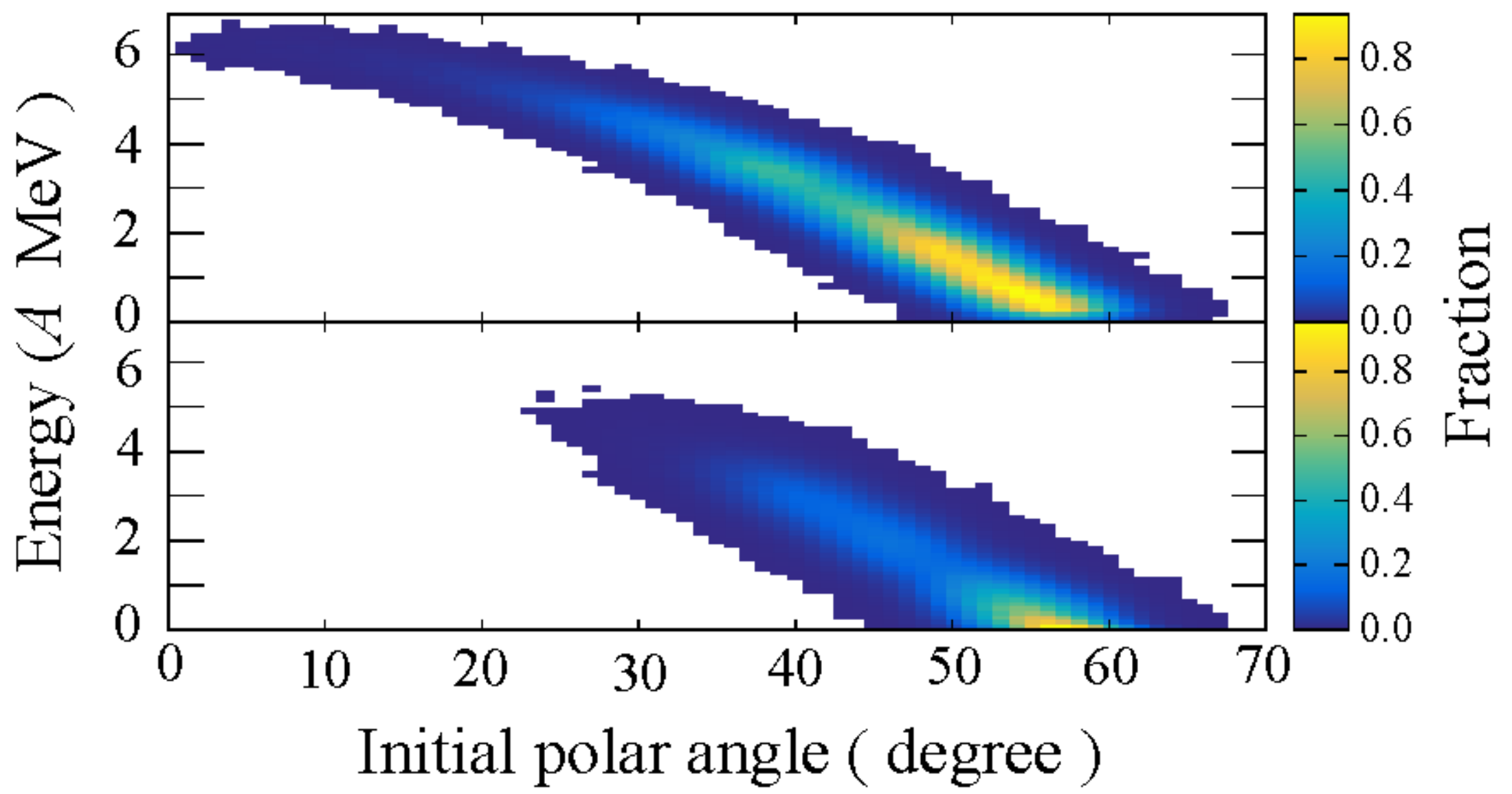

Figure 6-5. Initial $E$–$\theta_{lab}$ distribution of the TLF $^{250}$U. Upper panel: $^{238}$U($^{248}$Cm, $X$) case. Lower panel: $^{238}$U($^{238}$U, $X$) case.

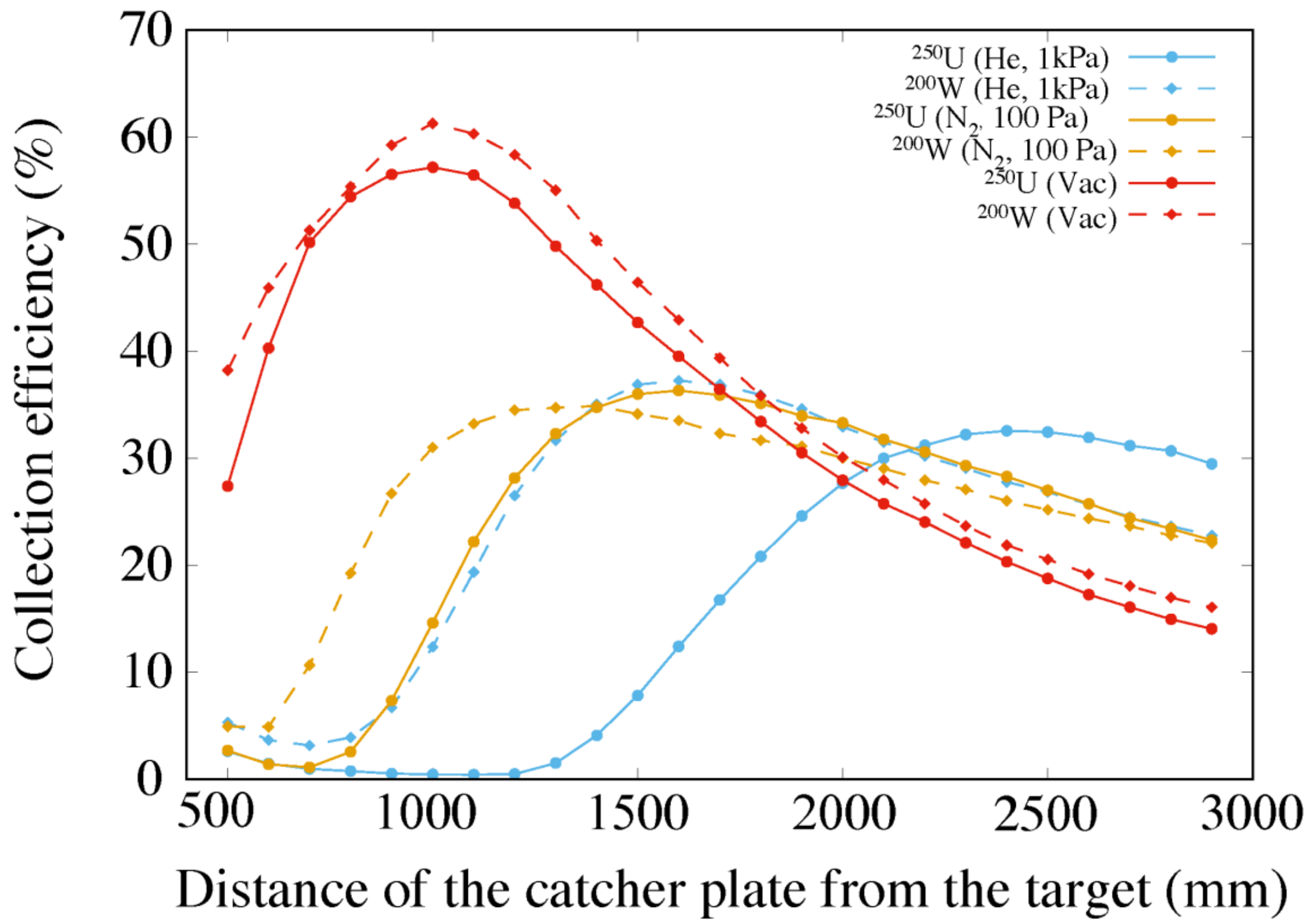

Figure 6-6. Collection efficiencies as a function of the distance from the target. Circles with solid lines indicate the results of the TLF $^{250}$U produced via $^{238}$U($^{248}$Cm, $X$) and diamonds with dashed lines show the case of the TLF $^{200}$W produced via $^{238}$U($^{198}$Pt, $X$). The 500 mm diameter catcher plate was assumed.

The performance of separation of the unwanted particle, such as, unreacted primary beam, target recoil, and other light ions, is a crucial property of the solenoid filter in terms of the capability of the multi-segmented gas cell as mentioned in Sec. 4-4. In the MNT reactions, the kinetic energy distribution of the reaction products is similar to that of the elastically scattered particles of primary beam and target material. This situation results in there being only a small difference between the trajectories of the MNT reaction products and scattered primary beams and/or emitted. This feature will be emphasized as one goes to more symmetric reaction system. Only their different $\rho_{max}$ in the solenoid magnet can be utilized for the separation. And smaller resolution in Fig. 6-3 would result better separation of TLF from other background particles.

Figure 6-7 shows $\rho_{max}$-distributions with $^{238}U(^{238}U, X)$ reaction under different conditions. Each histogram shows $\rho_{max}$-distribution of ions collected by the 500 mm diameter catcher plate. The distance between the catcher plate and the target were set to be 1000, 1300, and 2400 mm for vacuum, $N_2$, and He gas conditions, which give maximum collection efficiencies as shown in Fig. 6-4. And for the cases of the scattered primary beam ("Proj" in Fig. 6-7) and target recoils ("Recoil" in Fig. 6-7) the initial polar angular distributions were limited to 30°-70° because we are interested in the separation of the MNT reaction products from unwanted particles emitted with similar angles. The numbers of trials were the same in all cases, thus the integrated count of each histogram (each line) represents the collection efficiency for each

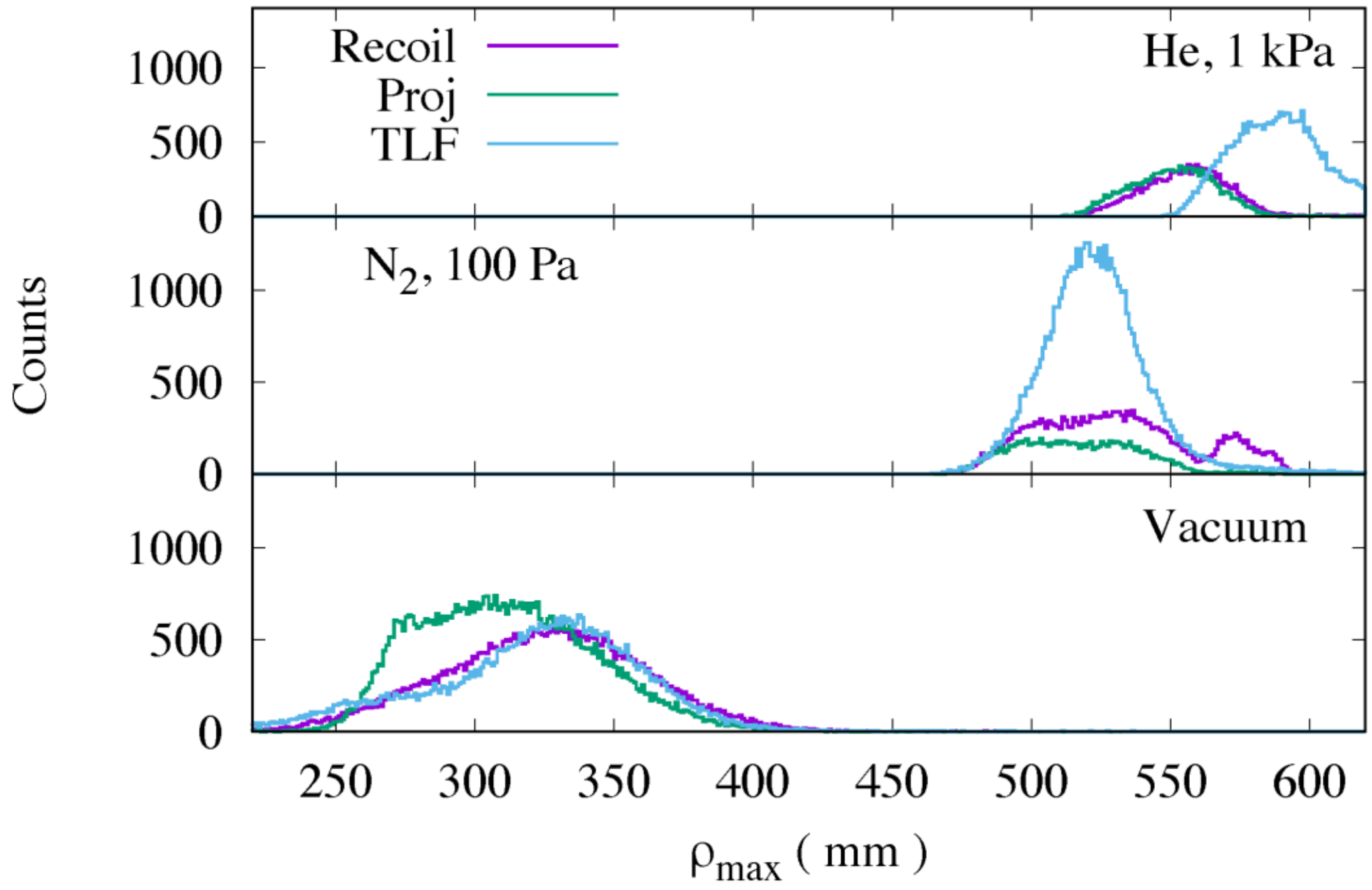

Figure 6-7. Simulated $\rho_{max}$-distribution of ions, which can be collected in a catcher plate with 500 mm in diameter, for the TLFs, Proj, and Recoil in $^{238}U(^{238}U, X)$ reaction. Top panel: 1 kPa of He gas. Middle panel: 100 Pa of $N_2$ gas. Bottom panel: vacuum.

case. Under the He gas condition (top panel of Fig. 6-7), the distributions of TLF and Proj/Recoil partially overlap and there is a room to cut Proj and Recoil ions inside the solenoid filter while maintaining the collection efficiency.

In contrast to this, the three histograms are fully overlap in the vacuum condition (bottom panel of Fig. 6-7). It is difficult to suppress the scattered projectile and elastic recoil components by using any type of particle stopper without loss of the TLFs. Same as the case of vacuum condition, in the $N_2$ gas filling case, the three histograms all overlap but the peak heights of "Proj" and "Recoil" are lower than in the case of the TLF's. It means some part of the scattered projectile and elastic recoil ions are lost in the solenoid filter.

These features are clearly shown in the values of transmission probability ratio between the TLF and "Proj"(or "Recoil") as shown in Table 6-1. In the He and $N_2$ gas conditions, in order to suppress the unwanted components in the solenoid filter, the values are in the range of 2 ~ 4. In contrast, the ratio remains nearly 1 in the vacuum. There is a room to improve the ratio by using the elastic stopper plate in the He gas condition. If one set the stopper plate radius to be 565 mm, the ratios are enhanced by a factor of 2~4 to be about 10. This result indicates that one order of magnitude intense beam can be utilized at the He gas condition than the vacuum condition. Thus, even if the vacuum condition is the most efficient for collecting the TLFs, He gas condition can be powerful and be ultimate for the use of intense $^{238}$U primary beam of more than 1 pμA.

Table 6-1. Ratios of the transmission probability $P_x$ (X=TLF, Proj, and Recoil).

| | He gas, 1 kPa | | $N_2$ gas 0.1 kPa | Vacuum |
|---|---|---|---|---|
| | No gate | $\rho_{max} > 565$ mm | | |
| $P_{TLF}/P_{Proj,\ 30^\circ-70^\circ}$ | 2.2 | 10 | 4.7 | 0.82 |
| $P_{TLF}/P_{Recoil,\ 30^\circ-70^\circ}$ | 2.3 | 7.3 | 2.2 | 1.0 |

### 6-1-4-2. Fusion reaction products

Here, we will discuss on a possible applying the solenoid filter to researches with using the fusion-evaporation reactions. Fusion-evaporation reactions to produce (super)heavy nuclei can be sorted into two types: cold and hot fusion. Cold fusion (CF) reactions typically use lead and/or bismuth elements as target materials and are characterized by a few-neutrons evaporation channel in the deexcitation process of compound nuclei. In the case of hot fusion (HF) reactions, actinoid elements are employed as target materials and the dominant deexcitation channel becomes at or above the 5 neutrons evaporation channel. The large number of evaporation particles in HF results a wide spread in velocity and angular distributions of the evaporation residues (ERs), which are the final products of the fusion reactions.

The gas-filled separators with dipole magnetic field are known to be the powerful tools for collecting and separating ERs, because they can collect ER's having large velocity spread by the effects of filling gas. But the angular acceptance of the gas-filled separator is limited and would have much room for improvement, when we apply this device to the HF reactions. Because ER's of HF reactions have wider spreads in velocity and angular distributions compared to ones of CF reactions as mentioned before.

It has been discussed that one of possible reactions for producing unknown (super)heavy nuclei would be the HF reactions with charged-particle evaporation channels together with neutron evaporations. To study such kind of HF reactions, a separator having an extremely large angular acceptance is required and the solenoid filter could be a good candidate.

In order to confirm the capability of the solenoid filter for HF reactions, the same simulation technique used for MNT reactions was employed. As a typical example of HF reaction, $^{238}$U($^{22}$Ne, $X$)($E_{lab}$ = 113.8 MeV) was considered. The experimental parameters, such as, target material, target thickness, and filling gas pressure, were adopted from the experiments reported in Ref. [10].

Figure 6-8 shows expected polar angular distributions of fusion evaporation residues, $^{255}$No and $^{253}$Fm, which are produced through (5n) and (α3n) channels in the HF reactions of

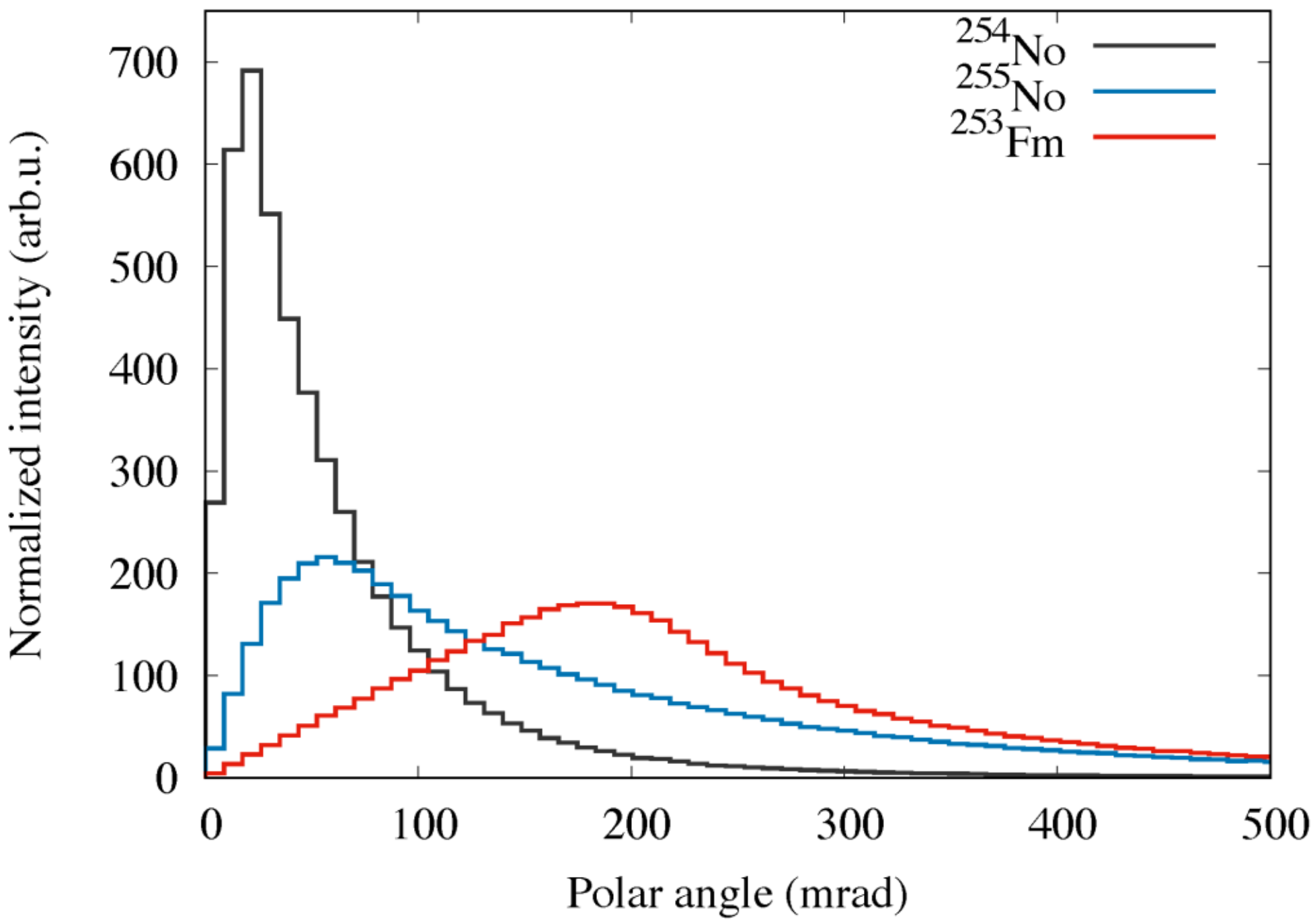

Figure 6-8. Estimated angular distributions of evaporation residues, $^{254,255}$No, and $^{253}$Fm. Residues, $^{255}$No and $^{253}$Fm are produced through HF reactions, while $^{254}$No is produced through CF reaction. Multiple scattering of ER's in the production target has been taken into account. Details are written in the text.

$^{238}U$ and $^{22}Ne$ system as mentioned above. In addition, the yield distribution of $^{254}No$ is estimated as a typical distribution of ER's in CF reaction, where we assume the reaction of $^{208}Pb(^{48}Ca,2n)^{254}No$($E_{lab}$ = 216 MeV and 450 μg/cm$^2$ Pb-target). As can be found in this figure, angular distributions of ER's in HF reactions are larger than the CF reaction. The ER's with (α$x$n) channel would have larger distribution around 200 mrad in this case.

Figure 6-9 shows the simulated trajectories of $^{255}No$ (5n-channel) and $^{253}Fm$ (α3n-channel) without stopper plates and Faraday cup. The relatively small angular distributions of ERs compared with the MNT cases give longer $z_0$ according to the Eq. (6-5) and trajectories do not return to the solenoid axis. The collection efficiency when the catcher plate was placed at 2400 mm from the target are 73.5% and 60.3% for $^{255}No$ and $^{253}Fm$, respectively. The initial angular distribution of $^{253}Fm$ was wider than that of $^{255}No$ due to recoiling effect of the α particle evaporation process as shown in Fig. 6-8. It results in a little bit small collection efficiency relative to the $^{255}No$ case.

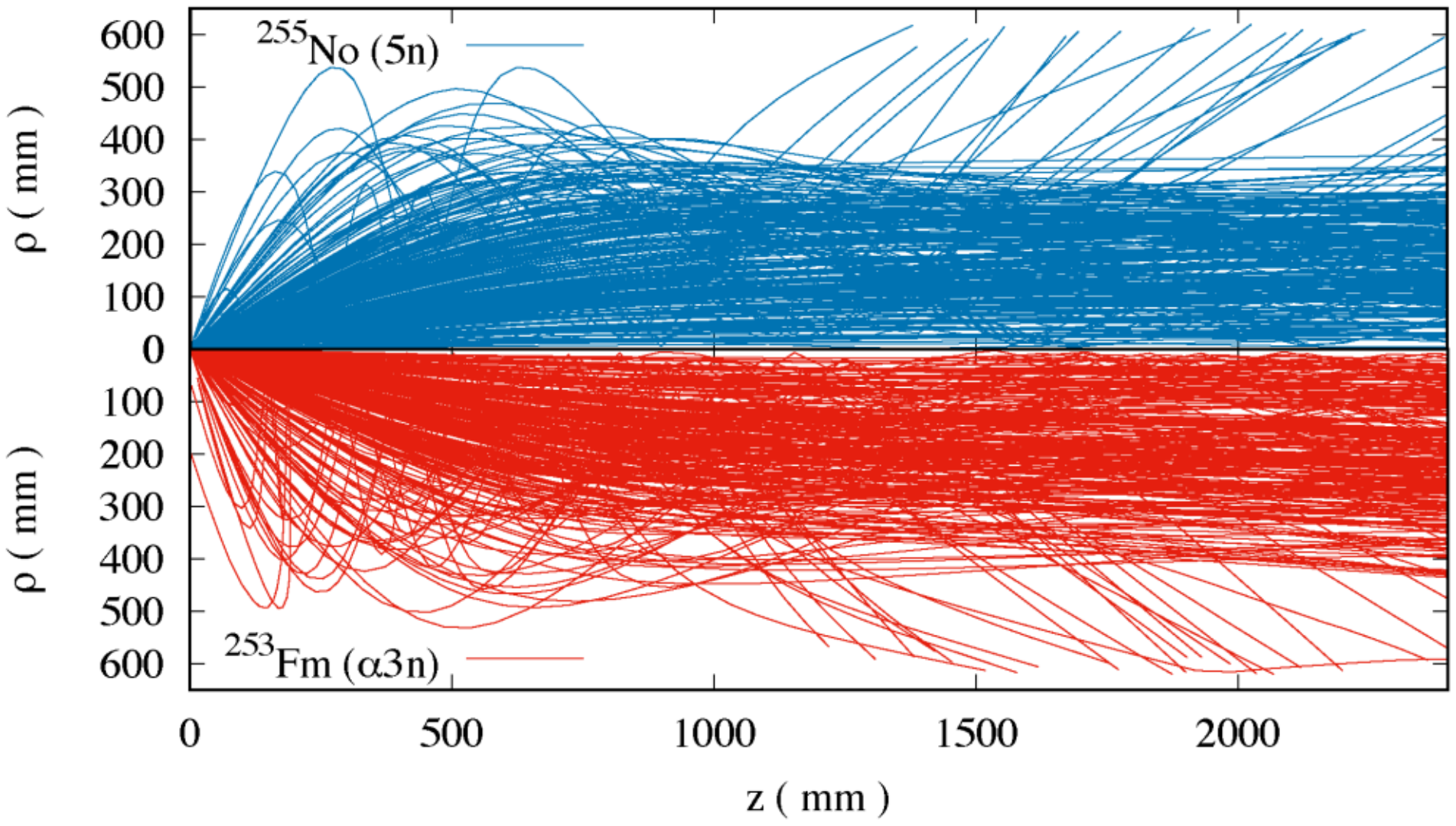

Figure 6-9. Ion trajectories of $^{255}No$ and $^{253}Fm$ in the solenoid filter. He gas pressure was 37 Pa and the thickness of $^{238}U_3O_8$ target was 370 μg/cm$^2$.

Further optimization is needed for starting researches with using HF reactions by the solenoid filter. At least, a Faraday cup and stopper plates will also be necessary when the solenoid filter is applied to HF reactions. Here, the influence of the disk-type stopper plate (as shown in Fig. 4-4) to the collection efficiency was evaluated at first by using the simulation for finding out the merit of the gas-filled solenoid.

The expected efficiencies as a function of the radius of the stopper plate are shown in Fig. 6-10. In the simulation, the stopper plate was located at a distance of 250 mm from the target. It covers the polar angle in the rage from 0° to 7°, for example, with its radius of 30 mm. Collection efficiencies would be gradually reduced to about 50% at this condition. It is noted this efficiency can be achieved even under such exaggerated stopper geometry with as large as 120 mrad, which is similar to the vertical angular acceptance of GARIS-II as mentioned below. Hence, the gas-filled solenoid filter may offer very clean circumstance to the downstream devices such as the helium gas-cell with keeping reasonable collection efficiency.

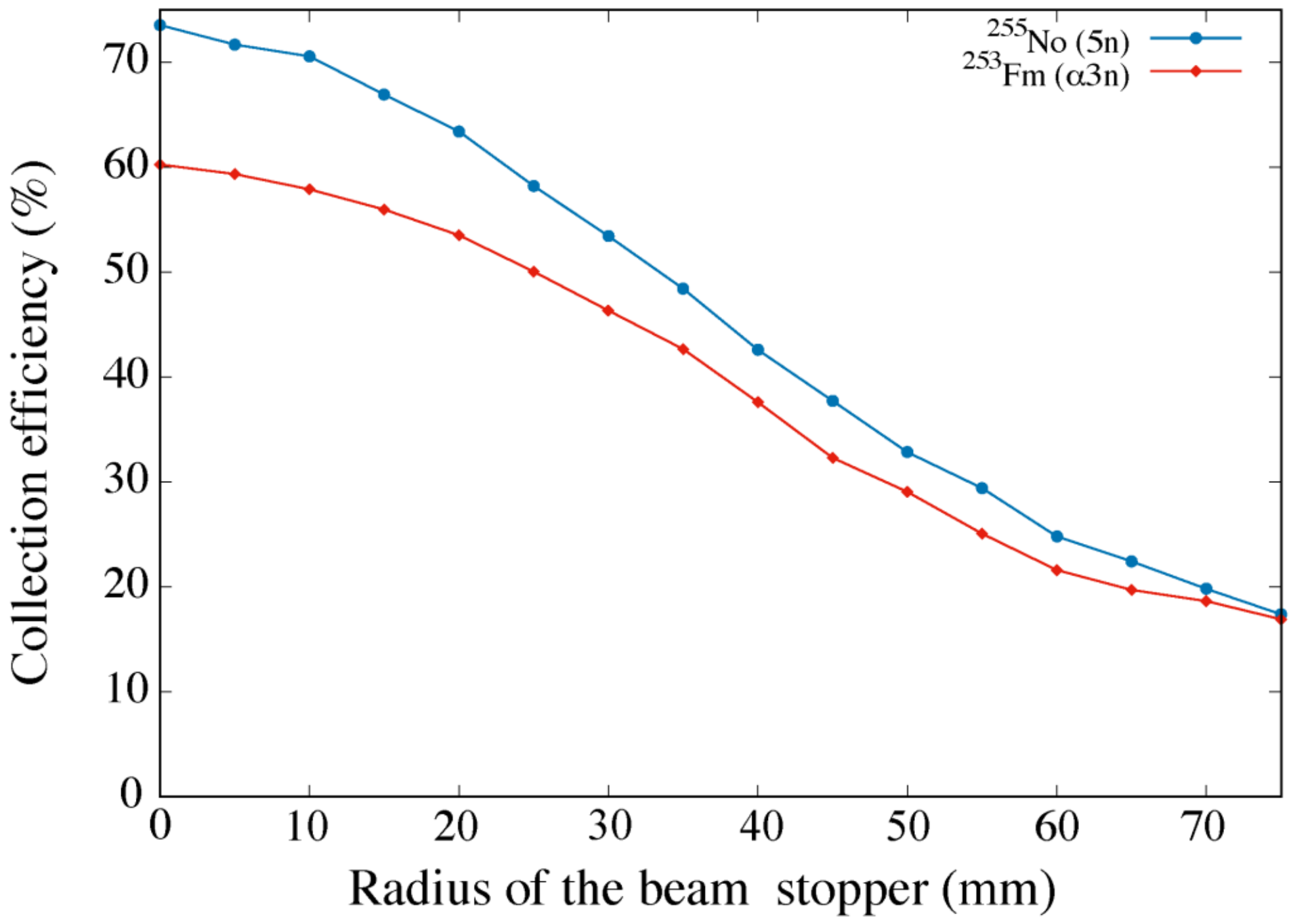

Figure 6-10. The collection efficiencies calculated for the 5n and α3n evaporation channels as functions of the beam stopper radius. As expected from the larger angular and momentum spreads, the charged particle evaporation channel has a reduced efficiency.

### 6-1-4-3. Comparison with GARIS-II

Herein, we will discuss the expected performance of the solenoid filter in comparison with an existing gas-filled recoil ion separator in RIKEN, GARIS-II [11], which is based on dipole magnetic field. The GARIS-II has been used for studies of nuclear physics and chemistry of the (super)heavy nuclei. The angular acceptances of GARIS-II are ±47 mrad and ±110 mrad in the horizontal and vertical directions, respectively. Those values are relatively large for this type of recoil ion separators in the world.

The collection efficiency of GARIS-II was evaluated with the same simulation applied to the gas-filled solenoid filter by replacing the ion-optical configuration of the GARIS-II. The results are tabulated in Table 6-2 with ones of the gas-filled solenoid.

In the case for collecting the TLF $^{250}$U produced via $^{238}$U($^{238}$U, $X$) with GARIS-II, the simulation presumed the primary beam impinged on the target at an angle of 55 degrees for the optical axis of the GARIS-II, although such beam deflection is not realistic for the existing system. While, the simulation can evaluate a maximum fraction of TLFs being accepted within the limited solid angle by setting the deflection angle of the GARISS-II to the angle of the intense TLF yield region as shown in Fig. 6-5. Despite such favorable condition, the efficiency of GARIS-II for the $^{250}$U nuclei was only 0.6%. In contrast with this, the collection efficiency for the solenoid filter under He gas condition reaches 28.9% by the help of azimuth integration characteristic at large polar angle. The solenoid filter will be a powerful tool for researches with the use of MNT reactions.

The simulated efficiency for $^{255}$No ER produced via $^{238}$U($^{22}$Ne, 5n) of 11.7% for GARIS-II setup is in well agreement with the experimentally obtained value 15% [12]. It has been verified the accuracy of this simulation. The collection efficiencies of the solenoid filter for the HF reaction products, $^{255}$No and $^{253}$Fm, are also higher than those of GARIS-II by a factor of 5 to 30, respectively. These results indicate a great advantage of the gas-filled SC solenoid filter to collect the HF reaction products compared with existing gas-filled separators.

Table 6-2. Results of the simulations. For the details, see the text.

| Reaction | Efficiency (%) | | |
|---|---|---|---|
| | Solenoid | GARIS-II | |
| | | Sim. | Exp. |
| $^{238}$U($^{238}$U, *X*)$^{250}$U | 28.9 | 0.6 | - |
| $^{238}$U($^{22}$Ne, 5n)$^{255}$No | 73.5 | 11.7 | 15 |
| $^{238}$U($^{22}$Ne, α3n)$^{253}$Fm | 60.3 | 2.0 | - |

## 6-1-5. Summary

The requirements for the solenoid filter and its performance evaluation via the simulation have been discussed. A solenoid magnet with a field of around 5.5 T and a 1.4 m diameter bore is suitable in order to efficiently collect the MNT reaction products. The collection efficiencies of various MNT reaction products in the gas-filled conditions are evaluated to be in the range of 30-50% depending on the reaction kinematics. The separation of unwanted particles has also been investigated. Comparison of these results makes it clear that the optimal filling gas condition would be helium gas of 1 kPa for accepting the use of the 1 puA primary beam of $^{238}$U.

In addition, the solenoid filter will be able to collect the HF reaction products with high efficiency of around 60%. The comparisons of the collection efficiencies between the solenoid filter and the presently existing GARIS-II separator indicate a superiority of the solenoid filter. These results show that the gas-filled SC solenoid filter will be an ideal tool for collecting the various reaction products having large emission angle.

# 6-2. Design of the SC solenoid filter, cryostat and yoke

## 6-2-1. Overview of the superconducting coil

The main parameters of the solenoid coils for KISS-II are summarized in Table 6-3.

Table 6-3. Basic parameters of SC solenoid coils for KISS-II.

| Item | Value |
|---|---|
| Magnet Parameters | |
| Coil inner diameter | 1380 mm |
| Coil outer diameter | 2042 mm |
| Coil length | 1000 mm |
| Cryostat inner bore diameter | 1260 mm |
| Cryostat outer diameter | 2260 mm |
| Cryostat length | 1150 mm |
| Central magnetic field | 5.0 T |
| Maximum magnetic field on conductor | 5.5 T |
| Operation current | 2600 A |
| Layers × turns | 20 × 105 |
| Inductance | 8.9 H |
| Stored energy | 30 MJ |
| Conductor Parameters | |
| Type | Aluminum clad NbTi/Cu cable |
| Cross sectional size w/o insulation | 8×15 $mm^2$ |
| Cross sectional size with insulation | 8.6×15.6 $mm^2$ |
| Insulation | Epoxy prepreg Polyimide film/Boron-free glass tape |
| Area ratio (NbTi/Cu/Al) | 1/1/17 |
| Critical current | 16,000 A (5 T), 13,000 A (6 T) and 9,500 A (7 T) at 4.3 K |

## 6-2-2. Conductor

The superconductor that will be used for the superconducting solenoid has a rectangular shape consisting of a Rutherford-type Nb-Ti cable located at the center of conductor and a stabilizer housing. The same superconductor is already used for the superconducting ring cyclotron at the RIBF facility in RIKEN. The conductors' cross-sectional area measures 8 mm by 15 mm. The stabilizer material is Al-alloy with 1000 ppm Ni, which gives a high 0.2%-yield strength of about 55 MPa at room temperature. The residual resistivity ratio is greater than 800-900. The cable is composed of ten strands of 1.18 mm in diameter, each being composed of about 960 filaments of 28 μm in diameter. The Cu/Nb-Ti ratio, the residual resistivity ratio of Cu stabilizer and the critical currents of the strand are 0.84, 100 and 16 000 A (5 T), 13 000 A (6 T) and 9 500 A (7 T) at 4.3 K, respectively. The conductor's cross section is shown in Fig. 6-11.

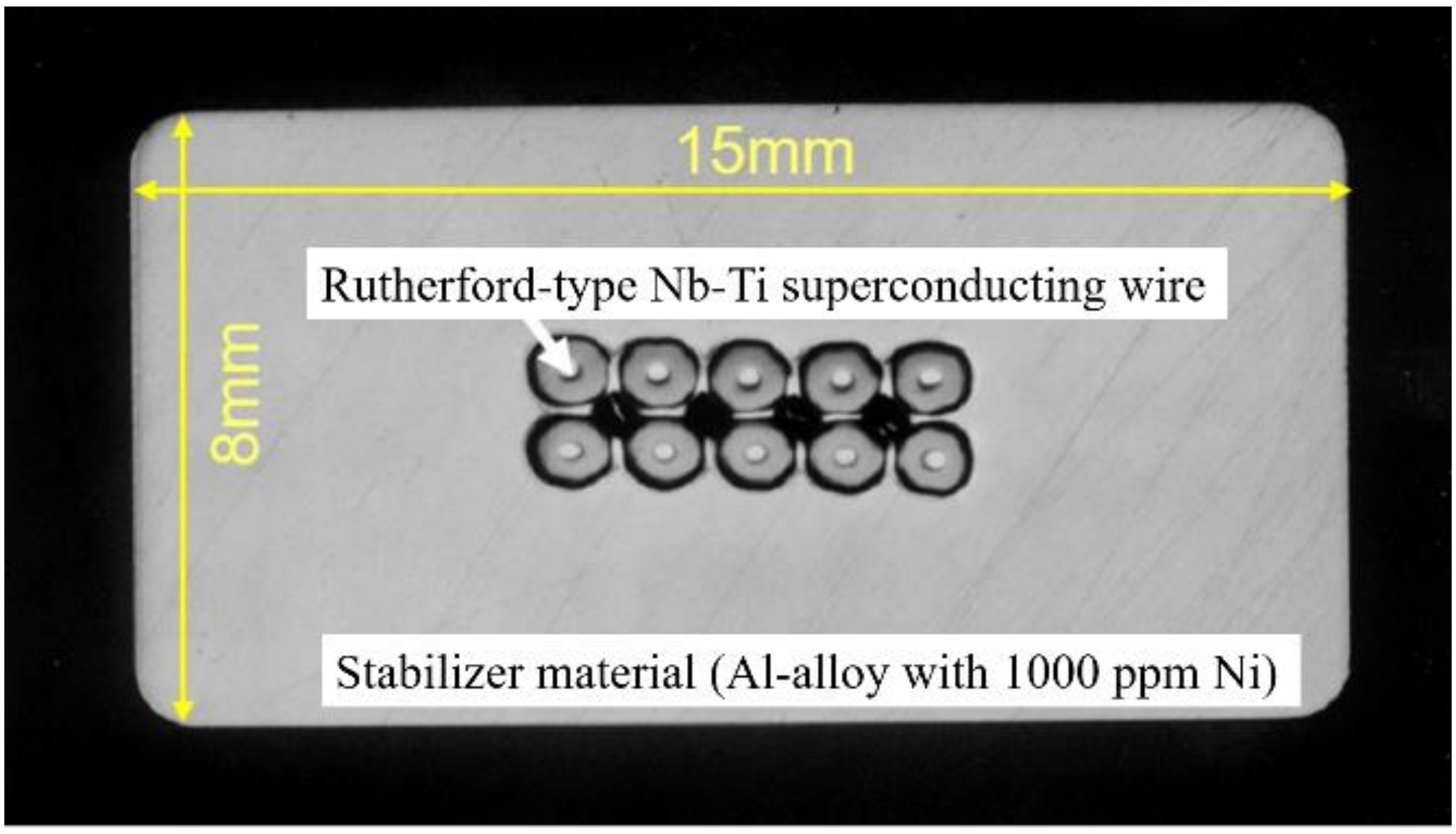

Figure 6-11. Superconducting wire.

## 6-2-3. Coil block

The coils will all be wound from the aluminum stabilized conductor insulated with a polyimide film covered by boron free glass with BT+Epoxy prepreg. The BT+Epoxy prepreg was utilized to ensure the radiation resistivity of the coils. Another blend of BT+Epoxy resin with silica filler will be smeared over the insulated conductor as it is wound up on a mandrel. The resin will be cured at 170~180℃. The proposed coil structure is shown in Fig. 6-12. The 20-layer 1000 mm long coil will be conductively cooled by 1 mm thick pure aluminum strips installed between each layer. The strips will be connected to cooling pipes that will be cooled by forced flow 2-phase helium. The connections will be made at the coil ends, at which some of the strips will be obstructed by the ramp and splice structures that are needed alternately at every other coil layer end.

The solenoid will have a total of 2100 turns (105 horizontal turns/layer × 20 vertical layers). The maximum current of the main coil will be 2600 A, giving a maximum magneto-motive force per sector of 5.46 mega-ampere-turn. The cross-sectional area of the coil will measure 300 mm by 1000 mm. The maximum current density will then be 19 A/mm$^2$. The maximum magnetic field at the coil will be 5.5 T. The temperature margin of the coil is estimated to be around 2 K. The maximum total stored energy of the main coils will be about 30 MJ.

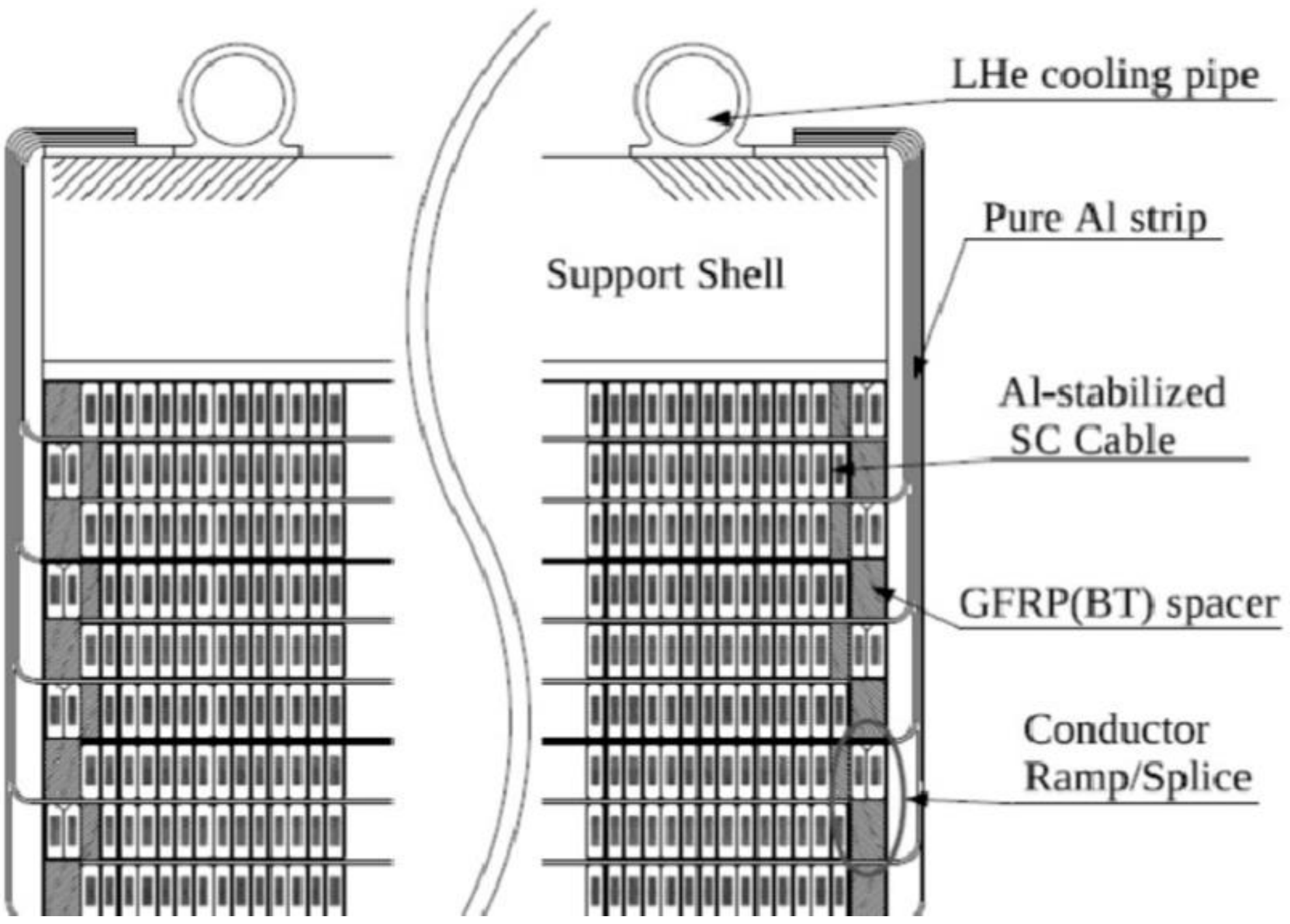

Figure 6-12. Coil block for the COMET.

### 6-2-4. Cryostat

The 4.5 K cold mass covered with a total area of ~21 m$^2$ of 70 K thermal radiation shields. The thermal shields will be composed of copper plates and multi-layer insulators. Gas He (GHe) cooling tubes are attached to the surfaces of the plates. The cold mass will be supported by a total of 7 thermal-insulation support links as shown in Fig. 6-13. The support system will be composed of four vertical supports, two horizontal supports and one longitudinal support. These supports will be made of CFRP rods. The 4.5 K cold mass and the thermal shields will be installed in the cryostat. The thickness of the plates for cryostats and the dimensions of the gaps between the 4.5 K cold mass and the cryostat are listed in Table 6-4. Total dimensions for the cold mass and cryostat are shown in Fig. 6-14.

Table 6-4. Dimensions for the cryostat.

| Item | | Value |
|---|---|---|
| Vacuum vessel | | |
| | Outer | 35 mm |
| | Inner | 15 mm |
| | End | 15 mm |
| Gap for thermal insulation | | |
| | Outer | 75 mm |
| | Inner | 25 mm |
| | End | 45 mm |

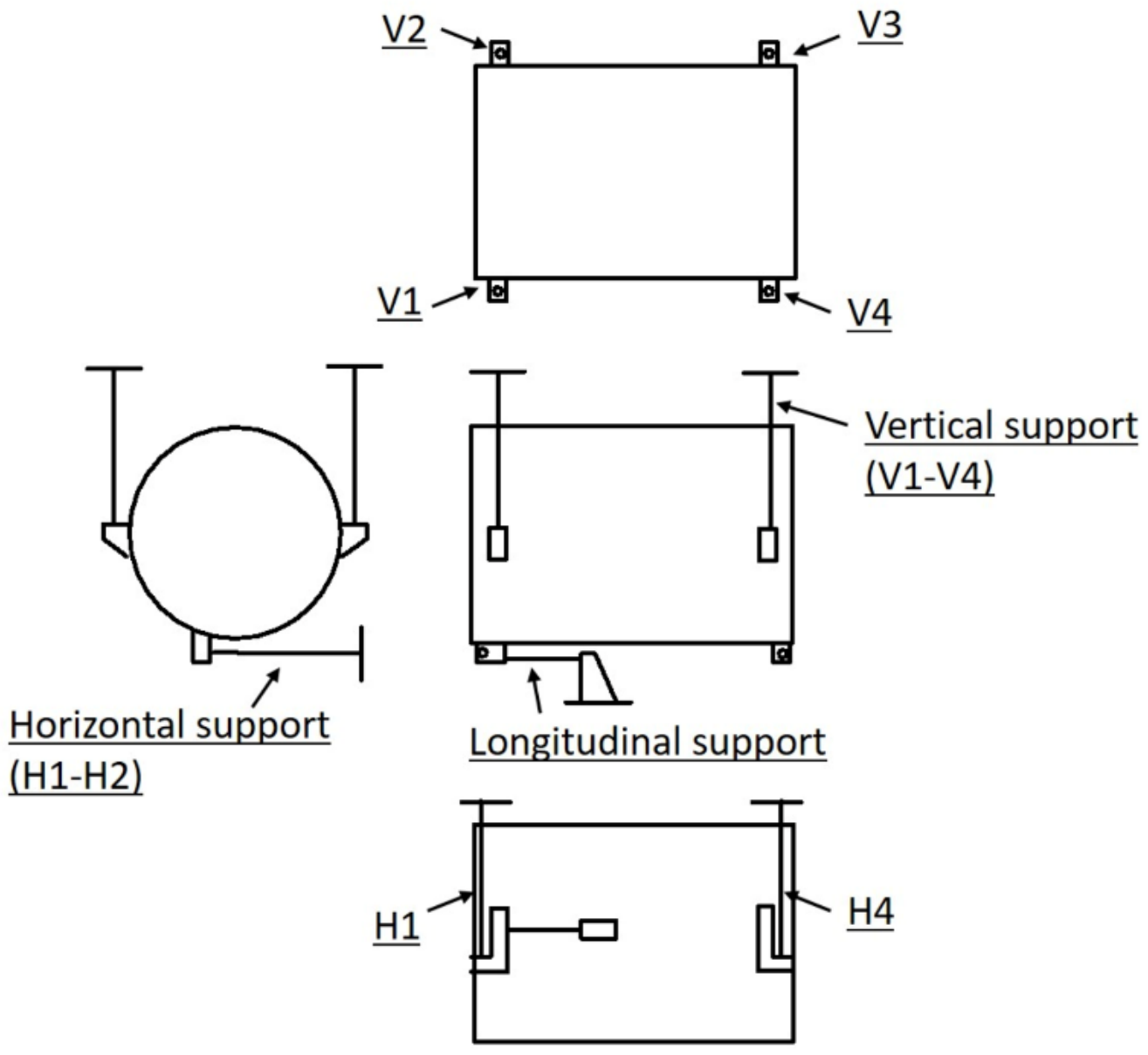

Figure 6-13. Proposed structure of thermal insulation supports for the solenoid.

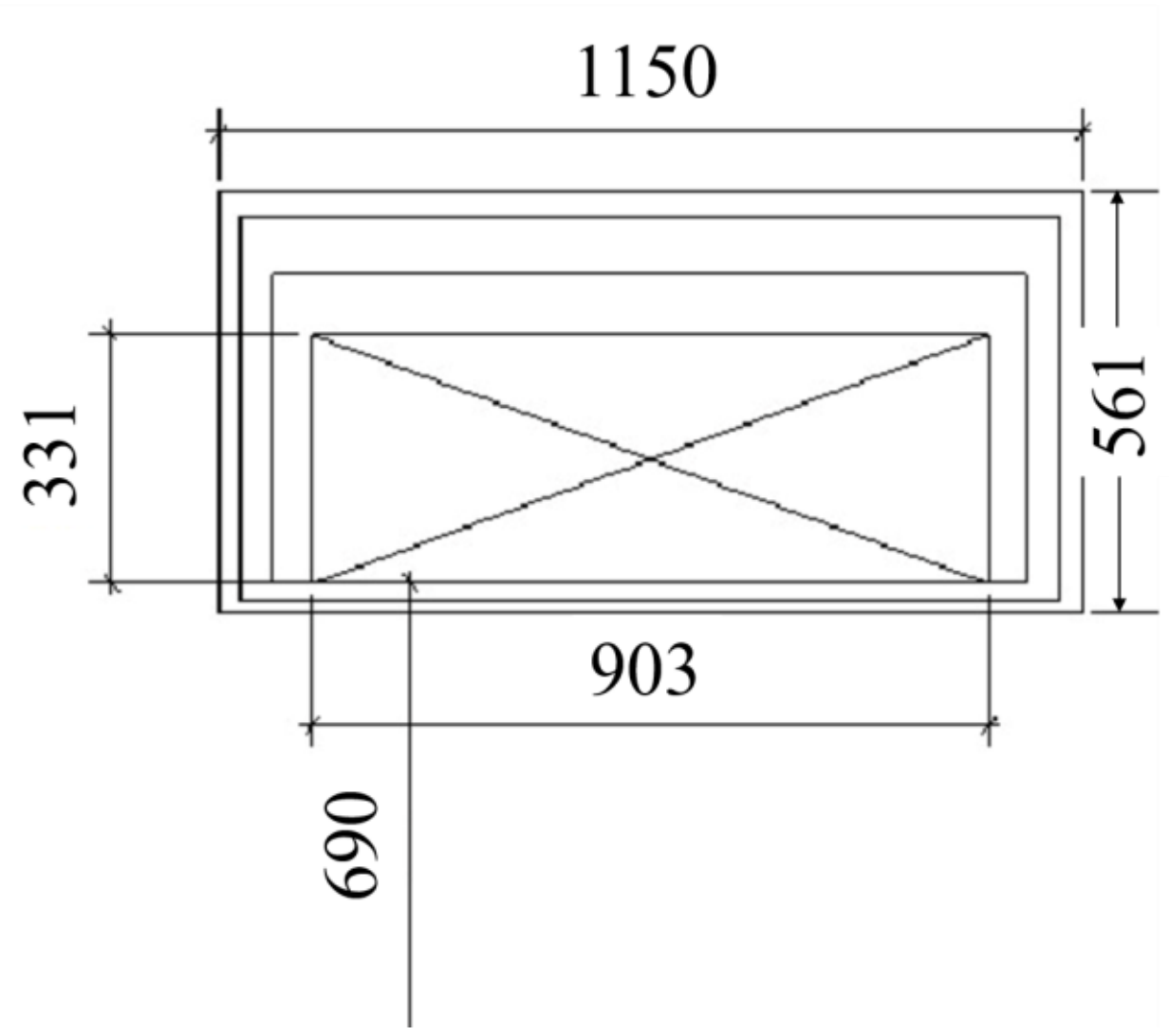

Figure 6-14. Dimension of the proposed coil and the cryostat.

### 6-2-5. Current leads in the service port

To induce a 5.5 T field will require injecting 2500 A into the superconducting solenoid coil. Doing so will require a pair of current leads that must carry up to 2700 A from terminals at room temperature to the coil at 4.5 K. The current leads will be also cooled by a set of Gifford-McMahon type cryocoolers (hereafter, GM cryocooler) conductively coupled to the coil, which is a challenging technical issue. Fortunately, this design challenge has already been met by the 3000 A current leads developed for the COMET beam line at J-PARC. We will adapt that design for use with the proposed KISS-II solenoid filter. Figure 6-15 shows a side view of a current lead used in the COMET magnet. It is composed of a feedthrough at a vacuum wall, a phosphorous copper lead from the feedthrough to a 60 K stage and a high temperature superconducting lead with a SUS backup lead from the 60 K stage to a 4 K stage. The 60 K stage is cooled by two to three single-stage GM cryocoolers (CH-110LT by SHI) and the 4 K stage is cooled by two-stage GM cryocooler with a capacity of 1.0 W at 4 K. A service port where the current leads are installed also has double-stage GM cryocoolers for coil. These cryocoolers, the current leads and other utilities are distributed on the service port with a diameter about 1.5 m as shown in Fig. 6-16. The number of double-stage cryocoolers depends on the heat load into the superconducting coil. Fortunately, radiative heat load from the target will be negligible, so we only need to consider heat leak through the magnet cryostat. Radiative heating at the coil (1.1 W), conductive heating through the coil supports (0.8 W), and radiative heating from the service ports (0.9 W) add to less than 3 W. Consequently, three cryocoolers may have sufficient cooling capacity.

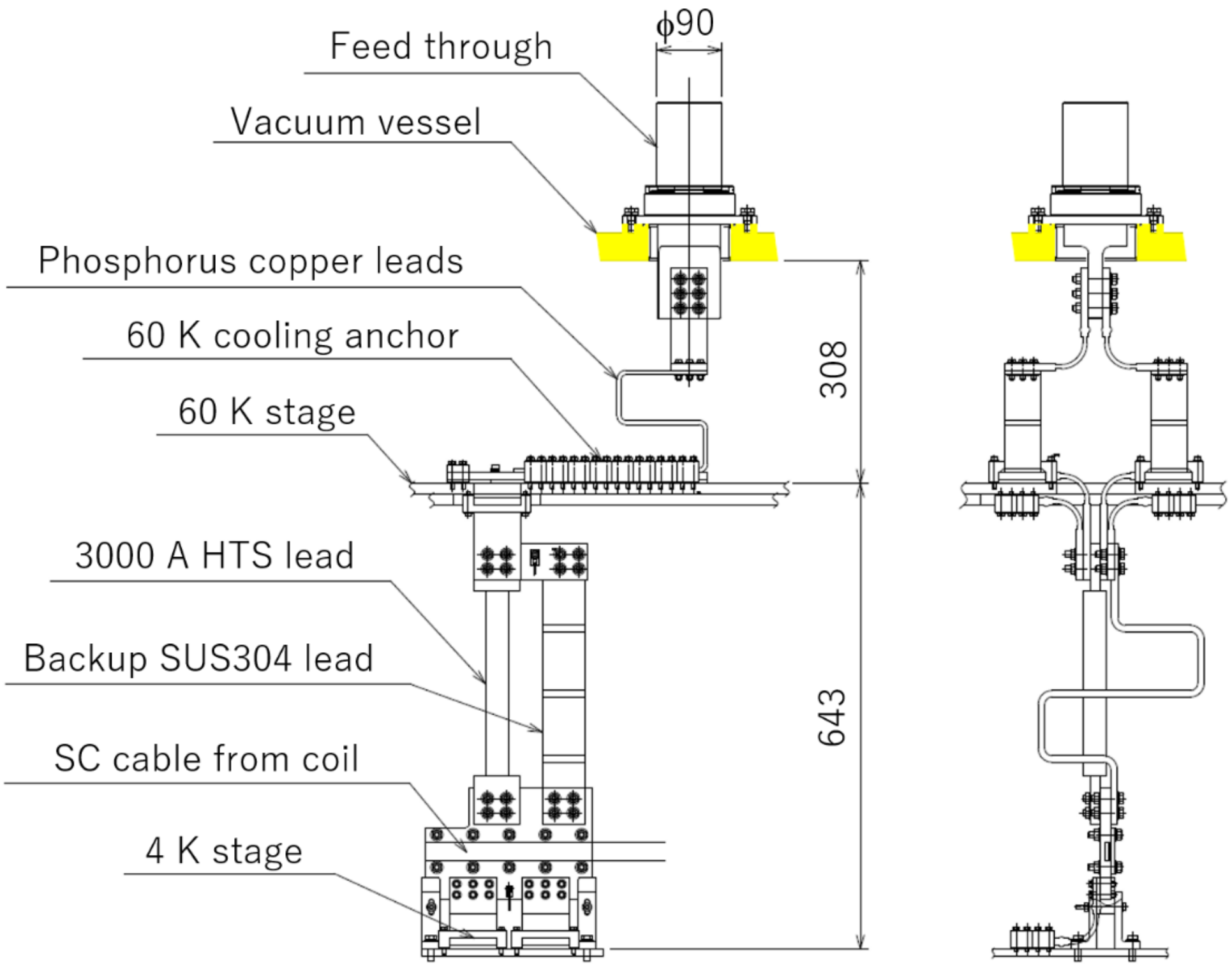

Figure 6-15. Current lead for the COMET magnet.

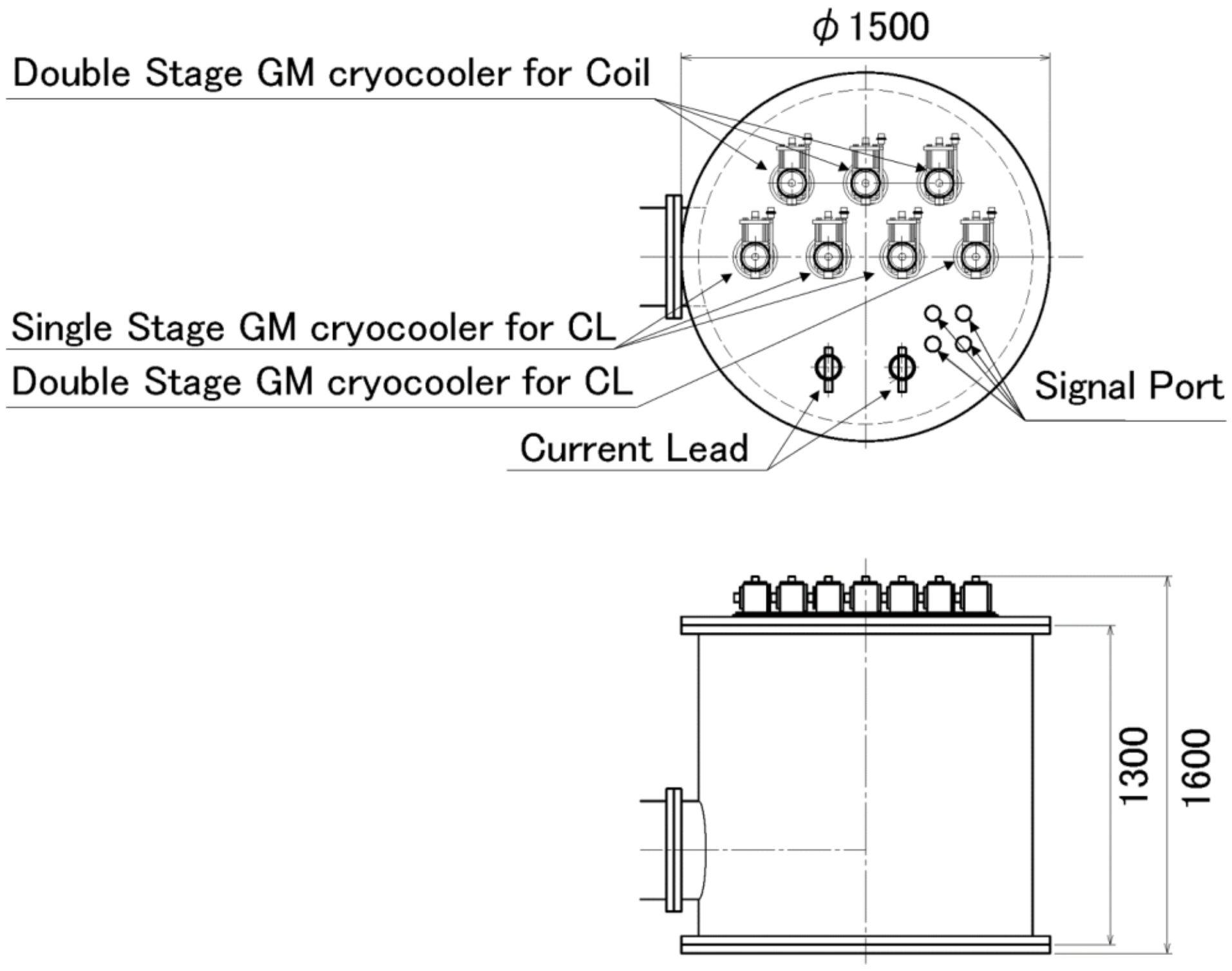

Figure 6-16. Top flange of the service port. CL refers to current leads.

## 6-2-6. Quench safety

A simple check concerning about coil quench safety is carried out by estimating the hot spot temperature at the point where the quench originates. The function shown in Fig. 6-17 indicates that the temperature at the quench origin increase adiabatically by joule heating until the current decays to zero. The time constant of current decay can be set with an external dump resistor, which must be included in the coil excitation circuit for quick current decay. Considering the limitations of the ground electric insulation, the maximum voltage over the dump resistor must be kept below 1 kV, so its resistivity is set about 0.2 Ω due to the excitation current of 2600 A. As the inductance of this magnet will be 8.88 H, the time constant for current decay $\tau_d$ will be about 44.4 s. Then according to the function curve described in Fig. 6-17, a hot spot temperature of 60 K is estimated, which will be low enough to ensure the coil safety.

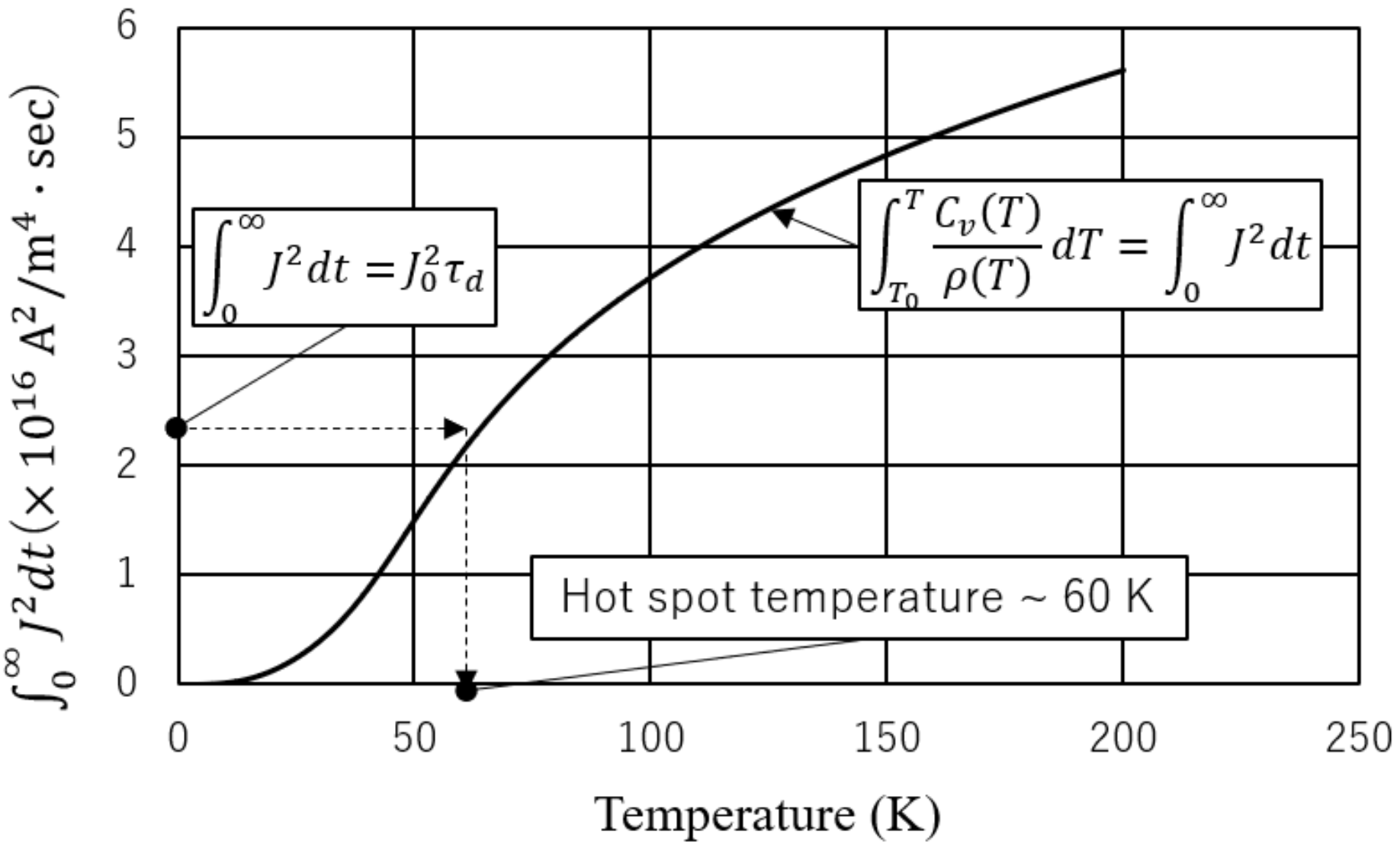

Figure 6-17. Hot spot temperature curve.

# 6-3. Fringing field to the other devices within the KISS-II facility

## 6-3-1. Optimization of yoke shape

In the design of a solenoid magnet, the shape of the yoke must be determined with consideration for the magnitude of the magnetic field leakage from the solenoid to the outside.

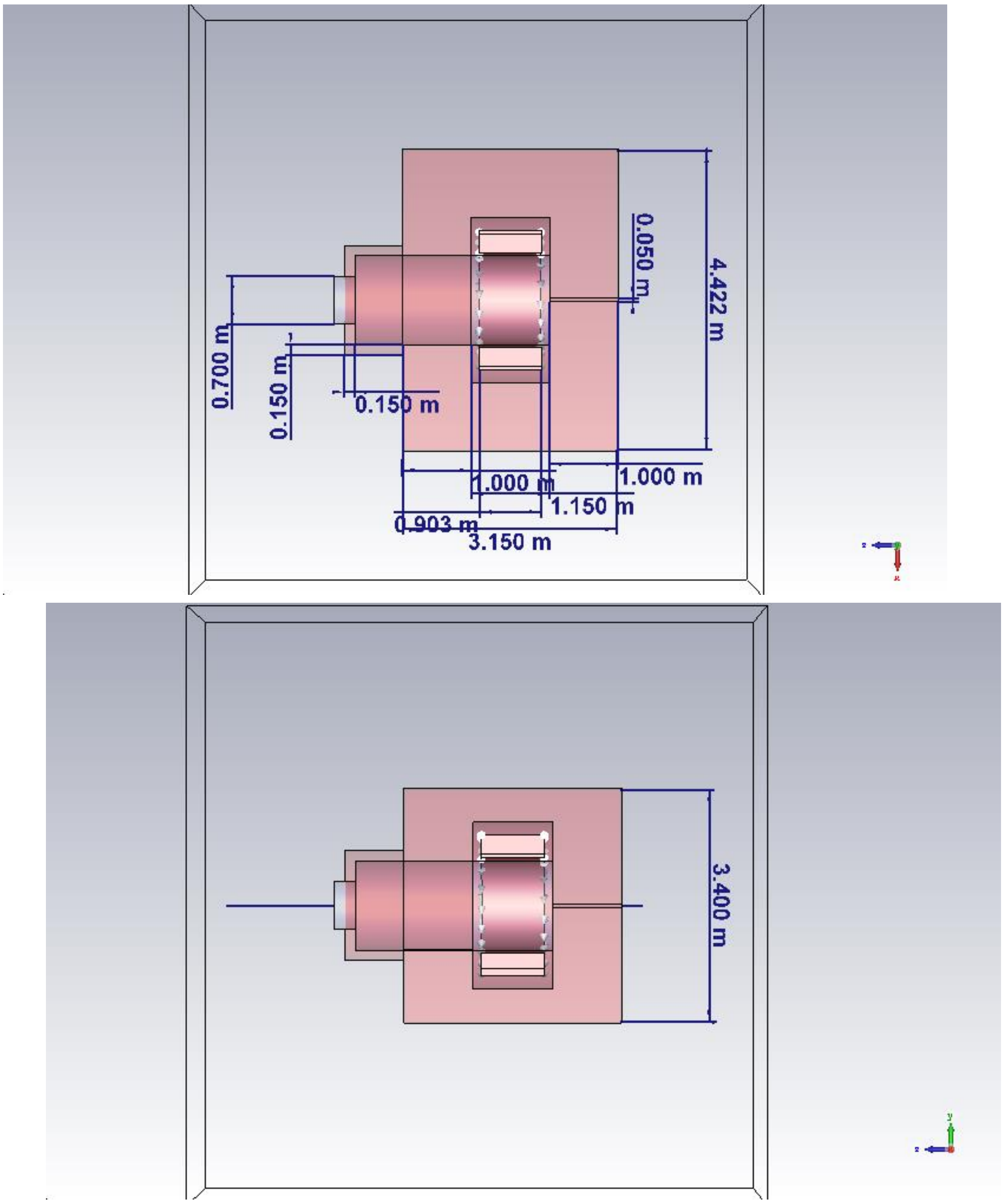

Figure 6-18. Input dimensions for field calculation. Upper (lower) figure is horizontal (vertical) view.

The solenoid magnet consists of an annular superconducting coil, a yoke that encases the coil and constitutes a magnetic circuit, and a field clump that suppresses the magnetic field leakage to the front of the beam direction. In this section, we present the optimization of the yoke in order to reduce the leakage field. The magnetic field was calculated using three-dimensional static magnetic field solver (M solver) of Dassault's CST [13]. Three cases were compared, all under the assumption that the beam line would be 1.7 m above the floor. In all cases iron was placed to a thickness just enough to not interfere with the floor, and the effect was calculated for iron extending horizontally to maximum distances of 1.95 m, 2.05 m, and 2.15 m from the beam axis. The input geometry with a horizontal size of 3.9 m in the case of 1.95 m iron depth is shown in Fig. 6-18. The fringe field was calculated from the proposed solenoid design of a single superconducting coil having 2100 turns and supporting a maximum current of 2600 A. In particular, to evaluate the magnetic field at the surface of the poles, which is important from the viewpoint of magnetic shielding, eight straight lines were defined at the outside with a distance of 1 mm from the yoke surface, and the distribution of the magnetic field along each line was considered. The evaluation lines are horizontal and vertical lines along the upstream and downstream end faces of the solenoid magnet, parallel to the beam direction through the top and side faces of the solenoid magnet, and horizontal and vertical lines through the field clamp end face connected to the gas catcher (see Fig. 6-19).

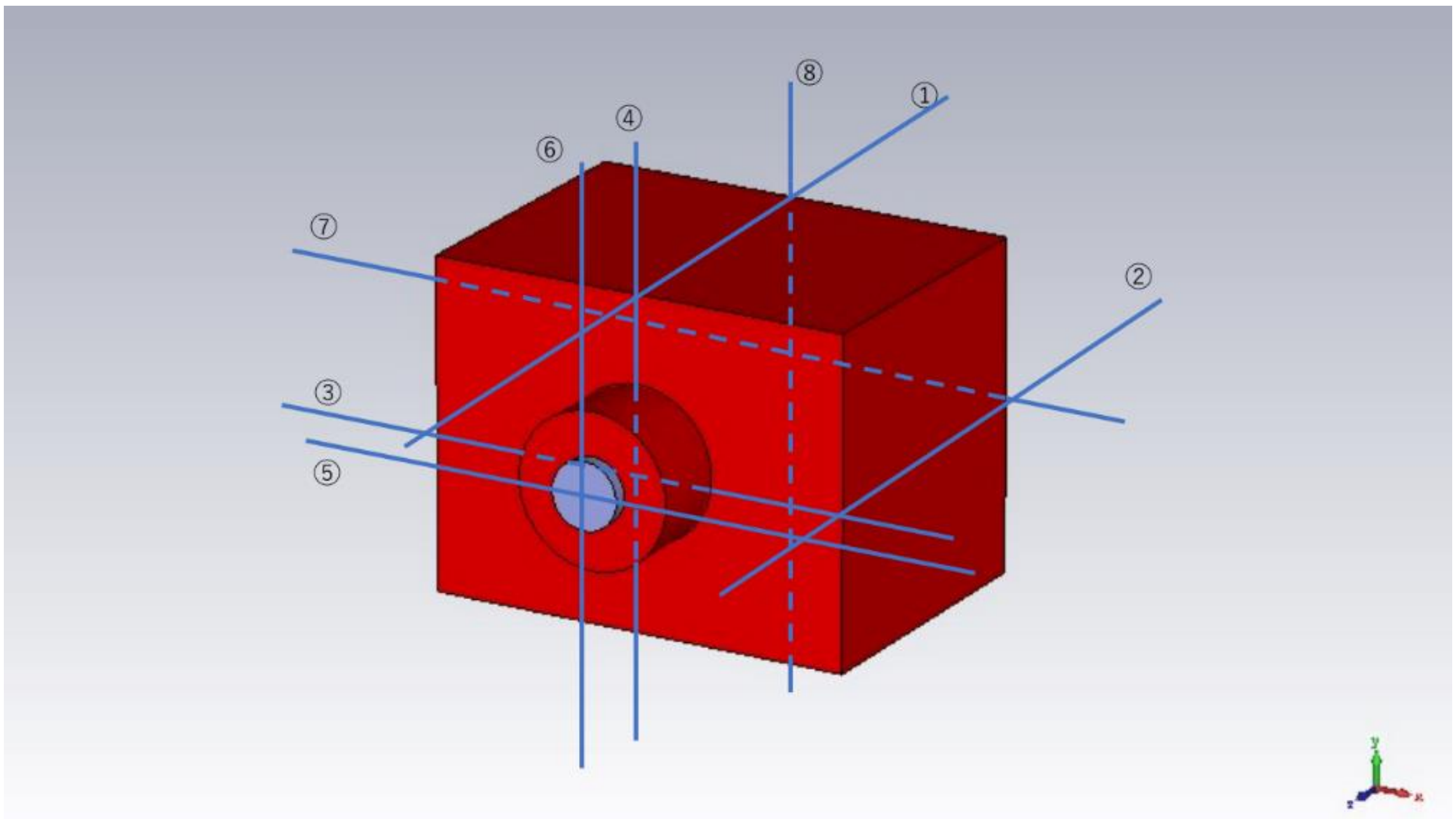

Figure 6-19. Lines along with the magnetic field was evaluated.

In the following, the results are presented in Cartesian coordinates, where the $Z$-axis is defined as parallel to the beam axis and in the same direction, and the origin is placed on the inner surface of the upstream pole; the *X*-axis is parallel to the horizontal direction and the *Y*-axis is parallel to the vertical direction. As a result of the magnetic field calculation, the magnetic field (in unit of T) distribution on *Z*-axis, shown in Fig. 6-20, was obtained. The maximum value of the magnetic field on the *Z*-axis was larger than 4.8 T, regardless of the yoke dimensions. The magnetic field intensities along line 1 to line 8 at the surface of the poles obtained as a result of the magnetic field calculation are shown in Figs. 6-21 through 6-28. The green, blue, and red lines in the graphs correspond to the magnetic field intensities for maximum width of the iron yoke: 3.9 m, 4.1 m, 4.3 m. This result indicates that the magnetic field at the surface is reduced by increasing the thickness of the yoke due to magnetic shielding. Setting a criterion that the magnetic field leakage to the outside should not exceed 30 mT, the magnetic field strength at the end face upstream of the beam with a horizontal size of 3.9 m exceeds the criterion within a distance of about 1 m from the beam axis. However, considering that it is near the beam line, the implication is simply that we must avoid installing magnetic field sensitive equipment in this length of beam line. Therefore, it is concluded that the size of the solenoid magnet's yoke should be at least 3.9 m in the horizontal direction and 3.4 m in the vertical direction.

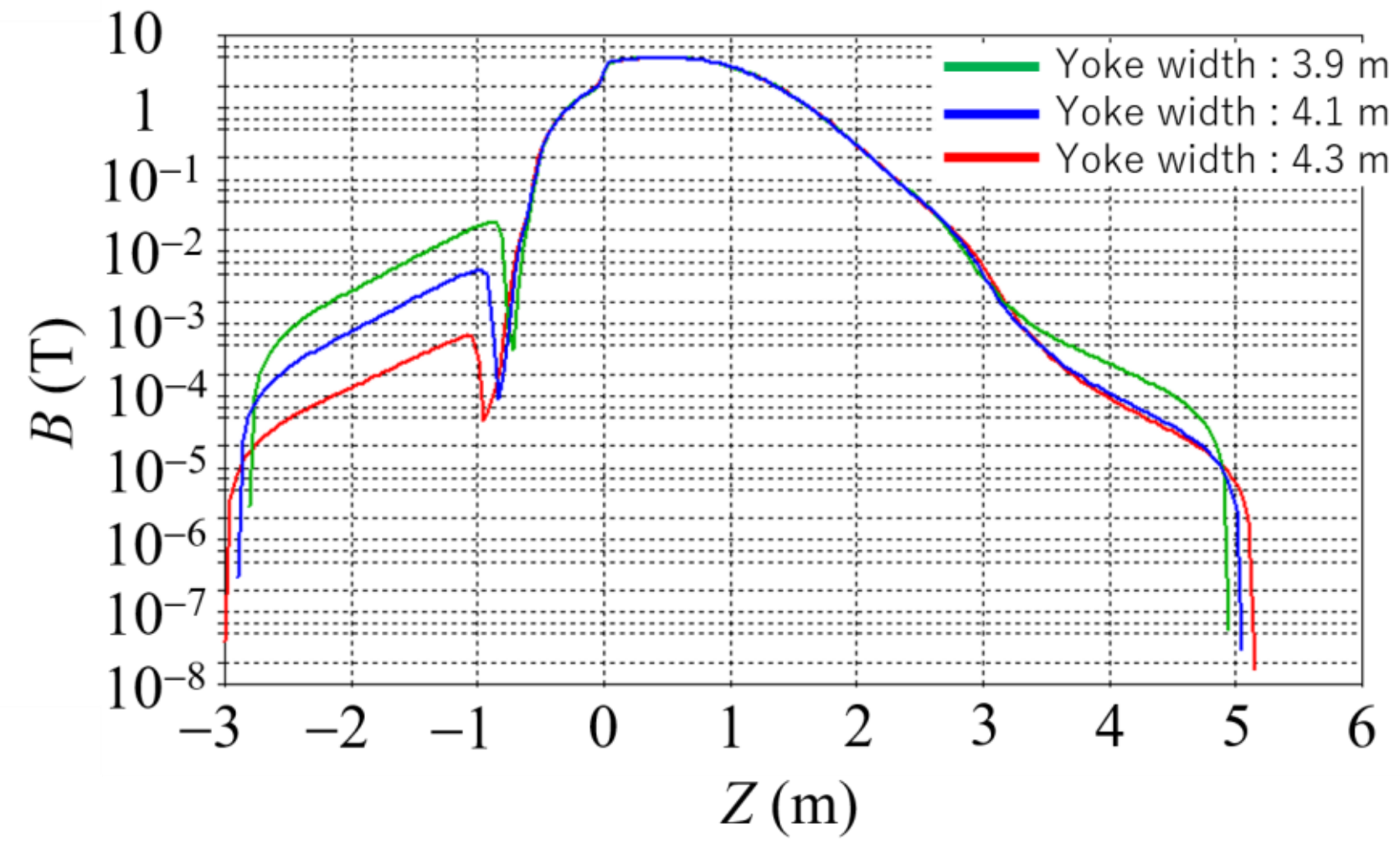

Figure 6-20. Magnetic field distribution on $Z$-Axis.

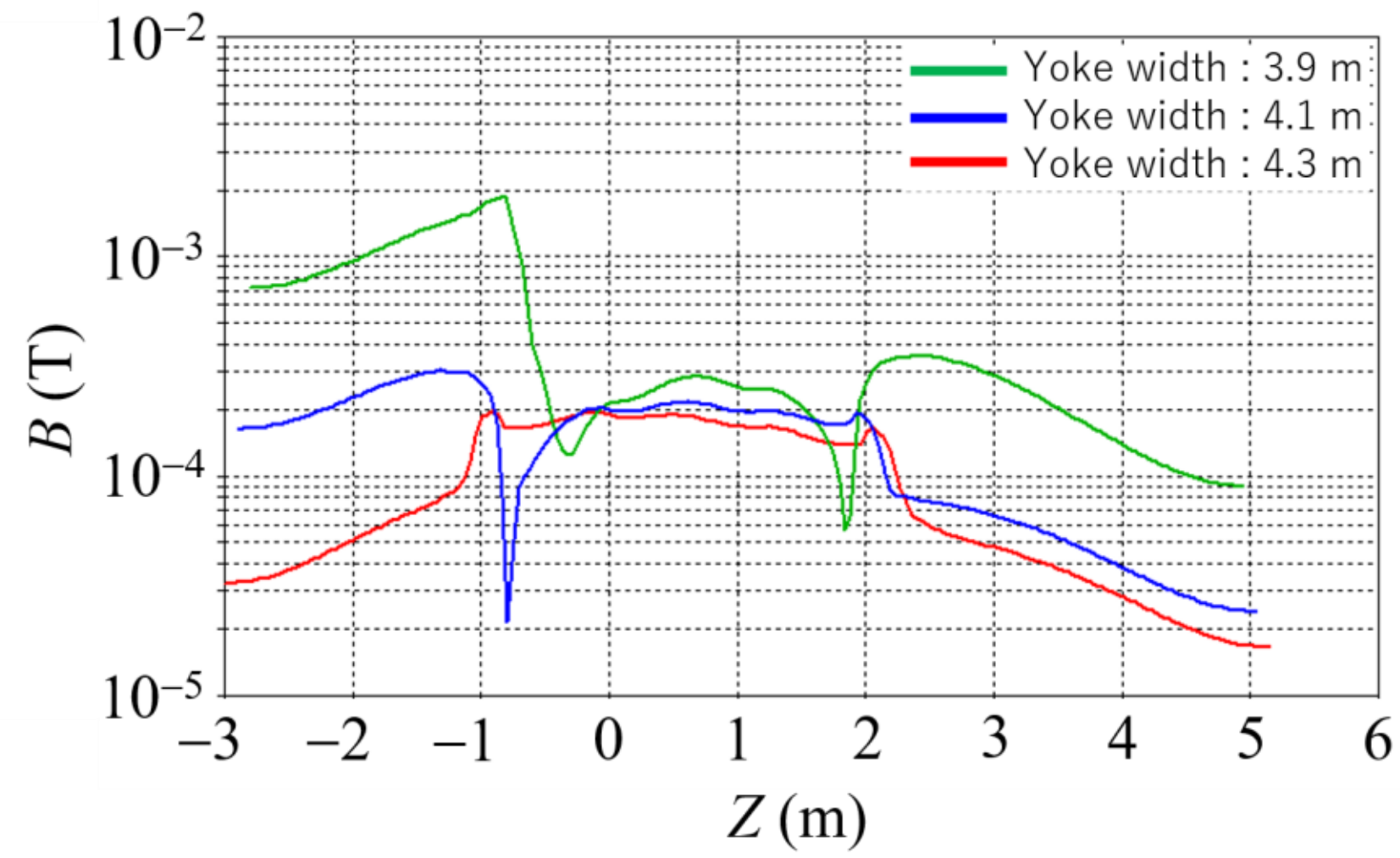

Figure 6-21. Magnetic field distribution on top surface of yoke (Line 1).

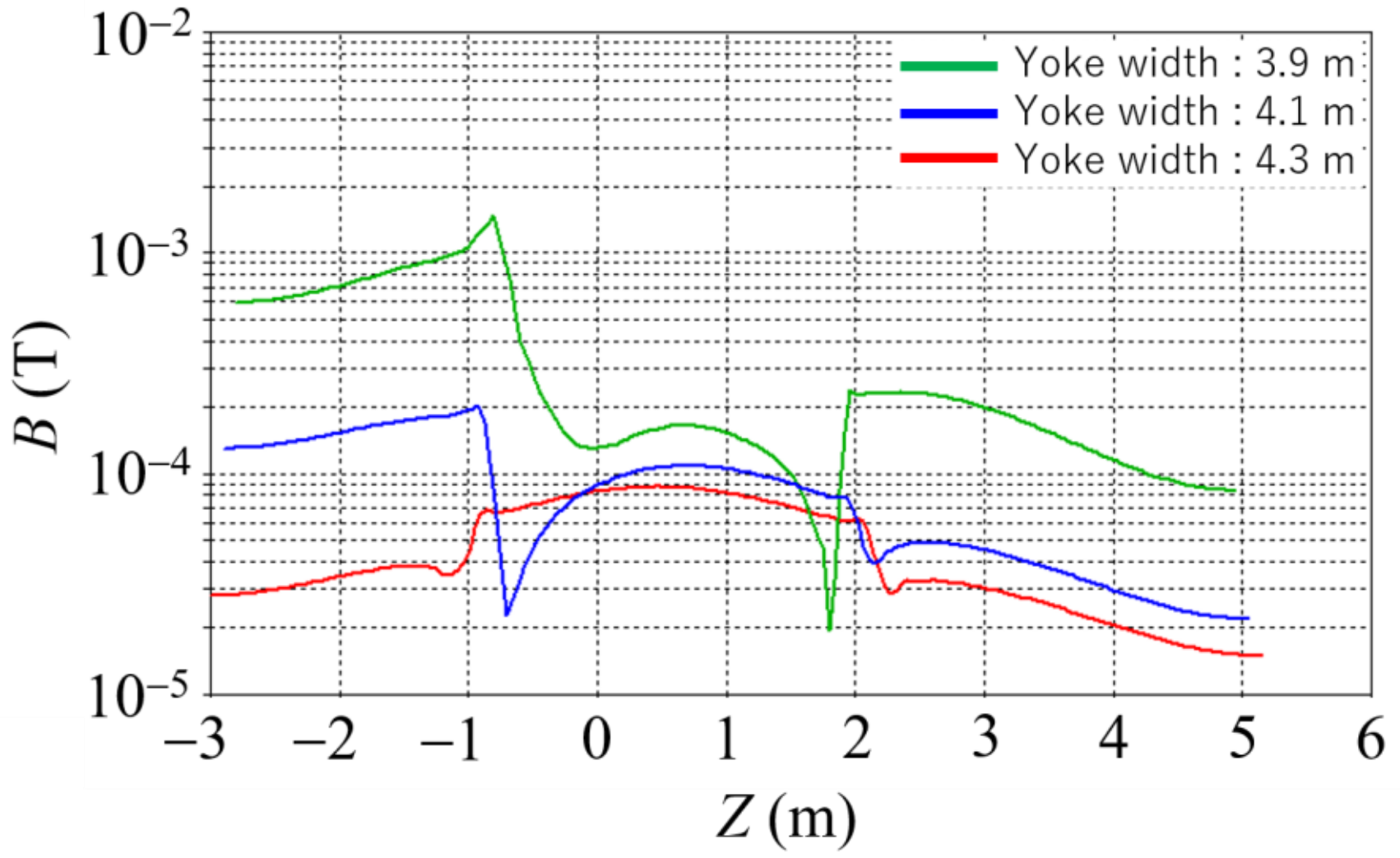

Figure 6-22. Magnetic field distribution on side surface of yoke (Line 2).

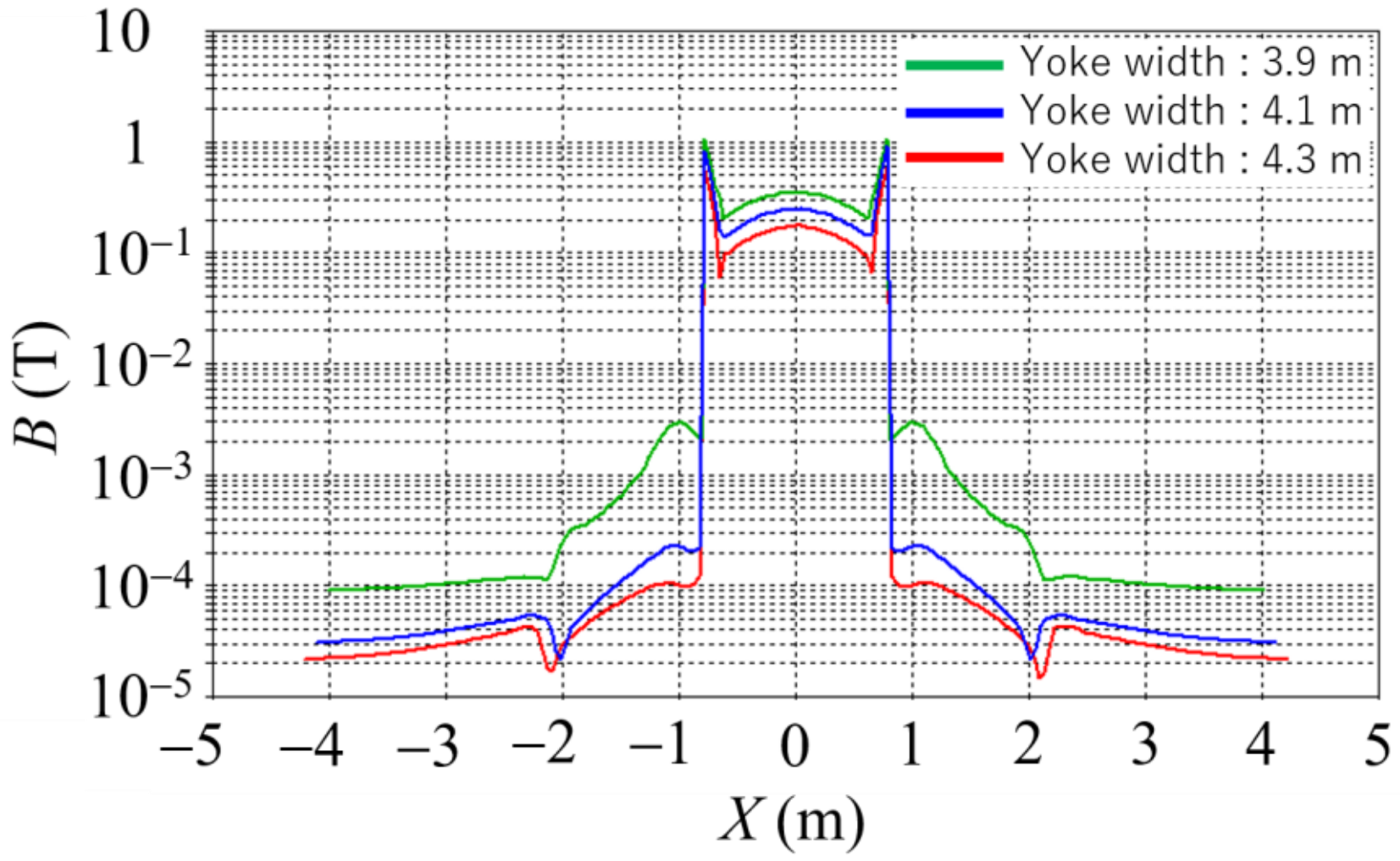

Figure 6-23. Magnetic field distribution on horizontal line at exit of yoke (Line 3).

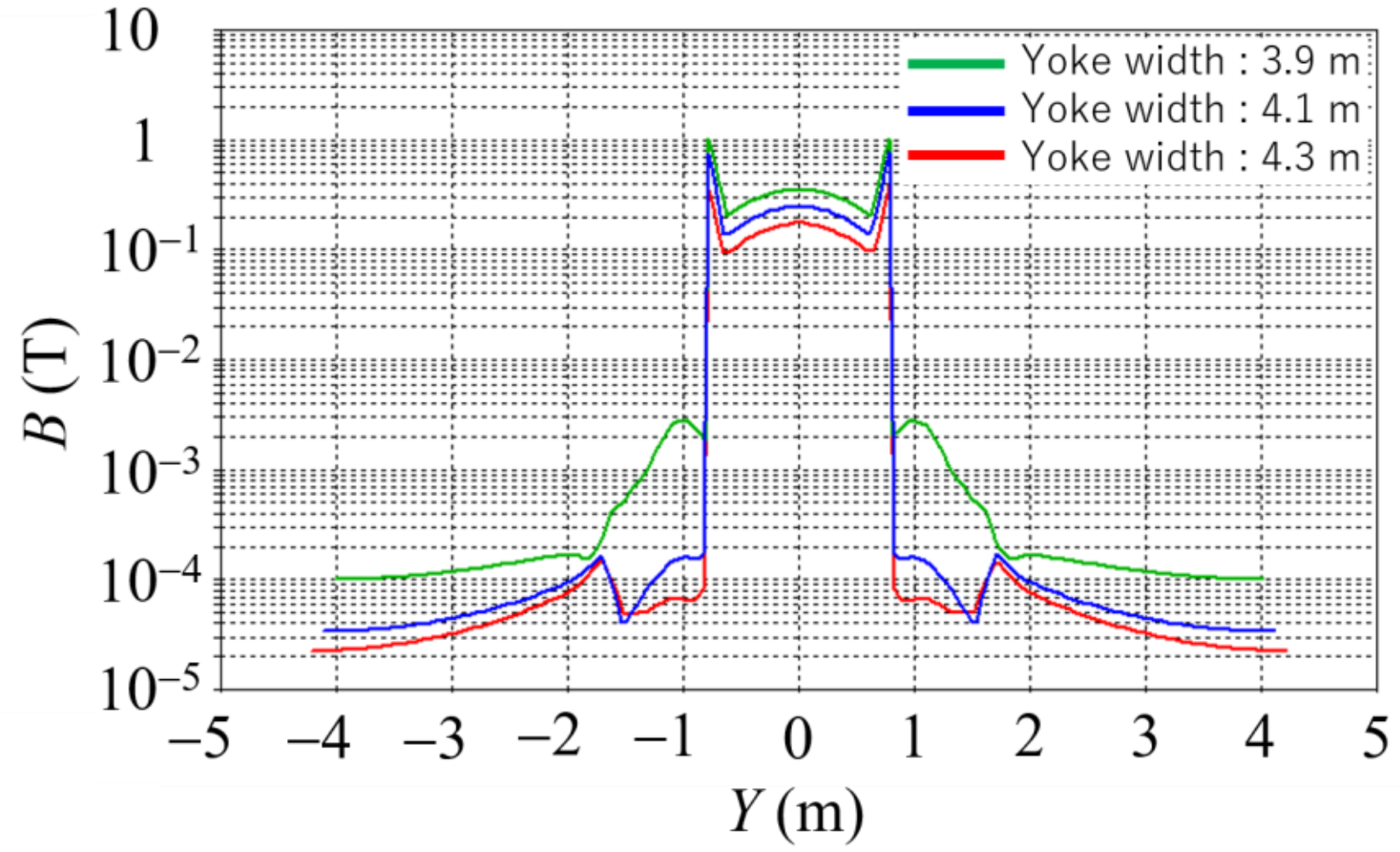

Figure 6-24. Magnetic field distribution on vertical line at exit of yoke (Line 4).

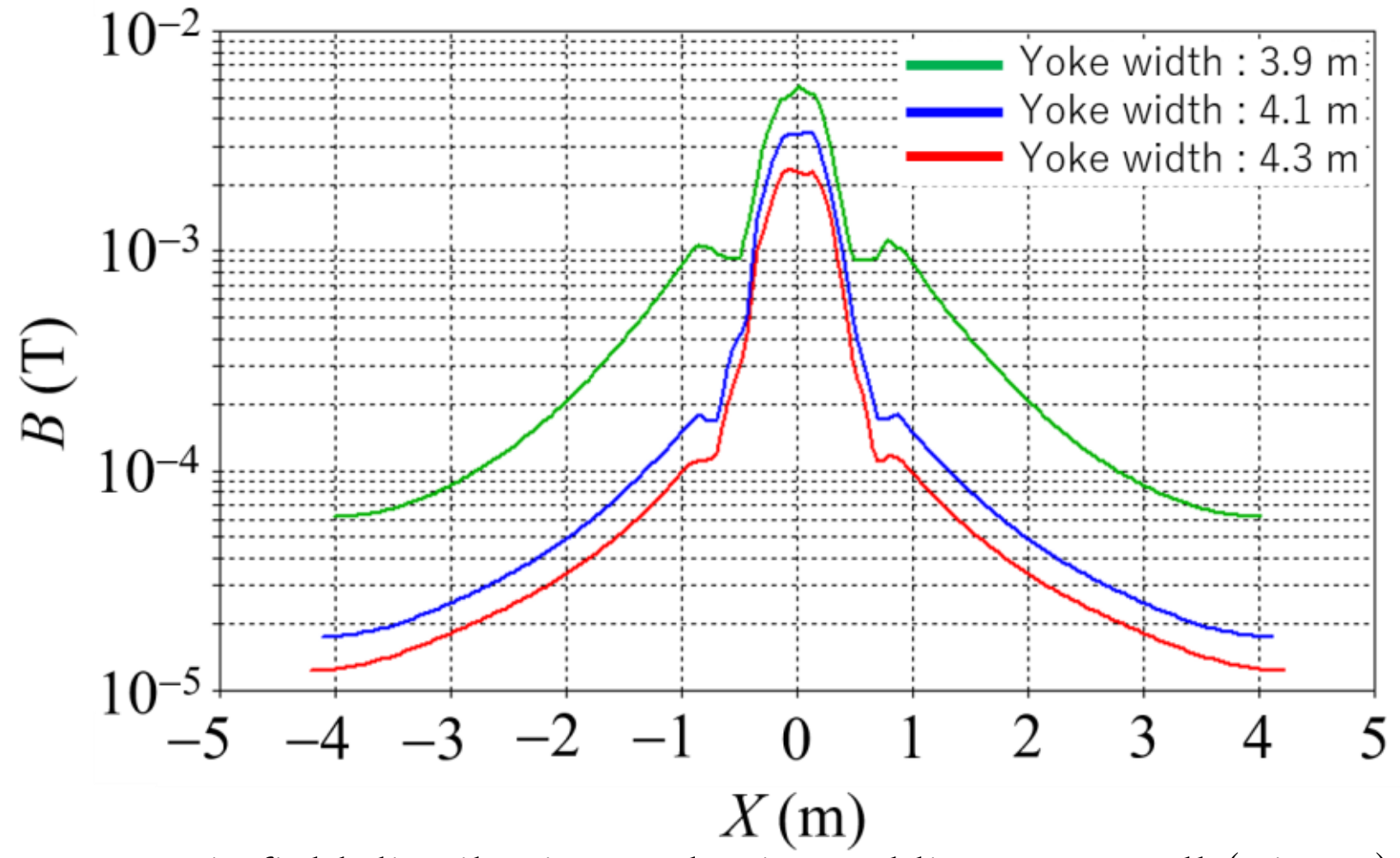

Figure 6-25. Magnetic field distribution on horizontal line at gas cell (Line 5).

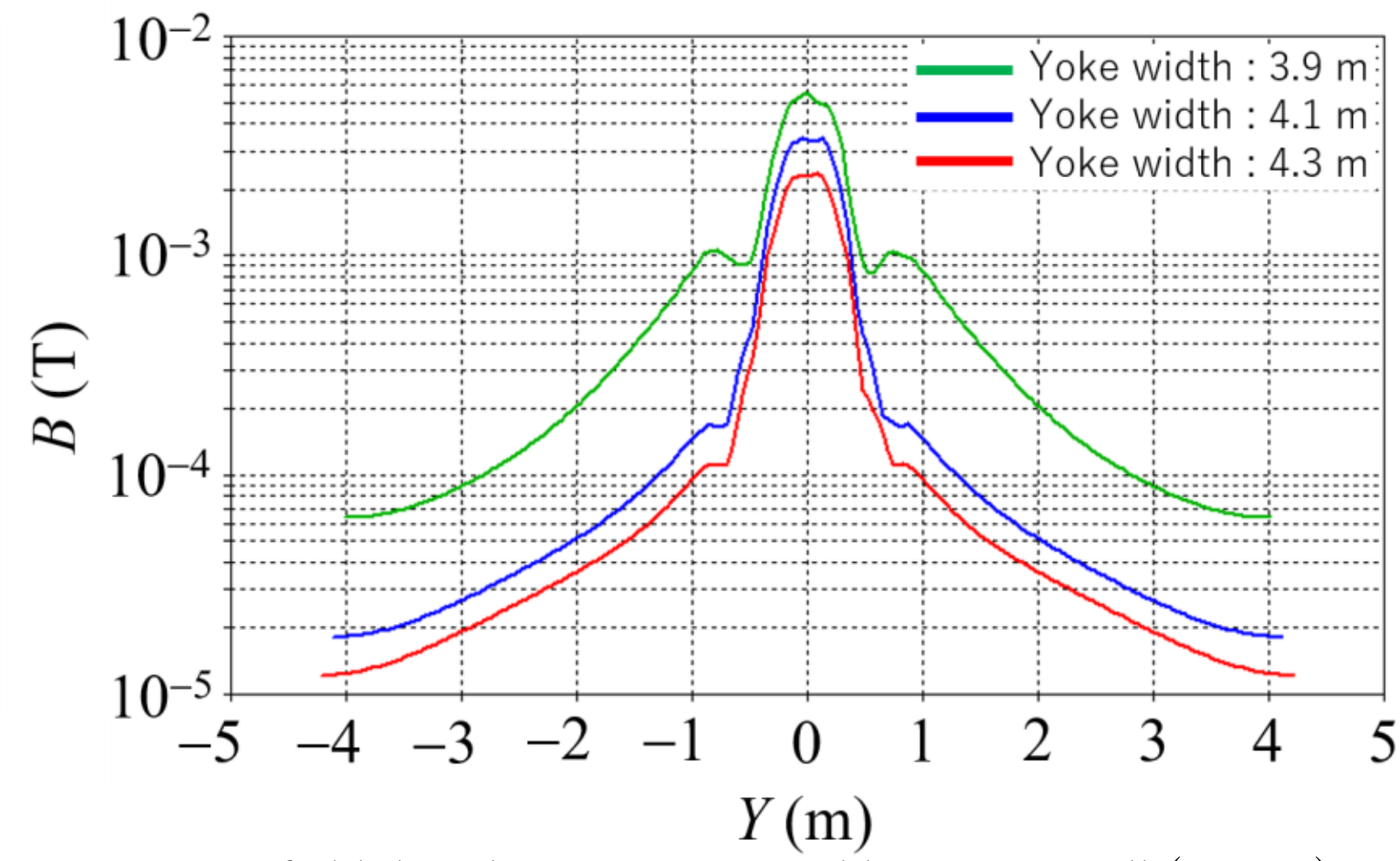

Figure 6-26. Magnetic field distribution on vertical line at gas cell (Line 6).

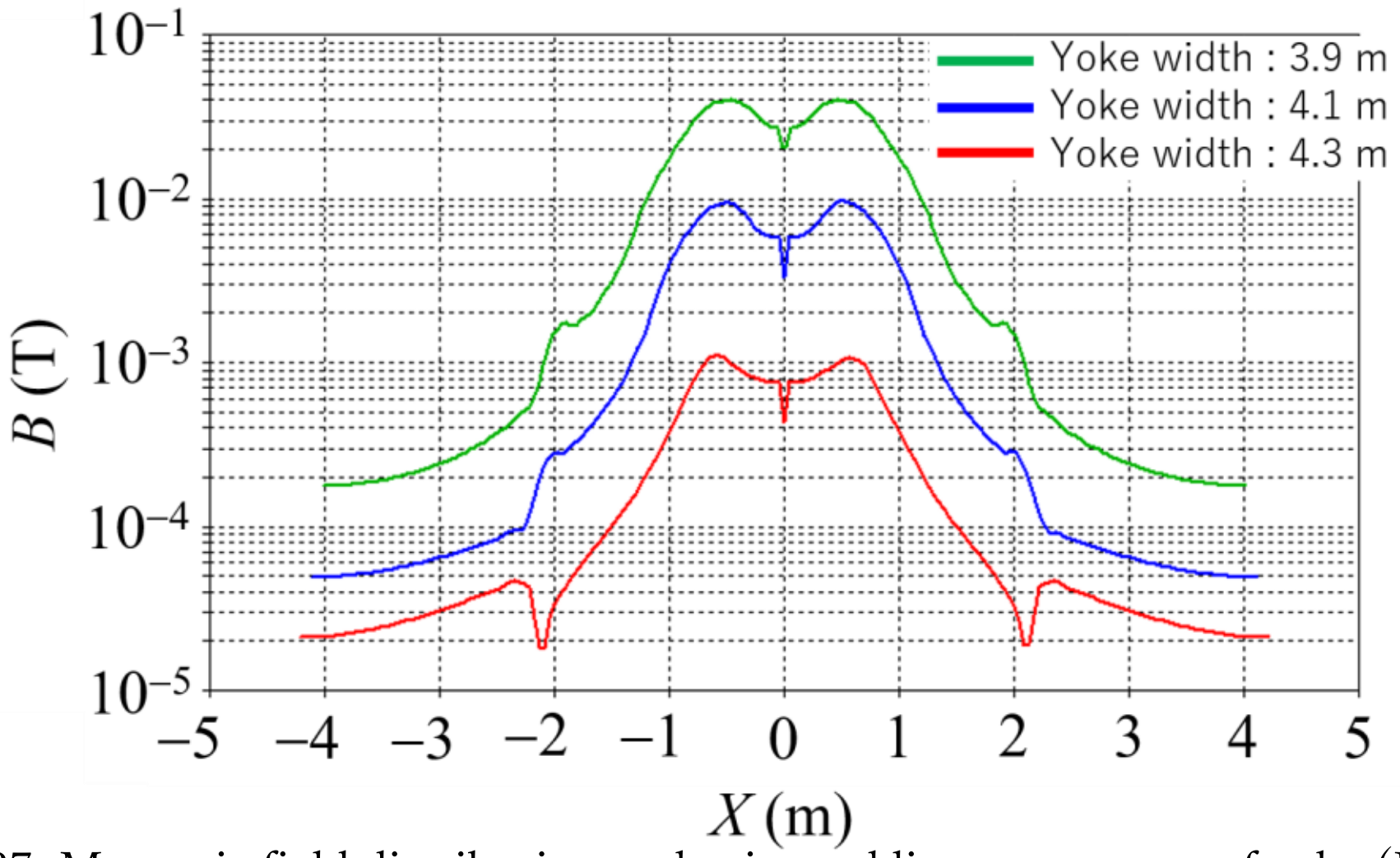

Figure 6-27. Magnetic field distribution on horizontal line at entrance of yoke (Line 7).

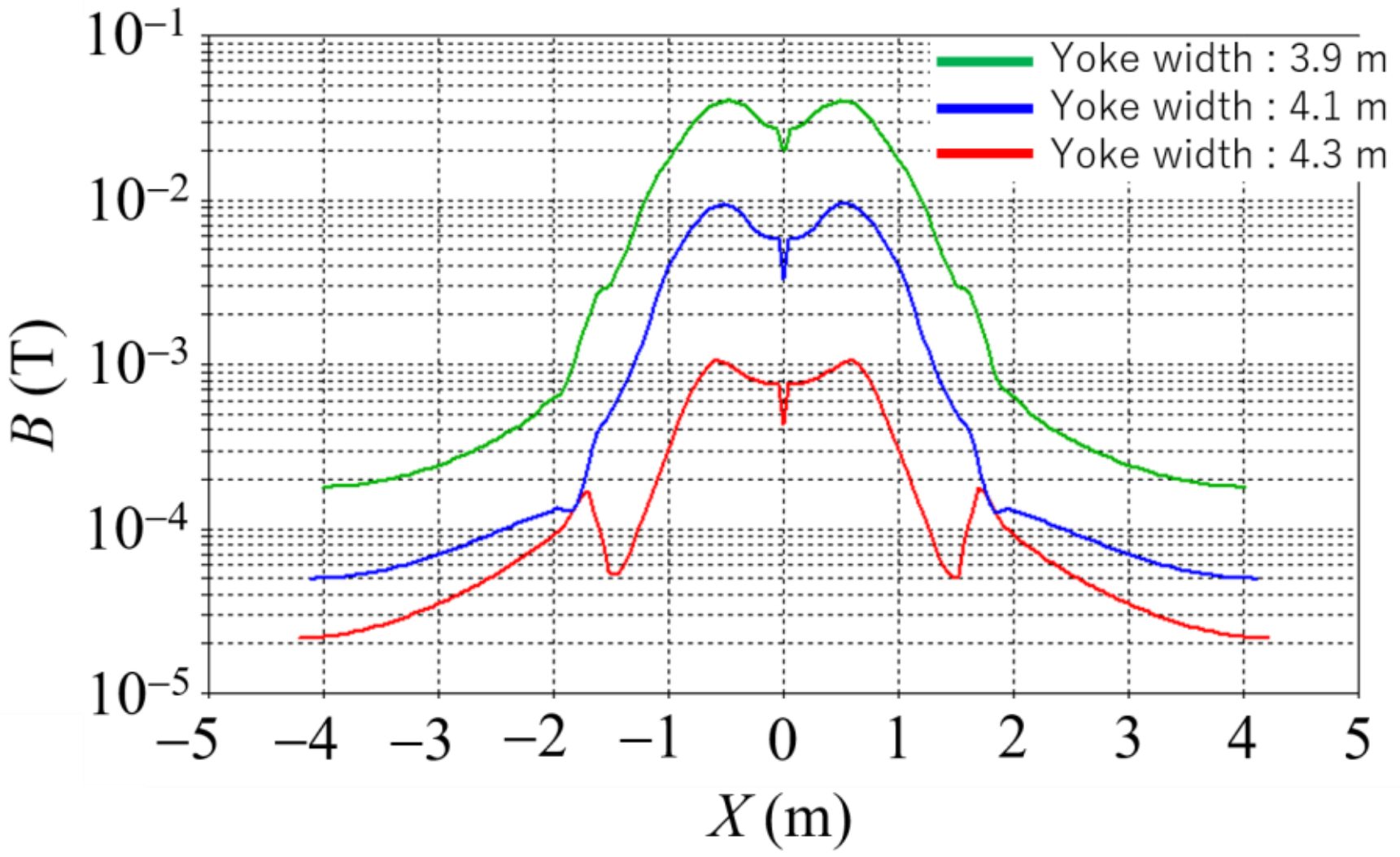

Figure 6-28. Magnetic field distribution on vertical line at entrance of yoke (Line 8).

### 6-3-2. Evaluation and Mitigation of Electromagnetic Force

In addition, the electromagnetic force on the coil was evaluated using the results of the three-dimensional magnetic field calculation. The electromagnetic force on the coil was calculated from the following equation.

$$F = \int mj \times B \, \mathrm{dV}$$

The current in the coil should always be orthogonal to the $Z$ axis, so the electromagnetic force in the $X$ and $Y$ directions will be zero. On the other hand, the electromagnetic force in Z direction can reach –5.02 MN due to the asymmetry of the magnetic field distribution. From this, the stresses on the coil can be calculated, and the angular stress will reach 76 MPa while the axial stress will reach 84 MPa when supported from the $Z < 0$ side.

This result suggests that mechanical supports for the coil are necessary to support the large $Z$-direction electromagnetic force. In this design, the heat input from the coil support to the low temperature region cannot be neglected, and the requirements for the refrigerator become more stringent. In order to reduce the cost of the refrigerator, we evaluated whether it is possible to cancel the electromagnetic force by changing the shape of the yoke around the coil. As a result of that study, reduction of the electromagnetic force was achieved by carving the magnetic material in the part of the upstream magnetic pole surface that faces the coil (see Fig. 6-29). By carving a 340 mm-deep groove in the area opposite the coil, the electromagnetic force under a current of 2600 A can be reduced to 0.12 MN. In this case, although the electromagnetic force is not proportional to the coil current due to the nonlinearity of iron permeability, the electromagnetic force can be reduced to the range of –0.38 MN ~ 0.52 MN for currents below 3000 A (see Fig. 6-30). The magnetic field on the $Z$-axis was reduced by about 5% compared to the initial geometry due to the decrease in the amount of magnetic material in the yoke, but this can be compensated by increasing the coil current up to 2700 A.

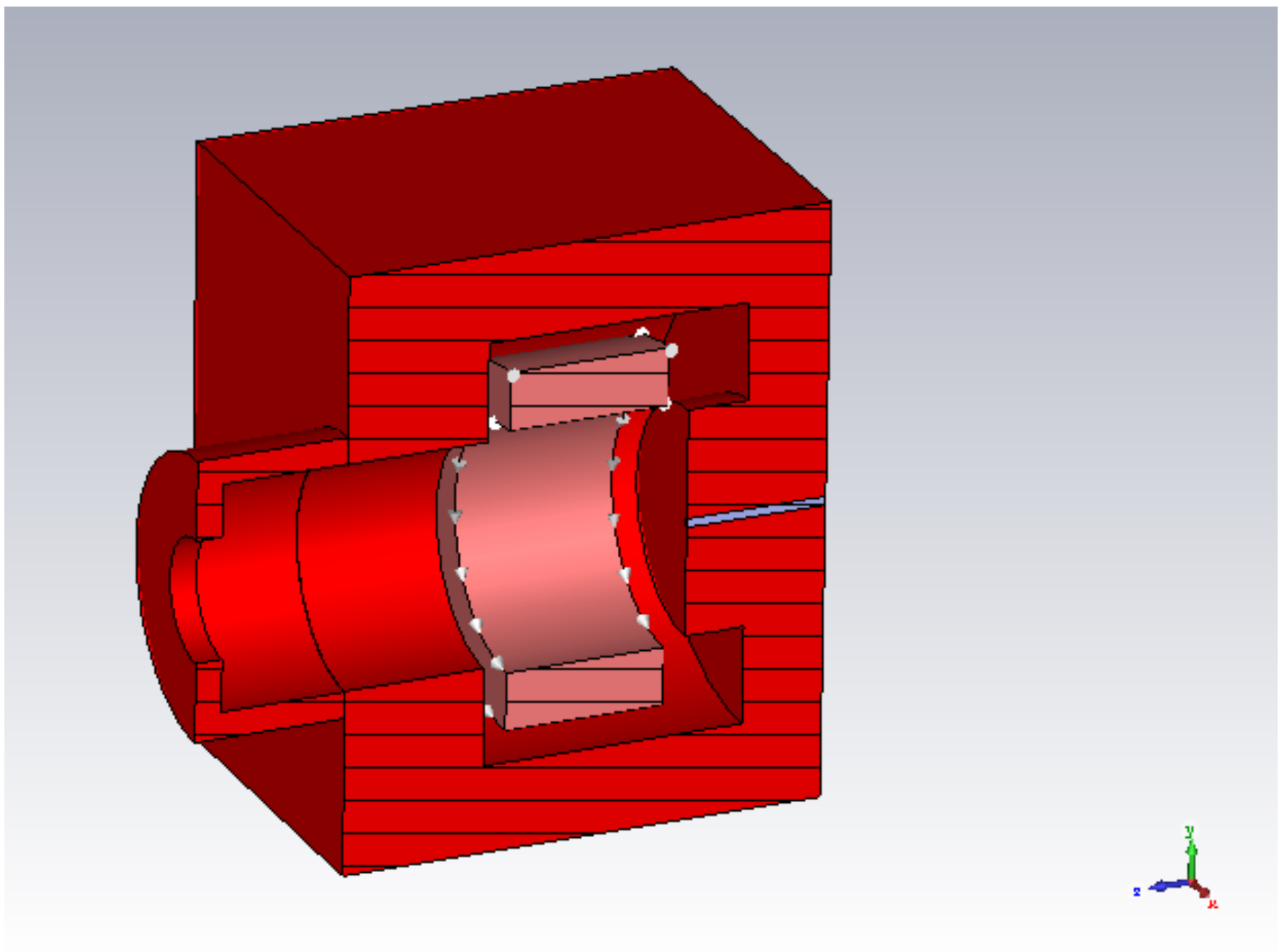

Figure 6-29. Geometry of digging into the yoke for cancellation of electromagnetic force.

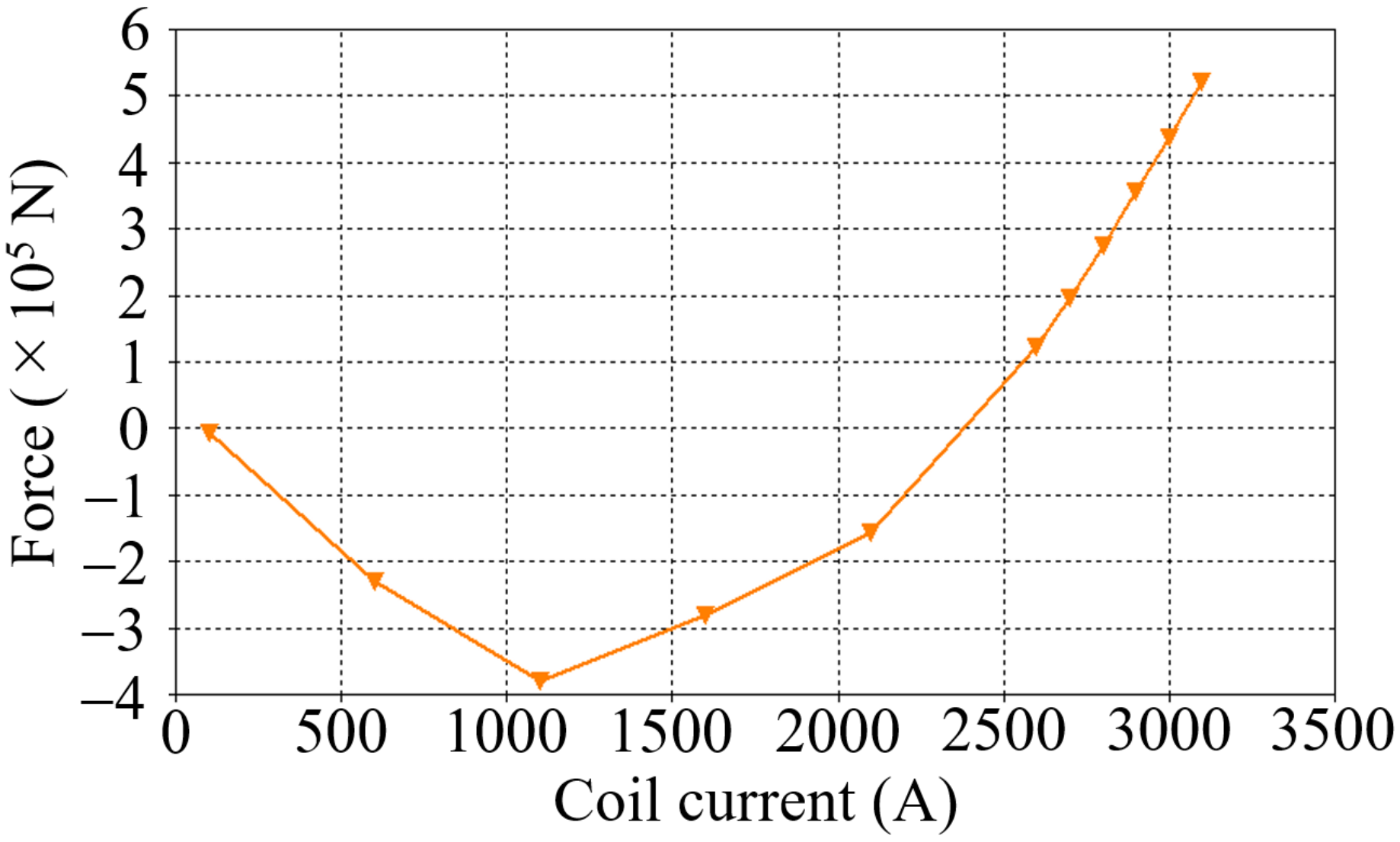

Figure 6-30. Dependence of the electromagnetic force on the coil current.

## 6-4. Radiation damage to the SC coils

Hadronic reactions of the primary beam with the target result in secondary radiation and activation of materials in the device. Assessment of the radiological impact to radiation workers who handle this device is essential for operational safety. In addition, prediction of radiation-induced heat is crucial to maintain the integrity of the device. Heating is particularly important for the superconducting magnet and the Faraday cup (FC). We have therefore conducted radiation transport simulation in and the vicinity of the device.

### 6-4-1. Methods

This calculation requires transport and reaction of heavy ions, neutrons, charged light particles, photons, electrons, and positrons. To perform such a complex calculation, general-purpose radiation transport calculation code PHITS (Particle and Heavy Ion Transport code

System) Ver. 3.26 was used.

The geometry used in the calculation is depicted as Fig. 6-31, which is consistent with the ones used in the previous sections. To calculate precise remnant dose taking into account the self-shielding effect, the yoke (indicated by pink around the green vacuum chamber in Fig. 6-31) was split into 10 virtual layers.

The other configuration parameters were as follows:

- A uniform magnetic field of 5.0 T directed to $+Z$ direction is applied to the cells filled with helium (or nitrogen) gas.
- Helium gas pressure of 1.013 kPa.
- Nitrogen gas pressure of 0.5 kPa.
- Primary beam ($^{238}$U or $^{136}$Xe) with intensity of 1 pμA and energy of 10$A$ MeV.
- The thickness of $^{238}$U target was 13 mg/cm$^2$.
- Periodical 30 days/year of irradiation is performed.
- After 20 years, the facility is shut down.

Simulations for the use of a primary beam $^{136}$Xe were performed in the case of test runs by using $^{136}$Xe beam.

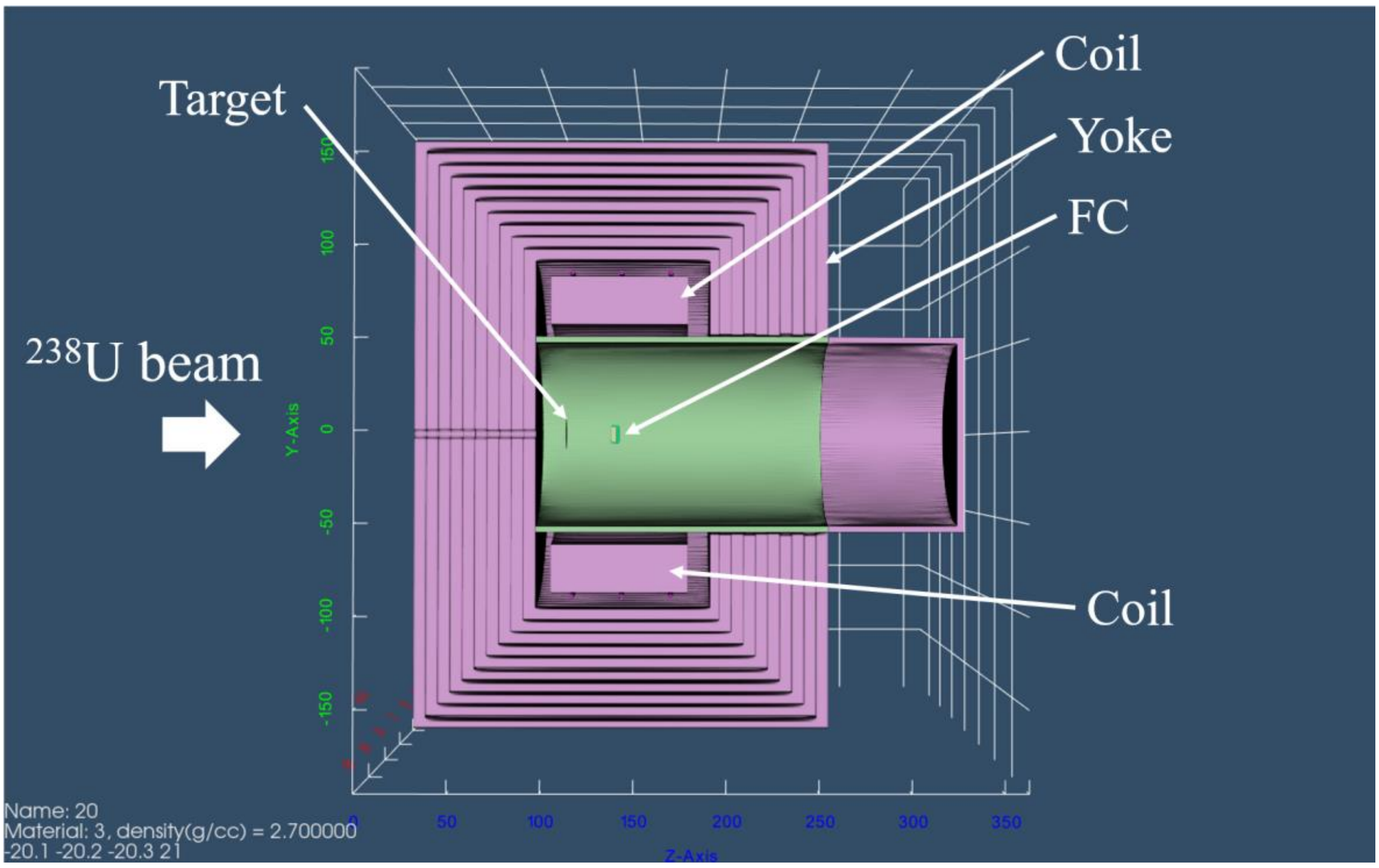

Figure 6-31. Schematic diagram of the calculation geometry. The unit of all axis is cm.

### 6-4-2. Results and discussion

Figures 6-32 and 6-33 show the energy deposited in the solenoid coil during the entire operational period. In the simulation, one solenoid coil was split into three sections to calculate the radiation doses. The data points in the figures represent the averaged doses of each section $Z$= 900 - 1200 mm, 1200 - 1500 mm, and 1500 - 1800 mm as the positions of $Z$ = 1050 mm, 1350 mm, and 1650 mm. The dose, about 150 Gy for both Xe and U beams, is enough lower than the safety limit of epoxy ($10^6$ Gy) which comprises the solenoid. One of the interesting features of this result is contribution of electrons and $\gamma$-rays. In cases of Xe and U beams, the majority of dose is attributed to electron-gamma cascade but without Al shielding, dose is more attributed to light charged particles (p5-group) as shown in Fig. 6-34. Comparison of Figs. 6-32 and 6-34 indicates that Al shielding converts neutrons to gammas which arrive at the solenoid but it blocks out light charged particle. To protect the solenoids this is advantageous because energy deposition by electron-gamma cascade is dispersed whereas that by light charged particles is particularly high near the surface, likely to reach or exceed safety limit locally.

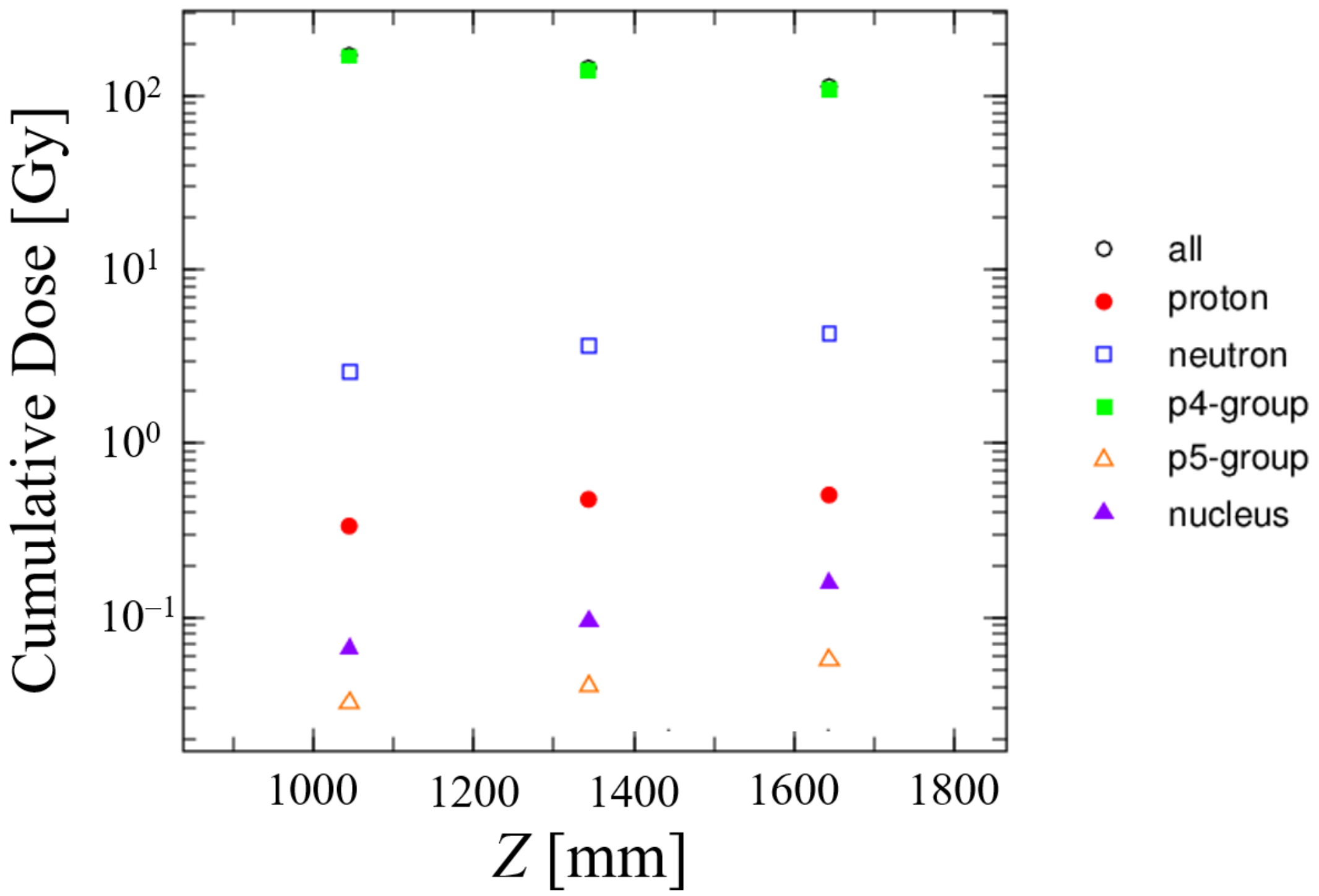

Figure 6-32. Energy deposited to solenoids by $^{136}$Xe beam irradiation throughout the whole operation period. In the legend, "p4-group" means photons and electrons. "p5-group" means deuteron, triton, $^3$He, and alpha. "nucleus" means nuclei heavier than $^4$He.

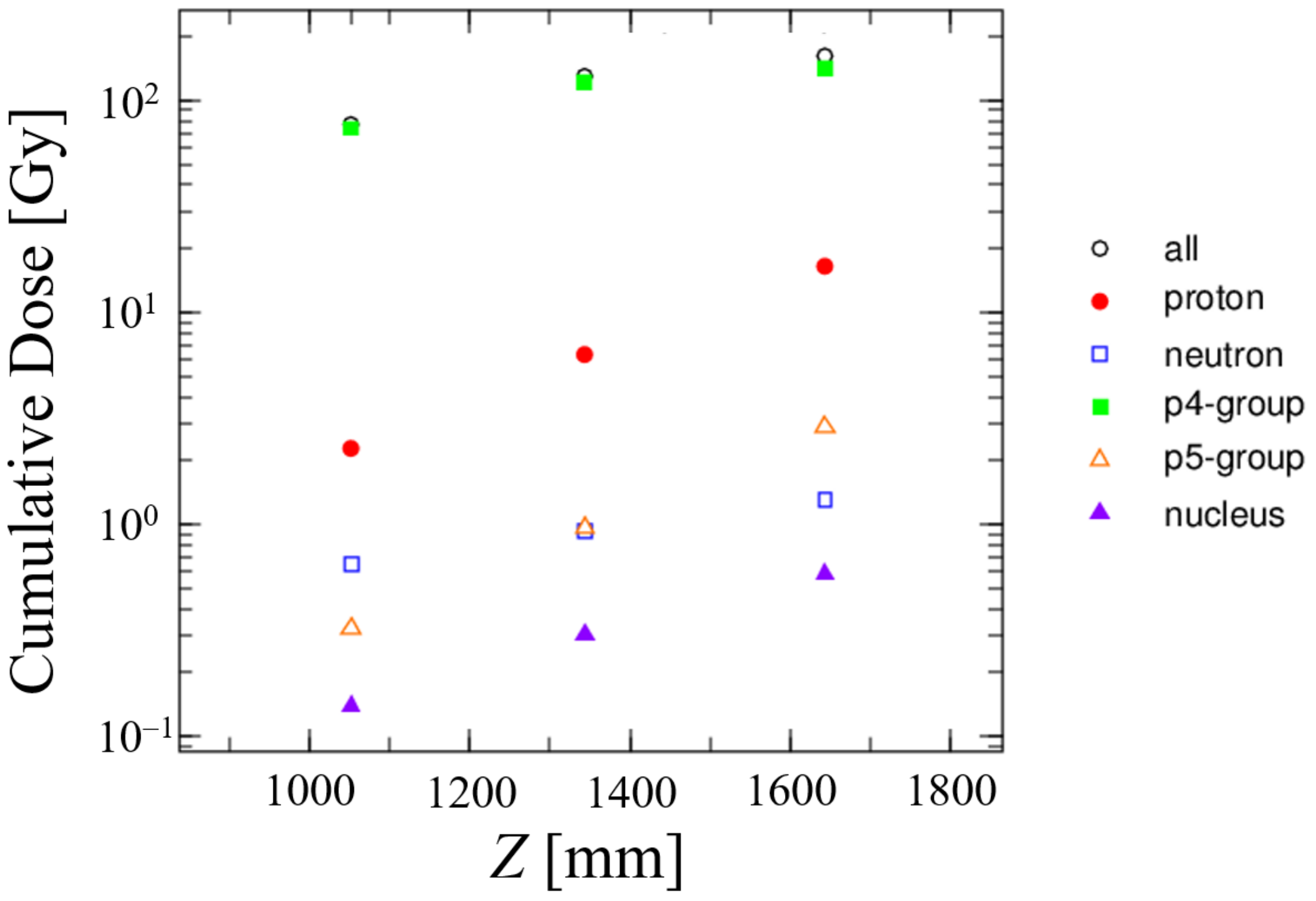

Figure 6-33. Same as Fig. 6-32 but by $^{238}$U beam.

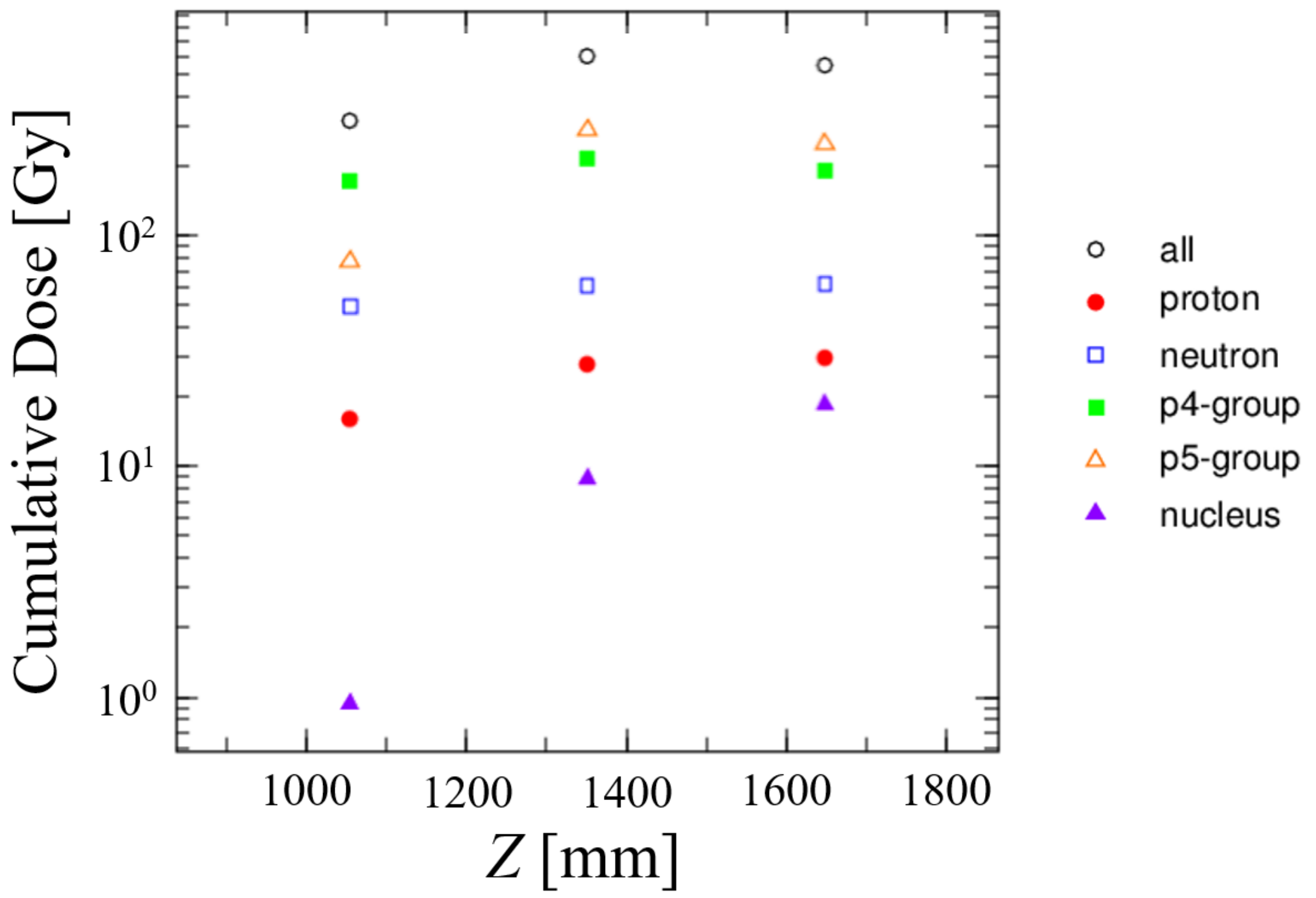

Figure 6-34. Same as Fig. 6-32 but without Al shielding.

Another radiological concern of this device, the remnant activation of the components, was also simulated. Figs. 6-35 and 6-36 show the time evolutions of induced activities in various components. Being dominated by a long-lived isotope $^{3}H$ or target-origin actinides, the activity in gas is cumulative. In contrast, the activity of metal parts declines quickly as soon as the cooling period starts owing to short-lived isotopes such as $^{24}Na$ (indicated as Na-24 in Tables 6-5, 6, 7, and 8 which show the residual nuclide important in each component of the device from the viewpoint of decay heat and dosimetric impact.), $^{52}Mn$, $^{64}Cu$, $^{92m}Nb$, $^{147}Nd$, and $^{237}U$. These isotopes disappear by the next beam period but metal parts also contain long-lived isotopes such as $^{22}Na$, $^{54}Mn$, $^{56}Co$, $^{59}Fe$, $^{60}Co$, and $^{65}Zn$ produced by spallation. Owing to these isotopes, the activity builds up over 20 years of operation. In view of the decommissioning, important isotopes are $^{26}Al$, $^{63}Ni$, and $^{238}U$, which remain even after 100 years of cooling.

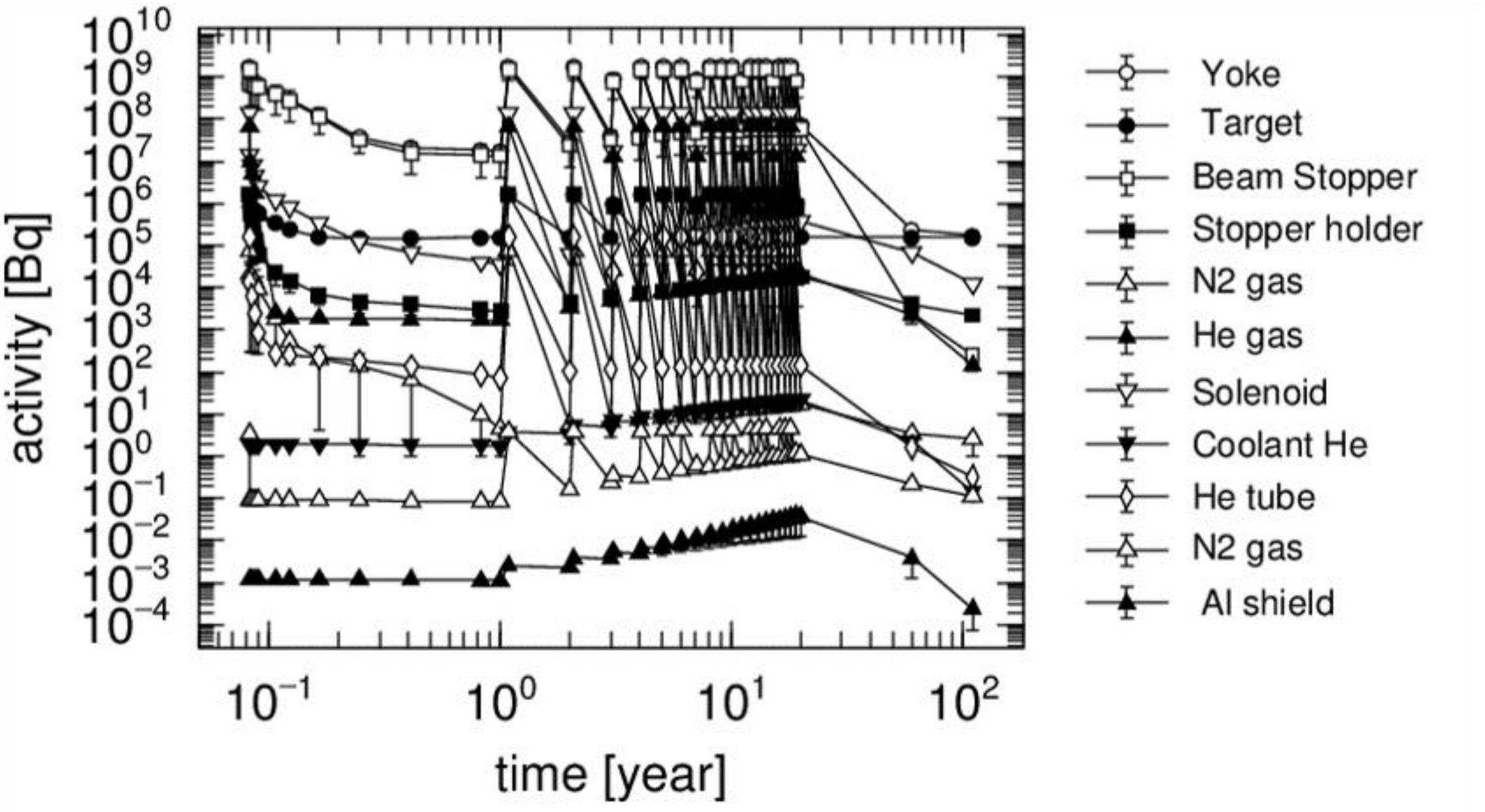

Figure 6-35. Time evolution of radioactivity in various components for $^{136}Xe$ beam.

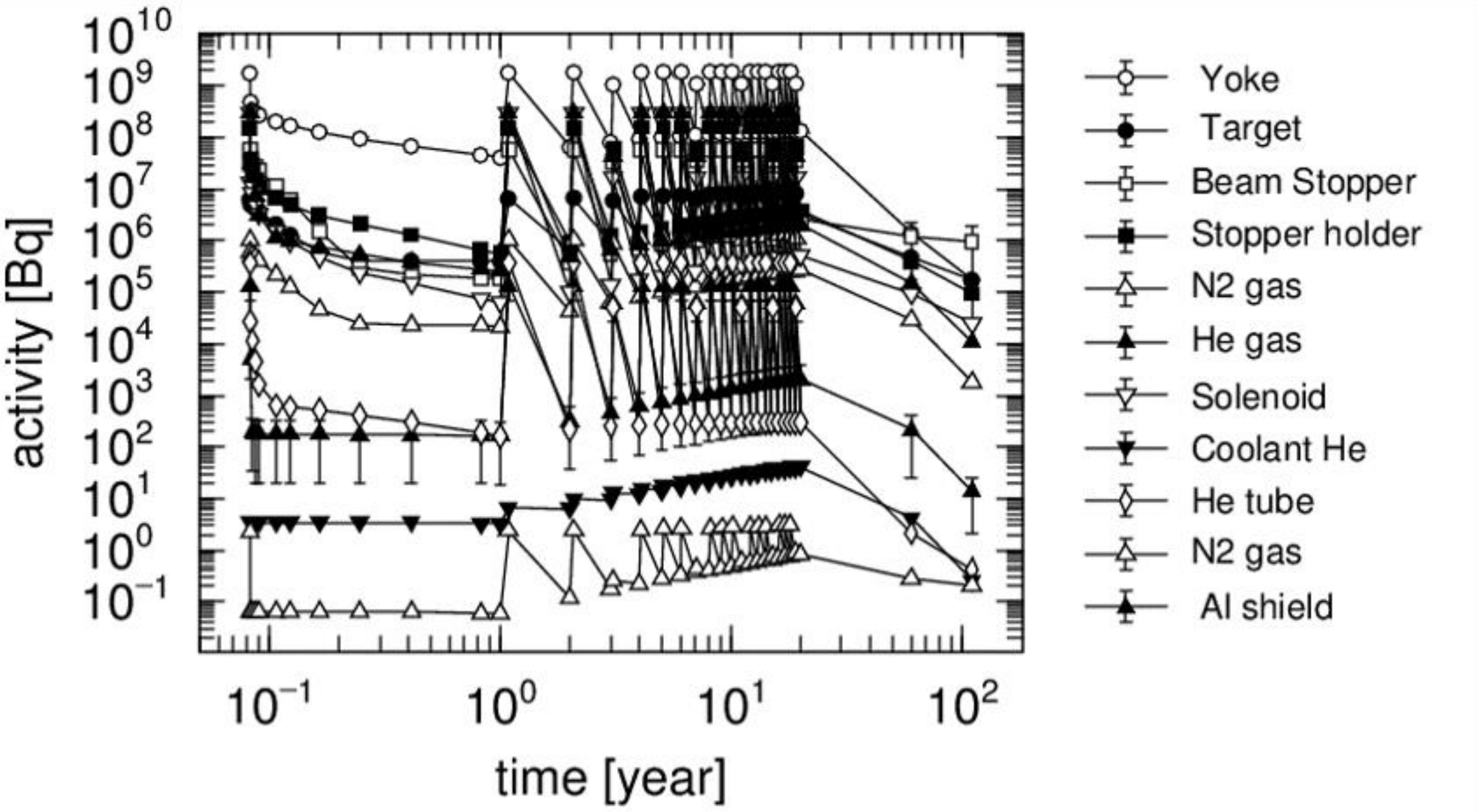

Figure 6-36. Time evolution of radioactivity in various components for $^{238}U$ beam.

Table 6-5. Decay heat source nuclide in the device components. In case the primary beam is $^{136}Xe$. "33 days" means the timing at the end of first beam run. "60 days" means the timing 30 days after the end of first beam run. "60 years" means the timing 60 years since the beginning of the first beam run, which is 40 years and 335 days after the beam shutdown. Nuclide whose contribution is more than 10% are listed.

| | 33 days | 60 days | 60 years |
|---|---|---|---|
| Yoke | Mn-54 (49.6%), Fe-59 (42.1%) | Mn-54 (57.9%), Fe-59 (34.4%) | H-3 (94.8%) |
| Target | U-238 (50.9%), U-237 (28.1%) | U-238 (82.4%), Pa-234m (13.0%) | U-238 (82.1%), Pa-234m (15.8%) |
| Beam stopper | Nd-147 (46.0%), Pm-148 (46.0%) | Nd-147 (83.9%), Pm-148 (14.1%) | Pu-238 (68.7%), Am-241 (18.0%) |
| Target holder | Cu-64 (61.0%), Co-60 (17.0%), U-237 (12.9%) | Co-60 (84.0%), Zn-65 (11.9%) | Co-60 (60.5%), Ni-63 (37.8%) |
| $N_2$ gas | U-237(54.8%), Np-239(45.1%) | U-237(96.4%) | Pu-239(97.5%) |
| Solenoid | Na-24(50.3%), Nb-92m(32.4%), Sc-48(12.2%) | Nb-92m(66.8%), Sc-46(22.8%) | Nb-94(82.8%), Nb-93m(16.8%) |
| Coolant He | H-3(100%) | H-3(100%) | H-3(100%) |
| He tube | Na-24(93.9%) | Zn-65(86.8%) | Al-26(45.7%), H-3(18.8%), Co-60(18.5%) |
| Al shield | Na-24(100.0%) | Al-26(68.5%), H-3(31.5%) | Al-26(97.2%) |

The activity in the nitrogen gas and coolant He is potentially released to the outside the device owing to the circulation between the device and external elements such as pump and filter. However, long-lived isotopes produced by one cycle of irradiation will be 22 kBq $^{3}H$ in the nitrogen gas, whereas 1.2 Bq $^{3}H$ in the coolant He in case of U beam. It should be noted that these radioactive nuclei will be collected by the filter.

Decay heat in the solenoid coil, target and beam stopper parts are much less than 1 mW in the use of intense primary beams of $^{238}U$ and $^{136}Xe$, and are negligible for the KISS-II experiments.

Table 6-6. Same as Table 6-5 but activity weighted by ambient dose conversion coefficient.

| | 33 days | 60 days | 60 years |
|---|---|---|---|
| Yoke | Mn-54 (53.8%), Fe-59 (39.1%) | Mn-54 (62.1%), Fe-59 (31.5%) | Fe-55 (55.1%), Co-60 (33.1%) |
| Target | U-237 (55.2%), Np-239 (19.8%) | U-237 (49.8%), Pa-234m (14.7%) | Pa-234m (41.6%), Pa-234 (35.2%) |
| Beam stopper | Pm-148 (55.3%), Nd-147 (44.0%) | Nd-147 (82.5%), Pm-148 (17.5%) | Pa-233 (50.7%), Np-236 (14.4%) |
| Target holder | Co-60 (64.3%), U-237 (14.7%) | Co-60 (86.1%), Zn-65 (12.5%) | Co-60 (100.0%) |
| $N_2$ | U-237(53.6%), Np-239(45.8%) | U-237(88.8%), Be-7(10.8%) | Pa-233(87.5%), Np-237(12.5%) |
| Solenoid | Na-24(43.3%), Nb-92m(38.9%), Sc-48(13.5%) | Nb-92m(69.1%), Sc-46(22.4%) | Nb-94(99.8%) |
| Coolant He | - | - | - |
| He tube | Na-24(93.9%) | Zn-65(86.4%) | Al-26(57.8%), Co-60(42.2%) |
| Al shield | Na-24(100.0%) | Al-26(100.0%) | Al-26(100.0%) |

Table 6-7. Same as Table 6-5 but the primary beam is $^{238}U$.

| | 33 days | 60 days | 60 years |
|---|---|---|---|
| Yoke | Mn-52 (54.5%) | Co-56 (36.6%), Mn-54 (27.6%), Fe-59 (18.8%) | H-3 (99.8%) |
| Target | U-237 (73.7%), U-238 (15.7%) | U-238 (67.0%), U-237 (19.6%), Pa-234m (10.6%) | U-238 (79.7%), Pa-234m (15.3%) |
| Beam stopper | U-237 (82.3%), Np-239 (10.1%) | Cf-249 (56.7%), U-237 (42.2%) | Cf-249 (99.5%) |
| Target holder | U-237 (17.5%), Fe-59 (17.2%), Co-56 (11.6%) | Fe-59 (30.6%), Co-56 (26.1%), Co-58 (17.8%) | Co-60 (16.8%), Sc-44 (11.3%) |
| $N_2$ | U-237(91.8%) | U-237(98.3%) | H-3(72.5%), Pu-240 (20.9%) |
| Solenoid | Na-24(35.0%), Sc-48(24.7%), Nb-92m(20.9%) | Sc-46(52.0%), Nb-92m(28.7%) | Nb-94(87.1%) |
| Coolant He | - | - | - |
| He tube | Na-24(76.4%), Zn-65(15.7%) | Zn-65(89.3%) | Al-26(56.9%), H-3(18.4%), Ni-63(13.8%) |
| Al shield | Na-24(90.8%) | Na-22(82.3%) | - |

Table 6-8. Same as Table 6-5 but activity weighted by ambient dose conversion coefficient and the primary beam is $^{238}U$.

| | 33 days | 60 days | 60 years |
|---|---|---|---|
| Yoke | Mn-52 (55.3%), Co-56 (14.7%), Mn-54 (11.5%) | Co-56 (34.2%), Mn-54 (32.3%), Fe-59 (18.8%) | Fe-55 (65.7%), Co-60 (23.0%), Mn-54 (11.2%) |
| Target | U-237 (89.3%), Np-239 (10.1%) | U-237 (92.9%) | Pa-234m (41.7%), Pa-234 (35.3%), Th-234 (22.5%) |
| Beam stopper | U-237 (86.4%), Np-239 (11.1%) | U-237 (77.0%), Cf-249 (20.0%) | Cf-249 (99.5%) |
| Target holder | Fe-59 (22.7%), Co-56 (14.3%), Mn-52 (11.0%) | Fe-59 (32.6%), Co-56 (26.1%), Co-58 (18.3%) | Co-60 (64.6%), Sc-44 (19.6%) |
| $N_2$ | U-237(91.6%) | U-237(100.0%) | Al-26(65.5%), Pa-233(30.0%) |
| Solenoid | Na-24(31.1%), Sc-48(28.3%), Nb-92m(26.0%) | Sc-46(56.8%), Nb-92m(32.9%) | Nb-94(98.3%) |
| Coolant He | - | - | - |
| He tube | Na-24(75.6%), Zn-65(20.5%) | Zn-65(89.4%) | Al-26(74.4%), Co-60(25.6%) |
| Al shield | Na-24(93.8%) | Na-22(88.5%) | - |

Figs. 6-37 and 6-38 show the remnant dose distribution after 30 days of beam run followed by 3 days of cooling period. This condition was taken as an example to know if one can approach the device soon after the beam shutdown. This result indicates that the dose is less than 1 μSv/h outside the device. To access the interior of the device, 1-10 μSv/h of dose is expected. It is not advisable to handle the target this soon because the ambient dose is more than 10 μSv/h in its close vicinity. For the safe access to the interior of the device, 30 days of cooling is recommended. Figs. 6-39, and 6-40 show the remnant dose distribution after 30 days of beam run followed by 30 days of cooling period. The dose inside the device becomes less than 1 μSv/h except for the close vicinity of the target.

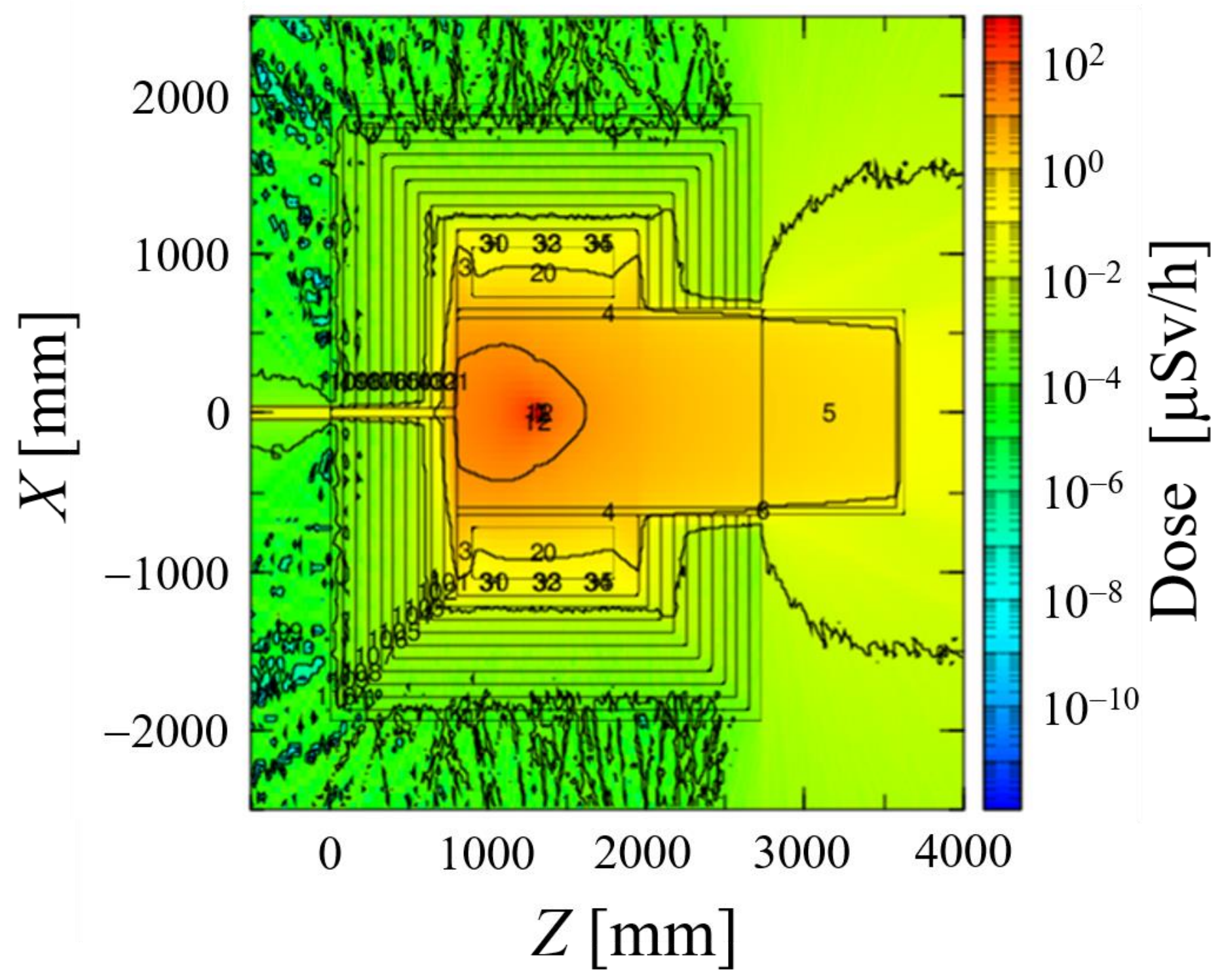

Figure 6-37. Remnant dose distribution after 30 days of $^{136}Xe$ beam run and 3 days of cooling period.

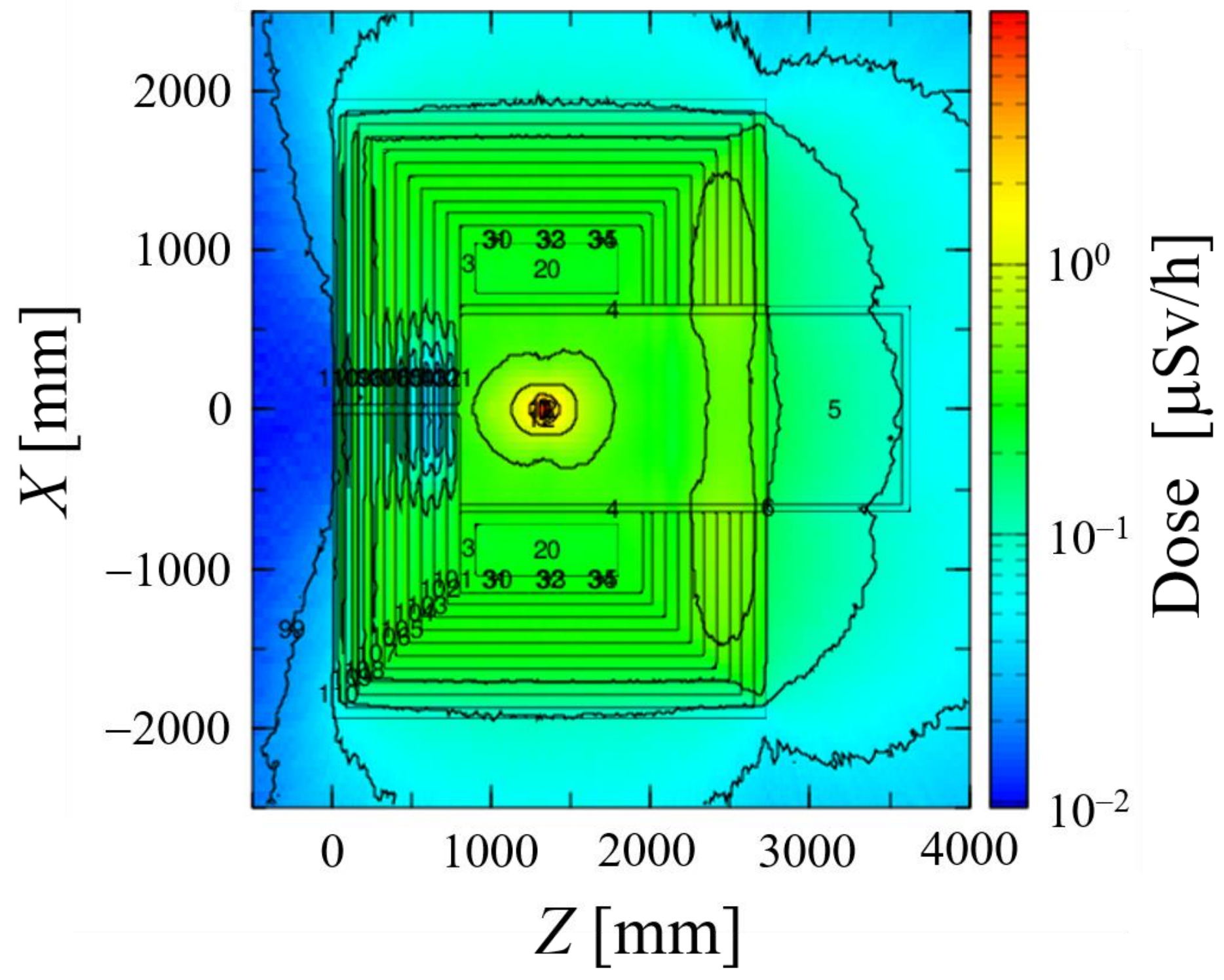

Figure 6-38. Same as Fig. 6-37 but the beam is $^{238}U$.

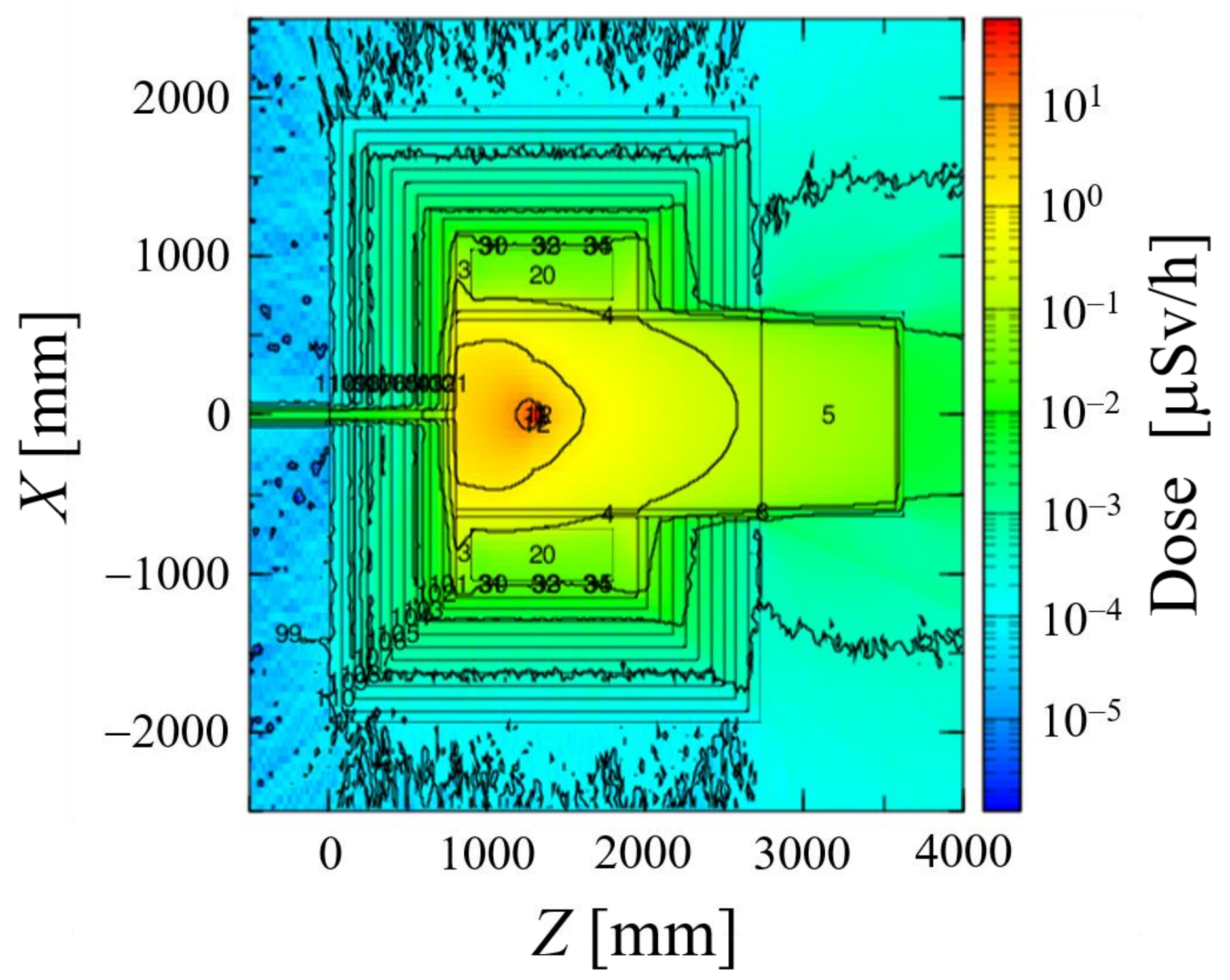

Figure 6-39. Same as Fig. 6-37 but the cooling period is 30 days long.

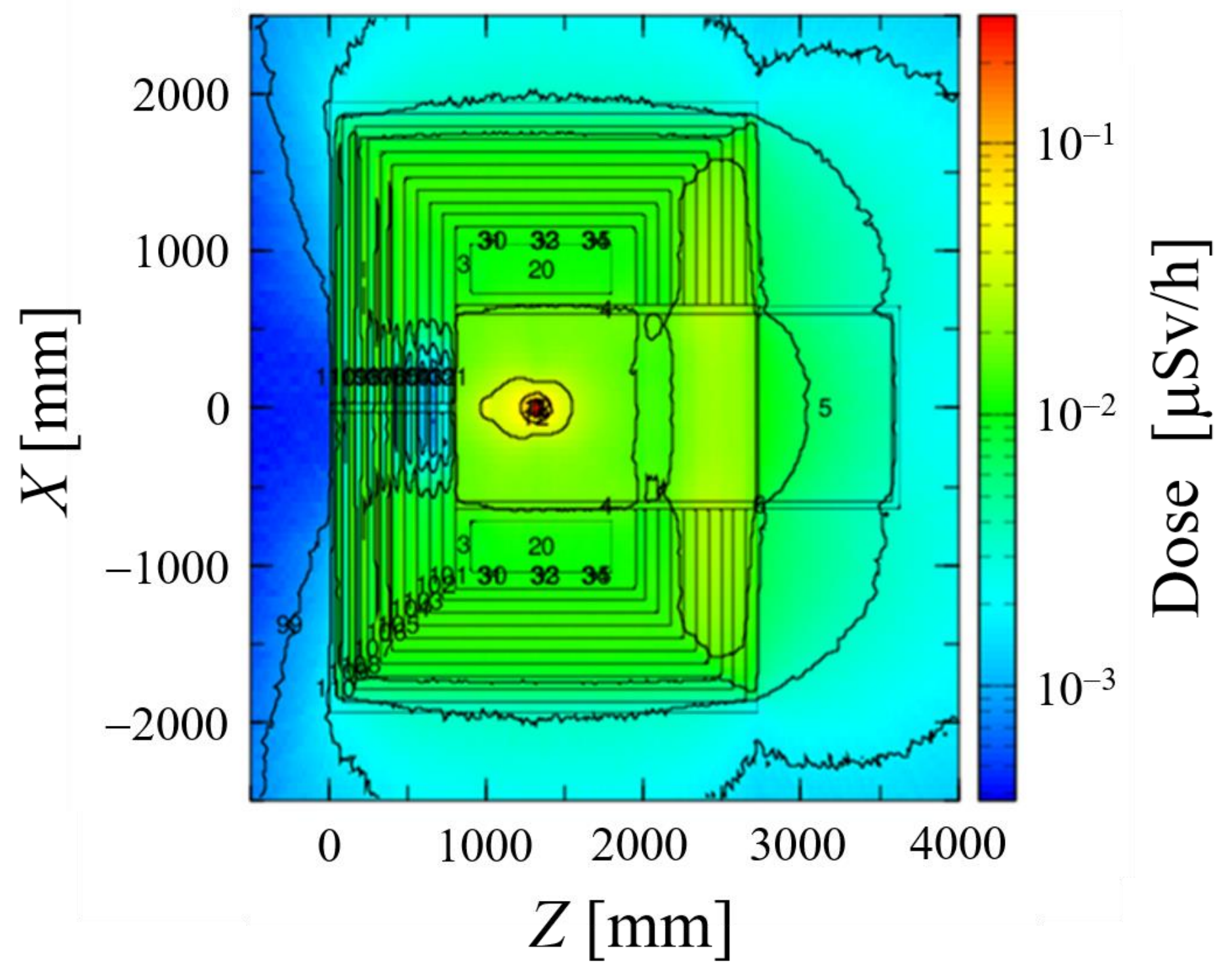

Figure 6-40. Same as Fig. 6-37 but the cooling period is 30 days long and the beam is $^{238}U$.

# 7. International context for the KISS-II facility

Many radioactive ion beam (RIB) facilities around the world, as shown in Table 7-1, have started to propose new projects to extend the nuclear spectroscopic works into the *"Terra Incognita"* of the nuclear chart to investigate the origin of uranium. The nuclei heavier than uranium can be produced by using fusion or MNT reactions with low-energy primary beam. To produce intense RIB in the heavy region, intense primary beam and the gas catcher system, which can handle the intense primary beams, are required, and, therefore, are under design and construction in these facilities.

The nuclear spectroscopy in the vicinity of $N = 126$ recently started by using helium gas ion-guide installed in the IGISOL facility [1] at the University of Jyväsyklä in Finland. The MNT reactions between $^{136}$Xe beam and $^{209}$Bi or $^{nat.}$Pb targets were tested to produce α-emitters as the commissioning. In principle, by replacing the production target to $^{238}$U target, they could have accessed the nuclei heavier than uranium. The INCREASE [2,3] project at GSI in Germany and the "$N = 126$ Factory" [4,5,6] at Argonne National Laboratory (ANL) in USA are under and close to commissioning, respectively. The ion catcher of the INCREASE project is a proto-type setup to study the feasibility of the RIB production by the MNT reactions and the extraction from the ion-catcher for the future low-energy-branch (LEB) of Super-FRS where much more intense primary beams will be used. The "$N = 126$ Factory" is intended for studying the nuclear region around $N = 126$ similar to the original KISS facility by using the MNT reaction with $^{198}$Pt production target, only with a helium gas catcher, and is under construction. They could have accessed the nuclei heavier than uranium by replacing the production target to $^{238}$U target.

Three other projects are under construction (design) stage. Helium gas catchers will be used in the LEB [7] at the HIAF facility in China and in the NEXT [8] at the Groningen facility [9,10] in Netherlands. At the Groningen facility, they plan to install a solenoid filter to remove the intense primary beam similar to the KISS-II facility, and have already received partial funding to promote the project. The HIAF facility has a significant budget to promote their project. However, they presently do not plan to use any primary beam separators. These two projects are strong competitors. In the S3-LEB project [11] at GANIL in France, they will use the S3 spectrometer, which has been designed and optimized to transport the fusion reaction products and in-flight fission products (however, not MNT reaction products), to transport the RIB and an argon gas cell as an ion-catcher for use with the laser resonance ionization technique. Their ability to only access heavy nuclei by using fusion reactions will limit their possible nuclear spectroscopy work of astrophysical interest.

Table 7-1 compares the KISS-II project with its leading global competitors. As noted, some of these facilities are under design or construction while others are near commissioning. In order to maintain a global competitive research position within Japan, it will be important to fund the KISS-II project. The expertise in MNT reactions, gas catchers, ion trapping and manipulation, and nuclear spectroscopy within the WNSC will offer one of the best opportunities for successfully mapping out the properties of the nuclear *terra incognita* and gaining valuable insight into the origins of the heaviest natural elements.

Table 7-1. Competitive projects in the world.

| Facility | Project | Nation | Beams | Intensity (pμA) | Production method | Status |
|---|---|---|---|---|---|---|
| RIBF | KISS-II | Japan | $^{238}U$ | 2(5) | MNT, fusion | Under design |
| RIBF | KISS | Japan | $^{136}Xe$ | 0.01 | MNT | Spectroscopy of neutron-rich refractory elements being progress |
| Jyväsyklä | IGISOL [1] | Finland | $^{136}Xe$ | 0.003(>0.1) | MNT | Experiments are on going in the region of around $N = 126$ |
| GSI | INCREASE [2,3] | Germany | $^{238}U$/RIB | $7\times10^8$ ions/spill | MNT | Pilot setup for LEB of Super-FRS Under commissioning |
| ANL | N =126 Factory [4,5,6] | USA | $^{136}Xe/^{238}U$ | (5/5) | MNT | Under construction<br>Pilot system (CARIBU, FP of $^{252}Cf$) |
| HIAF | LEB [7] | China | $^{238}U$ | (50) | MNT | Accelerator is under construction |
| Groningen | NEXT [8,9,10] | Netherlands | $^{208}Pb$ | 0.03 | MNT | Under construction |
| GANIL | SPIRAL-II S3-LEB [11] | France | $^{136}Xe/^{238}U$ | (18/2.5) | Fusion, MNT | Under construction of S3 spectrometer<br>Budget request for LEB |

Beam intensity in the bracket indicates the reported goal intensity.

# 8. Time line and budgetary

We plan to install the KISS-II facility in the E2 experimental room in RIKEN Nishina center as shown in Fig. 4-1. Three components – the gas-filled superconducting solenoid, helium gas cell, and variable mass-range separator (see Table 8-1 for the components) – will be newly constructed. The details of the specifications have been described in the preceding sections. The expected time line for each component along with the anticipated annual expenditure for each year are listed in Table 8-1. We will parallelly design and construct the components, and finish them by the end of the third year. In the fourth year, we will start with the installation of the solenoid, and then install the other equipment. After that, we can start offline tests (ion extraction and transport) of the KISS-II facility. We anticipate first online tests within the fifth year.

The superconducting solenoid coil, cryostat, and yoke will be designed and manufactured in the first three years. The refrigerator and service port will be constructed from the second year. All components will be installed in the fourth year. After that, the excitation test will start. After the installation of the solenoid, we can start the installation of the primary beam line, contents in the solenoid, and equipment located at the downstream of the solenoid such as the helium gas cell system, mass separator and beam lines.

The primary beam line with a differential pumping for the use of helium gas in the solenoid filter should be installed. For installing the production target, FC, and stopper plates from the upstream side of the solenoid filter, a part of the yoke and the primary beam line should be movable by using a rail system. These will be also designed and constructed by the end of the second year. The assembling and vacuum test will be done by the end of the third year.

The multi-segmented helium gas cell, ion extraction parts, and first MRTOF-MS will be manufactured by the middle of the second year. Their assembly and operation tests are scheduled from the second year. After installing the solenoid filter, all other components will be installed for the offline and online tests.

The variable mass-range separator and the beam line, which will be connected to the present KISS beam line in the E3 experimental room to share the experimental equipment, are parallelly manufactured and installed in the E2 and E3 experimental rooms.

At the beginning of the online tests and experiments, we will start with the use of low-intensity of $^{238}$U beam such as 10 pnA, and increase it up to more than 1 pμA finally by improving and optimizing the structures of the helium gas cell and buncher system, which should work under the high-plasma condition induced by the elastic particles of the primary beam and the production target. In order to access more heavier region, we will prepare radioactive actinoid target such as $^{243}$Am.

We can perform the mass measurement, decay spectroscopy, and laser spectroscopy to reveal the origin of the uranium produced in the r-process at the KISS-II facility, which will be open for external users to study the nuclear properties around the actinoid region from around the sixth year.

Table. 8-1. Time schedule and cost estimations for the three major new components which comprise the KISS-II project. The red numbers indicate the costs (unit : 1 M JPY)

| | Year-1 | Year-2 | Year-3 | Year-4 | Year-5 | Year-6 | Year-7 | Year-8 | Year-9 | Year-10 | Cost (1 M JPY) |
|---|---|---|---|---|---|---|---|---|---|---|---|
| **Gas filled Superconducting Solenoid** | 70 | 360 | 260 | 35 | | | | | | | 725 |
| Coil | Manufacture 20 | 180 | 100 | | | | | | | | |
| Cryostat | Manufacture 20 | 50 | 50 | | | | | | | | |
| Yoke | Manufacture 30 | 80 | 60 | | | | | | | | |
| Refrigerator & Service port | | Manufacture 50 | 50 | | | | | | | | |
| Install and excitation test | | | | Install 10, Excitation test 25 | | | | | | | |
| **Primary beam line Target, FC, Stopper plates (SPs)** | 25 | 60 | 5 | 10 | Offline test | | 100 | | | | 200 |
| Primary beam line | Manufacture 25 | 50 Assembling | Offline tests | Install :10 | | Start Online tests and Experiments | | | | | |
| Target, FC SPs | | Manufacture 10 | 5 Assembling | Install | | | Actinoid Target 100 | | | | |
| **He gas cell, Ion extraction parts MRTOF-MS** | 40 | 70 | 20 | | | | | | | | 130 |
| He gas cell | Manufacture 20 | 30 Assembling RF tests 20 | Offline test | Install | | | | | | | |
| Ion extraction parts | Manufacture 10 | 20 Assembling | Offline test | Install | | | | | | | |
| MRTOF-MS | Manufacture 10 | 20 Assembling | Offline test | Install | | | | | | | |
| **Mass separator & Beam line** | 100 | 100 | 20 | | | | | | | | 220 |
| Magnets | Manufacture 50 | 50 | Install 10, Offline test | | | | | | | | |
| Optical components | Manufacture 50 | 50 | Install 10, Offline test | | | | | | | | |
| **Consumables** | 10 | 10 | 20 | 30 | 30 | 30 | 30 | 30 | 30 | 30 | 250 |
| **Total costs** | 245 | 600 | 335 | 65 | 30 | 30 | 130 | 30 | 30 | 30 | 1,525 |

# Addendum A. The review report of the review committee

The present proposal was reviewed by an international review committee organized by Institute for Particle and Nuclear Studies (IPNS) in KEK.

The committee met online on January 12 and 21, 2022. After the introduction of the review by the director of IPNS, KEK, an overview, physics motivations, methodologies, and technical details were presented by the proponents, followed by questions and answers. The director of the RIKEN Nishina Center stressed the complementary aspects of research in RIKEN RIBF and KISS-II.

In the closed sessions by the committee members, the KISS-II project was evaluated and the outline of the review report was discussed. Final documentations were made through discussions by e-mail exchanges.

We would like to express our great thanks for the favorable review to the following committee members:

| | |
|---|---|
| Tohru Motobayashi | RIKEN Nishina Center, Chair |
| Robert Tribble | Brookhaven National Laboratory, Deputy Chair |
| Georg Bollen | FRIB, Michigan State University |
| Matthew Mumpower | Los Alamos National Laboratory |
| Alexandre Obertelli | Technische Universität Darmstadt |

The review report is composed of a following "Executive Summary" summarizing the basic conclusions of the review.
(*The "basic conclusions of the review" are not included in this volume by the editor's decision.*)

## Executive Summary

### Findings:

The KISS-II project will build a high-performance beam line for heavy neutron-rich nuclei coupled to the RIKEN RIBF (RI Beam Factory) with its strong $^{238}U$ beams. It affords the exploration of heavy nuclei thereby advancing our understanding of the origin of the elements in cooperation with research in other fields as multi-messenger observations of compact object mergers.

KISS-II is aimed at discovering hundreds of new nuclides, and obtaining precise mass values, in the $N$=126 and actinide regions. It will provide the first data towards progenitors of uranium from the rapid neutron capture process (r-process) and it will provide valuable information to clarify the r-process nucleosynthesis. KISS-II's high research potential will also strengthen nuclear models, leading to a deeper understanding of the nucleus, which strongly supports the study of heavy-element synthesis.

The physics program is of high priority, as the origin of the heaviest elements in the r-process nucleosynthesis is a fundamental scientific question. The information is needed also to better understand celestial signals from, for example, merger events.

By constructing various state-of-art devices, KISS-II will increase the capability of producing neutron-rich heavy nuclei by a factor of 10000 compared with its precursor KISS in operation since 2015.

The core team of KISS-II has very strong expertise accumulated by the successful KISS project. The visibility of WNSC (Wako Nuclear Science Center, IPNS, KEK) in the nuclear physics field is high.

KISS-II and RIKEN RIBF are complementary to each other in their coverage of nuclei relevant to the r-process: RIBF is for lighter nuclei.

The collaboration already includes 27 international renowned laboratories, with 68 expert users. This would substantially increase once the project is accepted to start, since KISS-II draws strong interest from the international community.

## Comments:

The science program from KISS-II would be world-leading in its ability to quantify the r-process. This is extremely important, and timely, due to the ability to see the astrophysical process that produce r-process nuclei through gravitational waves that pinpoint mergers of massive stars and the subsequent measurements of nuclei produced in the mergers.

The committee reviewed all parts of the KISS-II proposal and found no shortfalls in it. The technical details presented were of sufficient detail to lift any concerns about the outcome of the project.

The committee noted that if the estimates of improvements were not fully realized, the project would still be highly successful. A factor of 1000 improvement rather than 10000 over the successful KISS project would be enormous in the science reach. Thus, proceeding with KISS-II is low risk.

## Recommendation:

Further estimate of production rates assuming a more pessimistic evolution of the production cross section for the most neutron-rich nuclei should be performed to assess risks; It is expected that even in the most conservative scenario the physics outcome of KISS-II will remain very high.

KISS-II will be a major upgrade from KISS that will provide for a significantly broader experimental program in the future. Adding new dedicated research staff to take advantage of this opportunity is highly recommended. With the interest in the project from the international community, more collaborators will be attracted to work on KISS-II science as well.

KISS-II is the only realistic project in the world to access to the unknown regions of nuclei in the actinide region responsible for creating heavy natural elements including uranium. Its construction should proceed as soon as possible.

# Addendum B. Collaboration

This proposal has been endorsed by the steering committee of the Japan Nuclear Physics Forum (KAKUDAN), Stopped and Slow RI for Precise Nuclear Spectroscopy (SSRI-PNS) collaboration, and Japan Forum of Nuclear Astrophysics.

This proposal has been discussed at the collaboration meeting held in 2020 and 2021.

Collaborators (68 members) :

| | |
|---|---|
| Andrei Andreyev | *University of York* |
| Masato Asai | *Japan Atomic Energy Agency* |
| Giovanna Benzoni | *INFN sezione di Milano* |
| Seonho Choi | *Seoul National University* |
| Martha Liliana Cortes | *TU Darmstadt* |
| James Cubiss | *University of York* |
| Timo Dickel | *GSI Helmholtz Center for Heavy Ion Research* |
| Pieter Doornenbal | *RIKEN Nishina Center* |
| Alfredo Estrade | *Central Michigan University* |
| Yongde Fang | *Institute of Modern Physics, Chinese Academy of Sciences* |
| Rafael Ferrer | *IKS, KU Leuven* |
| Luis Mario Fraile | *Universidad Complutense de Madrid* |
| Ting Gao | *The University of Hong Kong* |
| Georgi Georgiev | *IJCLab, IN2P3/CNRS, Orsay* |
| Hiromitsu Haba | *RIKEN Nishina Center* |
| Takashi Hashimoto | *RISP, Institute for Basic Science* |
| Yoshikazu Hirayama | *WNSC, IPNS, KEK* |
| Wenxue Huang | *Institute of Modern Physics, Chinese Academy of Sciences* |
| Hideki Iimura | *Japan Atomic Energy Agency* |
| Shun Iimura | *RIKEN Nishina Center* |
| Hironobu Ishiyama | *RIKEN Nishina Center* |
| Yuta Ito | *Japan Atomic Energy Agency* |
| SunChan Jeong | *WNSC, IPNS, KEK* |
| Toshitaka Kajino | *Beihang University* |
| Sota Kimura | *RIKEN Nishina Center* |
| Filip G. Kondev | *Argonne National Laboratory* |
| Shigeru Kubono | *RIKEN Nishina Center* |
| Teresa Kurtukian-Nieto | *CNRS/IN2P3 – Université de Bordeaux, CENBG/LP2I* |
| Guang-shun Li | *Institute of Modern Physics, Chinese Academy of Sciences* |
| Pengjie Li | *Institute of Modern Physics, Chinese Academy of Sciences* |
| Yury A. Litvinov | *GSI Helmholtz Center for Heavy Ion Research* |
| Zhong Liu | *Institute of Modern Physics, Chinese Academy of Sciences* |
| Radomira Lozeva | *CNRS, IJCLab* |
| David Lunney | *CNRS – Université Paris – Saclay* |
| Yasuhiro Makida | *IPNS, KEK* |
| Hiroyuki Makii | *Japan Atomic Energy Agency* |
| Hiroari Miyatake | *WNSC, IPNS, KEK* |
| Jun-Young Moon | *RISP, Institute for Basic Science* |
| Ana Isabel Morales López | *Instituto de Física Corpuscular (IFIC, CSIC-Valencia)* |
| Momo Mukai | *RIKEN Nishina Center* |
| Hitoshi Nakada | *Chiba University* |
| Nobuya Nishimura | *RIKEN* |
| Shunji Nishimura | *RIKEN Nishina Center* |
| Katsuhisa Nishio | *Japan Atomic Energy Agency* |
| Toshitaka Niwase | *WNSC, IPNS, KEK* |
| David O'Donnell | *University of the West of Scotland* |
| Tatsuhiko Ogawa | *Japan Atomic Energy Agency* |
| Tetsuya Ohnishi | *RIKEN Nishina Center* |
| Akira Ozawa | *University of Tsukuba* |
| Vi Ho Phong | *RIKEN Nishina Center* |
| Zsolt Podolyák | *University of Surrey* |
| Marco Rosenbusch | *WNSC, IPNS, KEK* |
| Satoshi Sakaguchi | *Kyushu University* |
| Peter Schury | *WNSC, IPNS, KEK* |
| Taeksu Shin | *RISP, Institute for Basic Science* |
| Tetsu Sonoda | *RIKEN Nishina Center* |
| Baohua Sun | *Beihang University* |
| Minori Tajima | *RIKEN Nishina Center* |
| Aiko Takamine | *RIKEN Nishina Center* |
| Akihiro Taniguchi | *Institute for Integrated Radiation and Nuclear Science, Kyoto University* |
| Hideki Tomita | *Nagoya University* |
| Hideki Ueno | *RIKEN Nishina Center* |
| Piet Van Duppen | *IKS, KU Leuven* |
| Michiharu Wada | *WNSC, IPNS, KEK* |
| Philip Walker | *University of Surrey* |
| Hiroshi Watanabe | *Beihang University* |
| Yutaka Watanabe | *WNSC, IPNS, KEK* |
| Oliver Wieland | *INFN sezione di Milano* |
| Jinn Ming Yap | *The University of Hong Kong* |